%% file: review.tex
\documentstyle[aps,prb,floats,epsf,psfig,12pt]{revtex}
\tightenlines
\input{psfig}
\input{review.def}
\begin{document}

\title{Monte Carlo Optimization of Trial Wave Functions in Quantum
Mechanics and Statistical Mechanics}

\author{ M.P. Nightingale }
\address{Department of Physics, University of Rhode Island, \\
Kingston, Rhode Island, 02881. }

\author{C.J. Umrigar}
\address{Cornell Theory Center and Laboratory of Atomic
and Solid State Physics,\\ Cornell University, Ithaca, New York 14853.}

\maketitle
%\baselineskip 14pt
%\baselineskip = 0.5\baselineskip
\begin{abstract}

%\baselineskip = 0.5\baselineskip
This review covers applications of quantum Monte Carlo\index{quantum
Monte Carlo} methods to quantum mechanical problems in the study of
electronic\index{electronic structure} and atomic
structure\index{atomic structure}, as well as applications to
statistical mechanical\index{statistical mechanics} problems both of
static and dynamic nature.  The common thread in all these
applications is optimization\index{trial state optimization} of
many-parameter trial states\index{trial state}, which is done by
minimization of the variance\index{variance minimization} of the local
energy\index{local energy} or, more generally for arbitrary eigenvalue
problems, minimization of the variance of the configurational
eigenvalue\index{configurational eigenvalue}.

\end{abstract}

\section{introduction}

Many computational problems can be reduced to the computation of
eigenvalues of operators.  Examples of operators discussed in this
paper are quantum mechanical Hamiltonians \index{Hamiltonian},
transfer matrices\index{transfer matrix}, and stochastic (Markov)
matrices\index{stochastic matrix}\index{Markov matrix}.  The
applications discussed are: electronic eigenstates of
atoms\index{atom}, and molecules\index{molecule}; atomic eigenstates
of clusters\index{cluster}; critical point\index{critical point}
properties of classical statistical mechanical
systems\index{statistical mechanics}; and critical
slowing-down\index{critical slowing-down} of spin models.

In the review we assume that the reader is familiar with the Monte
Carlo techniques in general.  Readers who lack this familiarity may
wish to consult the literature listed in Ref.~\onlinecite{reviews}.

Numerical computation of the eigenvalue spectrum of the full operator,
or its approximate counterpart in a truncated representation of the
state space, has been employed traditionally for this purpose, but
more recently Monte Carlo eigenvalue methods have started to provide
practical alternatives.

A particularly appealing feature of Monte Carlo eigenvalue methods is
that the stochastic process can be used to estimate {\em just the
corrections} to already sophisticated approximations.  More
specifically, these methods can be designed so that statistical errors
vanish in the ideal case that the optimized trial states\index{trial
state}, which are pivotal in this approach, are exact eigenstates.  In
practice, one obviously cannot realize this ideal, but one can get
close since there is much flexibility in the choice of trial
functions, a flexibility which can be exploited to incorporate
important physics in the approximation from the outset.

For the quantum mechanical applications, the flexibility of the
functional form of the wave function is an important feature that can
be exploited by the quantum Monte Carlo method\index{quantum Monte
Carlo} and this distinguishes it from conventional quantum chemistry
methods, such as Configuration Interaction (CI)\index{Configuration
Interaction}.  It is possible to construct a compact and accurate wave
function if its functional form incorporates the singularities of the
true wave function, but CI, for example, uses an expansion in
determinants of single-particle orbitals, which is slowly convergent
because it cannot reproduce the cusps in the wave function at
electron-electron coincidences\cite{M}.  In quantum Monte Carlo,
however, functional forms can be used that are sufficiently flexible
to have the correct singular behavior at the electron-nucleus,
electron-electron and possibly higher-order particle coincidence
points.  This allows one to construct relatively compact wave
functions with 50-100 free parameters and of quality comparable to CI
wave functions with millions of determinants.

In applications, much effort is spent on the design and optimization
of these trial states\index{trial state optimization} and this paper
reviews various examples in which optimized trial states are employed,
{\it viz.} the computation of eigenstates (mostly ground states) of
atoms\index{atom}, molecules\index{molecule} and van der Waals
clusters\index{cluster}, and the computation of critical exponents of
static and dynamic equilibrium models with phase transitions in two
and three dimensions.  The examples discussed here are selected from
work performed by the authors with various collaborators
\cite{CyrusPRL88,CyrusAthens88,georgia.88,boston.88,NB.PRL.60,%
GraNigh93,MushNigh.94,NighGraKos95,%
MeiMushNigh96,NighBloeprl.96,NighBloeprb.96,FilippiUmrigar96} and this
paper is not intended as a comprehensive review of the field.

\section{trial state optimization}\index{trial state optimization}
\label{sec.optimization}

First we consider the general principle
\cite{CyrusPRL88,CyrusAthens88} of Monte Carlo optimization of a trial
vector \index{trial state optimization} to approximate an eigenstate
of an operator $\T$.  In principle, the method is applicable to an
arbitrary discrete state anywhere in the spectrum, but in practice it
is used for eigenvalues at the top or bottom of the spectral sector
compatible with the desired symmetry, unless one starts out with an
estimate of the energy of an excited state, that is accurate relative
to the gaps separating it from neighboring eigenvalues.  More
specifically, as mentioned, the applications to be discussed will be
drawn from quantum mechanics\index{quantum mechanics}, and static and
dynamic equilibrium statistical mechanics\index{statistical
mechanics}.

If the operator $\T$ is a quantum mechanical Hamiltonian
\index{Hamiltonian} $\cal H,$ the trial state\index{trial state}, in
most cases discussed below, is an approximation for the fermionic
ground state in the case of atoms\index{atom} and
molecules\index{molecule}, or the bosonic\index{boson} ground state in
the case of the van der Waals clusters\index{cluster}.  In the
statistical mechanics statics case, $\T$ is the transfer matrix
\index{transfer matrix}, and here the state of interest is the
dominant eigenstate, the state with the eigenvalue of largest
magnitude.  In the application to stochastic processes satisfying
detailed balance, the largest eigenvalue is unity and the associated
eigenstate is the Boltzmann\index{Boltzmann distribution}
distribution.  In this case, one is interested in the dominant
non-trivial state, i.e., the state with the second largest eigenvalue,
which in the case discussed below is the dominant antisymmetric state.

For simplicity of presentation we assume that the operator $\T$ is
hermitian.  In practice, the method is applied to non-hermitian
operators as well, since transfer matrices\index{transfer matrix} are
used for systems with helical boundary conditions.  Owing to the
helicity, the transfer matrices have the property
\beq
\T^{\dagger} =
\R \T \R,
\label{eq.tdag}
\eeq
where $\R$ is a reflection operator, i.e., $\R^2 = \openone$.  This
property implies that a simple relation exists between left and right
eigenstates, which is usually not the case for non-hermitian
operators.  As a consequence, the discussion below can be immediately
generalized to transfer matrices, but to simplify the presentation we
omit further technical details.

Consider a trial state\index{trial state optimization} $\psiTket$
depending on a set of optimization parameters.  Optimal values for the
parameters can, in principle, be found by minimization of the quantity
\beq
\sigma^2={\psiTbra (\T -\Tav)^2 \psiTket \over \psiTbr \psiTket},
\label{eq.sigma2}
\eeq
where
\beq
\Tav={\psiTbra \T \psiTket \over \psiTbr \psiTket}.
\label{eq.Tav}
\eeq
Note that $\sigma^2$ assumes its minimum value, zero, for {\it any
eigenstate} of the operator $\T$.

In practical applications, the variance $\sigma^2$ cannot be evaluated
directly, and a Monte Carlo estimate is used instead.  This can be
done by generating a sample $\S$ of configurations $s$ drawn from a
suitable relative probability distribution $\br{s}\psiTzket ^2$, as
given, for example, by a trial state\index{trial state optimization}
defined by initial guesses for the optimization parameters.  Given
this sample $\S$, one can evaluate
\beq
\sigma^2 \approx {\sumS{s} [t(s)-\bar{t}\,]^2 w_s^2 \over \sumS{s} w_s^2}.
\label{eq.sigma2est}
\eeq
Here the $w_s$ are weights defined by $w_s={\br s \psiTket / \br s
\psiTzket}$ and $\bar{t}$ is the corresponding weighted sample average
of the {\em configurational eigenvalue,}\index{configurational
eigenvalue} which is called the {\em local energy}\index{local energy}
in real-space quantum Monte Carlo applications\index{quantum Monte
Carlo},
\beq
t(s)= {\bra s \T \psiTket \over \br s \psiTket}.
\label{eq.conf_eigenvalue}
\eeq
To avoid spurious minima of $\sigma^2$, which may arise when very few
configurations acquire most of the weight, it is advisable in practice
to evaluate the expressions in Eqs.~(\ref{eq.sigma2est}) and
(\ref{eq.conf_eigenvalue}) using redefined weights $w_s$ that are
bounded from above.

For this optimization algorithm to be practical, the configurational
eigenvalue\index{configurational eigenvalue} $t(s)$ has to be
computable without explicit summation or integration over all states
of a complete set inserted between $\T$ and $\psiTket$, a condition
which is satisfied, {\it e.g.,} if $\T$ is sparse or near-local.  In
this case, optimal parameters can be found efficiently with a modified
version \cite{NighUmrLM} of the Levenberg-Marquardt
\index{Levenberg-Marquardt algorithm} algorithm, designed to minimize a
sum of squares, such as the expression given in
Eq.~(\ref{eq.sigma2est}).

The crux of the method is that this procedure is applied to a fixed,
small sample, which works efficiently as long as the sample is a good
representation of $\psiTket$, the trial state\index{trial state}
defined by the current parameter estimates.  How well the sample
generated by means of the state $\psiTzket$ represents the state
$\psiTket$, can be measured by the overlap
\beq
\label{eq.overlap}
c={\psiTbr \psiTzket \over \sqrt{\psiTzbr \psiTzket \psiTbr \psiTket}},
\eeq
or in more practical terms
\beq
\label{eq.overlap.estim}
c^2 \approx {(\sumS{s} w_s)^2 \over |S| \sumS{s} w_s^2}.
\eeq
Usually, one chooses for $\psiTzket$ the best available approximation
to the optimized trial state, in which case Eq.~(\ref{eq.overlap}) can
be used irrespective of the signs of the weights $w_s$, which will be
predominantly of one sign.  In some cases, however, it is convenient
to draw a sample from a state (almost) orthogonal to $\psiTket$.  For
example, in the case of the optimization\index{trial state
optimization} of the trial state for the second-largest eigenstate of
the Markov matrix \index{Markov matrix}, discussed in more detail in
Section \ref{sec.dynamics}, it is convenient to sample states from the
Boltzmann\index{Boltzmann distribution} distribution, a state of even
symmetry, while the desired state is odd.  Whether this process yields
a representative sample can be estimated if $w_s$ is replaced by
$|w_s|$ in Eq.~(\ref{eq.overlap.estim}). In any event, whenever the
overlap becomes too small, a new sample has to be generated, e.g.,
from the current, improved trial state, and this process is iterated
until it converges.

The above procedure deviates from the standard Rayleigh-Ritz
variational approach of optimizing the Rayleigh quotient defined in
Eq.~(\ref{eq.Tav}).  As mentioned above, minimization of the variance
\index{variance minimization} of the configurational
eigenvalue\index{configurational eigenvalue} has the advantage of
being applicable also to excited states.  Perhaps more important in a
Monte Carlo context is that minimization of the $\sigma^2$ is more
stable numerically.  That is, although the exact Rayleigh quotient is
bounded by the extremal eigenvalues, this property no longer holds for
the approximation involving sums over a finite sample of states, in
particular because this sample is necessarily extremely sparse when
one is dealing with high-dimensional problems, which is the case for
most problems of practical interest.  Since one is varying many
parameters, the estimate of the Rayleigh quotient in terms of finite
sums can easily become dominated by a single state, as can be seen by
considering the Monte Carlo estimate of the Rayleigh quotient used in
the minimization, {\it viz.} \beq \langle \T \rangle \approx {\sumS{s}
t(s) w_s^2 \over \sumS{s} w_s^2}.  \label{eq.RQest} \eeq an expression
in which $t(s)$ is not necessarily bounded from below for all parameter values.
The Monte Carlo estimate of $\sigma^2$, on the other hand, although
approximate, remains a sum of squares and cannot be made artificially
small as long as more states effectively contribute to the sum than
there are parameters.  An additional computational advantage is that
there are better methods, such as the Levenberg-Marquardt algorithm
\index{Levenberg-Marquardt algorithm}
mentioned above\cite{NighUmrLM}, to optimize a sum of squares than to
optimize a general function.

A possible source of instability of the above algorithm is the fact
that $\bar{t}$ can assume misleading values.  For instance, in the
optimization\index{trial state optimization} of a quantum mechanical
bound state, one can encounter a sample and parameter values such that
only states in the tail of the wave function carry considerable
weight.  This can produce an artificially small estimate of the
variance, resulting from the fact that frequently only a few
parameters have to be adjusted to obtain a local energy\index{local
energy} that is fairly constant over the contributing part of such
sample.  Simple variations of the object function can help cure this
problem.  First of all, one can replace the weighted sample average
$\bar{t}$ in Eq.~(\ref{eq.sigma2est}) by a constant estimate of the
eigenvalue~\cite{CyrusPRL88,CyrusAthens88}.  In this way, one can
maintain focus on the ground state and obtain an object function that
interpolates between minimization of the variance and minimization of
the Rayleigh quotient by choosing a fixed value for $\bar{t}$ below
the true eigenvalue (or above in case one is interested in the largest
eigenvalue).  An alternative method \cite{MeiMushNigh96} of
accomplishing this is to use as the object function
$\sigma^2/\bar{t}\,^2$.

\section{Atoms and Molecules}\index{atom}
\label{sec.atomsNmols}
\subsection{\it Trial wave functions}

Trial wave functions commonly used in \qMC\ applications to problems
in electronic structure are a sum of products of up- and down-spin
determinants of single-particle orbitals multiplied by a Jastrow
factor
\begin{eqnarray}
\Psi=J \sum_n d_n D^\uparrow_n D^\downarrow_n. \label{eq1}
\end{eqnarray}
$D^\uparrow_n$ and $D^\downarrow_n$ are the Slater determinants of
single particle orbitals for the up and down electrons respectively.
The orbitals are linear combinations of products of Slater basis
functions and spherical harmonics centered at the nuclei.

Filippi and Umrigar use a Feenberg or generalized Jastrow factor $J$,
which is a modification of the form introduced by Umrigar {\it et
al.}\cite{CyrusPRL88,CyrusAthens88,CyrusPeterKarl93} to account
explicitly for correlations between a nucleus and two electrons.  This
form is a generalization of the Boys and Handy~\cite{BH} form.
Subsequently, employing wave functions based on back-flow arguments,
Schmidt and Moskowitz~\cite{SM,SXM} attempted to reduce the number of
parameters in the wave functions used by Umrigar {\it et al.}.

The generalized Jastrow factor is written as a product of factors
describing two- and three-body interactions.  The notation is as
follows: electrons are labeled by $i$ and $j$, while $\alpha$ labels
the nuclei.  The electron-nucleus correlation is described by
$A_{\alpha i}$, the electron-electron correlation $B_{ij}$; and
$C_{\alpha ij}$ gives the correlation of two electrons and a nucleus.
Thus, the following form is obtained
\begin{eqnarray}
J=\prod_{\alpha, i} \,\exp{A_{\alpha i}}
\;\;\prod_{i,j}\,\exp{B_{ij}}
\;\;\prod_{\alpha ,i,j}\,\exp{C_{\alpha ij}}. \label{Jastrow}
\end{eqnarray}

The electron-nucleus contribution $A_{\alpha i}$ could in principle be
omitted from the Jastrow factor, provided that a sufficiently large
single-particle basis is used in the determinantal factor of the wave
function.  Three-electron correlations are not included in the Jastrow
factor, since the proximity of more than two electrons is rendered
unlikely by the exclusion principle, incorporated in the Slater
determinant.  The importance of three-electron and higher-order terms
is discussed in \onlinecite{HU}.

The exponents $A_{\alpha i}$, $B_{ij}$, and $C_{\alpha ij}$ of the
Jastrow factor are written as functions of the inter-electronic
distances $r_{ij}$ and electron-nucleus distances $r_{i\alpha}$.
These exponents are chosen to be either polynomials or rational
functions expanded in scaled variables $\hat r_{ij}$ and $\hat
r_{i\alpha}$, where $\hat r= [1-\exp(-\kappa r)]/\kappa$, which at
large inter-particle distances prevents domination of these
polynomials by their highest-order terms.  Typically, either a
5$^{th}$ order polynomial or a 4$^{th}$ order rational function is
used.  Increasing the order beyond this does not yield a significant
improvement in the wave function (except in the case of the
two-electron systems) since the bottleneck is due to the missing
higher body-order correlations in the Jastrow and the determinantal
parts of the wave function.  In addition to these analytic terms,
$C_{\alpha ij}$ contains non-analytic terms that suppress the
dependence of the local energy\index{local energy} on the shape of the
triangle formed by two electrons and a nucleus for configurations in
which two electrons simultaneously approach the nucleus.  The Fock
expansion\index{Fock expansion} motivates the detailed form of these
terms; we refer to Refs.~\onlinecite{FilippiUmrigar96} and
\onlinecite{MUSM} for details.

To guarantee that the local energy\index{local energy} remains finite
when two electrons, or an electron and a nucleus approach each other,
cusp conditions\index{cusp conditions} are imposed on the wave
function.  The resulting algebraic relations among the variational
parameters of the wave function significantly reduce the number of free
parameters.

To summarize, the parameters are of several different types: a) the
exponents of the Slater basis functions; b) the linear coefficients
used to construct orbitals from the basis functions; c) the linear
coefficients in the combination of determinants; d) the exponent
$\kappa$ used to define the scaled inter-particle distances; and
e) expansion coefficients of the polynomial or rational functions in
the Jastrow factor [{\it cf.} Eq.~(\ref{Jastrow})].  The last set
accounts for most of the parameters, the total number being on the
order of 60, even after taking into account the reduction in the
number of free parameters resulting from the imposition of symmetry
and cusp-condition constraints.  Although this is a substantial
number, it is orders of magnitude smaller than the number of
coefficients required in configuration interaction wave functions of
similar accuracy.  The freedom quantum Monte Carlo\index{quantum Monte
Carlo} methods provide in the choice of the form of the trial wave
functions pays off!

Optimization of a trial wave function of the complexity described
above is best performed in a step-wise fashion.  The starting point is
an approximate Hartree-Fock\index{Hartree-Fock} wave function composed
of the minimum number of determinants of the lowest single particle
orbitals necessary to obtain a state of the desired symmetry.  This
Hartree-Fock wave function is multiplied by the Jastrow factor
described above, whereupon all parameters are optimized.  Continuing
from this intermediate trial function, multi-determinantal trial
functions can be constructed by adding configuration state functions
corresponding to single and double excitations from the \HF\
configuration.  To select these additional configurations, a \MCSCF\
wave function is obtained with a standard quantum chemistry package
and the configurations with large weight are added to the best
previous wave function.  Finally, all parameters are re-optimized
simultaneously.

\subsection{\it Results for atoms and molecules}\index{atom}\index{molecule}

Accurate wave functions and energies for small atoms\index{atom} and
ions\cite{CyrusPRL88,CyrusAthens88,CyrusPeterKarl93}\index{ions} and
small molecules\cite{FilippiUmrigar96}\index{molecule} were calculated.
Three measures were used to judge the quality of the wave functions: the
percentages of the correlation energy recovered in variational and in
\dMC\ \index{diffusion Monte Carlo}, and the root-mean-square
fluctuation of the local energy\index{local energy} $\sigma_{\rm VMC}$
obtained by \vMC.  The results are contained in Table~\ref{tab.t1},
which also contains the variational and \dMC\ energy estimates $E_{\rm
VMC}$ and $E_{\rm DMC}$.  Some of the values of $E_{\rm VMC}$ and
$\sigma_{\rm VMC}$ are better than those presented in earlier papers,
the improvements being both the result of small modifications that
were made to the form of the wave functions and of further
optimization\index{trial state optimization}.  We note that the
results for the two-electron atoms\index{atom} and ions are
exceptionally accurate.  This is because the Jastrow factor includes
all-body correlations, i.e., the correlations of both electrons and
the nucleus.  Note however that simply increasing the order of the
polynomials or rational functions in the Jastrow factor yields almost
arbitrarily accurate wave functions only for node-less ground states or
states in which the nodal surface is determined by symmetry alone.

Included in Table~\ref{tab.t1} are the energies of the 1$^3$S He and
the 2$^3$S He excited states.  The 1$^3$S He state is the lowest state
of that symmetry but the 2$^3$S He is not.  The accurate energy
obtained for the 2$^3$S He illustrates that the variance minimization
\index{variance minimization} method can be used to obtain
accurate wave functions not only for the lowest states of a given
symmetry, but for true excited states as well, provided that the
energy gap between the desired state and neighboring states of the
same symmetry is larger than the fluctuations of the local
energy\index{local energy} for the desired trial state\index{trial
state}.  The energies obtained for all the two-electron states from
\vMC\ \index{variational Monte Carlo} is so good that it is not
necessary to perform \dMC\ \index{diffusion Monte Carlo} calculations.

The energies obtained for the four- and ten-electron systems are also
very accurate though of course not of accuracy comparable to those for
the two-electron systems.  We note that for the four-electron systems
it is essential to include the low-lying second configuration state
function, coming from the 2s$^2\to$2p$^2$ excitation, in order to
obtain an accurate energy.

\begin{table}[tbhp]
\caption[]{
Total energies for atoms\index{atom} and ions\index{ions}.
$E^{\rm VMC}_{c}$ and $E^{\rm DMC}_{c}$ are the percentages of
correlation energy recovered in \vMC\ \index{variational Monte
Carlo} and \dMC.\index{diffusion Monte Carlo}  $\sigma_{\rm VMC}$ is
the root-mean-square fluctuation of the local energy\index{local
energy}.  The numbers in parentheses are the statistical errors in the
last digit.  The second column lists the number of configuration state
functions (CSF) and the number of determinants ($D$) in the wave
function. Energies are in Hartree atomic units.  The percentages were
calculated using the Hartree Fock and virtually exact energies from the
references cited.  (Taken from
Ref.~\onlinecite{CyrusPRL88,CyrusAthens88,CyrusPeterKarl93} and
updated.)}

\label{tab.t1}
\squeezetable
\vskip 2mm
\begin{tabular}{lcdddddc}
%\baselineskip=0.5\baselineskip
atom/ion & CSF,D & $E_{\rm VMC}$ & $E_{\rm DMC}$
         & $E^{\rm VMC}_{c}$ ($\%$) & $E^{\rm DMC}_{c}$ ($\%$)
         & $\sigma_{\rm VMC}$ & \\
\tableline
1$^1$S H$^-$    &1,1 & $-$0.527,750,6(1)&                & 99.9990(2)\cite{Fisher,Freund} &         & 0.0007 & \\
1$^1$S He       &1,1 & $-$2.903,724,4(1)&                &100.0000(2)\cite{Fisher,Freund} &         & 0.0007 & \\
1$^3$S He       &1,1 & $-$2.175,229,3(1)&                &100.00  (1)\cite{Fisher,Drake}  &         & 0.0003 & \\
2$^3$S He       &1,1 & $-$2.068,688,7(1)&                & 99.80  (5)\cite{Fisher,Drake}  &         & 0.0011 & \\
1$^1$S Be$^{2+}$&1,1 &$-$13.655,566,2(1)&                &100.0000(2)\cite{Fisher,Freund} &         & 0.0014 & \\
1$^1$S Li$^-$   &2,4 & $-$7.500,30   (1)&  $-$7.500,69(1)& 99.45(1)\cite{Fisher,Fisher93} & 99.91(1)& 0.044  & \\
1$^1$S Be       &2,4 & $-$14.666,65  (1)& $-$14.667,19(1)& 99.25(1)\cite{Fisher,Davidson} & 99.82(1)& 0.084  & \\
1$^1$S Ne       &1,1 &$-$128.901,1   (1)&$-$128.923,6 (2)& 90.56(2)\cite{Fisher,Chakravor}& 96.32(4)& 0.89   & \\
\end{tabular}
\end{table}

\begin{table}[tbhp]
\caption[]
{Total energy of Li$_2$ for increasing number of configuration state
functions (CSF).  The configurations are listed without the doubly
occupied core molecular orbital $1\sigma_{\rm g}$.  $E_{\rm VMC}$ and
$E_{\rm DMC}$ are estimates of the energy obtained by
\vMC\ \index{variational Monte Carlo} and fixed node
\dMC\index{diffusion Monte Carlo}; $E^{\rm VMC}_{c}$ and $E^{\rm
DMC}_{c}$ are the respective percentages of correlation energy.
$\sigma_{\rm VMC}$ is the root-mean-square fluctuation of the local
energy\index{local energy} of the optimized trial
state\index{trial state}. The numbers in parentheses are the
statistical errors in the last digit.  Energies are in Hartree atomic
units.  (Taken from Ref.~\onlinecite{FilippiUmrigar96}.)}

\label{tab.t2}
\vskip 2mm
\begin{tabular}{ccddddd}
CSF/D & \ additional CSF  & $E_{\rm VMC}$ & $E_{\rm DMC}$
      & $E_c^{\rm VMC}$ (\%) & $E_c^{\rm DMC}$ (\%) & $\sigma_{\rm VMC}$ \\
\tableline
\\
1,1 & $2\sigma_{\rm g}^2$              & $-$14.97343(7) & $-$14.9911(1) & 82.26(5) &
96.5(1) & 0.112 \\
2,1 & $2\sigma_{\rm u}^2$              & $-$14.97745(6) & $-$14.9909(1) & 85.51(4) &
96.4(1) & 0.098 \\
3,4 & $1\pi_{\rm ux}^2+1\pi_{\rm uy}^2$& $-$14.98404(5) & $-$14.9923(1) & 90.83(4) &
97.5(1) & 0.086 \\
4,5 & $3\sigma_{\rm g}^2$              & $-$14.98850(4) & $-$14.9938(1) & 94.43(4) &
98.7(1) & 0.086
\end{tabular}
\end{table}

\begin{table}[tbh]

\caption[] {Total energies for the best single- and multi-configuration
wave functions.  As in Table~\ref{tab.t2}, $E^{\rm VMC}_{c}$ and
$E^{\rm DMC}_{c}$ are the percentages of correlation energy recovered
in \vMC\ \index{variational Monte Carlo} and \dMC\index{diffusion Monte
Carlo}. $\sigma_{\rm VMC}$ is the root-mean-square fluctuation of the
local energy\index{local energy} of the optimized trial
state\index{trial state}.  The numbers in parentheses are the
statistical errors in the last digit.  The second column lists the
number of configuration state functions (CSF) and the number of
different determinants ($D$) in the wave function.  Energies are in
Hartree atomic units. (Taken from Ref.~\onlinecite{FilippiUmrigar96}.)}

\label{tab.t3}
\vskip 2mm
\begin{tabular}{lcdddddc}\index{molecule}
molecule & CSF,D & $E_{\rm VMC}$ & $E_{\rm DMC}$
         & $E^{\rm VMC}_{c}$ ($\%$) & $E^{\rm DMC}_{c}$ ($\%$)
         & $\sigma_{\rm VMC}$ & \\
\tableline
Li$_2$ & 1,1 & $-$14.97343(7) & $-$14.9911(1) &  82.26(5) &  96.5(1) & 0.112 & \\
  &4,5       & $-$14.98850(4) & $-$14.9938(1) &  94.43(4) &  98.7(1) & 0.086 & \\[.9mm]
Be$_2$ & 1,1 & $-$29.2782 (1) & $-$29.3176(4) &  70.70(7) &  89.8(2) & 0.242 & \\
  &5,16      & $-$29.3129 (1) & $-$29.3301(2) &  87.56(6) &  95.9(1) & 0.215 & \\[.9mm]
B$_2$  & 1,1 & $-$49.3115 (3) & $-$49.3778(8) &  68.06(8) &  88.5(2) & 0.432 & \\
 & 6,11      & $-$49.3602 (2) & $-$49.3979(6) &  83.10(7) &  94.7(2) & 0.408 & \\[.9mm]
C$_2$  & 1,1 & $-$75.7567 (5) & $-$75.8613(8) &  67.82(9) &  88.1(2) & 0.707 & \\
 & 4,16      & $-$75.8282 (4) & $-$75.8901(7) &  81.66(7) &  93.6(1) & 0.641 & \\[.9mm]
N$_2$  & 1,1 & $-$109.3756(6) & $-$109.487(1) &  69.7 (1) &  89.9(2) & 0.935 & \\
 & 4,17      & $-$109.4376(5) & $-$109.505(1) &  80.94(8) &  93.1(2) & 0.863 & \\[.9mm]
O$_2$  & 1,1 & $-$150.1507(6) & $-$150.268(1) &  73.4 (1) &  91.0(2) & 1.09  & \\
 & 4,7       & $-$150.1885(5) & $-$150.277(1) &  79.08(8) &  92.5(2) & 1.05  & \\[.9mm]
F$_2$  & 1,1 & $-$199.3647(7) & $-$199.478(2) &  78.26(9) &  93.2(2) & 1.23  & \\
 & 2,2       & $-$199.4101(6) & $-$199.487(1) &  84.23(8) &  94.3(1) & 1.19  & \\[.9mm]
\end{tabular}
\end{table}

\begin{figure}[tbhp]
\centering
\centerline{\psfig{figure=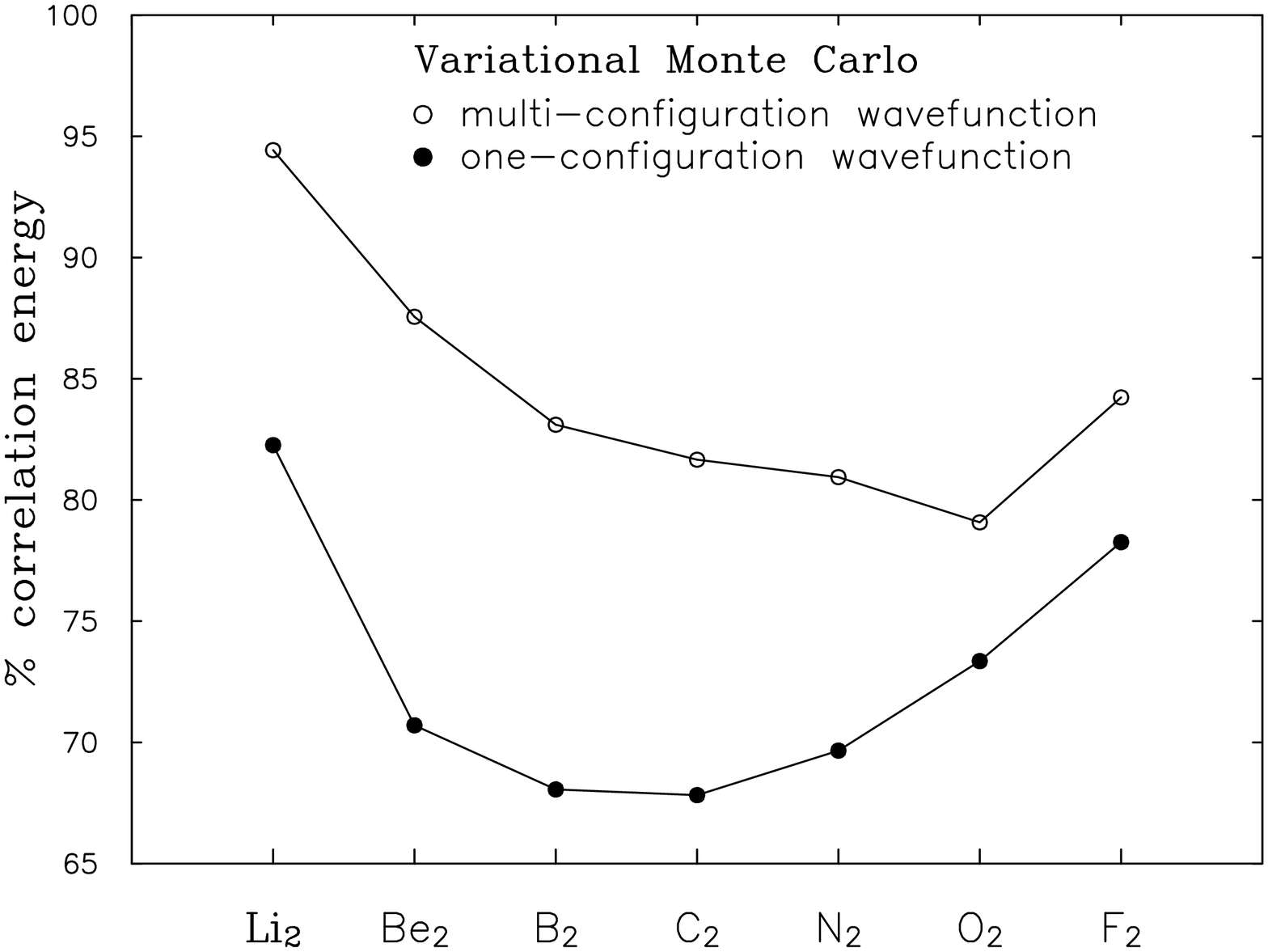,width=9cm,height=6.5cm}}
\vspace{.5cm}
\centerline{\psfig{figure=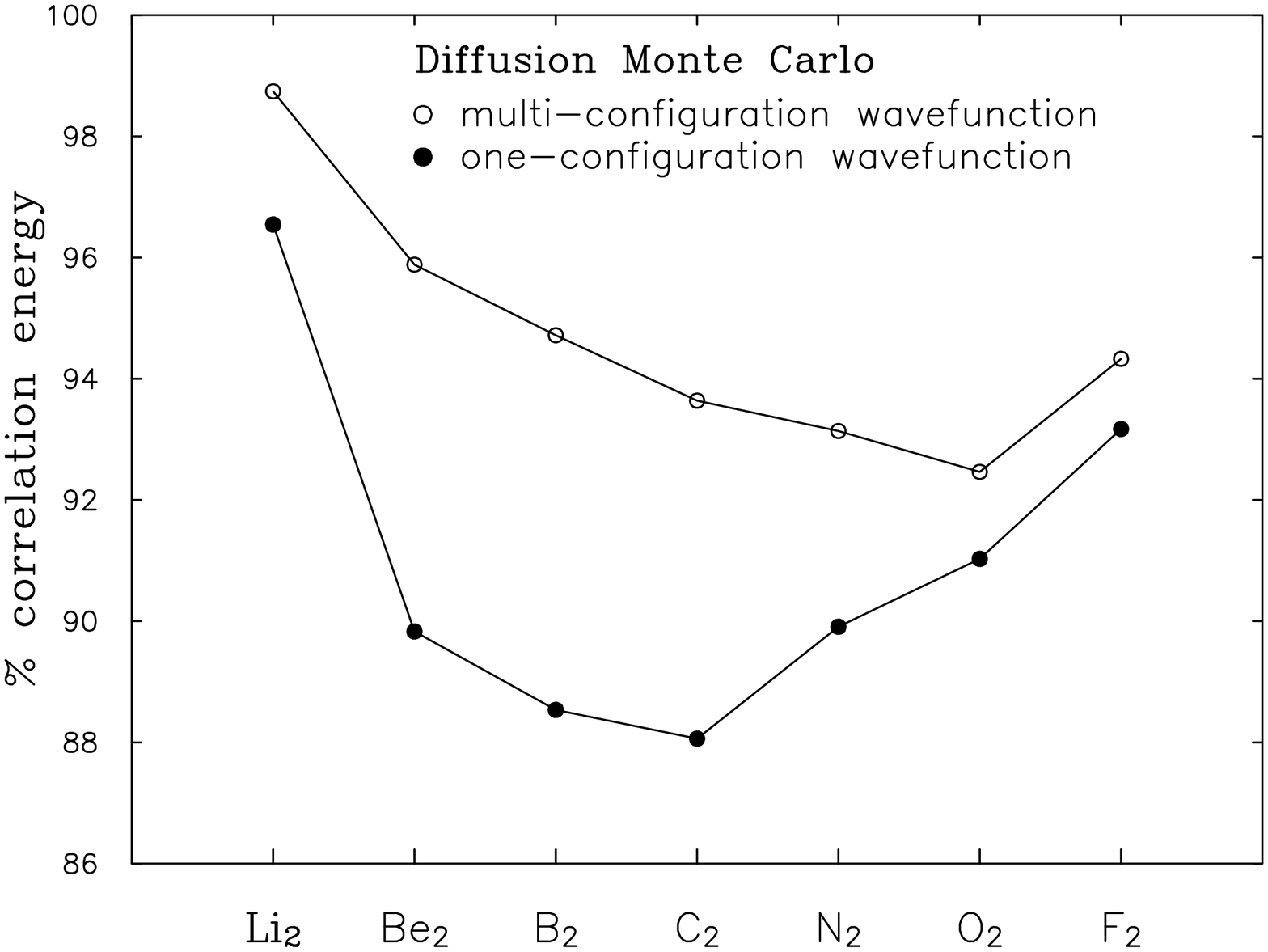,width=9cm,height=6.5cm}}
\vspace{.5cm}
\centerline{\psfig{figure=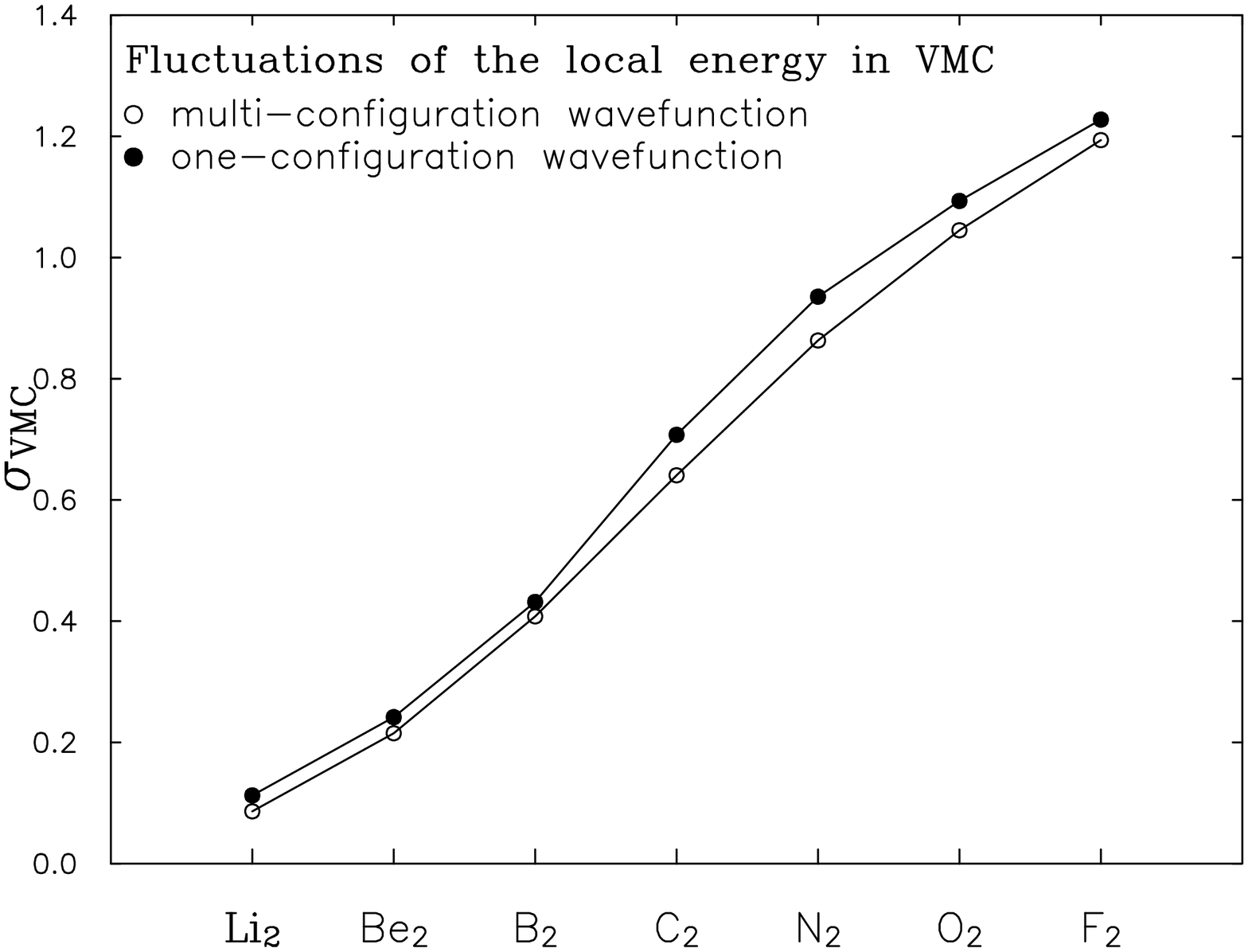,width=9cm,height=6.5cm}}
\vspace{.5cm}

\caption[]{Percentage of correlation energy in
\vMC\ \index{variational Monte Carlo} (upper plot) and
\dMC\ \index{diffusion Monte Carlo} (middle plot) and
root-mean-square fluctuation $\sigma_{\rm VMC}$ (lower plot) with
one-configuration (full circle) and multi-configuration (empty circle)
wave functions.  (Taken from Ref.~\onlinecite{FilippiUmrigar96}.)}

\label{f1}
\end{figure}

%\begin{figure}[tbhp]
%\centering
%\epsfxsize=9 cm
%\epsfysize=6.5cm
%\centerline{\epsfbox{vmc.ps}}
%\vspace{.5cm}
%\epsfxsize=9 cm
%\epsfysize=6.5cm
%\centerline{\epsfbox{dmc.ps}}
%\vspace{.5cm}
%\epsfxsize=9 cm
%\epsfysize=6.5cm
%\centerline{\epsfbox{sigma.ps}}
%\vspace{.5cm}
%
%\caption[]{Percentage of correlation energy in
%\vMC\ \index{variational Monte Carlo} (upper plot) and
%\dMC\ \index{diffusion Monte Carlo} (middle plot) and
%root-mean-square fluctuation $\sigma_{\rm VMC}$ (lower plot) with
%one-configuration (full circle) and multi-configuration (empty circle)
%wave functions.  (Taken from Ref.~\onlinecite{FilippiUmrigar96}.)}

%\label{f1}
%\end{figure}

Filippi and Umrigar computed both single and multi-configuration wave
functions for the first-row homo-nuclear diatomic
molecules\index{molecule}, Li$_2$, Be$_2$, B$_2$, C$_2$, N$_2$,
O$_2$ and F$_2$ with trial wave functions optimized at the experimental
bond length.  Results illustrating the effect of including additional
configuration state functions are shown in Table~\ref{tab.t2} for
Li$_2$.  As configurations are added, there is an improvement in all
three quantities of interest, $\sigma_{\rm VMC}$, E$_{\rm VMC}$ and
E$_{\rm DMC}$ (except that the one- and two-configuration E$_{\rm DMC}$
and the three- and four-configuration $\sigma_{\rm VMC}$ are the same
within the statistical errors), but the best result achieved for Li$_2$
is not as good as for the two-configuration Be wave function.  In the
case of Be there is a single low-lying configuration that mixes in
strongly, whereas for Li$_2$ there are several configurations that make
smaller contributions.

Table \ref{tab.t3} shows the total energies obtained in
\vMC\ \index{variational Monte Carlo} and
\dMC\ \index{diffusion Monte Carlo} and $\sigma_{\rm VMC}$
obtained with single-determinant and multi-determinant wave functions.
The percentage of correlation energy recovered is plotted in
Figure~\ref{f1}.  It ranges from 68-82\% for single-configuration and
79-94\% for multi-configuration wave functions in \vMC. The
corresponding numbers are 88.1-96.5\% and 92.5-98.7\% in \dMC.  For
the single-configuration wave functions, the smallest percentage of
correlation energy recovered is not for the heaviest
molecules\index{molecule} but rather for the molecules in the middle
of the row owing to the strong multi-configurational nature of their
true ground state.  For the multi-configuration wave functions, the
smallest percentage moves more to the right of the row where,
possibly, excitations to the next shell become important.  Since the
wave functions are obtained by non-linear optimization, it is possible
to get stuck in a local minimum.  From the shape of the curves it
seems likely that this is in fact the case for O$_2$.  There is
considerable resemblance in the shapes of the \vMC\ and \dMC\ curves
and the multi-configuration energies are consistently better than
their single-configuration counterparts. Therefore, it is clear that
when more configurations are added to the HF one, not only do
$\sigma_{\rm VMC}$ and E$_{\rm VMC}$ improve but there also is an
improvement in the shape of the nodal surface, flaws in which limit
the accuracy of the fixed-node \dMC\ energies.

These results shed some light on the question how the computational
complexity of \qMC\ scales with atomic number $Z$.  In going from
Li$_2$ to F$_2$, the root-mean-square fluctuation of the local
energy\index{local energy}, $\sigma_{\rm VMC}$, increases by more
than a factor of ten for both the single and the multi-configuration
wave functions.  The dependence of $\sigma_{\rm VMC}$ on atomic number
$Z$ appears to be considerably faster than linear for small $Z$ but
slower than linear for large $Z$.  This factor should be taken into
account in figuring the scaling of the computational cost.  In order to
estimate this scaling, it is necessary to have a systematic study of
several molecules\index{molecule}.  It would in fact be useful to
also have results on some second-row homo-nuclear diatomic molecules.

\section{Bosonic van der Waals clusters}\index{cluster}
\subsection{\it Introduction}

As a next example, we discuss bosonic\index{boson} van der Waals
clusters\index{cluster}.  Again, in this case most of the work
focuses on trial wave functions for the ground state.

Mushinski {\it et al.} \cite{MushNigh.94,MeiMushNigh96} study
clusters\index{cluster} consisting of $N$ Lennard-Jones atoms
interacting via the pair potential $v(r)=4\epsilon [ (r/\sigma)^{-12}-
(r/\sigma)^{-6}]$.  In reduced units, the potential takes the form
$v(r)=r^{-12}-2 r^{-6}$ and the only independent parameter in the
Schr\"odinger equation is the reduced inverse mass $m^{-1}$, which is
proportional to the square of the de Boer parameter, $h/\sigma \sqrt{m
\epsilon}$, a dimensionless quantity measuring the relative importance
of quantum mechanical effects.

Constructing accurate trial wave functions is particularly challenging
for large values of the de Boer parameter, and part of the work to be
discussed here deals with clusters\index{cluster} in the extreme case,
{\it viz.}  the unbinding limit, where the zero-point energy is
sufficiently strong to destroy the bound-state nature of the ground
state of a cluster.  The corresponding value of the de Boer parameter
plays the role of a critical point\index{critical point}, and in fact
the unbinding transition has many features in common with a critical
wetting transition \cite{wetting}.  For bosonic\index{boson} clusters,
the unbinding is of theoretical interest, but the transition is not
experimentally observable.  That is, in the first place, the critical
de Boer parameter is an increasing function of the geometric size of
the cluster, or in other words, the binding energy per atom increases
with the size of the cluster.  Secondly, for the bosonic atom with
largest de Boer parameter, $^4$He, the dimer is believed to have in a
bound state, so that one expects the same for clusters of all sizes.
For fermion clusters, however, it is estimated that $^3$He clusters
only have bound states for more than 30 atoms\cite{fermion.unbinding}.
{}From this point of view the boson\index{boson} computations are to
be viewed as a test case for the experimentally more interesting
fermion case.

Simple arguments\cite{MeiMushNigh96} applied to this unbinding
transition predict the way in which the energy vanishes as the de Boer
parameter approaches its critical value, and the nature of the
divergence of the size of the clusters\index{cluster}.  More
specifically, the following critical behavior is expected for the
ground state energy $E_0$ and the average size $\langle r\rangle$ (as
defined below) of the cluster
\begin{eqnarray}
E_0 & \sim & (\Delta m)^2, \nonumber \\
\langle r \rangle & \sim & (\Delta m)^{-1},
\label{eq.critical}
\end{eqnarray}
for ${m \downarrow m_{\rm c}}$ where $\Delta m = m-m_{\rm c}$ with
$m_{\rm c}$ the critical value of the dimensionless mass.  Results
corroborating these predictions will be discussed.  In addition, this
section discusses a graphical method of inspecting the quality of the
trial wave functions, in particular to identify regions that contribute
excessively to the variance of the local energy\index{local energy}.

\subsection{\it Cluster trial functions}

The construction of the trial wave functions for
clusters\index{cluster} closely parallels the discussion above in
section \ref{sec.atomsNmols}, i.e., a trial wave function for the
ground state of a bosonic\index{boson} cluster should be invariant
under translation, rotation and particle permutation.  The first two
of these requirements can be satisfied by choosing as coordinates the
inter-atomic distances $r_{ij},\ 1\le i<j \le N$.  The third
condition, particle exchange symmetry, then can be imposed explicitly
by considering only functions that are invariant under the $N!$
permutations of the indices of the $r_{ij}$.  The technical problems
associated with this permutation symmetry can be dealt with
efficiently by employing an approach based on the theory of algebraic
invariants \index{algebraic invariants} and we refer to
Ref.~\onlinecite{MushNigh.94} for details.

In addition to having to satisfy these symmetry restrictions, the ground
state trial function has to be strictly positive.  In principle, $n$-body
correlations with all $n\le N$, should be incorporated in the trial
wave function, but these many-body effects are expected to become
progressively less important as $n$ increases.  Positivity and
many-body correlations suggest that the trial function be written as
 \begin{equation} \label{Psi_thru_fF} \log \psiT = \sum_{(i,j)}
u^{(2)}(r_{ij}) +
	     \sum_{(i,j,k)} u^{(3)}(r_{ij},~r_{jk},~r_{ki})+\cdots +
	     \sum_{(i_{1},\dots ,i_{N})} u^{(N)}(r_{i_1 i_2},\dots),
\end{equation} where the $u^{(n)}$ are real-valued, $n$-body
functions.

The design of trial wave functions, described in detail in
Ref.~\onlinecite{MushNigh.94}, used the procedure developed by Umrigar
{\it et al.} \cite{CyrusPRL88,CyrusAthens88}.  That is, the trial
functions satisfy boundary conditions associated with (a) the collision
of two atoms and (b) one atom going off to infinity.  The behavior
of the wave function for the most likely configurations, which involve
intermediate distances and require most of the variational freedom of
the trial wave functions, is described by expanding the $u^{(n)}$ in
polynomials of variables $\hat r_{ij}=\hat r(r_{ij})$, where $\hat r$
is an optimizable function that approaches a constant for $r \to
\infty$.  Note that $u^{(2)}$ is exceptional in that in addition to
these polynomial terms, it has terms dictated by the short-distance and
long-distance boundary conditions.

Wave functions of this type can be used to study the importance of
many-body effects for the quality of the trial wave function.  The
exact ground state energy $E_{0}$ and its
variational estimate $\tilde E_{0}$ satisfy the inequality
\begin{equation}
\tilde E_{0} - \sigma \leq  E_0 \leq \tilde E_{0}.
\label{eq.sigmabound}
\end{equation}
This suggests that the quality of the trial wave function $\psiT$ be
measured by
\begin{equation}
\label{Quality}
Q = -\log_{10}{ \sigma \over |E_0| }.
\end{equation}
The limit $Q \to\infty$ corresponds to an {\em exact} solution of the
time-independent Schr\"odinger equation.  The quantity $Q$ is a
conservative measure of the accuracy of the wave function since in fact,
there exists a tighter bound, linear in $\sigma^2$ rather than
$\sigma$; we refer to Refs.~\onlinecite{MushNigh.94} and
\onlinecite{MeiMushNigh96} for details.

\begin{figure}[tbhp]
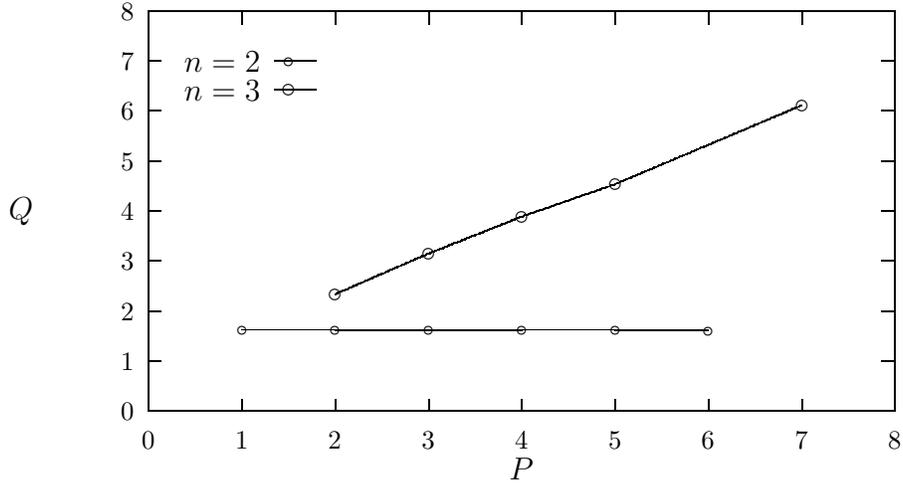

\centerline{\input argon3}
\vskip 0.5cm
\caption[f1]{
Quality $Q$, a measure of the accuracy of the optimized trial function
as defined in Eq.~(\ref{Quality}), as a function of the power $P$ of
two- and three-body polynomials, labeled by $n$, for
$\mbox{Ar}_3$. From Ref.~\onlinecite{MushNigh.94}.}
\label{fig.Ar3}
\end{figure}

With these wave functions containing an arbitrary number of many-body
correlations Mushinski and Nightingale \cite{MushNigh.94} address the question
how the quality of the trial wave function improves as $n$-body terms with
progressively larger values of $n$ are included.  Figures~\ref{fig.Ar3}
through \ref{fig.Ar5} display results for Ar clusters\index{cluster}
of sizes three, four, and five.  The estimate $Q$ of the quality of
the wave functions is plotted versus the power $P$ of the polynomials
used in the expansion of Eq.~(\ref{Psi_thru_fF}) for different values
of $n$.  The bottom curve, starting at $P=1$, is for the case in which
only two-body interactions are present ($n=2$).  As shown, the quality
levels off at a fixed value of $Q$ with increasing $P$, an indication
that the absence of three-body terms is the dominant source of
variance of the local energy.  In the next curve segment, three-body
terms are added ($n=3$).  Here $P$ is redefined to denote the order of
the three-body polynomial, which starts at $P=2$; in the second curve
segment the order of the polynomial that describes two-body effects is
kept constant at the highest $P$-value used in the previous segment of
the curve.  This process is repeated for $n$-body terms with
increasing $n$, until finally the complete $N$-body polynomial for the
$N$-atom cluster is included in the trial function.  At this point,
the quality starts to go up roughly linearly with the order of the
polynomial.  Similar results were obtained for systems made of lighter
atoms\cite{MushNigh.94}. Here we only note that the order $P$ of the
four-body polynomials is not sufficiently high in any of the figures
for the quality to have leveled off, as ultimately it must for
five-atom clusters.

\begin{figure}[tbhp]
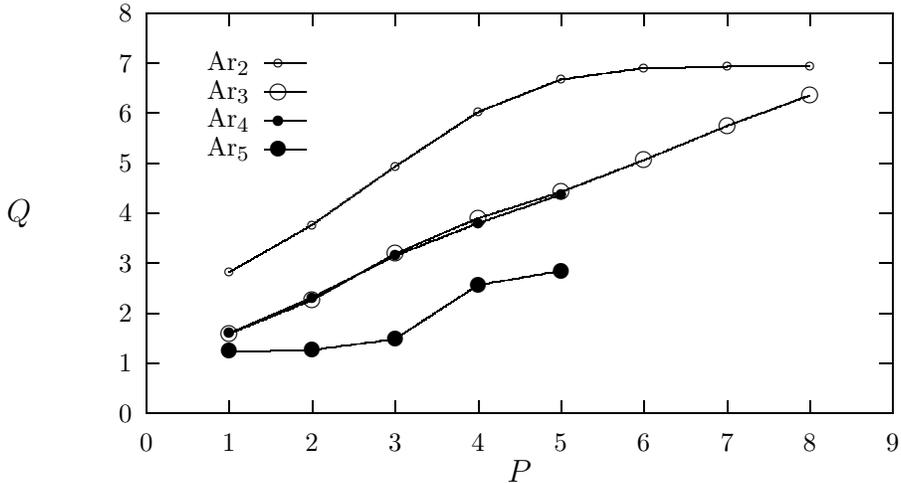

\centerline{\input argon} \vskip 0.5cm
\caption[f1]
{ Quality $Q$, a measure of the accuracy of the optimized trial
function as defined in Eq.~(\ref{Quality}), as a function of power $P$
of the complete $N$-body polynomial for argon clusters\index{cluster}
of sizes two through five.  The curve for Ar$_2$ levels off because of
noisy behavior of the numerical derivatives used in the computations.
Apparent erratic behavior of the Ar$_5$ curve is presumably due to
occasional suboptimal parameter choices for the trial wave function.
(Taken from Ref.~\onlinecite{MushNigh.94}.)}
\label{fig.argon}
\end{figure}

\begin{figure}[tbhp]
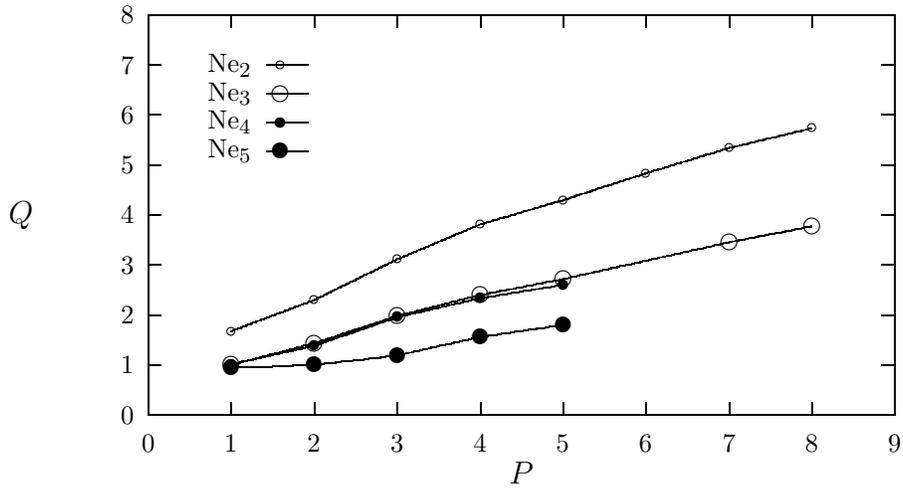

\centerline{\input neon} \vskip 0.5cm
\caption[f1]{
Quality $Q$, a measure of the accuracy of the optimized trial function
as defined in Eq.~(\ref{Quality}), as a function of power $P$ of the
complete $N$-body polynomial for neon clusters\index{cluster} of sizes
two through five. (Taken from Ref.~\onlinecite{MushNigh.94}.)}
\label{fig.neon}
\end{figure}

\begin{figure}[tbhp]
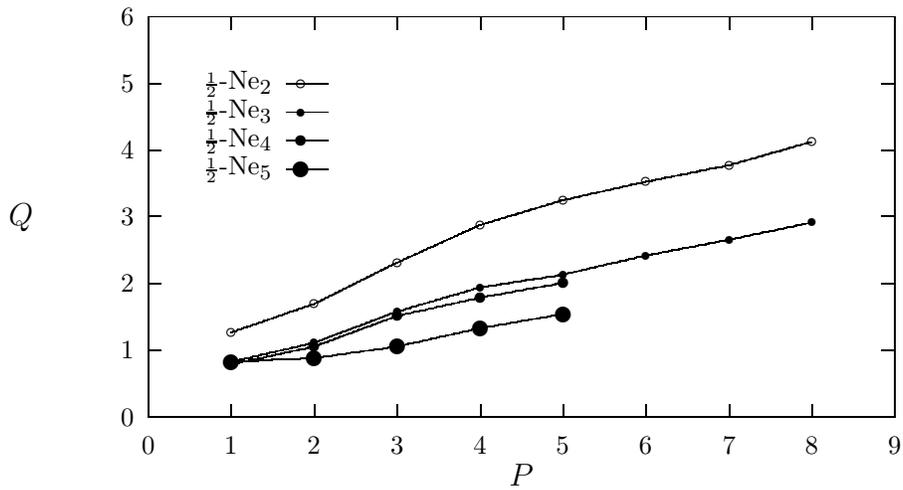

\centerline{\input half} \vskip 0.5cm
\caption[f1]{
Quality $Q$, a measure of the accuracy of the optimized trial function
as defined in Eq.~(\ref{Quality}), as a function of power $P$ of the
complete $N$-body polynomial for clusters\index{cluster} of sizes two
through five; the clusters consist of hypothetical ``{\tiny ${1\over
2}$}-Ne'' atoms with a reduced
mass equal to half that of neon . (Taken from
Ref.~\onlinecite{MushNigh.94}.)}
\label{fig.half}
\end{figure}

Figures \ref{fig.argon}, \ref{fig.neon}, and \ref{fig.half} are plots of
the quality $Q$ vs. the power $P$ for each type and size of
cluster\index{cluster} for optimized wave functions constructed of the
full many-body polynomials.  An interesting feature of these plots is
that the $Q$ vs. $P$ curves for three- and four-atom ($N=3,4$)
clusters almost coincide but that they are distinct from the curves
for $N=2$ and $N=5$.  The $N=2$ clusters are unique in that the
short-distance divergences in the local energy\index{local energy}
have been fully removed, while also the large-distance asymptotic
properties of the trial function is superior in this case.  Apart from
this, the effect probably is geometric in nature: only for sizes $N
\le 4$ are clusters fully symmetric in the classical configuration of
minimum energy and are all atoms are at the bottom of the individual
pair potentials.  The same is true for $N=5$ in four dimensions and
indeed the accuracy of the wave functions in that case is comparable
to that in the $N=3$ and $N=4$ cases in three dimensions.

Finally, Table~\ref{tab.cluster_energies} contains ground state
energies obtained with the help of the optimized trial functions
discussed above.

%1
%2
\begin{figure}[tbhp]
\centerline{\input argon4}
\vskip 0.5cm
\caption[f1]{
Quality $Q$, a measure of the accuracy of the optimized trial function
as defined in Eq.~(\ref{Quality}), as a function of the power $P$ of
two- , three- and four-body polynomials, labeled by $n$, for
$\mbox{Ar}_4$. (Taken from Ref.~\onlinecite{MushNigh.94}.)}
\label{fig.Ar4}
\end{figure}
%3
\begin{figure}[tbhp]
\centerline{\input argon5}
\vskip 0.5cm
\caption[f1]{
Quality $Q$, a measure of the accuracy of the optimized trial function
as defined in Eq.~(\ref{Quality}), as a function of the power $P$ of
two-, three-, four- and five-body polynomials, labeled by $n$, for
$\mbox{Ar}_5$. (Taken from Ref.~\onlinecite{MushNigh.94}.)}
\label{fig.Ar5}
\end{figure}
%9
%10
%11

\subsection{\it Visualization of flaws of the trial wave function}

In the process of construction of the trial functions, it is important
to know what regions of configuration space contribute most to the
variance $\sigma^2$, as defined in Eq.~(\ref{eq.sigma2}).  For
instance, it is useful to know if the quality of the wave function is
limited by poorly satisfied boundary conditions (either at infinity or
when two particles get close) or if the quality can be improved by
adding more variational parameters.  A wave function subjected to the
optimization\index{trial state optimization} procedure described in
Section~\ref{sec.optimization} might have too much variational freedom
relative to the size of the sample over which it is optimized.  This
might show up in the form of unphysical peaks, which might
remain invisible in the variance of the local energy\index{local
energy} estimated from the relatively small samples used to optimize
the trial function.

To help answer such questions, Meierovich {\it et
al.}\cite{MeiMushNigh96} made density plots of the local error, the
deviation of the local energy\index{local energy} from its average
and, as an illustration, they discuss a five-atom
cluster\index{cluster}.  In fact, it is useful to superimpose color
density plots of both the wave function and the local error, which
contain more information than can be reproduced by the grey-scale
plots reproduced in this paper. See
Ref.~\onlinecite{Marks.internet.visualization} for the color graphics.

Obviously, the fact that the ground state wave function depends on
$3N-6$ independent coordinates, seriously limits any graphical
approach.  For five-atom clusters\index{cluster} the following planar
cut through configuration space is informative: four atoms are fixed
at the vertices of a regular tetrahedron, while the fifth particle is
located in a plane that contains two of these vertices and bisects the
edge connecting the two remaining atoms.

%1
\begin{figure}[tbh]
\centerline{\psfig{figure=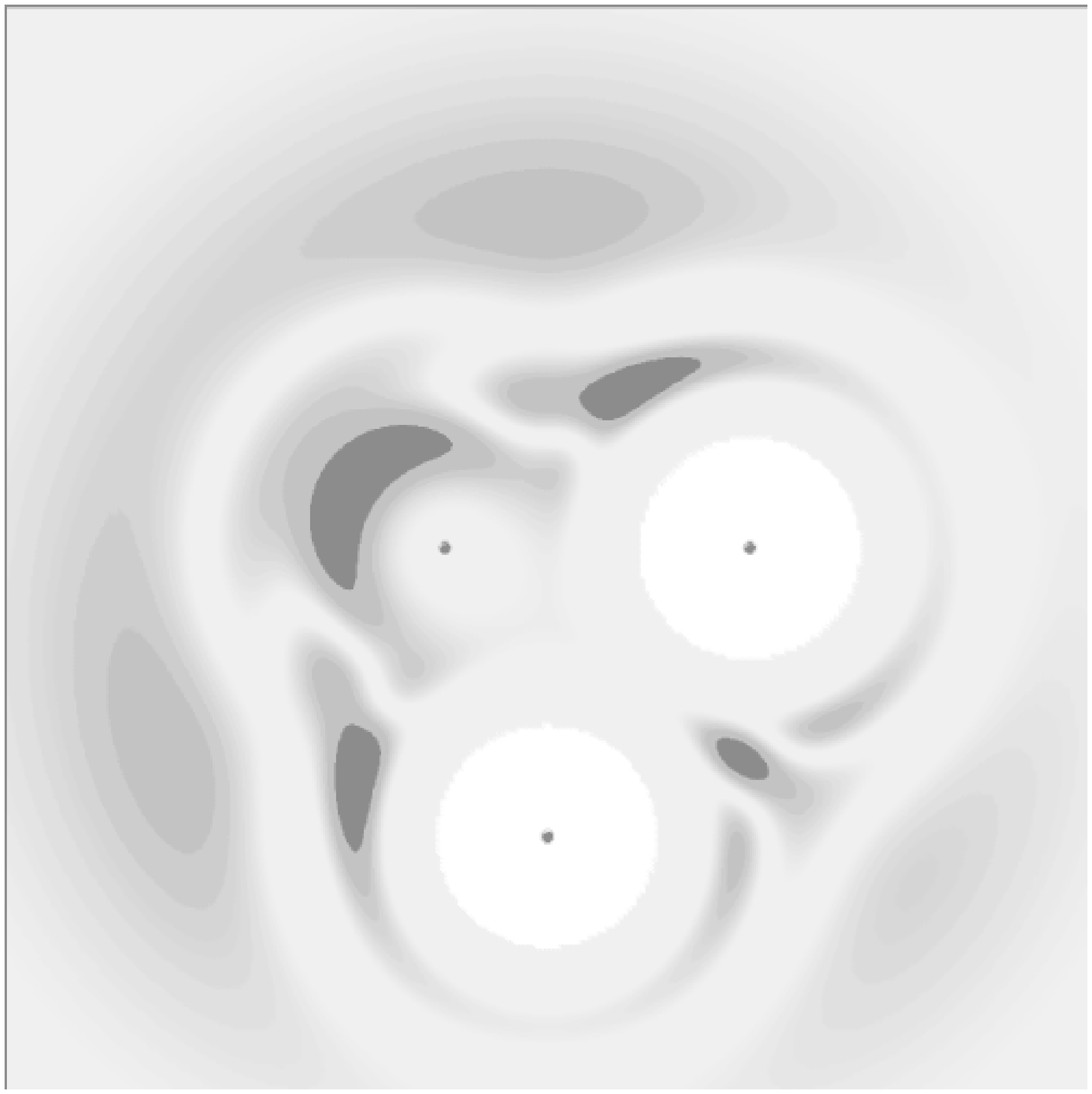,width=8cm,height=8cm}
\psfig{figure=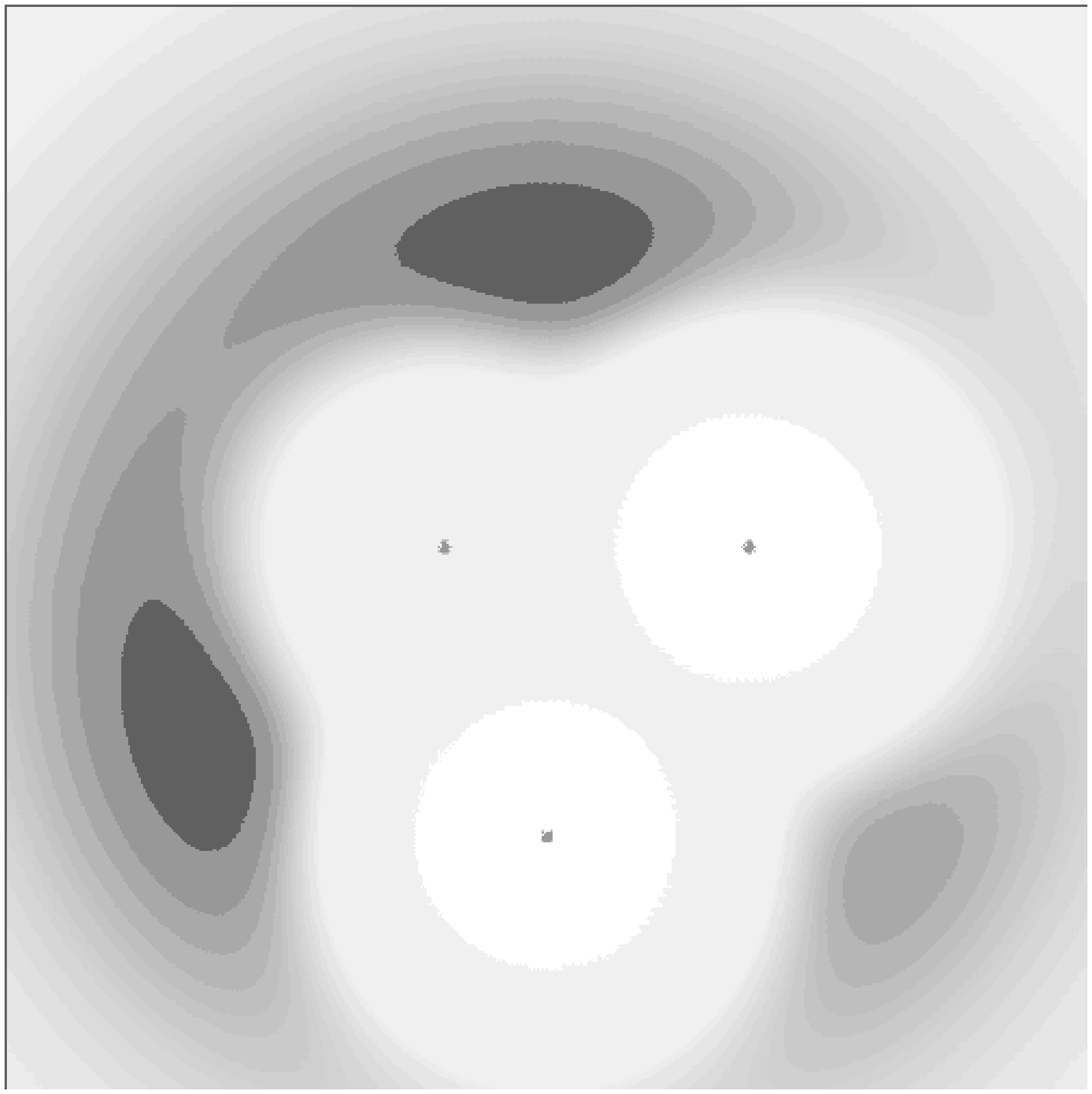,width=8cm,height=8cm}}
\vskip 0.5cm

\caption[Local error for $N=5$, $m^{-1}=0.16$]{
Left: Density plot of the ``local error'' in the geometry described in the
text.  The two dots in the lower right-hand corner are the two
in-plane vertices of the tetrahedron; the one in the upper left corner
is the projection of the {\it two} out-of-plane vertices.  The length
of the tetrahedron edges is 1.3 and $m^{-1}=0.16$.  The darker the
region, the more it contributes to $\sigma^2$.  Note that the dark
region in the lower right is a cut through the banana-shaped dark
region in the upper left.  The regions of the largest local error are
the two symmetrically located regions of collision of four atoms.
White lines are are cuts through nodal surfaces of the local error and
have no physical significance.\\

% (Taken from Ref.~\onlinecite{MeiMushNigh96}.) }

%\label{fig.n5.e.16.ps}
%\end{figure}
%2
%\begin{figure}[tbhp]
%\centerline{\psfig{figure=n5.16.L13.psi.ps,width=6cm,height=6cm}}
%\vskip 0.5cm
%\caption[$\psi^2$ for $N=5$, $m^{-1}=0.16$]{

Right: Density plot of $|\psi_T|^2$ as a function of the position of
the fifth particle in same geometry Dark regions correspond to high
probability density.  Note that the regions with the largest local
error are contained in the region where the wave function becomes small
because of repulsive potential at short range. (Taken from
Ref.~\onlinecite{MeiMushNigh96}.)}

\label{fig.n5.epsi.16.ps}
\end{figure}

Figures \ref{fig.n5.epsi.16.ps} strongly suggest which regions of
configuration space contribute most to $\sigma^2$.  The density plots
should be interpreted according to the following convention: zero
intensity (white) corresponds to a minimum, while full intensity
(black) corresponds to a maximum of the plotted function.  The picture
on the left in Figure~\ref{fig.n5.epsi.16.ps} represents the density
plot of the weighted {\it local error}, defined as $|(\Eloc -
\Etr)\psi_{\/\rm T}|$, as a function of the position of the fifth,
wandering atom.  [Note that the {\it local energy}\index{local energy}
$\Eloc$ is defined by Eq.~(\ref{eq.conf_eigenvalue}) with $\T$ replaced
by the Hamiltonian.  \index{Hamiltonian}] The picture on the right in
Figure~\ref{fig.n5.epsi.16.ps} shows the dependence of $\psi_{\/\rm
T}^2$, on the position of the fifth atom while the other four atoms are
fixed in the geometry mentioned above.

The conclusion drawn from inspection of the density plots is that the
trial function fails particularly in regions where more than two atoms
collide and we see that, of these, the four-body collisions are worst at
least in a local sense.

\begin{table}[tbhp]

\caption[] {$E_{\rm T}$ and $E_0$, variational and
\dMC\ \index{diffusion Monte Carlo}estimates of the ground state
energies of noble gases Ar and Ne, and hypothetical ``{\tiny ${1\over
2}$}-Ne'' for clusters\index{cluster} of $N$ atoms, from
Refs.~\onlinecite{MushNigh.94} and \onlinecite{MeiMushNigh96}.
Standard errors in the last digits are given in parentheses.  The
relative numerical error of these estimates is on the order of
$10^{-7}$, which in some cases exceeds the statistical error.  (Taken
from Ref.~\onlinecite{MeiMushNigh96}.)}

\input{cluster_energies}

\label{tab.cluster_energies}
\end{table}

\subsection{\it Unbinding transition}

\begin{figure}[tbhp]
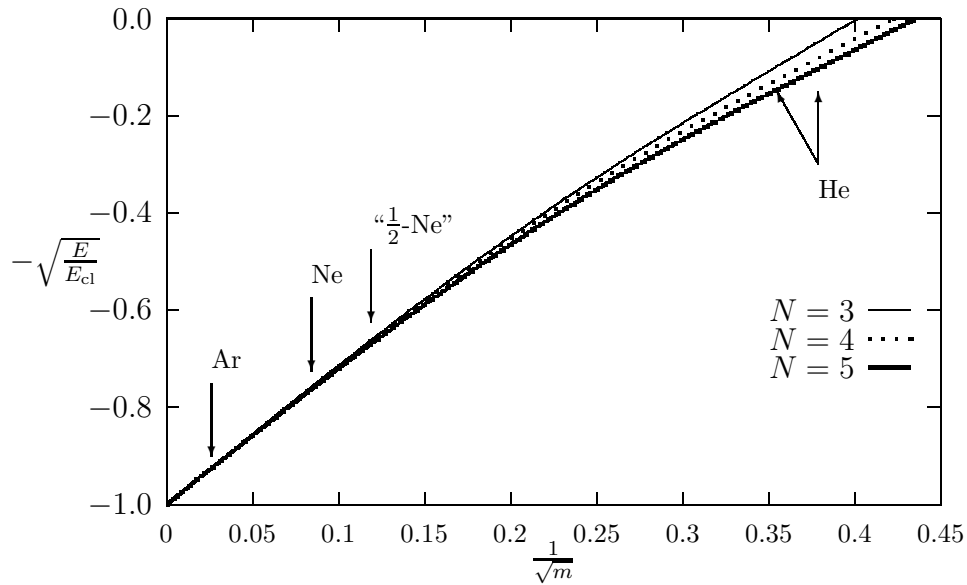

\centerline{\input ./e}
\vskip 0.5cm

\caption[f1]{Curves fit to \dMC\ \index{diffusion Monte
Carlo}estimates of the ground state energy for clusters\index{cluster}
of sizes $N = 3,4$ and $5$.  (Taken from
Ref.~\onlinecite{MeiMushNigh96}.)}

\label{fig.comb}
\end{figure}

Figure \ref{fig.comb} shows the behavior of the energy vs. the de Boer
parameter for clusters\index{cluster} of sizes $N=3$,4, and 5.  The
figure shows the square root of the scaled energy plotted vs.
$1/\sqrt m$.  The former is chosen because the energy vanishes
quadratically at the unbinding transition, {\it cf.}
Eq.~(\ref{eq.critical}).  The $1/\sqrt m$ produces a linear curve at
large mass, according to the harmonic approximation, which becomes
exact in this limit.  Plotted in this way, the curves are remarkably
dull over the whole range from the classical to the quantum regime.
The reader is referred to Ref.~\onlinecite{MeiMushNigh96} for a more
detailed discussion of this topic.  Finally, Figure~\ref{fig.mc}
contains a plot of the value of the critical mass vs. the inverse
cluster size.  The behavior again is remarkably linear, but obviously
there are not enough points to allow a reliable extrapolation to the
infinite system limit.

\begin{figure}[tbhp]
\centerline{\input ./mc}
\vskip 0.5cm
\caption[$m_c$ vs. $1/N$]{Critical mass $m_{\rm c}$ versus
cluster\index{cluster} size $N$ plotted on a $1/N$ scale for $N = 2,
3, 4$ and $5$.  (Taken from Ref.~\onlinecite{MeiMushNigh96}.)}
\label{fig.mc}
\end{figure}

\section{statistical mechanics}\index{statistical mechanics}
\subsection{\it Connection with quantum mechanics}\index{quantum mechanics}

In the remainder of this paper we discuss applications of optimized
trial functions to statistical mechanics\index{statistical mechanics}.
Compared to the problems discussed in the previous sections, these
applications involve very different physics.  Yet, computationally
they are quite similar. In the first set of examples, we discuss the
use of optimized trial functions to compute the dominant eigenvalue of
a transfer matrix\index{transfer matrix}, the largest eigenvalue of
which is of immediate physical significance, since its logarithm is
proportional to the free energy.  A final example deals with the
computation of the second-largest eigenvalue of a Markov
matrix\index{Markov matrix} which defines the single-flip dynamics of
a stochastic process commonly used to sample the
Boltzmann\index{Boltzmann distribution} distribution by Monte Carlo.

The Markov matrix\index{Markov matrix} $\P$ generates the stochastic
time evolution of a system and $\P$ is the immediate analog of the
imaginary-time evolution operator in quantum mechanics.  On the other
hand, the transfer matrix\index{transfer matrix} $\T$ has the
following interpretation.  If one thinks of a $d$-dimensional lattice
system as a one-dimensional array of $d-1$-dimensional slices, the
transfer matrix can be viewed as the evolution operator from one slice
to the next. In other words, one has the following correspondences:
$\T \leftrightarrow \exp(-t{\H}) \leftrightarrow \P$.  In all cases,
optimized trial states \index{trial state} can be used to reduce the
statistical variance of unbiased Monte Carlo methods, such as
\dMC\ \index{diffusion Monte Carlo} and transfer-matrix Monte
Carlo\index{transfer-matrix Monte Carlo}, or to or to both reduce the
statistical variance and obtain less biased results with
\vMC\ methods\index{variational Monte Carlo}.

\subsection{\it Statics at the critical point}\index{critical point}
\label{sec.statics}

We give a brief sketch of how one can design trial states\index{trial
state} associated with the dominant transfer matrix\index{transfer
matrix} eigenstate of lattice systems, and we use as an illustration
the two-dimensional Ising model.  The reader is referred for further
details to Ref.~\onlinecite{NighBloeprb.96}.

\begin{figure}[tbh]
\centerline{\psfig{figure=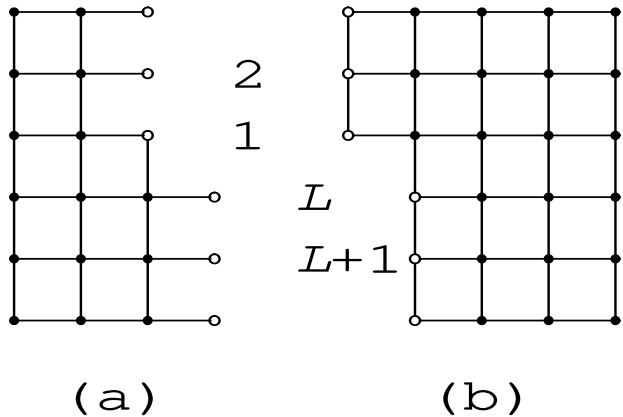,height=3.2truein,width=3.2in,%
rheight=3.2truein,rwidth=3.2truein}}
\caption[c1]{
Illustration of left and right eigenvectors of the transfer matrix.
Open circles are the sites of fixed spins; closed circles have been
summed over in the conditional partition function.  The lattice
segments are to be considered semi infinite.  (Taken from
Ref.~\onlinecite{NighBloeprb.96}.)}
\label{lreivecs}
\end{figure}

Consider a simple-quadratic lattice of $M$ sites, wrapped on a
cylinder with a circumference of $L$ lattice units. If helical
boundary conditions are used, the one-spin transfer
matrix\index{transfer matrix} is
\begin{equation}
T_{S,R} =e^{K(s_1 r_1 + s_1 r_L)}
\prod_{i=1}^{L-1}\delta_{s_i,r_{i+1}},
\label{tm.ising}
\end{equation}
with $S=(s_1,s_2,\dots,s_L)$ and $R\,=(r_1,r_2,\dots,r_L)$, where the
$s_i=\pm 1$ and $r_i=\pm 1$.  The conditional partition function of
the lattice of $M$ sites, subject to the condition that the spins on
the left-hand edge are in state $R$, as illustrated in
Figure~\ref{lreivecs}, is denoted $Z_{M}(R)$.  One has
\begin{equation}
Z_{M+1}(S)=\sum_{R}T_{S,R}Z_{M}(R).
\label{Z}
\end{equation}

For $M\to\infty$ the vector of conditional partition sums $Z_{M}(R)$
forms the dominant right eigenvector of the transfer
matrix\index{transfer matrix}.  An element of this eigenvector is
represented by the graph on the right in Figure~\ref{lreivecs}, at
least if the lattice segments are viewed as semi-infinite. Full
circles indicate spins that have been summed over in the partition
sum; the fixed surface spins are represented by the open circles.
Each bond represents a factor $\exp(K s_i s_j)$.  The left
eigenvector, which is the one that has to be approximated by an
optimized trial vector, is represented by the graph on the left.  In
constructing a trial state\index{trial state}, an effective
Hamiltonian\index{Hamiltonian} is required, to describe the
interactions of the surface spins (open circles) in equilibrium with
the bulk (full circles); the eigenvector is the corresponding vector
of Boltzmann\index{Boltzmann distribution} weights.  The following is
an obvious choice for a trial vector
\begin{equation}
\psi_{\/\rm T} =\exp\sum_{i,j}^\ast K_{ij} s_i s_j.
\label{eq.trial.ising}
\end{equation}
Here the asterisk indicates that the sum over pair interactions has to
be truncated for reasons of numerical efficiency, but of course even
with all pair interactions included, Eq.~(\ref{eq.trial.ising}) is
only an approximation.  The coupling constants are the variational
parameters.  Note that the $K_{ij}$ lack translational invariance,
because of the step on the surface, apparent from
Figure~\ref{lreivecs}.  This causes technical problems requiring a
solution outside the scope of this review.

As an illustration of the efficiency and flexibility of these trial
vectors, we discuss the $XY$-Ising model.  It consists of coupled
Ising and planar rotator degrees of freedom on a simple-quadratic
lattice. On each lattice site there are two variables: $s_i = \pm 1$ and
${\rm\bf n}_i$, a two-component unit vector.  The Hamiltonian
\index{Hamiltonian} of this model is given by
\begin{equation}
H=-k_{\rm B}T\sum_{(i,j)}\left( A\,{\rm\bf n}_i \cdot {\rm\bf n}_j +
B\, {\rm\bf n}_i \cdot {\rm\bf n}_j s_i s_j+ C s_i s_j\right),
\label{XY.Ising.ham}
\end{equation}
where the summation is over nearest-neighbor pairs.

To illustrate the performance of transfer-matrix\index{transfer
matrix} Monte Carlo algorithm we consider the special case $A=B$. For
a discussion of the physics of this model the reader is referred to
Ref.~\onlinecite{NighGraKos95}.  The trial vectors discussed above for
the Ising model have an immediate generalization
\begin{equation}
\psi_{\/\rm T} =\exp\sum_{i,j}^\ast
\left( A_{ij}\,{\rm\bf n}_i \cdot {\rm\bf n}_j  +
B_{ij}\, {\rm\bf n}_i \cdot {\rm\bf n}_j s_i s_j+ C_{ij}
s_i s_j\right).
\label{trial.xy.ising}
\end{equation}

Table~\ref{tab.xy-ising} shows estimates of the dominant eigenvalue of
the $XY$-Ising model for trial vectors truncated at various maximum
values of $d_{ij}$, which essentially is the distance of the pairs
$(i,j)$ appearing in the sum over pair interactions.  As can be seen
by comparing the first and last lines of the table, the variance in
the estimate of the eigenvalue is reduced by a factor 300 for a fixed
number of Monte Carlo steps.  Taking into account that the computer
time per step doubles, this constitutes a speed-up by a factor of 150.

\begin{table}[tbhp]
\caption[tc1]{
Estimated eigenvalue and standard errors for the $XY$-Ising model.
These data apply to the point $(A=B=1.005,C=-0.2285)$ [{\it cf.}
Eq.~(\ref{XY.Ising.ham})] on the line where Ising and $XY$ transitions
coincide \cite{NighGraKos95}. Results are shown for various values of
$d_{\/\rm max}$, the maximum path length of the cutoff in
Eq.~(\ref{trial.xy.ising}).  The results \cite{NighBloeprb.96} are for
a strip of width $L=20$.  }
\label{tab.xy-ising}
\bigskip
\begin{tabular}{dddd}
$\lambda_0$ & $\sigma$ & $d_{\/\rm max}$ &time (arbitrary units)\\
\tableline
34.17406   &   0.0071  & 0  & 15\\
34.20875   &   0.0052  & 2  & 15\\
34.21658   &   0.0015  & 3  & 17\\
34.21418   &   0.00083 & 4  & 19\\
34.21384   &   0.00052 & 5  & 21\\
34.21366   &   0.00049 & 6  & 23\\
34.21379   &   0.00041 & 7  & 26\\
\end{tabular}
\end{table}

The truncation scheme introduced above for the Ising model is purely
geometrical, and therefore carries over with only the obvious
modifications to the $XY$-Ising model.  The same is true for the
three-dimensional models discussed below.  It should, however, be
noted that there are models and choices of transfer
matrices\index{transfer matrix} to which this scheme does not apply.
Ref.~\onlinecite{GraNigh93} contains a discussion and an example of
such a case.

As a further illustration of the trial function
optimization\index{trial state optimization} technique, we discuss
applications to three-dimensional, simple-cubic O($n$) models for
$n=1$, 2 and 3, i.e., the Ising, planar and Heisenberg models.  As
mentioned above, transfer-matrix Monte Carlo\index{transfer-matrix
Monte Carlo} is designed to compute the dominant eigenvalue
$\lambda_0$ of the transfer matrix.  The reduced free energy per site
is $f=-\ln \lambda_0$.  One can calculate the interface free energy as
the difference in free energy of two systems: one with ferromagnetic
and the other with antiferromagnetic interactions, if the dimensions
are chosen so as to force an interface in the antiferromagnetic
system.  For $L\times L\times\infty$ systems with helical boundary
conditions, this means that $L$ has to be even.

Renormalization group theory predicts that the values of $\Delta$, the
reduced interface free energy per lattice site, as a function of
coupling $K$ and system sizes $L$ collapse onto a single curve, at
least close to the critical point\index{critical point} $K_{\/\rm c}$
and for sufficiently big systems.  In terms of the non-linear thermal
scaling field
\begin{equation}
 u(K)=K-K_{\/\rm c}+a(K-K_{\/\rm c})^2+\dots, \label{scalingfield}
\end{equation}
this curve
$\Sigma(x)$ is of the form
\begin{equation} \Delta (u,L) = L^{1-d} \Sigma
(L^{y_{\/\rm T}}u), \label{Deltahomog}
\end{equation}
for a $d$-dimensional system with a thermal scaling
exponent\index{scaling exponent} $y_{\/\rm T}$. The function $\Sigma$
can be expanded in a series:
\begin{equation}
\Sigma(x)=\sum_{l=0}^{\infty}\sigma_l x^l.
\end{equation}

\begin{figure}[tbhp]
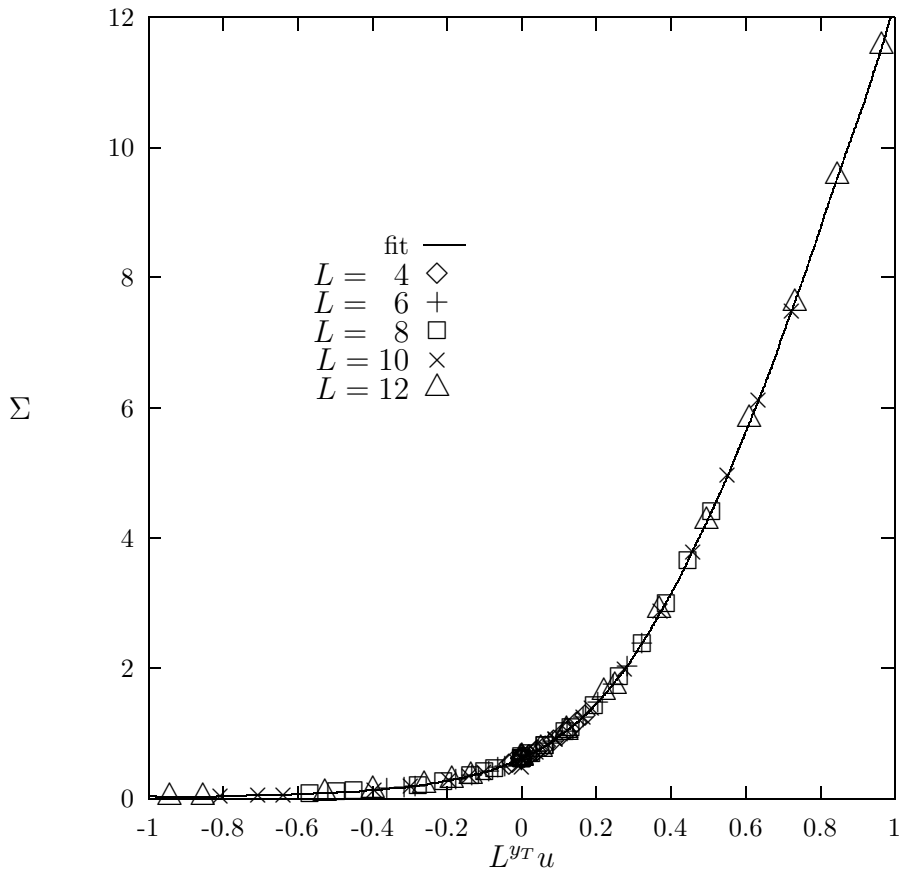

\centerline{\input O1ScalingPlot}
\caption[is]{Finite-size scaling\index{finite-size scaling} plot for the
interface free energy of the three dimensional Ising model.  (Taken
from Ref.~\onlinecite{NighBloeprb.96}.)}
\label{scaling.plot.n1}
\end{figure}

The scaling plot is obtained by treating $K_{\/\rm c}$, $y_{\/\rm T}$,
and the $\sigma_l$ as fitting parameters and the result for the
three-dimensional Ising model is shown in
Figure~\ref{scaling.plot.n1}.
To check if the system sizes are in the asymptotic, finite-size scaling
\index{finite-size scaling}
regime given the statistical accuracy of the Monte Carlo data, fits
can be made both with and without the $6\times 6\times \infty$ data.
Nightingale and Bl\"ote\cite{NighBloeprb.96} found the following
results: $K_{\/\rm c}=0.22162\pm 0.00002$ and $y_{\/\rm T}=1.584\pm
0.004$ using data with $L=6$ through $12$; and $K_{\/\rm c}=0.22167\pm
0.00004$ and $y_{\/\rm T}=1.584\pm 0.014$ if the $L=6$ are omitted.
These results agree well with those of other methods (see e.g.
Ref.~\onlinecite{BLH,BHHSM,GT} and references therein) which are in
the vicinity of $K_{\/\rm c}=0.221655$ (with a margin of error of
about $10^{-6}$) and $y_{\/\rm T}=1.586$ (with a precision of a few
times $10^{-3}$).  Analysis of the interface free energy of $L \times
L \times \infty$ systems, as illustrated in
Figure~\ref{scaling.plot.n1}, suggests that corrections to scaling are
remarkably small compared to the corrections that haunt standard Monte
Carlo analyses \cite{FerLan} for $L \times L \times L$ systems.

A similar analysis can be performed for the three-dimensional planar
model and yields the critical coupling $K_{\/\rm c}=0.45410\pm
0.00003$ for system sizes $L=6$ to 12, and $K_{\/\rm c}=0.45413 \pm
0.00005$ for $L=8$ to 12.  These values are close to results from
series expansions and standard Monte Carlo calculations.  Also the
estimates for the temperature exponent, {\it viz.} $y_{\/\rm T}=1.491
\pm 0.003 $ for $L\geq 6$ and $y_{\/\rm T}=1.487 \pm 0.006 $ for
$L\geq 8$ agree well with other work and the reader is referred to
Ref.~\onlinecite{NighBloeprb.96} for further details.

The calculations for the Heisenberg case were concentrated in a narrow
range of temperatures around the critical point\index{critical point},
and no attempt was made to determine the thermal exponent accurately
and independently; instead, the coupling-constant-expansion estimate
\cite{LGZJ} $y_{\/\rm T}=1.418$ was used.  This yields the following
results: $K_{\/\rm c}=0.69291\pm 0.00004$ for system sizes $L=6$ to
12, Again, this agrees well with other work.

\subsection{\it Dynamics at the critical point}\index{critical point}
\label{sec.dynamics}

The onset of criticality is marked by a divergence of both the
correlation length $\xi$ and the correlation time $\tau$.  The dynamic
exponent $z$ \index{dynamic exponent} links the divergences of length
and time scales: $\tau \sim \xi^z$.  The computation of $z$ to an
accuracy sufficient to address meaningfully questions of universality
has been problematic even for systems as simple as the two-dimensional
Ising model until very recently.

Optimized trial vectors have made it possible to perform finite-size
studies quite analogous to the ones discussed in the previous section.
For the dynamic problem the analog of the transfer
matrix\index{transfer matrix} is the stochastic (Markov)
matrix\index{Markov matrix}\index{stochastic matrix} ${\bf P}$
governing the dynamics; the interface free energy has as its analog
the inverse auto-correlation time. The only difference with the static
case is that the dominant eigenvalue and eigenvector are known: the
eigenvalue is unity and the corresponding eigenvector is the
Boltzmann\index{Boltzmann distribution} distribution, both by
construction.

The correlation time $\tau_L$ (in units of one flip per spin, i.e.,
$L^2$ single-spin flips) is determined by the second-largest
eigenvalue $\lambda_L$ of the Markov matrix\index{Markov matrix} ${\bf
P}$ by the relation
\begin{equation}
\tau_L = -\frac{1}{ L^2 \ln \lambda_L}.
\label{taulab}
\end{equation}
For a system symmetric under spin inversion, the corresponding
eigenvector is expected to be antisymmetric.

The element $P(s'|s)$ of the stochastic matrix\index{stochastic
matrix} denotes the probability of a single-spin flip transition from
configuration $s$ to $s'$.  The matrix ${\bf P}$ satisfies detailed
balance, and consequently, denoting by $\psi_{\/\rm B}(s)^2$ the
Boltzmann\index{Boltzmann distribution} weight of configuration $s$,
\begin{equation}
\hat P(s'|s)= {1 \over \psi_{\/\rm B}(s')} P(s'|s) \psi_{\/\rm B}(s)
\end{equation}
defines a symmetric matrix $\bf \hat P$.  The results discussed below
were obtained by Nightingale and Bl\"ote \cite{NighBloeprl.96}
for the heat bath or Yang~\cite{Yang} transition
probabilities with random site selection.

For an arbitrary trial state\index{trial state} $\psiTket$ and
time $t$ an effective eigenvalue $\lambda^{(t)}_L$ can be defined by
\begin{equation}
\lambda^{(t)}_L =
{\langle {\bf \hat P}^{t+1} \rangle_\psiT \over
 \langle {\bf \hat P}^t\rangle_{\psiT}},
\label{eq.lambdat}
\end{equation}
where $\langle \cdot \rangle_{\psiT}$ denotes the expectation value in
the state $\psiTket$.  In the generic case, the effective eigenvalue
converges for $t \to \infty$ to the dominant eigenvalue with the same
symmetry as the trial state $\psiTket$; for finite projection time
$t$, the effective eigenvalue has an exponentially vanishing,
systematic error.  It turns out that for any given trial state one can
compute the right-hand side of Eq.~(\ref{eq.lambdat}) with standard
Monte Carlo methods, since it can be written as the ratio of two
correlation functions.

A first guiding principle for the construction of trial
states\index{trial state} is that long-wavelength fluctuations of the
magnetization have the longest decay time.  This explains why
straightforward generalization of the ideas presented in Section
\ref{sec.statics}, {\it viz.} a short-distance expansion, fails to
produce a high-quality optimized trial vector, even when higher-order
local spin interactions are included.  A second heuristic ingredient
for this construction is analysis of the exact eigenvectors ${\bf P}$
for $L \leq 5$ systems.  It was found that the eigenvector elements
are given approximately by the Boltzmann\index{Boltzmann distribution}
factor times a function of the magnetization. Thus, one is led to
trial functions defined in terms of the energy, and long-wavelength
components of the Fourier transform of the spin configuration, the
zero momentum component $m_0$ of which is just the magnetization per
site.  Finally, constructions of this sort always require much
experimentation, and Nightingale and Bl\"ote selected the following
form
\begin{equation}
\psiT(s) = \tilde{\psi}_{\/\rm{B}}(s) \; \psi^{{(+)}}(s) \; \psi^{{(-)}}(s),
\end{equation}
where $\psi^{(\pm)}\to \pm \psi^{(\pm)}$ under spin inversion, so that
the trial function is antisymmetric under this transformation.  The
tilde in $\tilde{\psi}_{\/\rm{B}}$ indicates that the temperature is a
variational parameter, but the optimal value of this variational
temperature turns out to be virtually indistinguishable from the true
temperature.  The $\psi^{(\pm)}$ were chosen to be of the form
\begin{eqnarray}
\psi^{{(+)}}=\sum_{\bf k} a_{\bf k}(m_0^2) m_{\bf k}^{{(+)}} +
 m_0 \sum_{\bf k} b_{\bf k}(m_0^2) m_{\bf k}^{{(-)}} \\
\psi^{{(-)}}=m_0 \sum_{\bf k} c_{\bf k}(m_0^2) m_{\bf k}^{{(+)}} +
 \sum_{\bf k} d_{\bf k}(m_0^2) m_{\bf k}^{{(-)}},
\end{eqnarray}
where the index ${\bf k}$ runs through a small set of multiplets of
four or less long-wavelength wave vectors defining the $m_{\bf
k}^{(\pm)},$ translation and rotation symmetric sums of products of
Fourier transforms of the local magnetization; the values of ${\bf k}$
are selected so that $m_{\bf k}^{(-)}$ is odd and $m_{\bf k}^{(+)}$ is
even under spin inversion; the coefficients $a_{\bf k},\;b_{\bf
k},\;c_{\bf k}$ and $d_{\bf k}$ are polynomials of second order or
less in $m_0^2$.  The degrees of these polynomials were chosen so that
no terms occur of higher degree than four in the spin variables.  It
suffices to optimize approximately forty parameters for the trial
functions used in this example.  Table \ref{tab.tau} shows the values
and standard errors of the single-spin-flip auto-correlation time
computed with these trial functions.

To obtain an estimate for the critical exponent $z$, only a
finite-size scaling analysis\index{finite-size scaling} {\em at} the
critical point\index{critical point} is required.  This can be done by
fitting the data for the auto-correlation time to the form
\beq
\tau_L=L^z\sum_{k=0}^{n_{\/\rm c}} \alpha_k L^{2k},
\label{eq.zfit}
\eeq
with $z$ and the $\alpha_k$ as fitting parameters, and $n_{\/\rm c}$
as a cutoff which can be varied to check the convergence.  There is no
compelling theoretical justification for these particular correction
terms other than that they apply to the static two-dimensional Ising
model and provide a conservative extrapolation scheme.  The result
obtained from this analysis is $z=2.1665 \pm 0.0012$ where the error
quoted is the a two-sigma error as estimated from the least-squares
fit.  This appears to be the most precise estimate to date, and the
reader is referred to Ref.\onlinecite{NighBloeprl.96} for a more
detailed discussion of the analysis and an explicit comparison with
other work.

\begin{table}[tbhp]
\caption[auto]{
Estimated auto-correlation time $\tau_L$ for $L \times L$ Ising systems.}
\vskip 2mm
\begin{tabular}{rdd}
$L$&$\tau_1$&error\\
\tableline
5& 137.411& 0.0028\\
6& 206.85& 0.007\\
7& 291.414& 0.013\\
8& 391.453& 0.022\\
9& 507.274& 0.036\\
10& 639.287& 0.057\\
11& 787.507& 0.075\\
12& 952.687& 0.104\\
13& 1134.42& 0.13\\
14& 1333.38& 0.17\\
15& 1549.34& 0.27\\
\end{tabular}
\label{tab.tau}
\end{table}

\acknowledgments
This research was supported by the (US) National Science Foundation
through Grant \# DMR-9214669 and by the Office of Naval Research.
This research was conducted in part using the resources of the Cornell
Theory Center, which receives major funding from the National Science
Foundation (NSF) and New York State, with additional support from the
Advanced Research Projects Agency (ARPA), the National Center for
Research Resources at the National Institutes of Health (NIH), IBM
Corporation, and other members of the center's Corporate Research
Institute.

\end{document}

%% file: argon3.tex
% GNUPLOT: LaTeX picture
\setlength{\unitlength}{0.240900pt}
\ifx\plotpoint\undefined\newsavebox{\plotpoint}\fi
\begin{picture}(1500,900)(0,0)
\tenrm
\sbox{\plotpoint}{\rule[-0.175pt]{0.350pt}{0.350pt}}%
\put(264,158){\rule[-0.175pt]{282.335pt}{0.350pt}}
\put(264,158){\rule[-0.175pt]{0.350pt}{151.526pt}}
\put(264,158){\rule[-0.175pt]{4.818pt}{0.350pt}}
\put(242,158){\makebox(0,0)[r]{0}}
\put(1416,158){\rule[-0.175pt]{4.818pt}{0.350pt}}
\put(264,237){\rule[-0.175pt]{4.818pt}{0.350pt}}
\put(242,237){\makebox(0,0)[r]{1}}
\put(1416,237){\rule[-0.175pt]{4.818pt}{0.350pt}}
\put(264,315){\rule[-0.175pt]{4.818pt}{0.350pt}}
\put(242,315){\makebox(0,0)[r]{2}}
\put(1416,315){\rule[-0.175pt]{4.818pt}{0.350pt}}
\put(264,394){\rule[-0.175pt]{4.818pt}{0.350pt}}
\put(242,394){\makebox(0,0)[r]{3}}
\put(1416,394){\rule[-0.175pt]{4.818pt}{0.350pt}}
\put(264,473){\rule[-0.175pt]{4.818pt}{0.350pt}}
\put(242,473){\makebox(0,0)[r]{4}}
\put(1416,473){\rule[-0.175pt]{4.818pt}{0.350pt}}
\put(264,551){\rule[-0.175pt]{4.818pt}{0.350pt}}
\put(242,551){\makebox(0,0)[r]{5}}
\put(1416,551){\rule[-0.175pt]{4.818pt}{0.350pt}}
\put(264,630){\rule[-0.175pt]{4.818pt}{0.350pt}}
\put(242,630){\makebox(0,0)[r]{6}}
\put(1416,630){\rule[-0.175pt]{4.818pt}{0.350pt}}
\put(264,708){\rule[-0.175pt]{4.818pt}{0.350pt}}
\put(242,708){\makebox(0,0)[r]{7}}
\put(1416,708){\rule[-0.175pt]{4.818pt}{0.350pt}}
\put(264,787){\rule[-0.175pt]{4.818pt}{0.350pt}}
\put(242,787){\makebox(0,0)[r]{8}}
\put(1416,787){\rule[-0.175pt]{4.818pt}{0.350pt}}
\put(264,158){\rule[-0.175pt]{0.350pt}{4.818pt}}
\put(264,113){\makebox(0,0){0}}
\put(264,767){\rule[-0.175pt]{0.350pt}{4.818pt}}
\put(411,158){\rule[-0.175pt]{0.350pt}{4.818pt}}
\put(411,113){\makebox(0,0){1}}
\put(411,767){\rule[-0.175pt]{0.350pt}{4.818pt}}
\put(557,158){\rule[-0.175pt]{0.350pt}{4.818pt}}
\put(557,113){\makebox(0,0){2}}
\put(557,767){\rule[-0.175pt]{0.350pt}{4.818pt}}
\put(704,158){\rule[-0.175pt]{0.350pt}{4.818pt}}
\put(704,113){\makebox(0,0){3}}
\put(704,767){\rule[-0.175pt]{0.350pt}{4.818pt}}
\put(850,158){\rule[-0.175pt]{0.350pt}{4.818pt}}
\put(850,113){\makebox(0,0){4}}
\put(850,767){\rule[-0.175pt]{0.350pt}{4.818pt}}
\put(997,158){\rule[-0.175pt]{0.350pt}{4.818pt}}
\put(997,113){\makebox(0,0){5}}
\put(997,767){\rule[-0.175pt]{0.350pt}{4.818pt}}
\put(1143,158){\rule[-0.175pt]{0.350pt}{4.818pt}}
\put(1143,113){\makebox(0,0){6}}
\put(1143,767){\rule[-0.175pt]{0.350pt}{4.818pt}}
\put(1290,158){\rule[-0.175pt]{0.350pt}{4.818pt}}
\put(1290,113){\makebox(0,0){7}}
\put(1290,767){\rule[-0.175pt]{0.350pt}{4.818pt}}
\put(1436,158){\rule[-0.175pt]{0.350pt}{4.818pt}}
\put(1436,113){\makebox(0,0){8}}
\put(1436,767){\rule[-0.175pt]{0.350pt}{4.818pt}}
\put(264,158){\rule[-0.175pt]{282.335pt}{0.350pt}}
\put(1436,158){\rule[-0.175pt]{0.350pt}{151.526pt}}
\put(264,787){\rule[-0.175pt]{282.335pt}{0.350pt}}
\put(45,472){\makebox(0,0)[l]{\shortstack{$Q$}}}
\put(850,68){\makebox(0,0){$P$}}
\put(264,158){\rule[-0.175pt]{0.350pt}{151.526pt}}
\put(440,708){\makebox(0,0)[r]{$n=2$}}
\put(462,708){\rule[-0.175pt]{15.899pt}{0.350pt}}
\put(411,286){\usebox{\plotpoint}}
\put(411,286){\rule[-0.175pt]{35.171pt}{0.350pt}}
\put(557,285){\rule[-0.175pt]{70.584pt}{0.350pt}}
\put(850,286){\rule[-0.175pt]{35.412pt}{0.350pt}}
\put(484,708){\circle{12}}
\put(411,286){\circle{12}}
\put(557,285){\circle{12}}
\put(704,285){\circle{12}}
\put(850,286){\circle{12}}
\put(997,285){\circle{12}}
\put(1143,284){\circle{12}}
\put(997,285){\rule[-0.175pt]{35.171pt}{0.350pt}}
\put(440,663){\makebox(0,0)[r]{$n=3$}}
\put(462,663){\rule[-0.175pt]{15.899pt}{0.350pt}}
\put(557,342){\usebox{\plotpoint}}
\put(557,342){\rule[-0.175pt]{0.553pt}{0.350pt}}
\put(559,343){\rule[-0.175pt]{0.553pt}{0.350pt}}
\put(561,344){\rule[-0.175pt]{0.553pt}{0.350pt}}
\put(563,345){\rule[-0.175pt]{0.553pt}{0.350pt}}
\put(566,346){\rule[-0.175pt]{0.553pt}{0.350pt}}
\put(568,347){\rule[-0.175pt]{0.553pt}{0.350pt}}
\put(570,348){\rule[-0.175pt]{0.553pt}{0.350pt}}
\put(573,349){\rule[-0.175pt]{0.553pt}{0.350pt}}
\put(575,350){\rule[-0.175pt]{0.553pt}{0.350pt}}
\put(577,351){\rule[-0.175pt]{0.553pt}{0.350pt}}
\put(579,352){\rule[-0.175pt]{0.553pt}{0.350pt}}
\put(582,353){\rule[-0.175pt]{0.553pt}{0.350pt}}
\put(584,354){\rule[-0.175pt]{0.553pt}{0.350pt}}
\put(586,355){\rule[-0.175pt]{0.553pt}{0.350pt}}
\put(589,356){\rule[-0.175pt]{0.553pt}{0.350pt}}
\put(591,357){\rule[-0.175pt]{0.553pt}{0.350pt}}
\put(593,358){\rule[-0.175pt]{0.553pt}{0.350pt}}
\put(596,359){\rule[-0.175pt]{0.553pt}{0.350pt}}
\put(598,360){\rule[-0.175pt]{0.553pt}{0.350pt}}
\put(600,361){\rule[-0.175pt]{0.553pt}{0.350pt}}
\put(602,362){\rule[-0.175pt]{0.553pt}{0.350pt}}
\put(605,363){\rule[-0.175pt]{0.553pt}{0.350pt}}
\put(607,364){\rule[-0.175pt]{0.553pt}{0.350pt}}
\put(609,365){\rule[-0.175pt]{0.553pt}{0.350pt}}
\put(612,366){\rule[-0.175pt]{0.553pt}{0.350pt}}
\put(614,367){\rule[-0.175pt]{0.553pt}{0.350pt}}
\put(616,368){\rule[-0.175pt]{0.553pt}{0.350pt}}
\put(619,369){\rule[-0.175pt]{0.553pt}{0.350pt}}
\put(621,370){\rule[-0.175pt]{0.553pt}{0.350pt}}
\put(623,371){\rule[-0.175pt]{0.553pt}{0.350pt}}
\put(625,372){\rule[-0.175pt]{0.553pt}{0.350pt}}
\put(628,373){\rule[-0.175pt]{0.553pt}{0.350pt}}
\put(630,374){\rule[-0.175pt]{0.553pt}{0.350pt}}
\put(632,375){\rule[-0.175pt]{0.553pt}{0.350pt}}
\put(635,376){\rule[-0.175pt]{0.553pt}{0.350pt}}
\put(637,377){\rule[-0.175pt]{0.553pt}{0.350pt}}
\put(639,378){\rule[-0.175pt]{0.553pt}{0.350pt}}
\put(641,379){\rule[-0.175pt]{0.553pt}{0.350pt}}
\put(644,380){\rule[-0.175pt]{0.553pt}{0.350pt}}
\put(646,381){\rule[-0.175pt]{0.553pt}{0.350pt}}
\put(648,382){\rule[-0.175pt]{0.553pt}{0.350pt}}
\put(651,383){\rule[-0.175pt]{0.553pt}{0.350pt}}
\put(653,384){\rule[-0.175pt]{0.553pt}{0.350pt}}
\put(655,385){\rule[-0.175pt]{0.553pt}{0.350pt}}
\put(658,386){\rule[-0.175pt]{0.553pt}{0.350pt}}
\put(660,387){\rule[-0.175pt]{0.553pt}{0.350pt}}
\put(662,388){\rule[-0.175pt]{0.553pt}{0.350pt}}
\put(664,389){\rule[-0.175pt]{0.553pt}{0.350pt}}
\put(667,390){\rule[-0.175pt]{0.553pt}{0.350pt}}
\put(669,391){\rule[-0.175pt]{0.553pt}{0.350pt}}
\put(671,392){\rule[-0.175pt]{0.553pt}{0.350pt}}
\put(674,393){\rule[-0.175pt]{0.553pt}{0.350pt}}
\put(676,394){\rule[-0.175pt]{0.553pt}{0.350pt}}
\put(678,395){\rule[-0.175pt]{0.553pt}{0.350pt}}
\put(681,396){\rule[-0.175pt]{0.553pt}{0.350pt}}
\put(683,397){\rule[-0.175pt]{0.553pt}{0.350pt}}
\put(685,398){\rule[-0.175pt]{0.553pt}{0.350pt}}
\put(687,399){\rule[-0.175pt]{0.553pt}{0.350pt}}
\put(690,400){\rule[-0.175pt]{0.553pt}{0.350pt}}
\put(692,401){\rule[-0.175pt]{0.553pt}{0.350pt}}
\put(694,402){\rule[-0.175pt]{0.553pt}{0.350pt}}
\put(697,403){\rule[-0.175pt]{0.553pt}{0.350pt}}
\put(699,404){\rule[-0.175pt]{0.553pt}{0.350pt}}
\put(701,405){\rule[-0.175pt]{0.553pt}{0.350pt}}
\put(704,406){\rule[-0.175pt]{0.606pt}{0.350pt}}
\put(706,407){\rule[-0.175pt]{0.606pt}{0.350pt}}
\put(709,408){\rule[-0.175pt]{0.606pt}{0.350pt}}
\put(711,409){\rule[-0.175pt]{0.606pt}{0.350pt}}
\put(714,410){\rule[-0.175pt]{0.606pt}{0.350pt}}
\put(716,411){\rule[-0.175pt]{0.606pt}{0.350pt}}
\put(719,412){\rule[-0.175pt]{0.606pt}{0.350pt}}
\put(721,413){\rule[-0.175pt]{0.606pt}{0.350pt}}
\put(724,414){\rule[-0.175pt]{0.606pt}{0.350pt}}
\put(726,415){\rule[-0.175pt]{0.606pt}{0.350pt}}
\put(729,416){\rule[-0.175pt]{0.606pt}{0.350pt}}
\put(731,417){\rule[-0.175pt]{0.606pt}{0.350pt}}
\put(734,418){\rule[-0.175pt]{0.606pt}{0.350pt}}
\put(736,419){\rule[-0.175pt]{0.606pt}{0.350pt}}
\put(739,420){\rule[-0.175pt]{0.606pt}{0.350pt}}
\put(741,421){\rule[-0.175pt]{0.606pt}{0.350pt}}
\put(744,422){\rule[-0.175pt]{0.606pt}{0.350pt}}
\put(746,423){\rule[-0.175pt]{0.606pt}{0.350pt}}
\put(749,424){\rule[-0.175pt]{0.606pt}{0.350pt}}
\put(751,425){\rule[-0.175pt]{0.606pt}{0.350pt}}
\put(754,426){\rule[-0.175pt]{0.606pt}{0.350pt}}
\put(756,427){\rule[-0.175pt]{0.606pt}{0.350pt}}
\put(759,428){\rule[-0.175pt]{0.606pt}{0.350pt}}
\put(761,429){\rule[-0.175pt]{0.606pt}{0.350pt}}
\put(764,430){\rule[-0.175pt]{0.606pt}{0.350pt}}
\put(766,431){\rule[-0.175pt]{0.606pt}{0.350pt}}
\put(769,432){\rule[-0.175pt]{0.606pt}{0.350pt}}
\put(771,433){\rule[-0.175pt]{0.606pt}{0.350pt}}
\put(774,434){\rule[-0.175pt]{0.606pt}{0.350pt}}
\put(776,435){\rule[-0.175pt]{0.606pt}{0.350pt}}
\put(779,436){\rule[-0.175pt]{0.606pt}{0.350pt}}
\put(782,437){\rule[-0.175pt]{0.606pt}{0.350pt}}
\put(784,438){\rule[-0.175pt]{0.606pt}{0.350pt}}
\put(787,439){\rule[-0.175pt]{0.606pt}{0.350pt}}
\put(789,440){\rule[-0.175pt]{0.606pt}{0.350pt}}
\put(792,441){\rule[-0.175pt]{0.606pt}{0.350pt}}
\put(794,442){\rule[-0.175pt]{0.606pt}{0.350pt}}
\put(797,443){\rule[-0.175pt]{0.606pt}{0.350pt}}
\put(799,444){\rule[-0.175pt]{0.606pt}{0.350pt}}
\put(802,445){\rule[-0.175pt]{0.606pt}{0.350pt}}
\put(804,446){\rule[-0.175pt]{0.606pt}{0.350pt}}
\put(807,447){\rule[-0.175pt]{0.606pt}{0.350pt}}
\put(809,448){\rule[-0.175pt]{0.606pt}{0.350pt}}
\put(812,449){\rule[-0.175pt]{0.606pt}{0.350pt}}
\put(814,450){\rule[-0.175pt]{0.606pt}{0.350pt}}
\put(817,451){\rule[-0.175pt]{0.606pt}{0.350pt}}
\put(819,452){\rule[-0.175pt]{0.606pt}{0.350pt}}
\put(822,453){\rule[-0.175pt]{0.606pt}{0.350pt}}
\put(824,454){\rule[-0.175pt]{0.606pt}{0.350pt}}
\put(827,455){\rule[-0.175pt]{0.606pt}{0.350pt}}
\put(829,456){\rule[-0.175pt]{0.606pt}{0.350pt}}
\put(832,457){\rule[-0.175pt]{0.606pt}{0.350pt}}
\put(834,458){\rule[-0.175pt]{0.606pt}{0.350pt}}
\put(837,459){\rule[-0.175pt]{0.606pt}{0.350pt}}
\put(839,460){\rule[-0.175pt]{0.606pt}{0.350pt}}
\put(842,461){\rule[-0.175pt]{0.606pt}{0.350pt}}
\put(844,462){\rule[-0.175pt]{0.606pt}{0.350pt}}
\put(847,463){\rule[-0.175pt]{0.606pt}{0.350pt}}
\put(849,464){\rule[-0.175pt]{0.695pt}{0.350pt}}
\put(852,465){\rule[-0.175pt]{0.694pt}{0.350pt}}
\put(855,466){\rule[-0.175pt]{0.694pt}{0.350pt}}
\put(858,467){\rule[-0.175pt]{0.694pt}{0.350pt}}
\put(861,468){\rule[-0.175pt]{0.694pt}{0.350pt}}
\put(864,469){\rule[-0.175pt]{0.694pt}{0.350pt}}
\put(867,470){\rule[-0.175pt]{0.694pt}{0.350pt}}
\put(870,471){\rule[-0.175pt]{0.694pt}{0.350pt}}
\put(873,472){\rule[-0.175pt]{0.694pt}{0.350pt}}
\put(875,473){\rule[-0.175pt]{0.694pt}{0.350pt}}
\put(878,474){\rule[-0.175pt]{0.694pt}{0.350pt}}
\put(881,475){\rule[-0.175pt]{0.694pt}{0.350pt}}
\put(884,476){\rule[-0.175pt]{0.694pt}{0.350pt}}
\put(887,477){\rule[-0.175pt]{0.694pt}{0.350pt}}
\put(890,478){\rule[-0.175pt]{0.694pt}{0.350pt}}
\put(893,479){\rule[-0.175pt]{0.694pt}{0.350pt}}
\put(896,480){\rule[-0.175pt]{0.694pt}{0.350pt}}
\put(898,481){\rule[-0.175pt]{0.694pt}{0.350pt}}
\put(901,482){\rule[-0.175pt]{0.694pt}{0.350pt}}
\put(904,483){\rule[-0.175pt]{0.694pt}{0.350pt}}
\put(907,484){\rule[-0.175pt]{0.694pt}{0.350pt}}
\put(910,485){\rule[-0.175pt]{0.694pt}{0.350pt}}
\put(913,486){\rule[-0.175pt]{0.694pt}{0.350pt}}
\put(916,487){\rule[-0.175pt]{0.694pt}{0.350pt}}
\put(919,488){\rule[-0.175pt]{0.694pt}{0.350pt}}
\put(922,489){\rule[-0.175pt]{0.694pt}{0.350pt}}
\put(924,490){\rule[-0.175pt]{0.694pt}{0.350pt}}
\put(927,491){\rule[-0.175pt]{0.694pt}{0.350pt}}
\put(930,492){\rule[-0.175pt]{0.694pt}{0.350pt}}
\put(933,493){\rule[-0.175pt]{0.694pt}{0.350pt}}
\put(936,494){\rule[-0.175pt]{0.694pt}{0.350pt}}
\put(939,495){\rule[-0.175pt]{0.694pt}{0.350pt}}
\put(942,496){\rule[-0.175pt]{0.694pt}{0.350pt}}
\put(945,497){\rule[-0.175pt]{0.694pt}{0.350pt}}
\put(947,498){\rule[-0.175pt]{0.694pt}{0.350pt}}
\put(950,499){\rule[-0.175pt]{0.694pt}{0.350pt}}
\put(953,500){\rule[-0.175pt]{0.694pt}{0.350pt}}
\put(956,501){\rule[-0.175pt]{0.694pt}{0.350pt}}
\put(959,502){\rule[-0.175pt]{0.694pt}{0.350pt}}
\put(962,503){\rule[-0.175pt]{0.694pt}{0.350pt}}
\put(965,504){\rule[-0.175pt]{0.694pt}{0.350pt}}
\put(968,505){\rule[-0.175pt]{0.694pt}{0.350pt}}
\put(971,506){\rule[-0.175pt]{0.694pt}{0.350pt}}
\put(973,507){\rule[-0.175pt]{0.694pt}{0.350pt}}
\put(976,508){\rule[-0.175pt]{0.694pt}{0.350pt}}
\put(979,509){\rule[-0.175pt]{0.694pt}{0.350pt}}
\put(982,510){\rule[-0.175pt]{0.694pt}{0.350pt}}
\put(985,511){\rule[-0.175pt]{0.694pt}{0.350pt}}
\put(988,512){\rule[-0.175pt]{0.694pt}{0.350pt}}
\put(991,513){\rule[-0.175pt]{0.694pt}{0.350pt}}
\put(994,514){\rule[-0.175pt]{0.694pt}{0.350pt}}
\put(996,515){\rule[-0.175pt]{0.570pt}{0.350pt}}
\put(999,516){\rule[-0.175pt]{0.569pt}{0.350pt}}
\put(1001,517){\rule[-0.175pt]{0.569pt}{0.350pt}}
\put(1004,518){\rule[-0.175pt]{0.569pt}{0.350pt}}
\put(1006,519){\rule[-0.175pt]{0.569pt}{0.350pt}}
\put(1008,520){\rule[-0.175pt]{0.569pt}{0.350pt}}
\put(1011,521){\rule[-0.175pt]{0.569pt}{0.350pt}}
\put(1013,522){\rule[-0.175pt]{0.569pt}{0.350pt}}
\put(1015,523){\rule[-0.175pt]{0.569pt}{0.350pt}}
\put(1018,524){\rule[-0.175pt]{0.569pt}{0.350pt}}
\put(1020,525){\rule[-0.175pt]{0.569pt}{0.350pt}}
\put(1022,526){\rule[-0.175pt]{0.569pt}{0.350pt}}
\put(1025,527){\rule[-0.175pt]{0.569pt}{0.350pt}}
\put(1027,528){\rule[-0.175pt]{0.569pt}{0.350pt}}
\put(1030,529){\rule[-0.175pt]{0.569pt}{0.350pt}}
\put(1032,530){\rule[-0.175pt]{0.569pt}{0.350pt}}
\put(1034,531){\rule[-0.175pt]{0.569pt}{0.350pt}}
\put(1037,532){\rule[-0.175pt]{0.569pt}{0.350pt}}
\put(1039,533){\rule[-0.175pt]{0.569pt}{0.350pt}}
\put(1041,534){\rule[-0.175pt]{0.569pt}{0.350pt}}
\put(1044,535){\rule[-0.175pt]{0.569pt}{0.350pt}}
\put(1046,536){\rule[-0.175pt]{0.569pt}{0.350pt}}
\put(1048,537){\rule[-0.175pt]{0.569pt}{0.350pt}}
\put(1051,538){\rule[-0.175pt]{0.569pt}{0.350pt}}
\put(1053,539){\rule[-0.175pt]{0.569pt}{0.350pt}}
\put(1056,540){\rule[-0.175pt]{0.569pt}{0.350pt}}
\put(1058,541){\rule[-0.175pt]{0.569pt}{0.350pt}}
\put(1060,542){\rule[-0.175pt]{0.569pt}{0.350pt}}
\put(1063,543){\rule[-0.175pt]{0.569pt}{0.350pt}}
\put(1065,544){\rule[-0.175pt]{0.569pt}{0.350pt}}
\put(1067,545){\rule[-0.175pt]{0.569pt}{0.350pt}}
\put(1070,546){\rule[-0.175pt]{0.569pt}{0.350pt}}
\put(1072,547){\rule[-0.175pt]{0.569pt}{0.350pt}}
\put(1074,548){\rule[-0.175pt]{0.569pt}{0.350pt}}
\put(1077,549){\rule[-0.175pt]{0.569pt}{0.350pt}}
\put(1079,550){\rule[-0.175pt]{0.569pt}{0.350pt}}
\put(1082,551){\rule[-0.175pt]{0.569pt}{0.350pt}}
\put(1084,552){\rule[-0.175pt]{0.569pt}{0.350pt}}
\put(1086,553){\rule[-0.175pt]{0.569pt}{0.350pt}}
\put(1089,554){\rule[-0.175pt]{0.569pt}{0.350pt}}
\put(1091,555){\rule[-0.175pt]{0.569pt}{0.350pt}}
\put(1093,556){\rule[-0.175pt]{0.569pt}{0.350pt}}
\put(1096,557){\rule[-0.175pt]{0.569pt}{0.350pt}}
\put(1098,558){\rule[-0.175pt]{0.569pt}{0.350pt}}
\put(1100,559){\rule[-0.175pt]{0.569pt}{0.350pt}}
\put(1103,560){\rule[-0.175pt]{0.569pt}{0.350pt}}
\put(1105,561){\rule[-0.175pt]{0.569pt}{0.350pt}}
\put(1108,562){\rule[-0.175pt]{0.569pt}{0.350pt}}
\put(1110,563){\rule[-0.175pt]{0.569pt}{0.350pt}}
\put(1112,564){\rule[-0.175pt]{0.569pt}{0.350pt}}
\put(1115,565){\rule[-0.175pt]{0.569pt}{0.350pt}}
\put(1117,566){\rule[-0.175pt]{0.569pt}{0.350pt}}
\put(1119,567){\rule[-0.175pt]{0.569pt}{0.350pt}}
\put(1122,568){\rule[-0.175pt]{0.569pt}{0.350pt}}
\put(1124,569){\rule[-0.175pt]{0.569pt}{0.350pt}}
\put(1126,570){\rule[-0.175pt]{0.569pt}{0.350pt}}
\put(1129,571){\rule[-0.175pt]{0.569pt}{0.350pt}}
\put(1131,572){\rule[-0.175pt]{0.569pt}{0.350pt}}
\put(1134,573){\rule[-0.175pt]{0.569pt}{0.350pt}}
\put(1136,574){\rule[-0.175pt]{0.569pt}{0.350pt}}
\put(1138,575){\rule[-0.175pt]{0.569pt}{0.350pt}}
\put(1141,576){\rule[-0.175pt]{0.569pt}{0.350pt}}
\put(1143,577){\rule[-0.175pt]{0.569pt}{0.350pt}}
\put(1145,578){\rule[-0.175pt]{0.569pt}{0.350pt}}
\put(1148,579){\rule[-0.175pt]{0.569pt}{0.350pt}}
\put(1150,580){\rule[-0.175pt]{0.569pt}{0.350pt}}
\put(1152,581){\rule[-0.175pt]{0.569pt}{0.350pt}}
\put(1155,582){\rule[-0.175pt]{0.569pt}{0.350pt}}
\put(1157,583){\rule[-0.175pt]{0.569pt}{0.350pt}}
\put(1160,584){\rule[-0.175pt]{0.569pt}{0.350pt}}
\put(1162,585){\rule[-0.175pt]{0.569pt}{0.350pt}}
\put(1164,586){\rule[-0.175pt]{0.569pt}{0.350pt}}
\put(1167,587){\rule[-0.175pt]{0.569pt}{0.350pt}}
\put(1169,588){\rule[-0.175pt]{0.569pt}{0.350pt}}
\put(1171,589){\rule[-0.175pt]{0.569pt}{0.350pt}}
\put(1174,590){\rule[-0.175pt]{0.569pt}{0.350pt}}
\put(1176,591){\rule[-0.175pt]{0.569pt}{0.350pt}}
\put(1178,592){\rule[-0.175pt]{0.569pt}{0.350pt}}
\put(1181,593){\rule[-0.175pt]{0.569pt}{0.350pt}}
\put(1183,594){\rule[-0.175pt]{0.569pt}{0.350pt}}
\put(1186,595){\rule[-0.175pt]{0.569pt}{0.350pt}}
\put(1188,596){\rule[-0.175pt]{0.569pt}{0.350pt}}
\put(1190,597){\rule[-0.175pt]{0.569pt}{0.350pt}}
\put(1193,598){\rule[-0.175pt]{0.569pt}{0.350pt}}
\put(1195,599){\rule[-0.175pt]{0.569pt}{0.350pt}}
\put(1197,600){\rule[-0.175pt]{0.569pt}{0.350pt}}
\put(1200,601){\rule[-0.175pt]{0.569pt}{0.350pt}}
\put(1202,602){\rule[-0.175pt]{0.569pt}{0.350pt}}
\put(1204,603){\rule[-0.175pt]{0.569pt}{0.350pt}}
\put(1207,604){\rule[-0.175pt]{0.569pt}{0.350pt}}
\put(1209,605){\rule[-0.175pt]{0.569pt}{0.350pt}}
\put(1212,606){\rule[-0.175pt]{0.569pt}{0.350pt}}
\put(1214,607){\rule[-0.175pt]{0.569pt}{0.350pt}}
\put(1216,608){\rule[-0.175pt]{0.569pt}{0.350pt}}
\put(1219,609){\rule[-0.175pt]{0.569pt}{0.350pt}}
\put(1221,610){\rule[-0.175pt]{0.569pt}{0.350pt}}
\put(1223,611){\rule[-0.175pt]{0.569pt}{0.350pt}}
\put(1226,612){\rule[-0.175pt]{0.569pt}{0.350pt}}
\put(1228,613){\rule[-0.175pt]{0.569pt}{0.350pt}}
\put(1230,614){\rule[-0.175pt]{0.569pt}{0.350pt}}
\put(1233,615){\rule[-0.175pt]{0.569pt}{0.350pt}}
\put(1235,616){\rule[-0.175pt]{0.569pt}{0.350pt}}
\put(1238,617){\rule[-0.175pt]{0.569pt}{0.350pt}}
\put(1240,618){\rule[-0.175pt]{0.569pt}{0.350pt}}
\put(1242,619){\rule[-0.175pt]{0.569pt}{0.350pt}}
\put(1245,620){\rule[-0.175pt]{0.569pt}{0.350pt}}
\put(1247,621){\rule[-0.175pt]{0.569pt}{0.350pt}}
\put(1249,622){\rule[-0.175pt]{0.569pt}{0.350pt}}
\put(1252,623){\rule[-0.175pt]{0.569pt}{0.350pt}}
\put(1254,624){\rule[-0.175pt]{0.569pt}{0.350pt}}
\put(1256,625){\rule[-0.175pt]{0.569pt}{0.350pt}}
\put(1259,626){\rule[-0.175pt]{0.569pt}{0.350pt}}
\put(1261,627){\rule[-0.175pt]{0.569pt}{0.350pt}}
\put(1264,628){\rule[-0.175pt]{0.569pt}{0.350pt}}
\put(1266,629){\rule[-0.175pt]{0.569pt}{0.350pt}}
\put(1268,630){\rule[-0.175pt]{0.569pt}{0.350pt}}
\put(1271,631){\rule[-0.175pt]{0.569pt}{0.350pt}}
\put(1273,632){\rule[-0.175pt]{0.569pt}{0.350pt}}
\put(1275,633){\rule[-0.175pt]{0.569pt}{0.350pt}}
\put(1278,634){\rule[-0.175pt]{0.569pt}{0.350pt}}
\put(1280,635){\rule[-0.175pt]{0.569pt}{0.350pt}}
\put(1282,636){\rule[-0.175pt]{0.569pt}{0.350pt}}
\put(1285,637){\rule[-0.175pt]{0.569pt}{0.350pt}}
\put(484,663){\circle{18}}
\put(557,342){\circle{18}}
\put(704,406){\circle{18}}
\put(850,464){\circle{18}}
\put(997,515){\circle{18}}
\put(1290,639){\circle{18}}
\put(1287,638){\rule[-0.175pt]{0.569pt}{0.350pt}}
\end{picture}

%% file: argon4.tex
% GNUPLOT: LaTeX picture
\setlength{\unitlength}{0.240900pt}
\ifx\plotpoint\undefined\newsavebox{\plotpoint}\fi
\begin{picture}(1500,900)(0,0)
\tenrm
\sbox{\plotpoint}{\rule[-0.175pt]{0.350pt}{0.350pt}}%
\put(264,158){\rule[-0.175pt]{282.335pt}{0.350pt}}
\put(264,158){\rule[-0.175pt]{0.350pt}{151.526pt}}
\put(264,158){\rule[-0.175pt]{4.818pt}{0.350pt}}
\put(242,158){\makebox(0,0)[r]{0}}
\put(1416,158){\rule[-0.175pt]{4.818pt}{0.350pt}}
\put(264,263){\rule[-0.175pt]{4.818pt}{0.350pt}}
\put(242,263){\makebox(0,0)[r]{1}}
\put(1416,263){\rule[-0.175pt]{4.818pt}{0.350pt}}
\put(264,368){\rule[-0.175pt]{4.818pt}{0.350pt}}
\put(242,368){\makebox(0,0)[r]{2}}
\put(1416,368){\rule[-0.175pt]{4.818pt}{0.350pt}}
\put(264,473){\rule[-0.175pt]{4.818pt}{0.350pt}}
\put(242,473){\makebox(0,0)[r]{3}}
\put(1416,473){\rule[-0.175pt]{4.818pt}{0.350pt}}
\put(264,577){\rule[-0.175pt]{4.818pt}{0.350pt}}
\put(242,577){\makebox(0,0)[r]{4}}
\put(1416,577){\rule[-0.175pt]{4.818pt}{0.350pt}}
\put(264,682){\rule[-0.175pt]{4.818pt}{0.350pt}}
\put(242,682){\makebox(0,0)[r]{5}}
\put(1416,682){\rule[-0.175pt]{4.818pt}{0.350pt}}
\put(264,787){\rule[-0.175pt]{4.818pt}{0.350pt}}
\put(242,787){\makebox(0,0)[r]{6}}
\put(1416,787){\rule[-0.175pt]{4.818pt}{0.350pt}}
\put(264,158){\rule[-0.175pt]{0.350pt}{4.818pt}}
\put(264,113){\makebox(0,0){0}}
\put(264,767){\rule[-0.175pt]{0.350pt}{4.818pt}}
\put(431,158){\rule[-0.175pt]{0.350pt}{4.818pt}}
\put(431,113){\makebox(0,0){1}}
\put(431,767){\rule[-0.175pt]{0.350pt}{4.818pt}}
\put(599,158){\rule[-0.175pt]{0.350pt}{4.818pt}}
\put(599,113){\makebox(0,0){2}}
\put(599,767){\rule[-0.175pt]{0.350pt}{4.818pt}}
\put(766,158){\rule[-0.175pt]{0.350pt}{4.818pt}}
\put(766,113){\makebox(0,0){3}}
\put(766,767){\rule[-0.175pt]{0.350pt}{4.818pt}}
\put(934,158){\rule[-0.175pt]{0.350pt}{4.818pt}}
\put(934,113){\makebox(0,0){4}}
\put(934,767){\rule[-0.175pt]{0.350pt}{4.818pt}}
\put(1101,158){\rule[-0.175pt]{0.350pt}{4.818pt}}
\put(1101,113){\makebox(0,0){5}}
\put(1101,767){\rule[-0.175pt]{0.350pt}{4.818pt}}
\put(1269,158){\rule[-0.175pt]{0.350pt}{4.818pt}}
\put(1269,113){\makebox(0,0){6}}
\put(1269,767){\rule[-0.175pt]{0.350pt}{4.818pt}}
\put(1436,158){\rule[-0.175pt]{0.350pt}{4.818pt}}
\put(1436,113){\makebox(0,0){7}}
\put(1436,767){\rule[-0.175pt]{0.350pt}{4.818pt}}
\put(264,158){\rule[-0.175pt]{282.335pt}{0.350pt}}
\put(1436,158){\rule[-0.175pt]{0.350pt}{151.526pt}}
\put(264,787){\rule[-0.175pt]{282.335pt}{0.350pt}}
\put(45,472){\makebox(0,0)[l]{\shortstack{$Q$}}}
\put(850,68){\makebox(0,0){$P$}}
\put(264,158){\rule[-0.175pt]{0.350pt}{151.526pt}}
\put(431,682){\makebox(0,0)[r]{$n=2$}}
\put(453,682){\rule[-0.175pt]{15.899pt}{0.350pt}}
\put(431,327){\usebox{\plotpoint}}
\put(431,327){\rule[-0.175pt]{40.471pt}{0.350pt}}
\put(599,328){\rule[-0.175pt]{40.230pt}{0.350pt}}
\put(766,327){\rule[-0.175pt]{40.471pt}{0.350pt}}
\put(475,682){\circle{12}}
\put(431,327){\circle{12}}
\put(599,328){\circle{12}}
\put(766,327){\circle{12}}
\put(934,328){\circle{12}}
\put(1101,328){\circle{12}}
\put(1269,328){\circle{12}}
\put(934,328){\rule[-0.175pt]{80.701pt}{0.350pt}}
\put(431,637){\makebox(0,0)[r]{$n=3$}}
\put(453,637){\rule[-0.175pt]{15.899pt}{0.350pt}}
\put(599,378){\usebox{\plotpoint}}
\put(599,378){\rule[-0.175pt]{5.747pt}{0.350pt}}
\put(622,379){\rule[-0.175pt]{5.747pt}{0.350pt}}
\put(646,380){\rule[-0.175pt]{5.747pt}{0.350pt}}
\put(670,381){\rule[-0.175pt]{5.747pt}{0.350pt}}
\put(694,382){\rule[-0.175pt]{5.747pt}{0.350pt}}
\put(718,383){\rule[-0.175pt]{5.747pt}{0.350pt}}
\put(742,384){\rule[-0.175pt]{5.747pt}{0.350pt}}
\put(765,385){\rule[-0.175pt]{80.702pt}{0.350pt}}
\put(1101,386){\rule[-0.175pt]{20.236pt}{0.350pt}}
\put(475,637){\circle{18}}
\put(599,378){\circle{18}}
\put(766,385){\circle{18}}
\put(934,385){\circle{18}}
\put(1101,386){\circle{18}}
\put(1269,384){\circle{18}}
\put(1185,385){\rule[-0.175pt]{20.236pt}{0.350pt}}
\put(431,592){\makebox(0,0)[r]{$n=4$}}
\put(453,592){\rule[-0.175pt]{15.899pt}{0.350pt}}
\put(599,441){\usebox{\plotpoint}}
\put(599,441){\rule[-0.175pt]{0.516pt}{0.350pt}}
\put(601,442){\rule[-0.175pt]{0.516pt}{0.350pt}}
\put(603,443){\rule[-0.175pt]{0.516pt}{0.350pt}}
\put(605,444){\rule[-0.175pt]{0.516pt}{0.350pt}}
\put(607,445){\rule[-0.175pt]{0.516pt}{0.350pt}}
\put(609,446){\rule[-0.175pt]{0.516pt}{0.350pt}}
\put(611,447){\rule[-0.175pt]{0.516pt}{0.350pt}}
\put(613,448){\rule[-0.175pt]{0.516pt}{0.350pt}}
\put(616,449){\rule[-0.175pt]{0.516pt}{0.350pt}}
\put(618,450){\rule[-0.175pt]{0.516pt}{0.350pt}}
\put(620,451){\rule[-0.175pt]{0.516pt}{0.350pt}}
\put(622,452){\rule[-0.175pt]{0.516pt}{0.350pt}}
\put(624,453){\rule[-0.175pt]{0.516pt}{0.350pt}}
\put(626,454){\rule[-0.175pt]{0.516pt}{0.350pt}}
\put(628,455){\rule[-0.175pt]{0.516pt}{0.350pt}}
\put(631,456){\rule[-0.175pt]{0.516pt}{0.350pt}}
\put(633,457){\rule[-0.175pt]{0.516pt}{0.350pt}}
\put(635,458){\rule[-0.175pt]{0.516pt}{0.350pt}}
\put(637,459){\rule[-0.175pt]{0.516pt}{0.350pt}}
\put(639,460){\rule[-0.175pt]{0.516pt}{0.350pt}}
\put(641,461){\rule[-0.175pt]{0.516pt}{0.350pt}}
\put(643,462){\rule[-0.175pt]{0.516pt}{0.350pt}}
\put(646,463){\rule[-0.175pt]{0.516pt}{0.350pt}}
\put(648,464){\rule[-0.175pt]{0.516pt}{0.350pt}}
\put(650,465){\rule[-0.175pt]{0.516pt}{0.350pt}}
\put(652,466){\rule[-0.175pt]{0.516pt}{0.350pt}}
\put(654,467){\rule[-0.175pt]{0.516pt}{0.350pt}}
\put(656,468){\rule[-0.175pt]{0.516pt}{0.350pt}}
\put(658,469){\rule[-0.175pt]{0.516pt}{0.350pt}}
\put(661,470){\rule[-0.175pt]{0.516pt}{0.350pt}}
\put(663,471){\rule[-0.175pt]{0.516pt}{0.350pt}}
\put(665,472){\rule[-0.175pt]{0.516pt}{0.350pt}}
\put(667,473){\rule[-0.175pt]{0.516pt}{0.350pt}}
\put(669,474){\rule[-0.175pt]{0.516pt}{0.350pt}}
\put(671,475){\rule[-0.175pt]{0.516pt}{0.350pt}}
\put(673,476){\rule[-0.175pt]{0.516pt}{0.350pt}}
\put(676,477){\rule[-0.175pt]{0.516pt}{0.350pt}}
\put(678,478){\rule[-0.175pt]{0.516pt}{0.350pt}}
\put(680,479){\rule[-0.175pt]{0.516pt}{0.350pt}}
\put(682,480){\rule[-0.175pt]{0.516pt}{0.350pt}}
\put(684,481){\rule[-0.175pt]{0.516pt}{0.350pt}}
\put(686,482){\rule[-0.175pt]{0.516pt}{0.350pt}}
\put(688,483){\rule[-0.175pt]{0.516pt}{0.350pt}}
\put(691,484){\rule[-0.175pt]{0.516pt}{0.350pt}}
\put(693,485){\rule[-0.175pt]{0.516pt}{0.350pt}}
\put(695,486){\rule[-0.175pt]{0.516pt}{0.350pt}}
\put(697,487){\rule[-0.175pt]{0.516pt}{0.350pt}}
\put(699,488){\rule[-0.175pt]{0.516pt}{0.350pt}}
\put(701,489){\rule[-0.175pt]{0.516pt}{0.350pt}}
\put(703,490){\rule[-0.175pt]{0.516pt}{0.350pt}}
\put(706,491){\rule[-0.175pt]{0.516pt}{0.350pt}}
\put(708,492){\rule[-0.175pt]{0.516pt}{0.350pt}}
\put(710,493){\rule[-0.175pt]{0.516pt}{0.350pt}}
\put(712,494){\rule[-0.175pt]{0.516pt}{0.350pt}}
\put(714,495){\rule[-0.175pt]{0.516pt}{0.350pt}}
\put(716,496){\rule[-0.175pt]{0.516pt}{0.350pt}}
\put(718,497){\rule[-0.175pt]{0.516pt}{0.350pt}}
\put(721,498){\rule[-0.175pt]{0.516pt}{0.350pt}}
\put(723,499){\rule[-0.175pt]{0.516pt}{0.350pt}}
\put(725,500){\rule[-0.175pt]{0.516pt}{0.350pt}}
\put(727,501){\rule[-0.175pt]{0.516pt}{0.350pt}}
\put(729,502){\rule[-0.175pt]{0.516pt}{0.350pt}}
\put(731,503){\rule[-0.175pt]{0.516pt}{0.350pt}}
\put(733,504){\rule[-0.175pt]{0.516pt}{0.350pt}}
\put(736,505){\rule[-0.175pt]{0.516pt}{0.350pt}}
\put(738,506){\rule[-0.175pt]{0.516pt}{0.350pt}}
\put(740,507){\rule[-0.175pt]{0.516pt}{0.350pt}}
\put(742,508){\rule[-0.175pt]{0.516pt}{0.350pt}}
\put(744,509){\rule[-0.175pt]{0.516pt}{0.350pt}}
\put(746,510){\rule[-0.175pt]{0.516pt}{0.350pt}}
\put(748,511){\rule[-0.175pt]{0.516pt}{0.350pt}}
\put(751,512){\rule[-0.175pt]{0.516pt}{0.350pt}}
\put(753,513){\rule[-0.175pt]{0.516pt}{0.350pt}}
\put(755,514){\rule[-0.175pt]{0.516pt}{0.350pt}}
\put(757,515){\rule[-0.175pt]{0.516pt}{0.350pt}}
\put(759,516){\rule[-0.175pt]{0.516pt}{0.350pt}}
\put(761,517){\rule[-0.175pt]{0.516pt}{0.350pt}}
\put(763,518){\rule[-0.175pt]{0.515pt}{0.350pt}}
\put(766,519){\rule[-0.175pt]{0.547pt}{0.350pt}}
\put(768,520){\rule[-0.175pt]{0.547pt}{0.350pt}}
\put(770,521){\rule[-0.175pt]{0.547pt}{0.350pt}}
\put(772,522){\rule[-0.175pt]{0.547pt}{0.350pt}}
\put(775,523){\rule[-0.175pt]{0.547pt}{0.350pt}}
\put(777,524){\rule[-0.175pt]{0.547pt}{0.350pt}}
\put(779,525){\rule[-0.175pt]{0.547pt}{0.350pt}}
\put(781,526){\rule[-0.175pt]{0.547pt}{0.350pt}}
\put(784,527){\rule[-0.175pt]{0.547pt}{0.350pt}}
\put(786,528){\rule[-0.175pt]{0.547pt}{0.350pt}}
\put(788,529){\rule[-0.175pt]{0.547pt}{0.350pt}}
\put(790,530){\rule[-0.175pt]{0.547pt}{0.350pt}}
\put(793,531){\rule[-0.175pt]{0.547pt}{0.350pt}}
\put(795,532){\rule[-0.175pt]{0.547pt}{0.350pt}}
\put(797,533){\rule[-0.175pt]{0.547pt}{0.350pt}}
\put(800,534){\rule[-0.175pt]{0.547pt}{0.350pt}}
\put(802,535){\rule[-0.175pt]{0.547pt}{0.350pt}}
\put(804,536){\rule[-0.175pt]{0.547pt}{0.350pt}}
\put(806,537){\rule[-0.175pt]{0.547pt}{0.350pt}}
\put(809,538){\rule[-0.175pt]{0.547pt}{0.350pt}}
\put(811,539){\rule[-0.175pt]{0.547pt}{0.350pt}}
\put(813,540){\rule[-0.175pt]{0.547pt}{0.350pt}}
\put(815,541){\rule[-0.175pt]{0.547pt}{0.350pt}}
\put(818,542){\rule[-0.175pt]{0.547pt}{0.350pt}}
\put(820,543){\rule[-0.175pt]{0.547pt}{0.350pt}}
\put(822,544){\rule[-0.175pt]{0.547pt}{0.350pt}}
\put(825,545){\rule[-0.175pt]{0.547pt}{0.350pt}}
\put(827,546){\rule[-0.175pt]{0.547pt}{0.350pt}}
\put(829,547){\rule[-0.175pt]{0.547pt}{0.350pt}}
\put(831,548){\rule[-0.175pt]{0.547pt}{0.350pt}}
\put(834,549){\rule[-0.175pt]{0.547pt}{0.350pt}}
\put(836,550){\rule[-0.175pt]{0.547pt}{0.350pt}}
\put(838,551){\rule[-0.175pt]{0.547pt}{0.350pt}}
\put(840,552){\rule[-0.175pt]{0.547pt}{0.350pt}}
\put(843,553){\rule[-0.175pt]{0.547pt}{0.350pt}}
\put(845,554){\rule[-0.175pt]{0.547pt}{0.350pt}}
\put(847,555){\rule[-0.175pt]{0.547pt}{0.350pt}}
\put(849,556){\rule[-0.175pt]{0.547pt}{0.350pt}}
\put(852,557){\rule[-0.175pt]{0.547pt}{0.350pt}}
\put(854,558){\rule[-0.175pt]{0.547pt}{0.350pt}}
\put(856,559){\rule[-0.175pt]{0.547pt}{0.350pt}}
\put(859,560){\rule[-0.175pt]{0.547pt}{0.350pt}}
\put(861,561){\rule[-0.175pt]{0.547pt}{0.350pt}}
\put(863,562){\rule[-0.175pt]{0.547pt}{0.350pt}}
\put(865,563){\rule[-0.175pt]{0.547pt}{0.350pt}}
\put(868,564){\rule[-0.175pt]{0.547pt}{0.350pt}}
\put(870,565){\rule[-0.175pt]{0.547pt}{0.350pt}}
\put(872,566){\rule[-0.175pt]{0.547pt}{0.350pt}}
\put(874,567){\rule[-0.175pt]{0.547pt}{0.350pt}}
\put(877,568){\rule[-0.175pt]{0.547pt}{0.350pt}}
\put(879,569){\rule[-0.175pt]{0.547pt}{0.350pt}}
\put(881,570){\rule[-0.175pt]{0.547pt}{0.350pt}}
\put(884,571){\rule[-0.175pt]{0.547pt}{0.350pt}}
\put(886,572){\rule[-0.175pt]{0.547pt}{0.350pt}}
\put(888,573){\rule[-0.175pt]{0.547pt}{0.350pt}}
\put(890,574){\rule[-0.175pt]{0.547pt}{0.350pt}}
\put(893,575){\rule[-0.175pt]{0.547pt}{0.350pt}}
\put(895,576){\rule[-0.175pt]{0.547pt}{0.350pt}}
\put(897,577){\rule[-0.175pt]{0.547pt}{0.350pt}}
\put(899,578){\rule[-0.175pt]{0.547pt}{0.350pt}}
\put(902,579){\rule[-0.175pt]{0.547pt}{0.350pt}}
\put(904,580){\rule[-0.175pt]{0.547pt}{0.350pt}}
\put(906,581){\rule[-0.175pt]{0.547pt}{0.350pt}}
\put(909,582){\rule[-0.175pt]{0.547pt}{0.350pt}}
\put(911,583){\rule[-0.175pt]{0.547pt}{0.350pt}}
\put(913,584){\rule[-0.175pt]{0.547pt}{0.350pt}}
\put(915,585){\rule[-0.175pt]{0.547pt}{0.350pt}}
\put(918,586){\rule[-0.175pt]{0.547pt}{0.350pt}}
\put(920,587){\rule[-0.175pt]{0.547pt}{0.350pt}}
\put(922,588){\rule[-0.175pt]{0.547pt}{0.350pt}}
\put(924,589){\rule[-0.175pt]{0.547pt}{0.350pt}}
\put(927,590){\rule[-0.175pt]{0.547pt}{0.350pt}}
\put(929,591){\rule[-0.175pt]{0.547pt}{0.350pt}}
\put(931,592){\rule[-0.175pt]{0.547pt}{0.350pt}}
\put(933,593){\rule[-0.175pt]{0.473pt}{0.350pt}}
\put(935,594){\rule[-0.175pt]{0.473pt}{0.350pt}}
\put(937,595){\rule[-0.175pt]{0.473pt}{0.350pt}}
\put(939,596){\rule[-0.175pt]{0.473pt}{0.350pt}}
\put(941,597){\rule[-0.175pt]{0.473pt}{0.350pt}}
\put(943,598){\rule[-0.175pt]{0.473pt}{0.350pt}}
\put(945,599){\rule[-0.175pt]{0.473pt}{0.350pt}}
\put(947,600){\rule[-0.175pt]{0.473pt}{0.350pt}}
\put(949,601){\rule[-0.175pt]{0.473pt}{0.350pt}}
\put(951,602){\rule[-0.175pt]{0.473pt}{0.350pt}}
\put(953,603){\rule[-0.175pt]{0.473pt}{0.350pt}}
\put(955,604){\rule[-0.175pt]{0.473pt}{0.350pt}}
\put(957,605){\rule[-0.175pt]{0.473pt}{0.350pt}}
\put(959,606){\rule[-0.175pt]{0.473pt}{0.350pt}}
\put(961,607){\rule[-0.175pt]{0.473pt}{0.350pt}}
\put(963,608){\rule[-0.175pt]{0.473pt}{0.350pt}}
\put(965,609){\rule[-0.175pt]{0.473pt}{0.350pt}}
\put(967,610){\rule[-0.175pt]{0.473pt}{0.350pt}}
\put(969,611){\rule[-0.175pt]{0.473pt}{0.350pt}}
\put(971,612){\rule[-0.175pt]{0.473pt}{0.350pt}}
\put(973,613){\rule[-0.175pt]{0.473pt}{0.350pt}}
\put(975,614){\rule[-0.175pt]{0.473pt}{0.350pt}}
\put(977,615){\rule[-0.175pt]{0.473pt}{0.350pt}}
\put(979,616){\rule[-0.175pt]{0.473pt}{0.350pt}}
\put(981,617){\rule[-0.175pt]{0.473pt}{0.350pt}}
\put(983,618){\rule[-0.175pt]{0.473pt}{0.350pt}}
\put(985,619){\rule[-0.175pt]{0.473pt}{0.350pt}}
\put(987,620){\rule[-0.175pt]{0.473pt}{0.350pt}}
\put(989,621){\rule[-0.175pt]{0.473pt}{0.350pt}}
\put(990,622){\rule[-0.175pt]{0.473pt}{0.350pt}}
\put(992,623){\rule[-0.175pt]{0.473pt}{0.350pt}}
\put(994,624){\rule[-0.175pt]{0.473pt}{0.350pt}}
\put(996,625){\rule[-0.175pt]{0.473pt}{0.350pt}}
\put(998,626){\rule[-0.175pt]{0.473pt}{0.350pt}}
\put(1000,627){\rule[-0.175pt]{0.473pt}{0.350pt}}
\put(1002,628){\rule[-0.175pt]{0.473pt}{0.350pt}}
\put(1004,629){\rule[-0.175pt]{0.473pt}{0.350pt}}
\put(1006,630){\rule[-0.175pt]{0.473pt}{0.350pt}}
\put(1008,631){\rule[-0.175pt]{0.473pt}{0.350pt}}
\put(1010,632){\rule[-0.175pt]{0.473pt}{0.350pt}}
\put(1012,633){\rule[-0.175pt]{0.473pt}{0.350pt}}
\put(1014,634){\rule[-0.175pt]{0.473pt}{0.350pt}}
\put(1016,635){\rule[-0.175pt]{0.473pt}{0.350pt}}
\put(1018,636){\rule[-0.175pt]{0.473pt}{0.350pt}}
\put(1020,637){\rule[-0.175pt]{0.473pt}{0.350pt}}
\put(1022,638){\rule[-0.175pt]{0.473pt}{0.350pt}}
\put(1024,639){\rule[-0.175pt]{0.473pt}{0.350pt}}
\put(1026,640){\rule[-0.175pt]{0.473pt}{0.350pt}}
\put(1028,641){\rule[-0.175pt]{0.473pt}{0.350pt}}
\put(1030,642){\rule[-0.175pt]{0.473pt}{0.350pt}}
\put(1032,643){\rule[-0.175pt]{0.473pt}{0.350pt}}
\put(1034,644){\rule[-0.175pt]{0.473pt}{0.350pt}}
\put(1036,645){\rule[-0.175pt]{0.473pt}{0.350pt}}
\put(1038,646){\rule[-0.175pt]{0.473pt}{0.350pt}}
\put(1040,647){\rule[-0.175pt]{0.473pt}{0.350pt}}
\put(1042,648){\rule[-0.175pt]{0.473pt}{0.350pt}}
\put(1044,649){\rule[-0.175pt]{0.473pt}{0.350pt}}
\put(1045,650){\rule[-0.175pt]{0.473pt}{0.350pt}}
\put(1047,651){\rule[-0.175pt]{0.473pt}{0.350pt}}
\put(1049,652){\rule[-0.175pt]{0.473pt}{0.350pt}}
\put(1051,653){\rule[-0.175pt]{0.473pt}{0.350pt}}
\put(1053,654){\rule[-0.175pt]{0.473pt}{0.350pt}}
\put(1055,655){\rule[-0.175pt]{0.473pt}{0.350pt}}
\put(1057,656){\rule[-0.175pt]{0.473pt}{0.350pt}}
\put(1059,657){\rule[-0.175pt]{0.473pt}{0.350pt}}
\put(1061,658){\rule[-0.175pt]{0.473pt}{0.350pt}}
\put(1063,659){\rule[-0.175pt]{0.473pt}{0.350pt}}
\put(1065,660){\rule[-0.175pt]{0.473pt}{0.350pt}}
\put(1067,661){\rule[-0.175pt]{0.473pt}{0.350pt}}
\put(1069,662){\rule[-0.175pt]{0.473pt}{0.350pt}}
\put(1071,663){\rule[-0.175pt]{0.473pt}{0.350pt}}
\put(1073,664){\rule[-0.175pt]{0.473pt}{0.350pt}}
\put(1075,665){\rule[-0.175pt]{0.473pt}{0.350pt}}
\put(1077,666){\rule[-0.175pt]{0.473pt}{0.350pt}}
\put(1079,667){\rule[-0.175pt]{0.473pt}{0.350pt}}
\put(1081,668){\rule[-0.175pt]{0.473pt}{0.350pt}}
\put(1083,669){\rule[-0.175pt]{0.473pt}{0.350pt}}
\put(1085,670){\rule[-0.175pt]{0.473pt}{0.350pt}}
\put(1087,671){\rule[-0.175pt]{0.473pt}{0.350pt}}
\put(1089,672){\rule[-0.175pt]{0.473pt}{0.350pt}}
\put(1091,673){\rule[-0.175pt]{0.473pt}{0.350pt}}
\put(1093,674){\rule[-0.175pt]{0.473pt}{0.350pt}}
\put(1095,675){\rule[-0.175pt]{0.473pt}{0.350pt}}
\put(1097,676){\rule[-0.175pt]{0.473pt}{0.350pt}}
\put(475,592){\circle{24}}
\put(599,441){\circle{24}}
\put(766,519){\circle{24}}
\put(934,593){\circle{24}}
\put(1101,678){\circle{24}}
\put(1099,677){\rule[-0.175pt]{0.473pt}{0.350pt}}
\end{picture}

%% file: argon5.tex
% GNUPLOT: LaTeX picture
\setlength{\unitlength}{0.240900pt}
\ifx\plotpoint\undefined\newsavebox{\plotpoint}\fi
\begin{picture}(1500,900)(0,0)
\tenrm
\sbox{\plotpoint}{\rule[-0.175pt]{0.350pt}{0.350pt}}%
\put(264,158){\rule[-0.175pt]{282.335pt}{0.350pt}}
\put(264,158){\rule[-0.175pt]{0.350pt}{151.526pt}}
\put(264,158){\rule[-0.175pt]{4.818pt}{0.350pt}}
\put(242,158){\makebox(0,0)[r]{0}}
\put(1416,158){\rule[-0.175pt]{4.818pt}{0.350pt}}
\put(264,237){\rule[-0.175pt]{4.818pt}{0.350pt}}
\put(242,237){\makebox(0,0)[r]{0.5}}
\put(1416,237){\rule[-0.175pt]{4.818pt}{0.350pt}}
\put(264,315){\rule[-0.175pt]{4.818pt}{0.350pt}}
\put(242,315){\makebox(0,0)[r]{1}}
\put(1416,315){\rule[-0.175pt]{4.818pt}{0.350pt}}
\put(264,394){\rule[-0.175pt]{4.818pt}{0.350pt}}
\put(242,394){\makebox(0,0)[r]{1.5}}
\put(1416,394){\rule[-0.175pt]{4.818pt}{0.350pt}}
\put(264,473){\rule[-0.175pt]{4.818pt}{0.350pt}}
\put(242,473){\makebox(0,0)[r]{2}}
\put(1416,473){\rule[-0.175pt]{4.818pt}{0.350pt}}
\put(264,551){\rule[-0.175pt]{4.818pt}{0.350pt}}
\put(242,551){\makebox(0,0)[r]{2.5}}
\put(1416,551){\rule[-0.175pt]{4.818pt}{0.350pt}}
\put(264,630){\rule[-0.175pt]{4.818pt}{0.350pt}}
\put(242,630){\makebox(0,0)[r]{3}}
\put(1416,630){\rule[-0.175pt]{4.818pt}{0.350pt}}
\put(264,708){\rule[-0.175pt]{4.818pt}{0.350pt}}
\put(242,708){\makebox(0,0)[r]{3.5}}
\put(1416,708){\rule[-0.175pt]{4.818pt}{0.350pt}}
\put(264,787){\rule[-0.175pt]{4.818pt}{0.350pt}}
\put(242,787){\makebox(0,0)[r]{4}}
\put(1416,787){\rule[-0.175pt]{4.818pt}{0.350pt}}
\put(264,158){\rule[-0.175pt]{0.350pt}{4.818pt}}
\put(264,113){\makebox(0,0){0}}
\put(264,767){\rule[-0.175pt]{0.350pt}{4.818pt}}
\put(431,158){\rule[-0.175pt]{0.350pt}{4.818pt}}
\put(431,113){\makebox(0,0){1}}
\put(431,767){\rule[-0.175pt]{0.350pt}{4.818pt}}
\put(599,158){\rule[-0.175pt]{0.350pt}{4.818pt}}
\put(599,113){\makebox(0,0){2}}
\put(599,767){\rule[-0.175pt]{0.350pt}{4.818pt}}
\put(766,158){\rule[-0.175pt]{0.350pt}{4.818pt}}
\put(766,113){\makebox(0,0){3}}
\put(766,767){\rule[-0.175pt]{0.350pt}{4.818pt}}
\put(934,158){\rule[-0.175pt]{0.350pt}{4.818pt}}
\put(934,113){\makebox(0,0){4}}
\put(934,767){\rule[-0.175pt]{0.350pt}{4.818pt}}
\put(1101,158){\rule[-0.175pt]{0.350pt}{4.818pt}}
\put(1101,113){\makebox(0,0){5}}
\put(1101,767){\rule[-0.175pt]{0.350pt}{4.818pt}}
\put(1269,158){\rule[-0.175pt]{0.350pt}{4.818pt}}
\put(1269,113){\makebox(0,0){6}}
\put(1269,767){\rule[-0.175pt]{0.350pt}{4.818pt}}
\put(1436,158){\rule[-0.175pt]{0.350pt}{4.818pt}}
\put(1436,113){\makebox(0,0){7}}
\put(1436,767){\rule[-0.175pt]{0.350pt}{4.818pt}}
\put(264,158){\rule[-0.175pt]{282.335pt}{0.350pt}}
\put(1436,158){\rule[-0.175pt]{0.350pt}{151.526pt}}
\put(264,787){\rule[-0.175pt]{282.335pt}{0.350pt}}
\put(45,472){\makebox(0,0)[l]{\shortstack{$Q$}}}
\put(850,68){\makebox(0,0){$P$}}
\put(264,158){\rule[-0.175pt]{0.350pt}{151.526pt}}
\put(431,630){\makebox(0,0)[r]{$n=2$}}
\put(453,630){\rule[-0.175pt]{15.899pt}{0.350pt}}
\put(431,376){\usebox{\plotpoint}}
\put(431,376){\rule[-0.175pt]{20.236pt}{0.350pt}}
\put(515,375){\rule[-0.175pt]{20.236pt}{0.350pt}}
\put(599,374){\rule[-0.175pt]{1.149pt}{0.350pt}}
\put(603,375){\rule[-0.175pt]{1.149pt}{0.350pt}}
\put(608,376){\rule[-0.175pt]{1.149pt}{0.350pt}}
\put(613,377){\rule[-0.175pt]{1.149pt}{0.350pt}}
\put(618,378){\rule[-0.175pt]{1.149pt}{0.350pt}}
\put(622,379){\rule[-0.175pt]{1.149pt}{0.350pt}}
\put(627,380){\rule[-0.175pt]{1.149pt}{0.350pt}}
\put(632,381){\rule[-0.175pt]{1.149pt}{0.350pt}}
\put(637,382){\rule[-0.175pt]{1.149pt}{0.350pt}}
\put(641,383){\rule[-0.175pt]{1.149pt}{0.350pt}}
\put(646,384){\rule[-0.175pt]{1.149pt}{0.350pt}}
\put(651,385){\rule[-0.175pt]{1.149pt}{0.350pt}}
\put(656,386){\rule[-0.175pt]{1.149pt}{0.350pt}}
\put(661,387){\rule[-0.175pt]{1.149pt}{0.350pt}}
\put(665,388){\rule[-0.175pt]{1.149pt}{0.350pt}}
\put(670,389){\rule[-0.175pt]{1.149pt}{0.350pt}}
\put(675,390){\rule[-0.175pt]{1.149pt}{0.350pt}}
\put(680,391){\rule[-0.175pt]{1.149pt}{0.350pt}}
\put(684,392){\rule[-0.175pt]{1.149pt}{0.350pt}}
\put(689,393){\rule[-0.175pt]{1.149pt}{0.350pt}}
\put(694,394){\rule[-0.175pt]{1.149pt}{0.350pt}}
\put(699,395){\rule[-0.175pt]{1.149pt}{0.350pt}}
\put(703,396){\rule[-0.175pt]{1.149pt}{0.350pt}}
\put(708,397){\rule[-0.175pt]{1.149pt}{0.350pt}}
\put(713,398){\rule[-0.175pt]{1.149pt}{0.350pt}}
\put(718,399){\rule[-0.175pt]{1.149pt}{0.350pt}}
\put(723,400){\rule[-0.175pt]{1.149pt}{0.350pt}}
\put(727,401){\rule[-0.175pt]{1.149pt}{0.350pt}}
\put(732,402){\rule[-0.175pt]{1.149pt}{0.350pt}}
\put(737,403){\rule[-0.175pt]{1.149pt}{0.350pt}}
\put(742,404){\rule[-0.175pt]{1.149pt}{0.350pt}}
\put(746,405){\rule[-0.175pt]{1.149pt}{0.350pt}}
\put(751,406){\rule[-0.175pt]{1.149pt}{0.350pt}}
\put(756,407){\rule[-0.175pt]{1.149pt}{0.350pt}}
\put(761,408){\rule[-0.175pt]{1.149pt}{0.350pt}}
\put(765,409){\rule[-0.175pt]{13.490pt}{0.350pt}}
\put(822,410){\rule[-0.175pt]{13.490pt}{0.350pt}}
\put(878,411){\rule[-0.175pt]{13.490pt}{0.350pt}}
\put(934,412){\rule[-0.175pt]{6.705pt}{0.350pt}}
\put(961,413){\rule[-0.175pt]{6.705pt}{0.350pt}}
\put(989,414){\rule[-0.175pt]{6.705pt}{0.350pt}}
\put(1017,415){\rule[-0.175pt]{6.705pt}{0.350pt}}
\put(1045,416){\rule[-0.175pt]{6.705pt}{0.350pt}}
\put(1073,417){\rule[-0.175pt]{6.705pt}{0.350pt}}
\put(475,630){\circle{12}}
\put(431,376){\circle{12}}
\put(599,374){\circle{12}}
\put(766,409){\circle{12}}
\put(934,412){\circle{12}}
\put(1101,418){\circle{12}}
\put(1269,419){\circle{12}}
\put(1101,418){\rule[-0.175pt]{40.471pt}{0.350pt}}
\put(431,585){\makebox(0,0)[r]{$n=3$}}
\put(453,585){\rule[-0.175pt]{15.899pt}{0.350pt}}
\put(599,430){\usebox{\plotpoint}}
\put(599,430){\rule[-0.175pt]{1.341pt}{0.350pt}}
\put(604,431){\rule[-0.175pt]{1.341pt}{0.350pt}}
\put(610,432){\rule[-0.175pt]{1.341pt}{0.350pt}}
\put(615,433){\rule[-0.175pt]{1.341pt}{0.350pt}}
\put(621,434){\rule[-0.175pt]{1.341pt}{0.350pt}}
\put(626,435){\rule[-0.175pt]{1.341pt}{0.350pt}}
\put(632,436){\rule[-0.175pt]{1.341pt}{0.350pt}}
\put(637,437){\rule[-0.175pt]{1.341pt}{0.350pt}}
\put(643,438){\rule[-0.175pt]{1.341pt}{0.350pt}}
\put(649,439){\rule[-0.175pt]{1.341pt}{0.350pt}}
\put(654,440){\rule[-0.175pt]{1.341pt}{0.350pt}}
\put(660,441){\rule[-0.175pt]{1.341pt}{0.350pt}}
\put(665,442){\rule[-0.175pt]{1.341pt}{0.350pt}}
\put(671,443){\rule[-0.175pt]{1.341pt}{0.350pt}}
\put(676,444){\rule[-0.175pt]{1.341pt}{0.350pt}}
\put(682,445){\rule[-0.175pt]{1.341pt}{0.350pt}}
\put(688,446){\rule[-0.175pt]{1.341pt}{0.350pt}}
\put(693,447){\rule[-0.175pt]{1.341pt}{0.350pt}}
\put(699,448){\rule[-0.175pt]{1.341pt}{0.350pt}}
\put(704,449){\rule[-0.175pt]{1.341pt}{0.350pt}}
\put(710,450){\rule[-0.175pt]{1.341pt}{0.350pt}}
\put(715,451){\rule[-0.175pt]{1.341pt}{0.350pt}}
\put(721,452){\rule[-0.175pt]{1.341pt}{0.350pt}}
\put(727,453){\rule[-0.175pt]{1.341pt}{0.350pt}}
\put(732,454){\rule[-0.175pt]{1.341pt}{0.350pt}}
\put(738,455){\rule[-0.175pt]{1.341pt}{0.350pt}}
\put(743,456){\rule[-0.175pt]{1.341pt}{0.350pt}}
\put(749,457){\rule[-0.175pt]{1.341pt}{0.350pt}}
\put(754,458){\rule[-0.175pt]{1.341pt}{0.350pt}}
\put(760,459){\rule[-0.175pt]{1.341pt}{0.350pt}}
\put(765,460){\rule[-0.175pt]{1.927pt}{0.350pt}}
\put(774,461){\rule[-0.175pt]{1.927pt}{0.350pt}}
\put(782,462){\rule[-0.175pt]{1.927pt}{0.350pt}}
\put(790,463){\rule[-0.175pt]{1.927pt}{0.350pt}}
\put(798,464){\rule[-0.175pt]{1.927pt}{0.350pt}}
\put(806,465){\rule[-0.175pt]{1.927pt}{0.350pt}}
\put(814,466){\rule[-0.175pt]{1.927pt}{0.350pt}}
\put(822,467){\rule[-0.175pt]{1.927pt}{0.350pt}}
\put(830,468){\rule[-0.175pt]{1.927pt}{0.350pt}}
\put(838,469){\rule[-0.175pt]{1.927pt}{0.350pt}}
\put(846,470){\rule[-0.175pt]{1.927pt}{0.350pt}}
\put(854,471){\rule[-0.175pt]{1.927pt}{0.350pt}}
\put(862,472){\rule[-0.175pt]{1.927pt}{0.350pt}}
\put(870,473){\rule[-0.175pt]{1.927pt}{0.350pt}}
\put(878,474){\rule[-0.175pt]{1.927pt}{0.350pt}}
\put(886,475){\rule[-0.175pt]{1.927pt}{0.350pt}}
\put(894,476){\rule[-0.175pt]{1.927pt}{0.350pt}}
\put(902,477){\rule[-0.175pt]{1.927pt}{0.350pt}}
\put(910,478){\rule[-0.175pt]{1.927pt}{0.350pt}}
\put(918,479){\rule[-0.175pt]{1.927pt}{0.350pt}}
\put(926,480){\rule[-0.175pt]{1.927pt}{0.350pt}}
\put(934,481){\rule[-0.175pt]{43.343pt}{0.350pt}}
\put(1113,482){\rule[-0.175pt]{3.113pt}{0.350pt}}
\put(1126,483){\rule[-0.175pt]{3.113pt}{0.350pt}}
\put(1139,484){\rule[-0.175pt]{3.113pt}{0.350pt}}
\put(1152,485){\rule[-0.175pt]{3.113pt}{0.350pt}}
\put(1165,486){\rule[-0.175pt]{3.113pt}{0.350pt}}
\put(1178,487){\rule[-0.175pt]{3.113pt}{0.350pt}}
\put(1191,488){\rule[-0.175pt]{3.113pt}{0.350pt}}
\put(1204,489){\rule[-0.175pt]{3.113pt}{0.350pt}}
\put(1217,490){\rule[-0.175pt]{3.113pt}{0.350pt}}
\put(1230,491){\rule[-0.175pt]{3.113pt}{0.350pt}}
\put(1243,492){\rule[-0.175pt]{3.113pt}{0.350pt}}
\put(475,585){\circle{18}}
\put(599,430){\circle{18}}
\put(766,460){\circle{18}}
\put(934,481){\circle{18}}
\put(1101,481){\circle{18}}
\put(1269,494){\circle{18}}
\put(1256,493){\rule[-0.175pt]{3.113pt}{0.350pt}}
\put(431,540){\makebox(0,0)[r]{$n=4$}}
\put(453,540){\rule[-0.175pt]{15.899pt}{0.350pt}}
\put(599,496){\usebox{\plotpoint}}
\put(599,496){\rule[-0.175pt]{2.682pt}{0.350pt}}
\put(610,497){\rule[-0.175pt]{2.682pt}{0.350pt}}
\put(621,498){\rule[-0.175pt]{2.682pt}{0.350pt}}
\put(632,499){\rule[-0.175pt]{2.682pt}{0.350pt}}
\put(643,500){\rule[-0.175pt]{2.682pt}{0.350pt}}
\put(654,501){\rule[-0.175pt]{2.682pt}{0.350pt}}
\put(665,502){\rule[-0.175pt]{2.682pt}{0.350pt}}
\put(676,503){\rule[-0.175pt]{2.682pt}{0.350pt}}
\put(688,504){\rule[-0.175pt]{2.682pt}{0.350pt}}
\put(699,505){\rule[-0.175pt]{2.682pt}{0.350pt}}
\put(710,506){\rule[-0.175pt]{2.682pt}{0.350pt}}
\put(721,507){\rule[-0.175pt]{2.682pt}{0.350pt}}
\put(732,508){\rule[-0.175pt]{2.682pt}{0.350pt}}
\put(743,509){\rule[-0.175pt]{2.682pt}{0.350pt}}
\put(754,510){\rule[-0.175pt]{2.682pt}{0.350pt}}
\put(766,511){\rule[-0.175pt]{0.880pt}{0.350pt}}
\put(769,512){\rule[-0.175pt]{0.880pt}{0.350pt}}
\put(773,513){\rule[-0.175pt]{0.880pt}{0.350pt}}
\put(776,514){\rule[-0.175pt]{0.880pt}{0.350pt}}
\put(780,515){\rule[-0.175pt]{0.880pt}{0.350pt}}
\put(784,516){\rule[-0.175pt]{0.880pt}{0.350pt}}
\put(787,517){\rule[-0.175pt]{0.880pt}{0.350pt}}
\put(791,518){\rule[-0.175pt]{0.880pt}{0.350pt}}
\put(795,519){\rule[-0.175pt]{0.880pt}{0.350pt}}
\put(798,520){\rule[-0.175pt]{0.880pt}{0.350pt}}
\put(802,521){\rule[-0.175pt]{0.880pt}{0.350pt}}
\put(806,522){\rule[-0.175pt]{0.880pt}{0.350pt}}
\put(809,523){\rule[-0.175pt]{0.880pt}{0.350pt}}
\put(813,524){\rule[-0.175pt]{0.880pt}{0.350pt}}
\put(817,525){\rule[-0.175pt]{0.880pt}{0.350pt}}
\put(820,526){\rule[-0.175pt]{0.880pt}{0.350pt}}
\put(824,527){\rule[-0.175pt]{0.880pt}{0.350pt}}
\put(828,528){\rule[-0.175pt]{0.880pt}{0.350pt}}
\put(831,529){\rule[-0.175pt]{0.880pt}{0.350pt}}
\put(835,530){\rule[-0.175pt]{0.880pt}{0.350pt}}
\put(839,531){\rule[-0.175pt]{0.880pt}{0.350pt}}
\put(842,532){\rule[-0.175pt]{0.880pt}{0.350pt}}
\put(846,533){\rule[-0.175pt]{0.880pt}{0.350pt}}
\put(849,534){\rule[-0.175pt]{0.880pt}{0.350pt}}
\put(853,535){\rule[-0.175pt]{0.880pt}{0.350pt}}
\put(857,536){\rule[-0.175pt]{0.880pt}{0.350pt}}
\put(860,537){\rule[-0.175pt]{0.880pt}{0.350pt}}
\put(864,538){\rule[-0.175pt]{0.880pt}{0.350pt}}
\put(868,539){\rule[-0.175pt]{0.880pt}{0.350pt}}
\put(871,540){\rule[-0.175pt]{0.880pt}{0.350pt}}
\put(875,541){\rule[-0.175pt]{0.880pt}{0.350pt}}
\put(879,542){\rule[-0.175pt]{0.880pt}{0.350pt}}
\put(882,543){\rule[-0.175pt]{0.880pt}{0.350pt}}
\put(886,544){\rule[-0.175pt]{0.880pt}{0.350pt}}
\put(890,545){\rule[-0.175pt]{0.880pt}{0.350pt}}
\put(893,546){\rule[-0.175pt]{0.880pt}{0.350pt}}
\put(897,547){\rule[-0.175pt]{0.880pt}{0.350pt}}
\put(901,548){\rule[-0.175pt]{0.880pt}{0.350pt}}
\put(904,549){\rule[-0.175pt]{0.880pt}{0.350pt}}
\put(908,550){\rule[-0.175pt]{0.880pt}{0.350pt}}
\put(912,551){\rule[-0.175pt]{0.880pt}{0.350pt}}
\put(915,552){\rule[-0.175pt]{0.880pt}{0.350pt}}
\put(919,553){\rule[-0.175pt]{0.880pt}{0.350pt}}
\put(923,554){\rule[-0.175pt]{0.880pt}{0.350pt}}
\put(926,555){\rule[-0.175pt]{0.880pt}{0.350pt}}
\put(930,556){\rule[-0.175pt]{0.880pt}{0.350pt}}
\put(933,557){\rule[-0.175pt]{1.257pt}{0.350pt}}
\put(939,558){\rule[-0.175pt]{1.257pt}{0.350pt}}
\put(944,559){\rule[-0.175pt]{1.257pt}{0.350pt}}
\put(949,560){\rule[-0.175pt]{1.257pt}{0.350pt}}
\put(954,561){\rule[-0.175pt]{1.257pt}{0.350pt}}
\put(960,562){\rule[-0.175pt]{1.257pt}{0.350pt}}
\put(965,563){\rule[-0.175pt]{1.257pt}{0.350pt}}
\put(970,564){\rule[-0.175pt]{1.257pt}{0.350pt}}
\put(975,565){\rule[-0.175pt]{1.257pt}{0.350pt}}
\put(980,566){\rule[-0.175pt]{1.257pt}{0.350pt}}
\put(986,567){\rule[-0.175pt]{1.257pt}{0.350pt}}
\put(991,568){\rule[-0.175pt]{1.257pt}{0.350pt}}
\put(996,569){\rule[-0.175pt]{1.257pt}{0.350pt}}
\put(1001,570){\rule[-0.175pt]{1.257pt}{0.350pt}}
\put(1007,571){\rule[-0.175pt]{1.257pt}{0.350pt}}
\put(1012,572){\rule[-0.175pt]{1.257pt}{0.350pt}}
\put(1017,573){\rule[-0.175pt]{1.257pt}{0.350pt}}
\put(1022,574){\rule[-0.175pt]{1.257pt}{0.350pt}}
\put(1027,575){\rule[-0.175pt]{1.257pt}{0.350pt}}
\put(1033,576){\rule[-0.175pt]{1.257pt}{0.350pt}}
\put(1038,577){\rule[-0.175pt]{1.257pt}{0.350pt}}
\put(1043,578){\rule[-0.175pt]{1.257pt}{0.350pt}}
\put(1048,579){\rule[-0.175pt]{1.257pt}{0.350pt}}
\put(1054,580){\rule[-0.175pt]{1.257pt}{0.350pt}}
\put(1059,581){\rule[-0.175pt]{1.257pt}{0.350pt}}
\put(1064,582){\rule[-0.175pt]{1.257pt}{0.350pt}}
\put(1069,583){\rule[-0.175pt]{1.257pt}{0.350pt}}
\put(1074,584){\rule[-0.175pt]{1.257pt}{0.350pt}}
\put(1080,585){\rule[-0.175pt]{1.257pt}{0.350pt}}
\put(1085,586){\rule[-0.175pt]{1.257pt}{0.350pt}}
\put(1090,587){\rule[-0.175pt]{1.257pt}{0.350pt}}
\put(475,540){\circle{24}}
\put(599,496){\circle{24}}
\put(766,511){\circle{24}}
\put(934,557){\circle{24}}
\put(1101,589){\circle{24}}
\put(1095,588){\rule[-0.175pt]{1.257pt}{0.350pt}}
\put(431,495){\makebox(0,0)[r]{$n=5$}}
\put(453,495){\rule[-0.175pt]{15.899pt}{0.350pt}}
\put(766,588){\usebox{\plotpoint}}
\put(766,588){\rule[-0.175pt]{2.130pt}{0.350pt}}
\put(774,589){\rule[-0.175pt]{2.130pt}{0.350pt}}
\put(783,590){\rule[-0.175pt]{2.130pt}{0.350pt}}
\put(792,591){\rule[-0.175pt]{2.130pt}{0.350pt}}
\put(801,592){\rule[-0.175pt]{2.130pt}{0.350pt}}
\put(810,593){\rule[-0.175pt]{2.130pt}{0.350pt}}
\put(819,594){\rule[-0.175pt]{2.130pt}{0.350pt}}
\put(827,595){\rule[-0.175pt]{2.130pt}{0.350pt}}
\put(836,596){\rule[-0.175pt]{2.130pt}{0.350pt}}
\put(845,597){\rule[-0.175pt]{2.130pt}{0.350pt}}
\put(854,598){\rule[-0.175pt]{2.130pt}{0.350pt}}
\put(863,599){\rule[-0.175pt]{2.130pt}{0.350pt}}
\put(872,600){\rule[-0.175pt]{2.130pt}{0.350pt}}
\put(880,601){\rule[-0.175pt]{2.130pt}{0.350pt}}
\put(889,602){\rule[-0.175pt]{2.130pt}{0.350pt}}
\put(898,603){\rule[-0.175pt]{2.130pt}{0.350pt}}
\put(907,604){\rule[-0.175pt]{2.130pt}{0.350pt}}
\put(916,605){\rule[-0.175pt]{2.130pt}{0.350pt}}
\put(925,606){\rule[-0.175pt]{2.130pt}{0.350pt}}
\put(933,607){\rule[-0.175pt]{1.916pt}{0.350pt}}
\put(941,608){\rule[-0.175pt]{1.916pt}{0.350pt}}
\put(949,609){\rule[-0.175pt]{1.916pt}{0.350pt}}
\put(957,610){\rule[-0.175pt]{1.916pt}{0.350pt}}
\put(965,611){\rule[-0.175pt]{1.916pt}{0.350pt}}
\put(973,612){\rule[-0.175pt]{1.916pt}{0.350pt}}
\put(981,613){\rule[-0.175pt]{1.916pt}{0.350pt}}
\put(989,614){\rule[-0.175pt]{1.916pt}{0.350pt}}
\put(997,615){\rule[-0.175pt]{1.916pt}{0.350pt}}
\put(1005,616){\rule[-0.175pt]{1.916pt}{0.350pt}}
\put(1013,617){\rule[-0.175pt]{1.916pt}{0.350pt}}
\put(1021,618){\rule[-0.175pt]{1.916pt}{0.350pt}}
\put(1029,619){\rule[-0.175pt]{1.916pt}{0.350pt}}
\put(1037,620){\rule[-0.175pt]{1.916pt}{0.350pt}}
\put(1045,621){\rule[-0.175pt]{1.916pt}{0.350pt}}
\put(1053,622){\rule[-0.175pt]{1.916pt}{0.350pt}}
\put(1061,623){\rule[-0.175pt]{1.916pt}{0.350pt}}
\put(1069,624){\rule[-0.175pt]{1.916pt}{0.350pt}}
\put(1077,625){\rule[-0.175pt]{1.916pt}{0.350pt}}
\put(1085,626){\rule[-0.175pt]{1.916pt}{0.350pt}}
\put(475,495){\circle*{24}}
\put(766,588){\circle*{24}}
\put(934,607){\circle*{24}}
\put(1101,628){\circle*{24}}
\put(1093,627){\rule[-0.175pt]{1.916pt}{0.350pt}}
\end{picture}

%% file: neon.tex
% GNUPLOT: LaTeX picture
\setlength{\unitlength}{0.240900pt}
\ifx\plotpoint\undefined\newsavebox{\plotpoint}\fi
\begin{picture}(1500,900)(0,0)
\tenrm
\sbox{\plotpoint}{\rule[-0.175pt]{0.350pt}{0.350pt}}%
\put(264,158){\rule[-0.175pt]{282.335pt}{0.350pt}}
\put(264,158){\rule[-0.175pt]{0.350pt}{151.526pt}}
\put(264,158){\rule[-0.175pt]{4.818pt}{0.350pt}}
\put(242,158){\makebox(0,0)[r]{0}}
\put(1416,158){\rule[-0.175pt]{4.818pt}{0.350pt}}
\put(264,237){\rule[-0.175pt]{4.818pt}{0.350pt}}
\put(242,237){\makebox(0,0)[r]{1}}
\put(1416,237){\rule[-0.175pt]{4.818pt}{0.350pt}}
\put(264,315){\rule[-0.175pt]{4.818pt}{0.350pt}}
\put(242,315){\makebox(0,0)[r]{2}}
\put(1416,315){\rule[-0.175pt]{4.818pt}{0.350pt}}
\put(264,394){\rule[-0.175pt]{4.818pt}{0.350pt}}
\put(242,394){\makebox(0,0)[r]{3}}
\put(1416,394){\rule[-0.175pt]{4.818pt}{0.350pt}}
\put(264,473){\rule[-0.175pt]{4.818pt}{0.350pt}}
\put(242,473){\makebox(0,0)[r]{4}}
\put(1416,473){\rule[-0.175pt]{4.818pt}{0.350pt}}
\put(264,551){\rule[-0.175pt]{4.818pt}{0.350pt}}
\put(242,551){\makebox(0,0)[r]{5}}
\put(1416,551){\rule[-0.175pt]{4.818pt}{0.350pt}}
\put(264,630){\rule[-0.175pt]{4.818pt}{0.350pt}}
\put(242,630){\makebox(0,0)[r]{6}}
\put(1416,630){\rule[-0.175pt]{4.818pt}{0.350pt}}
\put(264,708){\rule[-0.175pt]{4.818pt}{0.350pt}}
\put(242,708){\makebox(0,0)[r]{7}}
\put(1416,708){\rule[-0.175pt]{4.818pt}{0.350pt}}
\put(264,787){\rule[-0.175pt]{4.818pt}{0.350pt}}
\put(242,787){\makebox(0,0)[r]{8}}
\put(1416,787){\rule[-0.175pt]{4.818pt}{0.350pt}}
\put(264,158){\rule[-0.175pt]{0.350pt}{4.818pt}}
\put(264,113){\makebox(0,0){0}}
\put(264,767){\rule[-0.175pt]{0.350pt}{4.818pt}}
\put(394,158){\rule[-0.175pt]{0.350pt}{4.818pt}}
\put(394,113){\makebox(0,0){1}}
\put(394,767){\rule[-0.175pt]{0.350pt}{4.818pt}}
\put(524,158){\rule[-0.175pt]{0.350pt}{4.818pt}}
\put(524,113){\makebox(0,0){2}}
\put(524,767){\rule[-0.175pt]{0.350pt}{4.818pt}}
\put(655,158){\rule[-0.175pt]{0.350pt}{4.818pt}}
\put(655,113){\makebox(0,0){3}}
\put(655,767){\rule[-0.175pt]{0.350pt}{4.818pt}}
\put(785,158){\rule[-0.175pt]{0.350pt}{4.818pt}}
\put(785,113){\makebox(0,0){4}}
\put(785,767){\rule[-0.175pt]{0.350pt}{4.818pt}}
\put(915,158){\rule[-0.175pt]{0.350pt}{4.818pt}}
\put(915,113){\makebox(0,0){5}}
\put(915,767){\rule[-0.175pt]{0.350pt}{4.818pt}}
\put(1045,158){\rule[-0.175pt]{0.350pt}{4.818pt}}
\put(1045,113){\makebox(0,0){6}}
\put(1045,767){\rule[-0.175pt]{0.350pt}{4.818pt}}
\put(1176,158){\rule[-0.175pt]{0.350pt}{4.818pt}}
\put(1176,113){\makebox(0,0){7}}
\put(1176,767){\rule[-0.175pt]{0.350pt}{4.818pt}}
\put(1306,158){\rule[-0.175pt]{0.350pt}{4.818pt}}
\put(1306,113){\makebox(0,0){8}}
\put(1306,767){\rule[-0.175pt]{0.350pt}{4.818pt}}
\put(1436,158){\rule[-0.175pt]{0.350pt}{4.818pt}}
\put(1436,113){\makebox(0,0){9}}
\put(1436,767){\rule[-0.175pt]{0.350pt}{4.818pt}}
\put(264,158){\rule[-0.175pt]{282.335pt}{0.350pt}}
\put(1436,158){\rule[-0.175pt]{0.350pt}{151.526pt}}
\put(264,787){\rule[-0.175pt]{282.335pt}{0.350pt}}
\put(45,472){\makebox(0,0)[l]{\shortstack{$Q$}}}
\put(850,68){\makebox(0,0){$P$}}
\put(264,158){\rule[-0.175pt]{0.350pt}{151.526pt}}
\put(427,708){\makebox(0,0)[r]{Ne$_2$}}
\put(449,708){\rule[-0.175pt]{15.899pt}{0.350pt}}
\put(394,290){\usebox{\plotpoint}}
\put(394,290){\rule[-0.175pt]{0.639pt}{0.350pt}}
\put(396,291){\rule[-0.175pt]{0.639pt}{0.350pt}}
\put(399,292){\rule[-0.175pt]{0.639pt}{0.350pt}}
\put(401,293){\rule[-0.175pt]{0.639pt}{0.350pt}}
\put(404,294){\rule[-0.175pt]{0.639pt}{0.350pt}}
\put(407,295){\rule[-0.175pt]{0.639pt}{0.350pt}}
\put(409,296){\rule[-0.175pt]{0.639pt}{0.350pt}}
\put(412,297){\rule[-0.175pt]{0.639pt}{0.350pt}}
\put(415,298){\rule[-0.175pt]{0.639pt}{0.350pt}}
\put(417,299){\rule[-0.175pt]{0.639pt}{0.350pt}}
\put(420,300){\rule[-0.175pt]{0.639pt}{0.350pt}}
\put(423,301){\rule[-0.175pt]{0.639pt}{0.350pt}}
\put(425,302){\rule[-0.175pt]{0.639pt}{0.350pt}}
\put(428,303){\rule[-0.175pt]{0.639pt}{0.350pt}}
\put(431,304){\rule[-0.175pt]{0.639pt}{0.350pt}}
\put(433,305){\rule[-0.175pt]{0.639pt}{0.350pt}}
\put(436,306){\rule[-0.175pt]{0.639pt}{0.350pt}}
\put(439,307){\rule[-0.175pt]{0.639pt}{0.350pt}}
\put(441,308){\rule[-0.175pt]{0.639pt}{0.350pt}}
\put(444,309){\rule[-0.175pt]{0.639pt}{0.350pt}}
\put(447,310){\rule[-0.175pt]{0.639pt}{0.350pt}}
\put(449,311){\rule[-0.175pt]{0.639pt}{0.350pt}}
\put(452,312){\rule[-0.175pt]{0.639pt}{0.350pt}}
\put(455,313){\rule[-0.175pt]{0.639pt}{0.350pt}}
\put(457,314){\rule[-0.175pt]{0.639pt}{0.350pt}}
\put(460,315){\rule[-0.175pt]{0.639pt}{0.350pt}}
\put(462,316){\rule[-0.175pt]{0.639pt}{0.350pt}}
\put(465,317){\rule[-0.175pt]{0.639pt}{0.350pt}}
\put(468,318){\rule[-0.175pt]{0.639pt}{0.350pt}}
\put(470,319){\rule[-0.175pt]{0.639pt}{0.350pt}}
\put(473,320){\rule[-0.175pt]{0.639pt}{0.350pt}}
\put(476,321){\rule[-0.175pt]{0.639pt}{0.350pt}}
\put(478,322){\rule[-0.175pt]{0.639pt}{0.350pt}}
\put(481,323){\rule[-0.175pt]{0.639pt}{0.350pt}}
\put(484,324){\rule[-0.175pt]{0.639pt}{0.350pt}}
\put(486,325){\rule[-0.175pt]{0.639pt}{0.350pt}}
\put(489,326){\rule[-0.175pt]{0.639pt}{0.350pt}}
\put(492,327){\rule[-0.175pt]{0.639pt}{0.350pt}}
\put(494,328){\rule[-0.175pt]{0.639pt}{0.350pt}}
\put(497,329){\rule[-0.175pt]{0.639pt}{0.350pt}}
\put(500,330){\rule[-0.175pt]{0.639pt}{0.350pt}}
\put(502,331){\rule[-0.175pt]{0.639pt}{0.350pt}}
\put(505,332){\rule[-0.175pt]{0.639pt}{0.350pt}}
\put(508,333){\rule[-0.175pt]{0.639pt}{0.350pt}}
\put(510,334){\rule[-0.175pt]{0.639pt}{0.350pt}}
\put(513,335){\rule[-0.175pt]{0.639pt}{0.350pt}}
\put(516,336){\rule[-0.175pt]{0.639pt}{0.350pt}}
\put(518,337){\rule[-0.175pt]{0.639pt}{0.350pt}}
\put(521,338){\rule[-0.175pt]{0.639pt}{0.350pt}}
\put(524,339){\rule[-0.175pt]{0.493pt}{0.350pt}}
\put(526,340){\rule[-0.175pt]{0.493pt}{0.350pt}}
\put(528,341){\rule[-0.175pt]{0.493pt}{0.350pt}}
\put(530,342){\rule[-0.175pt]{0.493pt}{0.350pt}}
\put(532,343){\rule[-0.175pt]{0.493pt}{0.350pt}}
\put(534,344){\rule[-0.175pt]{0.493pt}{0.350pt}}
\put(536,345){\rule[-0.175pt]{0.493pt}{0.350pt}}
\put(538,346){\rule[-0.175pt]{0.493pt}{0.350pt}}
\put(540,347){\rule[-0.175pt]{0.493pt}{0.350pt}}
\put(542,348){\rule[-0.175pt]{0.493pt}{0.350pt}}
\put(544,349){\rule[-0.175pt]{0.493pt}{0.350pt}}
\put(546,350){\rule[-0.175pt]{0.493pt}{0.350pt}}
\put(548,351){\rule[-0.175pt]{0.493pt}{0.350pt}}
\put(550,352){\rule[-0.175pt]{0.493pt}{0.350pt}}
\put(552,353){\rule[-0.175pt]{0.493pt}{0.350pt}}
\put(554,354){\rule[-0.175pt]{0.493pt}{0.350pt}}
\put(556,355){\rule[-0.175pt]{0.493pt}{0.350pt}}
\put(558,356){\rule[-0.175pt]{0.493pt}{0.350pt}}
\put(560,357){\rule[-0.175pt]{0.493pt}{0.350pt}}
\put(562,358){\rule[-0.175pt]{0.493pt}{0.350pt}}
\put(564,359){\rule[-0.175pt]{0.493pt}{0.350pt}}
\put(566,360){\rule[-0.175pt]{0.493pt}{0.350pt}}
\put(569,361){\rule[-0.175pt]{0.493pt}{0.350pt}}
\put(571,362){\rule[-0.175pt]{0.493pt}{0.350pt}}
\put(573,363){\rule[-0.175pt]{0.493pt}{0.350pt}}
\put(575,364){\rule[-0.175pt]{0.493pt}{0.350pt}}
\put(577,365){\rule[-0.175pt]{0.493pt}{0.350pt}}
\put(579,366){\rule[-0.175pt]{0.493pt}{0.350pt}}
\put(581,367){\rule[-0.175pt]{0.493pt}{0.350pt}}
\put(583,368){\rule[-0.175pt]{0.493pt}{0.350pt}}
\put(585,369){\rule[-0.175pt]{0.493pt}{0.350pt}}
\put(587,370){\rule[-0.175pt]{0.493pt}{0.350pt}}
\put(589,371){\rule[-0.175pt]{0.493pt}{0.350pt}}
\put(591,372){\rule[-0.175pt]{0.493pt}{0.350pt}}
\put(593,373){\rule[-0.175pt]{0.493pt}{0.350pt}}
\put(595,374){\rule[-0.175pt]{0.493pt}{0.350pt}}
\put(597,375){\rule[-0.175pt]{0.493pt}{0.350pt}}
\put(599,376){\rule[-0.175pt]{0.493pt}{0.350pt}}
\put(601,377){\rule[-0.175pt]{0.493pt}{0.350pt}}
\put(603,378){\rule[-0.175pt]{0.493pt}{0.350pt}}
\put(605,379){\rule[-0.175pt]{0.493pt}{0.350pt}}
\put(607,380){\rule[-0.175pt]{0.493pt}{0.350pt}}
\put(609,381){\rule[-0.175pt]{0.493pt}{0.350pt}}
\put(612,382){\rule[-0.175pt]{0.493pt}{0.350pt}}
\put(614,383){\rule[-0.175pt]{0.493pt}{0.350pt}}
\put(616,384){\rule[-0.175pt]{0.493pt}{0.350pt}}
\put(618,385){\rule[-0.175pt]{0.493pt}{0.350pt}}
\put(620,386){\rule[-0.175pt]{0.493pt}{0.350pt}}
\put(622,387){\rule[-0.175pt]{0.493pt}{0.350pt}}
\put(624,388){\rule[-0.175pt]{0.493pt}{0.350pt}}
\put(626,389){\rule[-0.175pt]{0.493pt}{0.350pt}}
\put(628,390){\rule[-0.175pt]{0.493pt}{0.350pt}}
\put(630,391){\rule[-0.175pt]{0.493pt}{0.350pt}}
\put(632,392){\rule[-0.175pt]{0.493pt}{0.350pt}}
\put(634,393){\rule[-0.175pt]{0.493pt}{0.350pt}}
\put(636,394){\rule[-0.175pt]{0.493pt}{0.350pt}}
\put(638,395){\rule[-0.175pt]{0.493pt}{0.350pt}}
\put(640,396){\rule[-0.175pt]{0.493pt}{0.350pt}}
\put(642,397){\rule[-0.175pt]{0.493pt}{0.350pt}}
\put(644,398){\rule[-0.175pt]{0.493pt}{0.350pt}}
\put(646,399){\rule[-0.175pt]{0.493pt}{0.350pt}}
\put(648,400){\rule[-0.175pt]{0.493pt}{0.350pt}}
\put(650,401){\rule[-0.175pt]{0.493pt}{0.350pt}}
\put(652,402){\rule[-0.175pt]{0.493pt}{0.350pt}}
\put(655,403){\rule[-0.175pt]{0.569pt}{0.350pt}}
\put(657,404){\rule[-0.175pt]{0.569pt}{0.350pt}}
\put(659,405){\rule[-0.175pt]{0.569pt}{0.350pt}}
\put(662,406){\rule[-0.175pt]{0.569pt}{0.350pt}}
\put(664,407){\rule[-0.175pt]{0.569pt}{0.350pt}}
\put(666,408){\rule[-0.175pt]{0.569pt}{0.350pt}}
\put(669,409){\rule[-0.175pt]{0.569pt}{0.350pt}}
\put(671,410){\rule[-0.175pt]{0.569pt}{0.350pt}}
\put(673,411){\rule[-0.175pt]{0.569pt}{0.350pt}}
\put(676,412){\rule[-0.175pt]{0.569pt}{0.350pt}}
\put(678,413){\rule[-0.175pt]{0.569pt}{0.350pt}}
\put(681,414){\rule[-0.175pt]{0.569pt}{0.350pt}}
\put(683,415){\rule[-0.175pt]{0.569pt}{0.350pt}}
\put(685,416){\rule[-0.175pt]{0.569pt}{0.350pt}}
\put(688,417){\rule[-0.175pt]{0.569pt}{0.350pt}}
\put(690,418){\rule[-0.175pt]{0.569pt}{0.350pt}}
\put(692,419){\rule[-0.175pt]{0.569pt}{0.350pt}}
\put(695,420){\rule[-0.175pt]{0.569pt}{0.350pt}}
\put(697,421){\rule[-0.175pt]{0.569pt}{0.350pt}}
\put(699,422){\rule[-0.175pt]{0.569pt}{0.350pt}}
\put(702,423){\rule[-0.175pt]{0.569pt}{0.350pt}}
\put(704,424){\rule[-0.175pt]{0.569pt}{0.350pt}}
\put(707,425){\rule[-0.175pt]{0.569pt}{0.350pt}}
\put(709,426){\rule[-0.175pt]{0.569pt}{0.350pt}}
\put(711,427){\rule[-0.175pt]{0.569pt}{0.350pt}}
\put(714,428){\rule[-0.175pt]{0.569pt}{0.350pt}}
\put(716,429){\rule[-0.175pt]{0.569pt}{0.350pt}}
\put(718,430){\rule[-0.175pt]{0.569pt}{0.350pt}}
\put(721,431){\rule[-0.175pt]{0.569pt}{0.350pt}}
\put(723,432){\rule[-0.175pt]{0.569pt}{0.350pt}}
\put(725,433){\rule[-0.175pt]{0.569pt}{0.350pt}}
\put(728,434){\rule[-0.175pt]{0.569pt}{0.350pt}}
\put(730,435){\rule[-0.175pt]{0.569pt}{0.350pt}}
\put(733,436){\rule[-0.175pt]{0.569pt}{0.350pt}}
\put(735,437){\rule[-0.175pt]{0.569pt}{0.350pt}}
\put(737,438){\rule[-0.175pt]{0.569pt}{0.350pt}}
\put(740,439){\rule[-0.175pt]{0.569pt}{0.350pt}}
\put(742,440){\rule[-0.175pt]{0.569pt}{0.350pt}}
\put(744,441){\rule[-0.175pt]{0.569pt}{0.350pt}}
\put(747,442){\rule[-0.175pt]{0.569pt}{0.350pt}}
\put(749,443){\rule[-0.175pt]{0.569pt}{0.350pt}}
\put(751,444){\rule[-0.175pt]{0.569pt}{0.350pt}}
\put(754,445){\rule[-0.175pt]{0.569pt}{0.350pt}}
\put(756,446){\rule[-0.175pt]{0.569pt}{0.350pt}}
\put(759,447){\rule[-0.175pt]{0.569pt}{0.350pt}}
\put(761,448){\rule[-0.175pt]{0.569pt}{0.350pt}}
\put(763,449){\rule[-0.175pt]{0.569pt}{0.350pt}}
\put(766,450){\rule[-0.175pt]{0.569pt}{0.350pt}}
\put(768,451){\rule[-0.175pt]{0.569pt}{0.350pt}}
\put(770,452){\rule[-0.175pt]{0.569pt}{0.350pt}}
\put(773,453){\rule[-0.175pt]{0.569pt}{0.350pt}}
\put(775,454){\rule[-0.175pt]{0.569pt}{0.350pt}}
\put(777,455){\rule[-0.175pt]{0.569pt}{0.350pt}}
\put(780,456){\rule[-0.175pt]{0.569pt}{0.350pt}}
\put(782,457){\rule[-0.175pt]{0.569pt}{0.350pt}}
\put(785,458){\rule[-0.175pt]{0.824pt}{0.350pt}}
\put(788,459){\rule[-0.175pt]{0.824pt}{0.350pt}}
\put(791,460){\rule[-0.175pt]{0.824pt}{0.350pt}}
\put(795,461){\rule[-0.175pt]{0.824pt}{0.350pt}}
\put(798,462){\rule[-0.175pt]{0.824pt}{0.350pt}}
\put(802,463){\rule[-0.175pt]{0.824pt}{0.350pt}}
\put(805,464){\rule[-0.175pt]{0.824pt}{0.350pt}}
\put(808,465){\rule[-0.175pt]{0.824pt}{0.350pt}}
\put(812,466){\rule[-0.175pt]{0.824pt}{0.350pt}}
\put(815,467){\rule[-0.175pt]{0.824pt}{0.350pt}}
\put(819,468){\rule[-0.175pt]{0.824pt}{0.350pt}}
\put(822,469){\rule[-0.175pt]{0.824pt}{0.350pt}}
\put(826,470){\rule[-0.175pt]{0.824pt}{0.350pt}}
\put(829,471){\rule[-0.175pt]{0.824pt}{0.350pt}}
\put(832,472){\rule[-0.175pt]{0.824pt}{0.350pt}}
\put(836,473){\rule[-0.175pt]{0.824pt}{0.350pt}}
\put(839,474){\rule[-0.175pt]{0.824pt}{0.350pt}}
\put(843,475){\rule[-0.175pt]{0.824pt}{0.350pt}}
\put(846,476){\rule[-0.175pt]{0.824pt}{0.350pt}}
\put(850,477){\rule[-0.175pt]{0.824pt}{0.350pt}}
\put(853,478){\rule[-0.175pt]{0.824pt}{0.350pt}}
\put(856,479){\rule[-0.175pt]{0.824pt}{0.350pt}}
\put(860,480){\rule[-0.175pt]{0.824pt}{0.350pt}}
\put(863,481){\rule[-0.175pt]{0.824pt}{0.350pt}}
\put(867,482){\rule[-0.175pt]{0.824pt}{0.350pt}}
\put(870,483){\rule[-0.175pt]{0.824pt}{0.350pt}}
\put(873,484){\rule[-0.175pt]{0.824pt}{0.350pt}}
\put(877,485){\rule[-0.175pt]{0.824pt}{0.350pt}}
\put(880,486){\rule[-0.175pt]{0.824pt}{0.350pt}}
\put(884,487){\rule[-0.175pt]{0.824pt}{0.350pt}}
\put(887,488){\rule[-0.175pt]{0.824pt}{0.350pt}}
\put(891,489){\rule[-0.175pt]{0.824pt}{0.350pt}}
\put(894,490){\rule[-0.175pt]{0.824pt}{0.350pt}}
\put(897,491){\rule[-0.175pt]{0.824pt}{0.350pt}}
\put(901,492){\rule[-0.175pt]{0.824pt}{0.350pt}}
\put(904,493){\rule[-0.175pt]{0.824pt}{0.350pt}}
\put(908,494){\rule[-0.175pt]{0.824pt}{0.350pt}}
\put(911,495){\rule[-0.175pt]{0.824pt}{0.350pt}}
\put(915,496){\rule[-0.175pt]{0.746pt}{0.350pt}}
\put(918,497){\rule[-0.175pt]{0.746pt}{0.350pt}}
\put(921,498){\rule[-0.175pt]{0.746pt}{0.350pt}}
\put(924,499){\rule[-0.175pt]{0.746pt}{0.350pt}}
\put(927,500){\rule[-0.175pt]{0.746pt}{0.350pt}}
\put(930,501){\rule[-0.175pt]{0.746pt}{0.350pt}}
\put(933,502){\rule[-0.175pt]{0.746pt}{0.350pt}}
\put(936,503){\rule[-0.175pt]{0.746pt}{0.350pt}}
\put(939,504){\rule[-0.175pt]{0.746pt}{0.350pt}}
\put(942,505){\rule[-0.175pt]{0.746pt}{0.350pt}}
\put(945,506){\rule[-0.175pt]{0.746pt}{0.350pt}}
\put(949,507){\rule[-0.175pt]{0.746pt}{0.350pt}}
\put(952,508){\rule[-0.175pt]{0.746pt}{0.350pt}}
\put(955,509){\rule[-0.175pt]{0.746pt}{0.350pt}}
\put(958,510){\rule[-0.175pt]{0.746pt}{0.350pt}}
\put(961,511){\rule[-0.175pt]{0.746pt}{0.350pt}}
\put(964,512){\rule[-0.175pt]{0.746pt}{0.350pt}}
\put(967,513){\rule[-0.175pt]{0.746pt}{0.350pt}}
\put(970,514){\rule[-0.175pt]{0.746pt}{0.350pt}}
\put(973,515){\rule[-0.175pt]{0.746pt}{0.350pt}}
\put(976,516){\rule[-0.175pt]{0.746pt}{0.350pt}}
\put(979,517){\rule[-0.175pt]{0.746pt}{0.350pt}}
\put(983,518){\rule[-0.175pt]{0.746pt}{0.350pt}}
\put(986,519){\rule[-0.175pt]{0.746pt}{0.350pt}}
\put(989,520){\rule[-0.175pt]{0.746pt}{0.350pt}}
\put(992,521){\rule[-0.175pt]{0.746pt}{0.350pt}}
\put(995,522){\rule[-0.175pt]{0.746pt}{0.350pt}}
\put(998,523){\rule[-0.175pt]{0.746pt}{0.350pt}}
\put(1001,524){\rule[-0.175pt]{0.746pt}{0.350pt}}
\put(1004,525){\rule[-0.175pt]{0.746pt}{0.350pt}}
\put(1007,526){\rule[-0.175pt]{0.746pt}{0.350pt}}
\put(1010,527){\rule[-0.175pt]{0.746pt}{0.350pt}}
\put(1014,528){\rule[-0.175pt]{0.746pt}{0.350pt}}
\put(1017,529){\rule[-0.175pt]{0.746pt}{0.350pt}}
\put(1020,530){\rule[-0.175pt]{0.746pt}{0.350pt}}
\put(1023,531){\rule[-0.175pt]{0.746pt}{0.350pt}}
\put(1026,532){\rule[-0.175pt]{0.746pt}{0.350pt}}
\put(1029,533){\rule[-0.175pt]{0.746pt}{0.350pt}}
\put(1032,534){\rule[-0.175pt]{0.746pt}{0.350pt}}
\put(1035,535){\rule[-0.175pt]{0.746pt}{0.350pt}}
\put(1038,536){\rule[-0.175pt]{0.746pt}{0.350pt}}
\put(1041,537){\rule[-0.175pt]{0.746pt}{0.350pt}}
\put(1044,538){\rule[-0.175pt]{0.789pt}{0.350pt}}
\put(1048,539){\rule[-0.175pt]{0.789pt}{0.350pt}}
\put(1051,540){\rule[-0.175pt]{0.789pt}{0.350pt}}
\put(1054,541){\rule[-0.175pt]{0.789pt}{0.350pt}}
\put(1058,542){\rule[-0.175pt]{0.789pt}{0.350pt}}
\put(1061,543){\rule[-0.175pt]{0.789pt}{0.350pt}}
\put(1064,544){\rule[-0.175pt]{0.789pt}{0.350pt}}
\put(1067,545){\rule[-0.175pt]{0.789pt}{0.350pt}}
\put(1071,546){\rule[-0.175pt]{0.789pt}{0.350pt}}
\put(1074,547){\rule[-0.175pt]{0.789pt}{0.350pt}}
\put(1077,548){\rule[-0.175pt]{0.789pt}{0.350pt}}
\put(1081,549){\rule[-0.175pt]{0.789pt}{0.350pt}}
\put(1084,550){\rule[-0.175pt]{0.789pt}{0.350pt}}
\put(1087,551){\rule[-0.175pt]{0.789pt}{0.350pt}}
\put(1090,552){\rule[-0.175pt]{0.789pt}{0.350pt}}
\put(1094,553){\rule[-0.175pt]{0.789pt}{0.350pt}}
\put(1097,554){\rule[-0.175pt]{0.789pt}{0.350pt}}
\put(1100,555){\rule[-0.175pt]{0.789pt}{0.350pt}}
\put(1103,556){\rule[-0.175pt]{0.789pt}{0.350pt}}
\put(1107,557){\rule[-0.175pt]{0.789pt}{0.350pt}}
\put(1110,558){\rule[-0.175pt]{0.789pt}{0.350pt}}
\put(1113,559){\rule[-0.175pt]{0.789pt}{0.350pt}}
\put(1117,560){\rule[-0.175pt]{0.789pt}{0.350pt}}
\put(1120,561){\rule[-0.175pt]{0.789pt}{0.350pt}}
\put(1123,562){\rule[-0.175pt]{0.789pt}{0.350pt}}
\put(1126,563){\rule[-0.175pt]{0.789pt}{0.350pt}}
\put(1130,564){\rule[-0.175pt]{0.789pt}{0.350pt}}
\put(1133,565){\rule[-0.175pt]{0.789pt}{0.350pt}}
\put(1136,566){\rule[-0.175pt]{0.789pt}{0.350pt}}
\put(1139,567){\rule[-0.175pt]{0.789pt}{0.350pt}}
\put(1143,568){\rule[-0.175pt]{0.789pt}{0.350pt}}
\put(1146,569){\rule[-0.175pt]{0.789pt}{0.350pt}}
\put(1149,570){\rule[-0.175pt]{0.789pt}{0.350pt}}
\put(1153,571){\rule[-0.175pt]{0.789pt}{0.350pt}}
\put(1156,572){\rule[-0.175pt]{0.789pt}{0.350pt}}
\put(1159,573){\rule[-0.175pt]{0.789pt}{0.350pt}}
\put(1162,574){\rule[-0.175pt]{0.789pt}{0.350pt}}
\put(1166,575){\rule[-0.175pt]{0.789pt}{0.350pt}}
\put(1169,576){\rule[-0.175pt]{0.789pt}{0.350pt}}
\put(1172,577){\rule[-0.175pt]{0.789pt}{0.350pt}}
\put(1176,578){\rule[-0.175pt]{1.010pt}{0.350pt}}
\put(1180,579){\rule[-0.175pt]{1.010pt}{0.350pt}}
\put(1184,580){\rule[-0.175pt]{1.010pt}{0.350pt}}
\put(1188,581){\rule[-0.175pt]{1.010pt}{0.350pt}}
\put(1192,582){\rule[-0.175pt]{1.010pt}{0.350pt}}
\put(1196,583){\rule[-0.175pt]{1.010pt}{0.350pt}}
\put(1201,584){\rule[-0.175pt]{1.010pt}{0.350pt}}
\put(1205,585){\rule[-0.175pt]{1.010pt}{0.350pt}}
\put(1209,586){\rule[-0.175pt]{1.010pt}{0.350pt}}
\put(1213,587){\rule[-0.175pt]{1.010pt}{0.350pt}}
\put(1217,588){\rule[-0.175pt]{1.010pt}{0.350pt}}
\put(1222,589){\rule[-0.175pt]{1.010pt}{0.350pt}}
\put(1226,590){\rule[-0.175pt]{1.010pt}{0.350pt}}
\put(1230,591){\rule[-0.175pt]{1.010pt}{0.350pt}}
\put(1234,592){\rule[-0.175pt]{1.010pt}{0.350pt}}
\put(1238,593){\rule[-0.175pt]{1.010pt}{0.350pt}}
\put(1243,594){\rule[-0.175pt]{1.010pt}{0.350pt}}
\put(1247,595){\rule[-0.175pt]{1.010pt}{0.350pt}}
\put(1251,596){\rule[-0.175pt]{1.010pt}{0.350pt}}
\put(1255,597){\rule[-0.175pt]{1.010pt}{0.350pt}}
\put(1259,598){\rule[-0.175pt]{1.010pt}{0.350pt}}
\put(1264,599){\rule[-0.175pt]{1.010pt}{0.350pt}}
\put(1268,600){\rule[-0.175pt]{1.010pt}{0.350pt}}
\put(1272,601){\rule[-0.175pt]{1.010pt}{0.350pt}}
\put(1276,602){\rule[-0.175pt]{1.010pt}{0.350pt}}
\put(1280,603){\rule[-0.175pt]{1.010pt}{0.350pt}}
\put(1285,604){\rule[-0.175pt]{1.010pt}{0.350pt}}
\put(1289,605){\rule[-0.175pt]{1.010pt}{0.350pt}}
\put(1293,606){\rule[-0.175pt]{1.010pt}{0.350pt}}
\put(1297,607){\rule[-0.175pt]{1.010pt}{0.350pt}}
\put(471,708){\circle{12}}
\put(394,290){\circle{12}}
\put(524,339){\circle{12}}
\put(655,403){\circle{12}}
\put(785,458){\circle{12}}
\put(915,496){\circle{12}}
\put(1045,538){\circle{12}}
\put(1176,578){\circle{12}}
\put(1306,609){\circle{12}}
\put(1301,608){\rule[-0.175pt]{1.010pt}{0.350pt}}
\put(427,663){\makebox(0,0)[r]{Ne$_3$}}
\put(449,663){\rule[-0.175pt]{15.899pt}{0.350pt}}
\put(394,237){\usebox{\plotpoint}}
\put(394,237){\rule[-0.175pt]{0.921pt}{0.350pt}}
\put(397,238){\rule[-0.175pt]{0.921pt}{0.350pt}}
\put(401,239){\rule[-0.175pt]{0.921pt}{0.350pt}}
\put(405,240){\rule[-0.175pt]{0.921pt}{0.350pt}}
\put(409,241){\rule[-0.175pt]{0.921pt}{0.350pt}}
\put(413,242){\rule[-0.175pt]{0.921pt}{0.350pt}}
\put(416,243){\rule[-0.175pt]{0.921pt}{0.350pt}}
\put(420,244){\rule[-0.175pt]{0.921pt}{0.350pt}}
\put(424,245){\rule[-0.175pt]{0.921pt}{0.350pt}}
\put(428,246){\rule[-0.175pt]{0.921pt}{0.350pt}}
\put(432,247){\rule[-0.175pt]{0.921pt}{0.350pt}}
\put(436,248){\rule[-0.175pt]{0.921pt}{0.350pt}}
\put(439,249){\rule[-0.175pt]{0.921pt}{0.350pt}}
\put(443,250){\rule[-0.175pt]{0.921pt}{0.350pt}}
\put(447,251){\rule[-0.175pt]{0.921pt}{0.350pt}}
\put(451,252){\rule[-0.175pt]{0.921pt}{0.350pt}}
\put(455,253){\rule[-0.175pt]{0.921pt}{0.350pt}}
\put(458,254){\rule[-0.175pt]{0.921pt}{0.350pt}}
\put(462,255){\rule[-0.175pt]{0.921pt}{0.350pt}}
\put(466,256){\rule[-0.175pt]{0.921pt}{0.350pt}}
\put(470,257){\rule[-0.175pt]{0.921pt}{0.350pt}}
\put(474,258){\rule[-0.175pt]{0.921pt}{0.350pt}}
\put(478,259){\rule[-0.175pt]{0.921pt}{0.350pt}}
\put(481,260){\rule[-0.175pt]{0.921pt}{0.350pt}}
\put(485,261){\rule[-0.175pt]{0.921pt}{0.350pt}}
\put(489,262){\rule[-0.175pt]{0.921pt}{0.350pt}}
\put(493,263){\rule[-0.175pt]{0.921pt}{0.350pt}}
\put(497,264){\rule[-0.175pt]{0.921pt}{0.350pt}}
\put(501,265){\rule[-0.175pt]{0.921pt}{0.350pt}}
\put(504,266){\rule[-0.175pt]{0.921pt}{0.350pt}}
\put(508,267){\rule[-0.175pt]{0.921pt}{0.350pt}}
\put(512,268){\rule[-0.175pt]{0.921pt}{0.350pt}}
\put(516,269){\rule[-0.175pt]{0.921pt}{0.350pt}}
\put(520,270){\rule[-0.175pt]{0.921pt}{0.350pt}}
\put(523,271){\rule[-0.175pt]{0.734pt}{0.350pt}}
\put(527,272){\rule[-0.175pt]{0.734pt}{0.350pt}}
\put(530,273){\rule[-0.175pt]{0.734pt}{0.350pt}}
\put(533,274){\rule[-0.175pt]{0.734pt}{0.350pt}}
\put(536,275){\rule[-0.175pt]{0.734pt}{0.350pt}}
\put(539,276){\rule[-0.175pt]{0.734pt}{0.350pt}}
\put(542,277){\rule[-0.175pt]{0.734pt}{0.350pt}}
\put(545,278){\rule[-0.175pt]{0.734pt}{0.350pt}}
\put(548,279){\rule[-0.175pt]{0.734pt}{0.350pt}}
\put(551,280){\rule[-0.175pt]{0.734pt}{0.350pt}}
\put(554,281){\rule[-0.175pt]{0.734pt}{0.350pt}}
\put(557,282){\rule[-0.175pt]{0.734pt}{0.350pt}}
\put(560,283){\rule[-0.175pt]{0.734pt}{0.350pt}}
\put(563,284){\rule[-0.175pt]{0.734pt}{0.350pt}}
\put(566,285){\rule[-0.175pt]{0.734pt}{0.350pt}}
\put(569,286){\rule[-0.175pt]{0.734pt}{0.350pt}}
\put(572,287){\rule[-0.175pt]{0.734pt}{0.350pt}}
\put(575,288){\rule[-0.175pt]{0.734pt}{0.350pt}}
\put(578,289){\rule[-0.175pt]{0.734pt}{0.350pt}}
\put(581,290){\rule[-0.175pt]{0.734pt}{0.350pt}}
\put(584,291){\rule[-0.175pt]{0.734pt}{0.350pt}}
\put(587,292){\rule[-0.175pt]{0.734pt}{0.350pt}}
\put(591,293){\rule[-0.175pt]{0.734pt}{0.350pt}}
\put(594,294){\rule[-0.175pt]{0.734pt}{0.350pt}}
\put(597,295){\rule[-0.175pt]{0.734pt}{0.350pt}}
\put(600,296){\rule[-0.175pt]{0.734pt}{0.350pt}}
\put(603,297){\rule[-0.175pt]{0.734pt}{0.350pt}}
\put(606,298){\rule[-0.175pt]{0.734pt}{0.350pt}}
\put(609,299){\rule[-0.175pt]{0.734pt}{0.350pt}}
\put(612,300){\rule[-0.175pt]{0.734pt}{0.350pt}}
\put(615,301){\rule[-0.175pt]{0.734pt}{0.350pt}}
\put(618,302){\rule[-0.175pt]{0.734pt}{0.350pt}}
\put(621,303){\rule[-0.175pt]{0.734pt}{0.350pt}}
\put(624,304){\rule[-0.175pt]{0.734pt}{0.350pt}}
\put(627,305){\rule[-0.175pt]{0.734pt}{0.350pt}}
\put(630,306){\rule[-0.175pt]{0.734pt}{0.350pt}}
\put(633,307){\rule[-0.175pt]{0.734pt}{0.350pt}}
\put(636,308){\rule[-0.175pt]{0.734pt}{0.350pt}}
\put(639,309){\rule[-0.175pt]{0.734pt}{0.350pt}}
\put(642,310){\rule[-0.175pt]{0.734pt}{0.350pt}}
\put(645,311){\rule[-0.175pt]{0.734pt}{0.350pt}}
\put(648,312){\rule[-0.175pt]{0.734pt}{0.350pt}}
\put(651,313){\rule[-0.175pt]{0.734pt}{0.350pt}}
\put(654,314){\rule[-0.175pt]{0.949pt}{0.350pt}}
\put(658,315){\rule[-0.175pt]{0.949pt}{0.350pt}}
\put(662,316){\rule[-0.175pt]{0.949pt}{0.350pt}}
\put(666,317){\rule[-0.175pt]{0.949pt}{0.350pt}}
\put(670,318){\rule[-0.175pt]{0.949pt}{0.350pt}}
\put(674,319){\rule[-0.175pt]{0.949pt}{0.350pt}}
\put(678,320){\rule[-0.175pt]{0.949pt}{0.350pt}}
\put(682,321){\rule[-0.175pt]{0.949pt}{0.350pt}}
\put(686,322){\rule[-0.175pt]{0.949pt}{0.350pt}}
\put(690,323){\rule[-0.175pt]{0.949pt}{0.350pt}}
\put(694,324){\rule[-0.175pt]{0.949pt}{0.350pt}}
\put(698,325){\rule[-0.175pt]{0.949pt}{0.350pt}}
\put(702,326){\rule[-0.175pt]{0.949pt}{0.350pt}}
\put(706,327){\rule[-0.175pt]{0.949pt}{0.350pt}}
\put(710,328){\rule[-0.175pt]{0.949pt}{0.350pt}}
\put(714,329){\rule[-0.175pt]{0.949pt}{0.350pt}}
\put(718,330){\rule[-0.175pt]{0.949pt}{0.350pt}}
\put(721,331){\rule[-0.175pt]{0.949pt}{0.350pt}}
\put(725,332){\rule[-0.175pt]{0.949pt}{0.350pt}}
\put(729,333){\rule[-0.175pt]{0.949pt}{0.350pt}}
\put(733,334){\rule[-0.175pt]{0.949pt}{0.350pt}}
\put(737,335){\rule[-0.175pt]{0.949pt}{0.350pt}}
\put(741,336){\rule[-0.175pt]{0.949pt}{0.350pt}}
\put(745,337){\rule[-0.175pt]{0.949pt}{0.350pt}}
\put(749,338){\rule[-0.175pt]{0.949pt}{0.350pt}}
\put(753,339){\rule[-0.175pt]{0.949pt}{0.350pt}}
\put(757,340){\rule[-0.175pt]{0.949pt}{0.350pt}}
\put(761,341){\rule[-0.175pt]{0.949pt}{0.350pt}}
\put(765,342){\rule[-0.175pt]{0.949pt}{0.350pt}}
\put(769,343){\rule[-0.175pt]{0.949pt}{0.350pt}}
\put(773,344){\rule[-0.175pt]{0.949pt}{0.350pt}}
\put(777,345){\rule[-0.175pt]{0.949pt}{0.350pt}}
\put(781,346){\rule[-0.175pt]{0.949pt}{0.350pt}}
\put(784,347){\rule[-0.175pt]{1.253pt}{0.350pt}}
\put(790,348){\rule[-0.175pt]{1.253pt}{0.350pt}}
\put(795,349){\rule[-0.175pt]{1.253pt}{0.350pt}}
\put(800,350){\rule[-0.175pt]{1.253pt}{0.350pt}}
\put(805,351){\rule[-0.175pt]{1.253pt}{0.350pt}}
\put(811,352){\rule[-0.175pt]{1.253pt}{0.350pt}}
\put(816,353){\rule[-0.175pt]{1.253pt}{0.350pt}}
\put(821,354){\rule[-0.175pt]{1.253pt}{0.350pt}}
\put(826,355){\rule[-0.175pt]{1.253pt}{0.350pt}}
\put(831,356){\rule[-0.175pt]{1.253pt}{0.350pt}}
\put(837,357){\rule[-0.175pt]{1.253pt}{0.350pt}}
\put(842,358){\rule[-0.175pt]{1.253pt}{0.350pt}}
\put(847,359){\rule[-0.175pt]{1.253pt}{0.350pt}}
\put(852,360){\rule[-0.175pt]{1.253pt}{0.350pt}}
\put(857,361){\rule[-0.175pt]{1.253pt}{0.350pt}}
\put(863,362){\rule[-0.175pt]{1.253pt}{0.350pt}}
\put(868,363){\rule[-0.175pt]{1.253pt}{0.350pt}}
\put(873,364){\rule[-0.175pt]{1.253pt}{0.350pt}}
\put(878,365){\rule[-0.175pt]{1.253pt}{0.350pt}}
\put(883,366){\rule[-0.175pt]{1.253pt}{0.350pt}}
\put(889,367){\rule[-0.175pt]{1.253pt}{0.350pt}}
\put(894,368){\rule[-0.175pt]{1.253pt}{0.350pt}}
\put(899,369){\rule[-0.175pt]{1.253pt}{0.350pt}}
\put(904,370){\rule[-0.175pt]{1.253pt}{0.350pt}}
\put(909,371){\rule[-0.175pt]{1.253pt}{0.350pt}}
\put(915,372){\rule[-0.175pt]{1.084pt}{0.350pt}}
\put(919,373){\rule[-0.175pt]{1.084pt}{0.350pt}}
\put(924,374){\rule[-0.175pt]{1.084pt}{0.350pt}}
\put(928,375){\rule[-0.175pt]{1.084pt}{0.350pt}}
\put(933,376){\rule[-0.175pt]{1.084pt}{0.350pt}}
\put(937,377){\rule[-0.175pt]{1.084pt}{0.350pt}}
\put(942,378){\rule[-0.175pt]{1.084pt}{0.350pt}}
\put(946,379){\rule[-0.175pt]{1.084pt}{0.350pt}}
\put(951,380){\rule[-0.175pt]{1.084pt}{0.350pt}}
\put(955,381){\rule[-0.175pt]{1.084pt}{0.350pt}}
\put(960,382){\rule[-0.175pt]{1.084pt}{0.350pt}}
\put(964,383){\rule[-0.175pt]{1.084pt}{0.350pt}}
\put(969,384){\rule[-0.175pt]{1.084pt}{0.350pt}}
\put(973,385){\rule[-0.175pt]{1.084pt}{0.350pt}}
\put(978,386){\rule[-0.175pt]{1.084pt}{0.350pt}}
\put(982,387){\rule[-0.175pt]{1.084pt}{0.350pt}}
\put(987,388){\rule[-0.175pt]{1.084pt}{0.350pt}}
\put(991,389){\rule[-0.175pt]{1.084pt}{0.350pt}}
\put(996,390){\rule[-0.175pt]{1.084pt}{0.350pt}}
\put(1000,391){\rule[-0.175pt]{1.084pt}{0.350pt}}
\put(1005,392){\rule[-0.175pt]{1.084pt}{0.350pt}}
\put(1009,393){\rule[-0.175pt]{1.084pt}{0.350pt}}
\put(1014,394){\rule[-0.175pt]{1.084pt}{0.350pt}}
\put(1018,395){\rule[-0.175pt]{1.084pt}{0.350pt}}
\put(1023,396){\rule[-0.175pt]{1.084pt}{0.350pt}}
\put(1027,397){\rule[-0.175pt]{1.084pt}{0.350pt}}
\put(1032,398){\rule[-0.175pt]{1.084pt}{0.350pt}}
\put(1036,399){\rule[-0.175pt]{1.084pt}{0.350pt}}
\put(1041,400){\rule[-0.175pt]{1.084pt}{0.350pt}}
\put(1045,401){\rule[-0.175pt]{1.084pt}{0.350pt}}
\put(1050,402){\rule[-0.175pt]{1.084pt}{0.350pt}}
\put(1054,403){\rule[-0.175pt]{1.084pt}{0.350pt}}
\put(1059,404){\rule[-0.175pt]{1.084pt}{0.350pt}}
\put(1063,405){\rule[-0.175pt]{1.084pt}{0.350pt}}
\put(1068,406){\rule[-0.175pt]{1.084pt}{0.350pt}}
\put(1072,407){\rule[-0.175pt]{1.084pt}{0.350pt}}
\put(1077,408){\rule[-0.175pt]{1.084pt}{0.350pt}}
\put(1081,409){\rule[-0.175pt]{1.084pt}{0.350pt}}
\put(1086,410){\rule[-0.175pt]{1.084pt}{0.350pt}}
\put(1090,411){\rule[-0.175pt]{1.084pt}{0.350pt}}
\put(1095,412){\rule[-0.175pt]{1.084pt}{0.350pt}}
\put(1099,413){\rule[-0.175pt]{1.084pt}{0.350pt}}
\put(1104,414){\rule[-0.175pt]{1.084pt}{0.350pt}}
\put(1108,415){\rule[-0.175pt]{1.084pt}{0.350pt}}
\put(1113,416){\rule[-0.175pt]{1.084pt}{0.350pt}}
\put(1117,417){\rule[-0.175pt]{1.084pt}{0.350pt}}
\put(1122,418){\rule[-0.175pt]{1.084pt}{0.350pt}}
\put(1126,419){\rule[-0.175pt]{1.084pt}{0.350pt}}
\put(1131,420){\rule[-0.175pt]{1.084pt}{0.350pt}}
\put(1135,421){\rule[-0.175pt]{1.084pt}{0.350pt}}
\put(1140,422){\rule[-0.175pt]{1.084pt}{0.350pt}}
\put(1144,423){\rule[-0.175pt]{1.084pt}{0.350pt}}
\put(1149,424){\rule[-0.175pt]{1.084pt}{0.350pt}}
\put(1153,425){\rule[-0.175pt]{1.084pt}{0.350pt}}
\put(1158,426){\rule[-0.175pt]{1.084pt}{0.350pt}}
\put(1162,427){\rule[-0.175pt]{1.084pt}{0.350pt}}
\put(1167,428){\rule[-0.175pt]{1.084pt}{0.350pt}}
\put(1171,429){\rule[-0.175pt]{1.084pt}{0.350pt}}
\put(1176,430){\rule[-0.175pt]{1.253pt}{0.350pt}}
\put(1181,431){\rule[-0.175pt]{1.253pt}{0.350pt}}
\put(1186,432){\rule[-0.175pt]{1.253pt}{0.350pt}}
\put(1191,433){\rule[-0.175pt]{1.253pt}{0.350pt}}
\put(1196,434){\rule[-0.175pt]{1.253pt}{0.350pt}}
\put(1201,435){\rule[-0.175pt]{1.253pt}{0.350pt}}
\put(1207,436){\rule[-0.175pt]{1.253pt}{0.350pt}}
\put(1212,437){\rule[-0.175pt]{1.253pt}{0.350pt}}
\put(1217,438){\rule[-0.175pt]{1.253pt}{0.350pt}}
\put(1222,439){\rule[-0.175pt]{1.253pt}{0.350pt}}
\put(1227,440){\rule[-0.175pt]{1.253pt}{0.350pt}}
\put(1233,441){\rule[-0.175pt]{1.253pt}{0.350pt}}
\put(1238,442){\rule[-0.175pt]{1.253pt}{0.350pt}}
\put(1243,443){\rule[-0.175pt]{1.253pt}{0.350pt}}
\put(1248,444){\rule[-0.175pt]{1.253pt}{0.350pt}}
\put(1253,445){\rule[-0.175pt]{1.253pt}{0.350pt}}
\put(1259,446){\rule[-0.175pt]{1.253pt}{0.350pt}}
\put(1264,447){\rule[-0.175pt]{1.253pt}{0.350pt}}
\put(1269,448){\rule[-0.175pt]{1.253pt}{0.350pt}}
\put(1274,449){\rule[-0.175pt]{1.253pt}{0.350pt}}
\put(1279,450){\rule[-0.175pt]{1.253pt}{0.350pt}}
\put(1285,451){\rule[-0.175pt]{1.253pt}{0.350pt}}
\put(1290,452){\rule[-0.175pt]{1.253pt}{0.350pt}}
\put(1295,453){\rule[-0.175pt]{1.253pt}{0.350pt}}
\put(1300,454){\rule[-0.175pt]{1.253pt}{0.350pt}}
\put(471,663){\circle{24}}
\put(394,237){\circle{24}}
\put(524,271){\circle{24}}
\put(655,314){\circle{24}}
\put(785,347){\circle{24}}
\put(915,372){\circle{24}}
\put(1176,430){\circle{24}}
\put(1306,455){\circle{24}}
\put(1305,455){\usebox{\plotpoint}}
\put(427,618){\makebox(0,0)[r]{Ne$_4$}}
\put(449,618){\rule[-0.175pt]{15.899pt}{0.350pt}}
\put(394,238){\usebox{\plotpoint}}
\put(394,238){\rule[-0.175pt]{1.080pt}{0.350pt}}
\put(398,239){\rule[-0.175pt]{1.080pt}{0.350pt}}
\put(402,240){\rule[-0.175pt]{1.080pt}{0.350pt}}
\put(407,241){\rule[-0.175pt]{1.080pt}{0.350pt}}
\put(411,242){\rule[-0.175pt]{1.080pt}{0.350pt}}
\put(416,243){\rule[-0.175pt]{1.080pt}{0.350pt}}
\put(420,244){\rule[-0.175pt]{1.080pt}{0.350pt}}
\put(425,245){\rule[-0.175pt]{1.080pt}{0.350pt}}
\put(429,246){\rule[-0.175pt]{1.080pt}{0.350pt}}
\put(434,247){\rule[-0.175pt]{1.080pt}{0.350pt}}
\put(438,248){\rule[-0.175pt]{1.080pt}{0.350pt}}
\put(443,249){\rule[-0.175pt]{1.080pt}{0.350pt}}
\put(447,250){\rule[-0.175pt]{1.080pt}{0.350pt}}
\put(452,251){\rule[-0.175pt]{1.080pt}{0.350pt}}
\put(456,252){\rule[-0.175pt]{1.080pt}{0.350pt}}
\put(461,253){\rule[-0.175pt]{1.080pt}{0.350pt}}
\put(465,254){\rule[-0.175pt]{1.080pt}{0.350pt}}
\put(470,255){\rule[-0.175pt]{1.080pt}{0.350pt}}
\put(474,256){\rule[-0.175pt]{1.080pt}{0.350pt}}
\put(479,257){\rule[-0.175pt]{1.080pt}{0.350pt}}
\put(483,258){\rule[-0.175pt]{1.080pt}{0.350pt}}
\put(488,259){\rule[-0.175pt]{1.080pt}{0.350pt}}
\put(492,260){\rule[-0.175pt]{1.080pt}{0.350pt}}
\put(497,261){\rule[-0.175pt]{1.080pt}{0.350pt}}
\put(501,262){\rule[-0.175pt]{1.080pt}{0.350pt}}
\put(506,263){\rule[-0.175pt]{1.080pt}{0.350pt}}
\put(510,264){\rule[-0.175pt]{1.080pt}{0.350pt}}
\put(515,265){\rule[-0.175pt]{1.080pt}{0.350pt}}
\put(519,266){\rule[-0.175pt]{1.080pt}{0.350pt}}
\put(524,267){\rule[-0.175pt]{0.701pt}{0.350pt}}
\put(526,268){\rule[-0.175pt]{0.701pt}{0.350pt}}
\put(529,269){\rule[-0.175pt]{0.701pt}{0.350pt}}
\put(532,270){\rule[-0.175pt]{0.701pt}{0.350pt}}
\put(535,271){\rule[-0.175pt]{0.701pt}{0.350pt}}
\put(538,272){\rule[-0.175pt]{0.701pt}{0.350pt}}
\put(541,273){\rule[-0.175pt]{0.701pt}{0.350pt}}
\put(544,274){\rule[-0.175pt]{0.701pt}{0.350pt}}
\put(547,275){\rule[-0.175pt]{0.701pt}{0.350pt}}
\put(550,276){\rule[-0.175pt]{0.701pt}{0.350pt}}
\put(553,277){\rule[-0.175pt]{0.701pt}{0.350pt}}
\put(556,278){\rule[-0.175pt]{0.701pt}{0.350pt}}
\put(558,279){\rule[-0.175pt]{0.701pt}{0.350pt}}
\put(561,280){\rule[-0.175pt]{0.701pt}{0.350pt}}
\put(564,281){\rule[-0.175pt]{0.701pt}{0.350pt}}
\put(567,282){\rule[-0.175pt]{0.701pt}{0.350pt}}
\put(570,283){\rule[-0.175pt]{0.701pt}{0.350pt}}
\put(573,284){\rule[-0.175pt]{0.701pt}{0.350pt}}
\put(576,285){\rule[-0.175pt]{0.701pt}{0.350pt}}
\put(579,286){\rule[-0.175pt]{0.701pt}{0.350pt}}
\put(582,287){\rule[-0.175pt]{0.701pt}{0.350pt}}
\put(585,288){\rule[-0.175pt]{0.701pt}{0.350pt}}
\put(588,289){\rule[-0.175pt]{0.701pt}{0.350pt}}
\put(590,290){\rule[-0.175pt]{0.701pt}{0.350pt}}
\put(593,291){\rule[-0.175pt]{0.701pt}{0.350pt}}
\put(596,292){\rule[-0.175pt]{0.701pt}{0.350pt}}
\put(599,293){\rule[-0.175pt]{0.701pt}{0.350pt}}
\put(602,294){\rule[-0.175pt]{0.701pt}{0.350pt}}
\put(605,295){\rule[-0.175pt]{0.701pt}{0.350pt}}
\put(608,296){\rule[-0.175pt]{0.701pt}{0.350pt}}
\put(611,297){\rule[-0.175pt]{0.701pt}{0.350pt}}
\put(614,298){\rule[-0.175pt]{0.701pt}{0.350pt}}
\put(617,299){\rule[-0.175pt]{0.701pt}{0.350pt}}
\put(620,300){\rule[-0.175pt]{0.701pt}{0.350pt}}
\put(622,301){\rule[-0.175pt]{0.701pt}{0.350pt}}
\put(625,302){\rule[-0.175pt]{0.701pt}{0.350pt}}
\put(628,303){\rule[-0.175pt]{0.701pt}{0.350pt}}
\put(631,304){\rule[-0.175pt]{0.701pt}{0.350pt}}
\put(634,305){\rule[-0.175pt]{0.701pt}{0.350pt}}
\put(637,306){\rule[-0.175pt]{0.701pt}{0.350pt}}
\put(640,307){\rule[-0.175pt]{0.701pt}{0.350pt}}
\put(643,308){\rule[-0.175pt]{0.701pt}{0.350pt}}
\put(646,309){\rule[-0.175pt]{0.701pt}{0.350pt}}
\put(649,310){\rule[-0.175pt]{0.701pt}{0.350pt}}
\put(652,311){\rule[-0.175pt]{0.701pt}{0.350pt}}
\put(655,312){\rule[-0.175pt]{1.044pt}{0.350pt}}
\put(659,313){\rule[-0.175pt]{1.044pt}{0.350pt}}
\put(663,314){\rule[-0.175pt]{1.044pt}{0.350pt}}
\put(667,315){\rule[-0.175pt]{1.044pt}{0.350pt}}
\put(672,316){\rule[-0.175pt]{1.044pt}{0.350pt}}
\put(676,317){\rule[-0.175pt]{1.044pt}{0.350pt}}
\put(680,318){\rule[-0.175pt]{1.044pt}{0.350pt}}
\put(685,319){\rule[-0.175pt]{1.044pt}{0.350pt}}
\put(689,320){\rule[-0.175pt]{1.044pt}{0.350pt}}
\put(693,321){\rule[-0.175pt]{1.044pt}{0.350pt}}
\put(698,322){\rule[-0.175pt]{1.044pt}{0.350pt}}
\put(702,323){\rule[-0.175pt]{1.044pt}{0.350pt}}
\put(706,324){\rule[-0.175pt]{1.044pt}{0.350pt}}
\put(711,325){\rule[-0.175pt]{1.044pt}{0.350pt}}
\put(715,326){\rule[-0.175pt]{1.044pt}{0.350pt}}
\put(719,327){\rule[-0.175pt]{1.044pt}{0.350pt}}
\put(724,328){\rule[-0.175pt]{1.044pt}{0.350pt}}
\put(728,329){\rule[-0.175pt]{1.044pt}{0.350pt}}
\put(732,330){\rule[-0.175pt]{1.044pt}{0.350pt}}
\put(737,331){\rule[-0.175pt]{1.044pt}{0.350pt}}
\put(741,332){\rule[-0.175pt]{1.044pt}{0.350pt}}
\put(745,333){\rule[-0.175pt]{1.044pt}{0.350pt}}
\put(750,334){\rule[-0.175pt]{1.044pt}{0.350pt}}
\put(754,335){\rule[-0.175pt]{1.044pt}{0.350pt}}
\put(758,336){\rule[-0.175pt]{1.044pt}{0.350pt}}
\put(763,337){\rule[-0.175pt]{1.044pt}{0.350pt}}
\put(767,338){\rule[-0.175pt]{1.044pt}{0.350pt}}
\put(771,339){\rule[-0.175pt]{1.044pt}{0.350pt}}
\put(776,340){\rule[-0.175pt]{1.044pt}{0.350pt}}
\put(780,341){\rule[-0.175pt]{1.044pt}{0.350pt}}
\put(784,342){\rule[-0.175pt]{1.491pt}{0.350pt}}
\put(791,343){\rule[-0.175pt]{1.491pt}{0.350pt}}
\put(797,344){\rule[-0.175pt]{1.491pt}{0.350pt}}
\put(803,345){\rule[-0.175pt]{1.491pt}{0.350pt}}
\put(809,346){\rule[-0.175pt]{1.491pt}{0.350pt}}
\put(815,347){\rule[-0.175pt]{1.491pt}{0.350pt}}
\put(822,348){\rule[-0.175pt]{1.491pt}{0.350pt}}
\put(828,349){\rule[-0.175pt]{1.491pt}{0.350pt}}
\put(834,350){\rule[-0.175pt]{1.491pt}{0.350pt}}
\put(840,351){\rule[-0.175pt]{1.491pt}{0.350pt}}
\put(846,352){\rule[-0.175pt]{1.491pt}{0.350pt}}
\put(853,353){\rule[-0.175pt]{1.491pt}{0.350pt}}
\put(859,354){\rule[-0.175pt]{1.491pt}{0.350pt}}
\put(865,355){\rule[-0.175pt]{1.491pt}{0.350pt}}
\put(871,356){\rule[-0.175pt]{1.491pt}{0.350pt}}
\put(877,357){\rule[-0.175pt]{1.491pt}{0.350pt}}
\put(884,358){\rule[-0.175pt]{1.491pt}{0.350pt}}
\put(890,359){\rule[-0.175pt]{1.491pt}{0.350pt}}
\put(896,360){\rule[-0.175pt]{1.491pt}{0.350pt}}
\put(902,361){\rule[-0.175pt]{1.491pt}{0.350pt}}
\put(471,618){\circle*{18}}
\put(394,238){\circle*{18}}
\put(524,267){\circle*{18}}
\put(655,312){\circle*{18}}
\put(785,342){\circle*{18}}
\put(915,363){\circle*{18}}
\put(908,362){\rule[-0.175pt]{1.491pt}{0.350pt}}
\put(427,573){\makebox(0,0)[r]{Ne$_5$}}
\put(449,573){\rule[-0.175pt]{15.899pt}{0.350pt}}
\put(394,233){\usebox{\plotpoint}}
\put(394,233){\rule[-0.175pt]{7.829pt}{0.350pt}}
\put(426,234){\rule[-0.175pt]{7.829pt}{0.350pt}}
\put(459,235){\rule[-0.175pt]{7.829pt}{0.350pt}}
\put(491,236){\rule[-0.175pt]{7.829pt}{0.350pt}}
\put(524,237){\rule[-0.175pt]{2.104pt}{0.350pt}}
\put(532,238){\rule[-0.175pt]{2.104pt}{0.350pt}}
\put(541,239){\rule[-0.175pt]{2.104pt}{0.350pt}}
\put(550,240){\rule[-0.175pt]{2.104pt}{0.350pt}}
\put(558,241){\rule[-0.175pt]{2.104pt}{0.350pt}}
\put(567,242){\rule[-0.175pt]{2.104pt}{0.350pt}}
\put(576,243){\rule[-0.175pt]{2.104pt}{0.350pt}}
\put(585,244){\rule[-0.175pt]{2.104pt}{0.350pt}}
\put(593,245){\rule[-0.175pt]{2.104pt}{0.350pt}}
\put(602,246){\rule[-0.175pt]{2.104pt}{0.350pt}}
\put(611,247){\rule[-0.175pt]{2.104pt}{0.350pt}}
\put(620,248){\rule[-0.175pt]{2.104pt}{0.350pt}}
\put(628,249){\rule[-0.175pt]{2.104pt}{0.350pt}}
\put(637,250){\rule[-0.175pt]{2.104pt}{0.350pt}}
\put(646,251){\rule[-0.175pt]{2.104pt}{0.350pt}}
\put(655,252){\rule[-0.175pt]{1.080pt}{0.350pt}}
\put(659,253){\rule[-0.175pt]{1.080pt}{0.350pt}}
\put(663,254){\rule[-0.175pt]{1.080pt}{0.350pt}}
\put(668,255){\rule[-0.175pt]{1.080pt}{0.350pt}}
\put(672,256){\rule[-0.175pt]{1.080pt}{0.350pt}}
\put(677,257){\rule[-0.175pt]{1.080pt}{0.350pt}}
\put(681,258){\rule[-0.175pt]{1.080pt}{0.350pt}}
\put(686,259){\rule[-0.175pt]{1.080pt}{0.350pt}}
\put(690,260){\rule[-0.175pt]{1.080pt}{0.350pt}}
\put(695,261){\rule[-0.175pt]{1.080pt}{0.350pt}}
\put(699,262){\rule[-0.175pt]{1.080pt}{0.350pt}}
\put(704,263){\rule[-0.175pt]{1.080pt}{0.350pt}}
\put(708,264){\rule[-0.175pt]{1.080pt}{0.350pt}}
\put(713,265){\rule[-0.175pt]{1.080pt}{0.350pt}}
\put(717,266){\rule[-0.175pt]{1.080pt}{0.350pt}}
\put(722,267){\rule[-0.175pt]{1.080pt}{0.350pt}}
\put(726,268){\rule[-0.175pt]{1.080pt}{0.350pt}}
\put(731,269){\rule[-0.175pt]{1.080pt}{0.350pt}}
\put(735,270){\rule[-0.175pt]{1.080pt}{0.350pt}}
\put(740,271){\rule[-0.175pt]{1.080pt}{0.350pt}}
\put(744,272){\rule[-0.175pt]{1.080pt}{0.350pt}}
\put(749,273){\rule[-0.175pt]{1.080pt}{0.350pt}}
\put(753,274){\rule[-0.175pt]{1.080pt}{0.350pt}}
\put(758,275){\rule[-0.175pt]{1.080pt}{0.350pt}}
\put(762,276){\rule[-0.175pt]{1.080pt}{0.350pt}}
\put(767,277){\rule[-0.175pt]{1.080pt}{0.350pt}}
\put(771,278){\rule[-0.175pt]{1.080pt}{0.350pt}}
\put(776,279){\rule[-0.175pt]{1.080pt}{0.350pt}}
\put(780,280){\rule[-0.175pt]{1.080pt}{0.350pt}}
\put(785,281){\rule[-0.175pt]{1.648pt}{0.350pt}}
\put(791,282){\rule[-0.175pt]{1.648pt}{0.350pt}}
\put(798,283){\rule[-0.175pt]{1.648pt}{0.350pt}}
\put(805,284){\rule[-0.175pt]{1.648pt}{0.350pt}}
\put(812,285){\rule[-0.175pt]{1.648pt}{0.350pt}}
\put(819,286){\rule[-0.175pt]{1.648pt}{0.350pt}}
\put(826,287){\rule[-0.175pt]{1.648pt}{0.350pt}}
\put(832,288){\rule[-0.175pt]{1.648pt}{0.350pt}}
\put(839,289){\rule[-0.175pt]{1.648pt}{0.350pt}}
\put(846,290){\rule[-0.175pt]{1.648pt}{0.350pt}}
\put(853,291){\rule[-0.175pt]{1.648pt}{0.350pt}}
\put(860,292){\rule[-0.175pt]{1.648pt}{0.350pt}}
\put(867,293){\rule[-0.175pt]{1.648pt}{0.350pt}}
\put(873,294){\rule[-0.175pt]{1.648pt}{0.350pt}}
\put(880,295){\rule[-0.175pt]{1.648pt}{0.350pt}}
\put(887,296){\rule[-0.175pt]{1.648pt}{0.350pt}}
\put(894,297){\rule[-0.175pt]{1.648pt}{0.350pt}}
\put(901,298){\rule[-0.175pt]{1.648pt}{0.350pt}}
\put(908,299){\rule[-0.175pt]{1.648pt}{0.350pt}}
\put(471,573){\circle*{24}}
\put(394,233){\circle*{24}}
\put(524,237){\circle*{24}}
\put(655,252){\circle*{24}}
\put(785,281){\circle*{24}}
\put(915,300){\circle*{24}}
\put(914,300){\usebox{\plotpoint}}
\end{picture}

%% file: half.tex
% GNUPLOT: LaTeX picture
\setlength{\unitlength}{0.240900pt}
\ifx\plotpoint\undefined\newsavebox{\plotpoint}\fi
\begin{picture}(1500,900)(0,0)
\tenrm
\sbox{\plotpoint}{\rule[-0.175pt]{0.350pt}{0.350pt}}%
\put(264,158){\rule[-0.175pt]{282.335pt}{0.350pt}}
\put(264,158){\rule[-0.175pt]{0.350pt}{151.526pt}}
\put(264,158){\rule[-0.175pt]{4.818pt}{0.350pt}}
\put(242,158){\makebox(0,0)[r]{0}}
\put(1416,158){\rule[-0.175pt]{4.818pt}{0.350pt}}
\put(264,263){\rule[-0.175pt]{4.818pt}{0.350pt}}
\put(242,263){\makebox(0,0)[r]{1}}
\put(1416,263){\rule[-0.175pt]{4.818pt}{0.350pt}}
\put(264,368){\rule[-0.175pt]{4.818pt}{0.350pt}}
\put(242,368){\makebox(0,0)[r]{2}}
\put(1416,368){\rule[-0.175pt]{4.818pt}{0.350pt}}
\put(264,473){\rule[-0.175pt]{4.818pt}{0.350pt}}
\put(242,473){\makebox(0,0)[r]{3}}
\put(1416,473){\rule[-0.175pt]{4.818pt}{0.350pt}}
\put(264,577){\rule[-0.175pt]{4.818pt}{0.350pt}}
\put(242,577){\makebox(0,0)[r]{4}}
\put(1416,577){\rule[-0.175pt]{4.818pt}{0.350pt}}
\put(264,682){\rule[-0.175pt]{4.818pt}{0.350pt}}
\put(242,682){\makebox(0,0)[r]{5}}
\put(1416,682){\rule[-0.175pt]{4.818pt}{0.350pt}}
\put(264,787){\rule[-0.175pt]{4.818pt}{0.350pt}}
\put(242,787){\makebox(0,0)[r]{6}}
\put(1416,787){\rule[-0.175pt]{4.818pt}{0.350pt}}
\put(264,158){\rule[-0.175pt]{0.350pt}{4.818pt}}
\put(264,113){\makebox(0,0){0}}
\put(264,767){\rule[-0.175pt]{0.350pt}{4.818pt}}
\put(394,158){\rule[-0.175pt]{0.350pt}{4.818pt}}
\put(394,113){\makebox(0,0){1}}
\put(394,767){\rule[-0.175pt]{0.350pt}{4.818pt}}
\put(524,158){\rule[-0.175pt]{0.350pt}{4.818pt}}
\put(524,113){\makebox(0,0){2}}
\put(524,767){\rule[-0.175pt]{0.350pt}{4.818pt}}
\put(655,158){\rule[-0.175pt]{0.350pt}{4.818pt}}
\put(655,113){\makebox(0,0){3}}
\put(655,767){\rule[-0.175pt]{0.350pt}{4.818pt}}
\put(785,158){\rule[-0.175pt]{0.350pt}{4.818pt}}
\put(785,113){\makebox(0,0){4}}
\put(785,767){\rule[-0.175pt]{0.350pt}{4.818pt}}
\put(915,158){\rule[-0.175pt]{0.350pt}{4.818pt}}
\put(915,113){\makebox(0,0){5}}
\put(915,767){\rule[-0.175pt]{0.350pt}{4.818pt}}
\put(1045,158){\rule[-0.175pt]{0.350pt}{4.818pt}}
\put(1045,113){\makebox(0,0){6}}
\put(1045,767){\rule[-0.175pt]{0.350pt}{4.818pt}}
\put(1176,158){\rule[-0.175pt]{0.350pt}{4.818pt}}
\put(1176,113){\makebox(0,0){7}}
\put(1176,767){\rule[-0.175pt]{0.350pt}{4.818pt}}
\put(1306,158){\rule[-0.175pt]{0.350pt}{4.818pt}}
\put(1306,113){\makebox(0,0){8}}
\put(1306,767){\rule[-0.175pt]{0.350pt}{4.818pt}}
\put(1436,158){\rule[-0.175pt]{0.350pt}{4.818pt}}
\put(1436,113){\makebox(0,0){9}}
\put(1436,767){\rule[-0.175pt]{0.350pt}{4.818pt}}
\put(264,158){\rule[-0.175pt]{282.335pt}{0.350pt}}
\put(1436,158){\rule[-0.175pt]{0.350pt}{151.526pt}}
\put(264,787){\rule[-0.175pt]{282.335pt}{0.350pt}}
\put(45,472){\makebox(0,0)[l]{\shortstack{$Q$}}}
\put(850,68){\makebox(0,0){$P$}}
\put(264,158){\rule[-0.175pt]{0.350pt}{151.526pt}}
\put(459,682){\makebox(0,0)[r]{{\tiny ${1\over 2}$}-Ne$_2$}}
\put(481,682){\rule[-0.175pt]{15.899pt}{0.350pt}}
\put(394,291){\usebox{\plotpoint}}
\put(394,291){\rule[-0.175pt]{0.696pt}{0.350pt}}
\put(396,292){\rule[-0.175pt]{0.696pt}{0.350pt}}
\put(399,293){\rule[-0.175pt]{0.696pt}{0.350pt}}
\put(402,294){\rule[-0.175pt]{0.696pt}{0.350pt}}
\put(405,295){\rule[-0.175pt]{0.696pt}{0.350pt}}
\put(408,296){\rule[-0.175pt]{0.696pt}{0.350pt}}
\put(411,297){\rule[-0.175pt]{0.696pt}{0.350pt}}
\put(414,298){\rule[-0.175pt]{0.696pt}{0.350pt}}
\put(417,299){\rule[-0.175pt]{0.696pt}{0.350pt}}
\put(419,300){\rule[-0.175pt]{0.696pt}{0.350pt}}
\put(422,301){\rule[-0.175pt]{0.696pt}{0.350pt}}
\put(425,302){\rule[-0.175pt]{0.696pt}{0.350pt}}
\put(428,303){\rule[-0.175pt]{0.696pt}{0.350pt}}
\put(431,304){\rule[-0.175pt]{0.696pt}{0.350pt}}
\put(434,305){\rule[-0.175pt]{0.696pt}{0.350pt}}
\put(437,306){\rule[-0.175pt]{0.696pt}{0.350pt}}
\put(440,307){\rule[-0.175pt]{0.696pt}{0.350pt}}
\put(443,308){\rule[-0.175pt]{0.696pt}{0.350pt}}
\put(445,309){\rule[-0.175pt]{0.696pt}{0.350pt}}
\put(448,310){\rule[-0.175pt]{0.696pt}{0.350pt}}
\put(451,311){\rule[-0.175pt]{0.696pt}{0.350pt}}
\put(454,312){\rule[-0.175pt]{0.696pt}{0.350pt}}
\put(457,313){\rule[-0.175pt]{0.696pt}{0.350pt}}
\put(460,314){\rule[-0.175pt]{0.696pt}{0.350pt}}
\put(463,315){\rule[-0.175pt]{0.696pt}{0.350pt}}
\put(466,316){\rule[-0.175pt]{0.696pt}{0.350pt}}
\put(469,317){\rule[-0.175pt]{0.696pt}{0.350pt}}
\put(471,318){\rule[-0.175pt]{0.696pt}{0.350pt}}
\put(474,319){\rule[-0.175pt]{0.696pt}{0.350pt}}
\put(477,320){\rule[-0.175pt]{0.696pt}{0.350pt}}
\put(480,321){\rule[-0.175pt]{0.696pt}{0.350pt}}
\put(483,322){\rule[-0.175pt]{0.696pt}{0.350pt}}
\put(486,323){\rule[-0.175pt]{0.696pt}{0.350pt}}
\put(489,324){\rule[-0.175pt]{0.696pt}{0.350pt}}
\put(492,325){\rule[-0.175pt]{0.696pt}{0.350pt}}
\put(495,326){\rule[-0.175pt]{0.696pt}{0.350pt}}
\put(497,327){\rule[-0.175pt]{0.696pt}{0.350pt}}
\put(500,328){\rule[-0.175pt]{0.696pt}{0.350pt}}
\put(503,329){\rule[-0.175pt]{0.696pt}{0.350pt}}
\put(506,330){\rule[-0.175pt]{0.696pt}{0.350pt}}
\put(509,331){\rule[-0.175pt]{0.696pt}{0.350pt}}
\put(512,332){\rule[-0.175pt]{0.696pt}{0.350pt}}
\put(515,333){\rule[-0.175pt]{0.696pt}{0.350pt}}
\put(518,334){\rule[-0.175pt]{0.696pt}{0.350pt}}
\put(521,335){\rule[-0.175pt]{0.696pt}{0.350pt}}
\put(524,336){\rule[-0.175pt]{0.486pt}{0.350pt}}
\put(526,337){\rule[-0.175pt]{0.486pt}{0.350pt}}
\put(528,338){\rule[-0.175pt]{0.486pt}{0.350pt}}
\put(530,339){\rule[-0.175pt]{0.486pt}{0.350pt}}
\put(532,340){\rule[-0.175pt]{0.486pt}{0.350pt}}
\put(534,341){\rule[-0.175pt]{0.486pt}{0.350pt}}
\put(536,342){\rule[-0.175pt]{0.486pt}{0.350pt}}
\put(538,343){\rule[-0.175pt]{0.486pt}{0.350pt}}
\put(540,344){\rule[-0.175pt]{0.486pt}{0.350pt}}
\put(542,345){\rule[-0.175pt]{0.486pt}{0.350pt}}
\put(544,346){\rule[-0.175pt]{0.486pt}{0.350pt}}
\put(546,347){\rule[-0.175pt]{0.486pt}{0.350pt}}
\put(548,348){\rule[-0.175pt]{0.486pt}{0.350pt}}
\put(550,349){\rule[-0.175pt]{0.486pt}{0.350pt}}
\put(552,350){\rule[-0.175pt]{0.486pt}{0.350pt}}
\put(554,351){\rule[-0.175pt]{0.486pt}{0.350pt}}
\put(556,352){\rule[-0.175pt]{0.486pt}{0.350pt}}
\put(558,353){\rule[-0.175pt]{0.486pt}{0.350pt}}
\put(560,354){\rule[-0.175pt]{0.486pt}{0.350pt}}
\put(562,355){\rule[-0.175pt]{0.486pt}{0.350pt}}
\put(564,356){\rule[-0.175pt]{0.486pt}{0.350pt}}
\put(566,357){\rule[-0.175pt]{0.486pt}{0.350pt}}
\put(568,358){\rule[-0.175pt]{0.486pt}{0.350pt}}
\put(570,359){\rule[-0.175pt]{0.486pt}{0.350pt}}
\put(572,360){\rule[-0.175pt]{0.486pt}{0.350pt}}
\put(574,361){\rule[-0.175pt]{0.486pt}{0.350pt}}
\put(576,362){\rule[-0.175pt]{0.486pt}{0.350pt}}
\put(578,363){\rule[-0.175pt]{0.486pt}{0.350pt}}
\put(580,364){\rule[-0.175pt]{0.486pt}{0.350pt}}
\put(582,365){\rule[-0.175pt]{0.486pt}{0.350pt}}
\put(584,366){\rule[-0.175pt]{0.486pt}{0.350pt}}
\put(586,367){\rule[-0.175pt]{0.486pt}{0.350pt}}
\put(588,368){\rule[-0.175pt]{0.486pt}{0.350pt}}
\put(590,369){\rule[-0.175pt]{0.486pt}{0.350pt}}
\put(592,370){\rule[-0.175pt]{0.486pt}{0.350pt}}
\put(594,371){\rule[-0.175pt]{0.486pt}{0.350pt}}
\put(596,372){\rule[-0.175pt]{0.486pt}{0.350pt}}
\put(598,373){\rule[-0.175pt]{0.486pt}{0.350pt}}
\put(600,374){\rule[-0.175pt]{0.486pt}{0.350pt}}
\put(602,375){\rule[-0.175pt]{0.486pt}{0.350pt}}
\put(604,376){\rule[-0.175pt]{0.486pt}{0.350pt}}
\put(606,377){\rule[-0.175pt]{0.486pt}{0.350pt}}
\put(608,378){\rule[-0.175pt]{0.486pt}{0.350pt}}
\put(610,379){\rule[-0.175pt]{0.486pt}{0.350pt}}
\put(612,380){\rule[-0.175pt]{0.486pt}{0.350pt}}
\put(614,381){\rule[-0.175pt]{0.486pt}{0.350pt}}
\put(616,382){\rule[-0.175pt]{0.486pt}{0.350pt}}
\put(618,383){\rule[-0.175pt]{0.486pt}{0.350pt}}
\put(620,384){\rule[-0.175pt]{0.486pt}{0.350pt}}
\put(622,385){\rule[-0.175pt]{0.486pt}{0.350pt}}
\put(624,386){\rule[-0.175pt]{0.486pt}{0.350pt}}
\put(626,387){\rule[-0.175pt]{0.486pt}{0.350pt}}
\put(628,388){\rule[-0.175pt]{0.486pt}{0.350pt}}
\put(630,389){\rule[-0.175pt]{0.486pt}{0.350pt}}
\put(632,390){\rule[-0.175pt]{0.486pt}{0.350pt}}
\put(634,391){\rule[-0.175pt]{0.486pt}{0.350pt}}
\put(636,392){\rule[-0.175pt]{0.486pt}{0.350pt}}
\put(638,393){\rule[-0.175pt]{0.486pt}{0.350pt}}
\put(640,394){\rule[-0.175pt]{0.486pt}{0.350pt}}
\put(642,395){\rule[-0.175pt]{0.486pt}{0.350pt}}
\put(644,396){\rule[-0.175pt]{0.486pt}{0.350pt}}
\put(646,397){\rule[-0.175pt]{0.486pt}{0.350pt}}
\put(648,398){\rule[-0.175pt]{0.486pt}{0.350pt}}
\put(650,399){\rule[-0.175pt]{0.486pt}{0.350pt}}
\put(652,400){\rule[-0.175pt]{0.486pt}{0.350pt}}
\put(654,401){\rule[-0.175pt]{0.531pt}{0.350pt}}
\put(657,402){\rule[-0.175pt]{0.531pt}{0.350pt}}
\put(659,403){\rule[-0.175pt]{0.531pt}{0.350pt}}
\put(661,404){\rule[-0.175pt]{0.531pt}{0.350pt}}
\put(663,405){\rule[-0.175pt]{0.531pt}{0.350pt}}
\put(666,406){\rule[-0.175pt]{0.531pt}{0.350pt}}
\put(668,407){\rule[-0.175pt]{0.531pt}{0.350pt}}
\put(670,408){\rule[-0.175pt]{0.531pt}{0.350pt}}
\put(672,409){\rule[-0.175pt]{0.531pt}{0.350pt}}
\put(674,410){\rule[-0.175pt]{0.531pt}{0.350pt}}
\put(677,411){\rule[-0.175pt]{0.531pt}{0.350pt}}
\put(679,412){\rule[-0.175pt]{0.531pt}{0.350pt}}
\put(681,413){\rule[-0.175pt]{0.531pt}{0.350pt}}
\put(683,414){\rule[-0.175pt]{0.531pt}{0.350pt}}
\put(685,415){\rule[-0.175pt]{0.531pt}{0.350pt}}
\put(688,416){\rule[-0.175pt]{0.531pt}{0.350pt}}
\put(690,417){\rule[-0.175pt]{0.531pt}{0.350pt}}
\put(692,418){\rule[-0.175pt]{0.531pt}{0.350pt}}
\put(694,419){\rule[-0.175pt]{0.531pt}{0.350pt}}
\put(696,420){\rule[-0.175pt]{0.531pt}{0.350pt}}
\put(699,421){\rule[-0.175pt]{0.531pt}{0.350pt}}
\put(701,422){\rule[-0.175pt]{0.531pt}{0.350pt}}
\put(703,423){\rule[-0.175pt]{0.531pt}{0.350pt}}
\put(705,424){\rule[-0.175pt]{0.531pt}{0.350pt}}
\put(707,425){\rule[-0.175pt]{0.531pt}{0.350pt}}
\put(710,426){\rule[-0.175pt]{0.531pt}{0.350pt}}
\put(712,427){\rule[-0.175pt]{0.531pt}{0.350pt}}
\put(714,428){\rule[-0.175pt]{0.531pt}{0.350pt}}
\put(716,429){\rule[-0.175pt]{0.531pt}{0.350pt}}
\put(718,430){\rule[-0.175pt]{0.531pt}{0.350pt}}
\put(721,431){\rule[-0.175pt]{0.531pt}{0.350pt}}
\put(723,432){\rule[-0.175pt]{0.531pt}{0.350pt}}
\put(725,433){\rule[-0.175pt]{0.531pt}{0.350pt}}
\put(727,434){\rule[-0.175pt]{0.531pt}{0.350pt}}
\put(729,435){\rule[-0.175pt]{0.531pt}{0.350pt}}
\put(732,436){\rule[-0.175pt]{0.531pt}{0.350pt}}
\put(734,437){\rule[-0.175pt]{0.531pt}{0.350pt}}
\put(736,438){\rule[-0.175pt]{0.531pt}{0.350pt}}
\put(738,439){\rule[-0.175pt]{0.531pt}{0.350pt}}
\put(740,440){\rule[-0.175pt]{0.531pt}{0.350pt}}
\put(743,441){\rule[-0.175pt]{0.531pt}{0.350pt}}
\put(745,442){\rule[-0.175pt]{0.531pt}{0.350pt}}
\put(747,443){\rule[-0.175pt]{0.531pt}{0.350pt}}
\put(749,444){\rule[-0.175pt]{0.531pt}{0.350pt}}
\put(751,445){\rule[-0.175pt]{0.531pt}{0.350pt}}
\put(754,446){\rule[-0.175pt]{0.531pt}{0.350pt}}
\put(756,447){\rule[-0.175pt]{0.531pt}{0.350pt}}
\put(758,448){\rule[-0.175pt]{0.531pt}{0.350pt}}
\put(760,449){\rule[-0.175pt]{0.531pt}{0.350pt}}
\put(762,450){\rule[-0.175pt]{0.531pt}{0.350pt}}
\put(765,451){\rule[-0.175pt]{0.531pt}{0.350pt}}
\put(767,452){\rule[-0.175pt]{0.531pt}{0.350pt}}
\put(769,453){\rule[-0.175pt]{0.531pt}{0.350pt}}
\put(771,454){\rule[-0.175pt]{0.531pt}{0.350pt}}
\put(773,455){\rule[-0.175pt]{0.531pt}{0.350pt}}
\put(776,456){\rule[-0.175pt]{0.531pt}{0.350pt}}
\put(778,457){\rule[-0.175pt]{0.531pt}{0.350pt}}
\put(780,458){\rule[-0.175pt]{0.531pt}{0.350pt}}
\put(782,459){\rule[-0.175pt]{0.531pt}{0.350pt}}
\put(784,460){\rule[-0.175pt]{0.803pt}{0.350pt}}
\put(788,461){\rule[-0.175pt]{0.803pt}{0.350pt}}
\put(791,462){\rule[-0.175pt]{0.803pt}{0.350pt}}
\put(794,463){\rule[-0.175pt]{0.803pt}{0.350pt}}
\put(798,464){\rule[-0.175pt]{0.803pt}{0.350pt}}
\put(801,465){\rule[-0.175pt]{0.803pt}{0.350pt}}
\put(804,466){\rule[-0.175pt]{0.803pt}{0.350pt}}
\put(808,467){\rule[-0.175pt]{0.803pt}{0.350pt}}
\put(811,468){\rule[-0.175pt]{0.803pt}{0.350pt}}
\put(814,469){\rule[-0.175pt]{0.803pt}{0.350pt}}
\put(818,470){\rule[-0.175pt]{0.803pt}{0.350pt}}
\put(821,471){\rule[-0.175pt]{0.803pt}{0.350pt}}
\put(824,472){\rule[-0.175pt]{0.803pt}{0.350pt}}
\put(828,473){\rule[-0.175pt]{0.803pt}{0.350pt}}
\put(831,474){\rule[-0.175pt]{0.803pt}{0.350pt}}
\put(834,475){\rule[-0.175pt]{0.803pt}{0.350pt}}
\put(838,476){\rule[-0.175pt]{0.803pt}{0.350pt}}
\put(841,477){\rule[-0.175pt]{0.803pt}{0.350pt}}
\put(844,478){\rule[-0.175pt]{0.803pt}{0.350pt}}
\put(848,479){\rule[-0.175pt]{0.803pt}{0.350pt}}
\put(851,480){\rule[-0.175pt]{0.803pt}{0.350pt}}
\put(854,481){\rule[-0.175pt]{0.803pt}{0.350pt}}
\put(858,482){\rule[-0.175pt]{0.803pt}{0.350pt}}
\put(861,483){\rule[-0.175pt]{0.803pt}{0.350pt}}
\put(864,484){\rule[-0.175pt]{0.803pt}{0.350pt}}
\put(868,485){\rule[-0.175pt]{0.803pt}{0.350pt}}
\put(871,486){\rule[-0.175pt]{0.803pt}{0.350pt}}
\put(874,487){\rule[-0.175pt]{0.803pt}{0.350pt}}
\put(878,488){\rule[-0.175pt]{0.803pt}{0.350pt}}
\put(881,489){\rule[-0.175pt]{0.803pt}{0.350pt}}
\put(884,490){\rule[-0.175pt]{0.803pt}{0.350pt}}
\put(888,491){\rule[-0.175pt]{0.803pt}{0.350pt}}
\put(891,492){\rule[-0.175pt]{0.803pt}{0.350pt}}
\put(894,493){\rule[-0.175pt]{0.803pt}{0.350pt}}
\put(898,494){\rule[-0.175pt]{0.803pt}{0.350pt}}
\put(901,495){\rule[-0.175pt]{0.803pt}{0.350pt}}
\put(904,496){\rule[-0.175pt]{0.803pt}{0.350pt}}
\put(908,497){\rule[-0.175pt]{0.803pt}{0.350pt}}
\put(911,498){\rule[-0.175pt]{0.803pt}{0.350pt}}
\put(914,499){\rule[-0.175pt]{1.080pt}{0.350pt}}
\put(919,500){\rule[-0.175pt]{1.080pt}{0.350pt}}
\put(923,501){\rule[-0.175pt]{1.080pt}{0.350pt}}
\put(928,502){\rule[-0.175pt]{1.080pt}{0.350pt}}
\put(932,503){\rule[-0.175pt]{1.080pt}{0.350pt}}
\put(937,504){\rule[-0.175pt]{1.080pt}{0.350pt}}
\put(941,505){\rule[-0.175pt]{1.080pt}{0.350pt}}
\put(946,506){\rule[-0.175pt]{1.080pt}{0.350pt}}
\put(950,507){\rule[-0.175pt]{1.080pt}{0.350pt}}
\put(955,508){\rule[-0.175pt]{1.080pt}{0.350pt}}
\put(959,509){\rule[-0.175pt]{1.080pt}{0.350pt}}
\put(964,510){\rule[-0.175pt]{1.080pt}{0.350pt}}
\put(968,511){\rule[-0.175pt]{1.080pt}{0.350pt}}
\put(973,512){\rule[-0.175pt]{1.080pt}{0.350pt}}
\put(977,513){\rule[-0.175pt]{1.080pt}{0.350pt}}
\put(982,514){\rule[-0.175pt]{1.080pt}{0.350pt}}
\put(986,515){\rule[-0.175pt]{1.080pt}{0.350pt}}
\put(991,516){\rule[-0.175pt]{1.080pt}{0.350pt}}
\put(995,517){\rule[-0.175pt]{1.080pt}{0.350pt}}
\put(1000,518){\rule[-0.175pt]{1.080pt}{0.350pt}}
\put(1004,519){\rule[-0.175pt]{1.080pt}{0.350pt}}
\put(1009,520){\rule[-0.175pt]{1.080pt}{0.350pt}}
\put(1013,521){\rule[-0.175pt]{1.080pt}{0.350pt}}
\put(1018,522){\rule[-0.175pt]{1.080pt}{0.350pt}}
\put(1022,523){\rule[-0.175pt]{1.080pt}{0.350pt}}
\put(1027,524){\rule[-0.175pt]{1.080pt}{0.350pt}}
\put(1031,525){\rule[-0.175pt]{1.080pt}{0.350pt}}
\put(1036,526){\rule[-0.175pt]{1.080pt}{0.350pt}}
\put(1040,527){\rule[-0.175pt]{1.080pt}{0.350pt}}
\put(1045,528){\rule[-0.175pt]{1.214pt}{0.350pt}}
\put(1050,529){\rule[-0.175pt]{1.214pt}{0.350pt}}
\put(1055,530){\rule[-0.175pt]{1.214pt}{0.350pt}}
\put(1060,531){\rule[-0.175pt]{1.214pt}{0.350pt}}
\put(1065,532){\rule[-0.175pt]{1.214pt}{0.350pt}}
\put(1070,533){\rule[-0.175pt]{1.214pt}{0.350pt}}
\put(1075,534){\rule[-0.175pt]{1.214pt}{0.350pt}}
\put(1080,535){\rule[-0.175pt]{1.214pt}{0.350pt}}
\put(1085,536){\rule[-0.175pt]{1.214pt}{0.350pt}}
\put(1090,537){\rule[-0.175pt]{1.214pt}{0.350pt}}
\put(1095,538){\rule[-0.175pt]{1.214pt}{0.350pt}}
\put(1100,539){\rule[-0.175pt]{1.214pt}{0.350pt}}
\put(1105,540){\rule[-0.175pt]{1.214pt}{0.350pt}}
\put(1110,541){\rule[-0.175pt]{1.214pt}{0.350pt}}
\put(1115,542){\rule[-0.175pt]{1.214pt}{0.350pt}}
\put(1120,543){\rule[-0.175pt]{1.214pt}{0.350pt}}
\put(1125,544){\rule[-0.175pt]{1.214pt}{0.350pt}}
\put(1130,545){\rule[-0.175pt]{1.214pt}{0.350pt}}
\put(1135,546){\rule[-0.175pt]{1.214pt}{0.350pt}}
\put(1140,547){\rule[-0.175pt]{1.214pt}{0.350pt}}
\put(1145,548){\rule[-0.175pt]{1.214pt}{0.350pt}}
\put(1150,549){\rule[-0.175pt]{1.214pt}{0.350pt}}
\put(1155,550){\rule[-0.175pt]{1.214pt}{0.350pt}}
\put(1160,551){\rule[-0.175pt]{1.214pt}{0.350pt}}
\put(1165,552){\rule[-0.175pt]{1.214pt}{0.350pt}}
\put(1170,553){\rule[-0.175pt]{1.214pt}{0.350pt}}
\put(1175,554){\rule[-0.175pt]{0.846pt}{0.350pt}}
\put(1179,555){\rule[-0.175pt]{0.846pt}{0.350pt}}
\put(1183,556){\rule[-0.175pt]{0.846pt}{0.350pt}}
\put(1186,557){\rule[-0.175pt]{0.846pt}{0.350pt}}
\put(1190,558){\rule[-0.175pt]{0.846pt}{0.350pt}}
\put(1193,559){\rule[-0.175pt]{0.846pt}{0.350pt}}
\put(1197,560){\rule[-0.175pt]{0.846pt}{0.350pt}}
\put(1200,561){\rule[-0.175pt]{0.846pt}{0.350pt}}
\put(1204,562){\rule[-0.175pt]{0.846pt}{0.350pt}}
\put(1207,563){\rule[-0.175pt]{0.846pt}{0.350pt}}
\put(1211,564){\rule[-0.175pt]{0.846pt}{0.350pt}}
\put(1214,565){\rule[-0.175pt]{0.846pt}{0.350pt}}
\put(1218,566){\rule[-0.175pt]{0.846pt}{0.350pt}}
\put(1221,567){\rule[-0.175pt]{0.846pt}{0.350pt}}
\put(1225,568){\rule[-0.175pt]{0.846pt}{0.350pt}}
\put(1228,569){\rule[-0.175pt]{0.846pt}{0.350pt}}
\put(1232,570){\rule[-0.175pt]{0.846pt}{0.350pt}}
\put(1235,571){\rule[-0.175pt]{0.846pt}{0.350pt}}
\put(1239,572){\rule[-0.175pt]{0.846pt}{0.350pt}}
\put(1242,573){\rule[-0.175pt]{0.846pt}{0.350pt}}
\put(1246,574){\rule[-0.175pt]{0.846pt}{0.350pt}}
\put(1249,575){\rule[-0.175pt]{0.846pt}{0.350pt}}
\put(1253,576){\rule[-0.175pt]{0.846pt}{0.350pt}}
\put(1256,577){\rule[-0.175pt]{0.846pt}{0.350pt}}
\put(1260,578){\rule[-0.175pt]{0.846pt}{0.350pt}}
\put(1263,579){\rule[-0.175pt]{0.846pt}{0.350pt}}
\put(1267,580){\rule[-0.175pt]{0.846pt}{0.350pt}}
\put(1270,581){\rule[-0.175pt]{0.846pt}{0.350pt}}
\put(1274,582){\rule[-0.175pt]{0.846pt}{0.350pt}}
\put(1277,583){\rule[-0.175pt]{0.846pt}{0.350pt}}
\put(1281,584){\rule[-0.175pt]{0.846pt}{0.350pt}}
\put(1284,585){\rule[-0.175pt]{0.846pt}{0.350pt}}
\put(1288,586){\rule[-0.175pt]{0.846pt}{0.350pt}}
\put(1291,587){\rule[-0.175pt]{0.846pt}{0.350pt}}
\put(1295,588){\rule[-0.175pt]{0.846pt}{0.350pt}}
\put(1298,589){\rule[-0.175pt]{0.846pt}{0.350pt}}
\put(503,682){\circle{12}}
\put(394,291){\circle{12}}
\put(524,336){\circle{12}}
\put(655,401){\circle{12}}
\put(785,460){\circle{12}}
\put(915,499){\circle{12}}
\put(1045,528){\circle{12}}
\put(1176,554){\circle{12}}
\put(1306,591){\circle{12}}
\put(1302,590){\rule[-0.175pt]{0.846pt}{0.350pt}}
\put(459,637){\makebox(0,0)[r]{{\tiny ${1\over 2}$}-Ne$_3$}}
\put(481,637){\rule[-0.175pt]{15.899pt}{0.350pt}}
\put(394,245){\usebox{\plotpoint}}
\put(394,245){\rule[-0.175pt]{1.044pt}{0.350pt}}
\put(398,246){\rule[-0.175pt]{1.044pt}{0.350pt}}
\put(402,247){\rule[-0.175pt]{1.044pt}{0.350pt}}
\put(407,248){\rule[-0.175pt]{1.044pt}{0.350pt}}
\put(411,249){\rule[-0.175pt]{1.044pt}{0.350pt}}
\put(415,250){\rule[-0.175pt]{1.044pt}{0.350pt}}
\put(420,251){\rule[-0.175pt]{1.044pt}{0.350pt}}
\put(424,252){\rule[-0.175pt]{1.044pt}{0.350pt}}
\put(428,253){\rule[-0.175pt]{1.044pt}{0.350pt}}
\put(433,254){\rule[-0.175pt]{1.044pt}{0.350pt}}
\put(437,255){\rule[-0.175pt]{1.044pt}{0.350pt}}
\put(441,256){\rule[-0.175pt]{1.044pt}{0.350pt}}
\put(446,257){\rule[-0.175pt]{1.044pt}{0.350pt}}
\put(450,258){\rule[-0.175pt]{1.044pt}{0.350pt}}
\put(454,259){\rule[-0.175pt]{1.044pt}{0.350pt}}
\put(459,260){\rule[-0.175pt]{1.044pt}{0.350pt}}
\put(463,261){\rule[-0.175pt]{1.044pt}{0.350pt}}
\put(467,262){\rule[-0.175pt]{1.044pt}{0.350pt}}
\put(472,263){\rule[-0.175pt]{1.044pt}{0.350pt}}
\put(476,264){\rule[-0.175pt]{1.044pt}{0.350pt}}
\put(480,265){\rule[-0.175pt]{1.044pt}{0.350pt}}
\put(485,266){\rule[-0.175pt]{1.044pt}{0.350pt}}
\put(489,267){\rule[-0.175pt]{1.044pt}{0.350pt}}
\put(493,268){\rule[-0.175pt]{1.044pt}{0.350pt}}
\put(498,269){\rule[-0.175pt]{1.044pt}{0.350pt}}
\put(502,270){\rule[-0.175pt]{1.044pt}{0.350pt}}
\put(506,271){\rule[-0.175pt]{1.044pt}{0.350pt}}
\put(511,272){\rule[-0.175pt]{1.044pt}{0.350pt}}
\put(515,273){\rule[-0.175pt]{1.044pt}{0.350pt}}
\put(519,274){\rule[-0.175pt]{1.044pt}{0.350pt}}
\put(524,275){\rule[-0.175pt]{0.644pt}{0.350pt}}
\put(526,276){\rule[-0.175pt]{0.644pt}{0.350pt}}
\put(529,277){\rule[-0.175pt]{0.644pt}{0.350pt}}
\put(532,278){\rule[-0.175pt]{0.644pt}{0.350pt}}
\put(534,279){\rule[-0.175pt]{0.644pt}{0.350pt}}
\put(537,280){\rule[-0.175pt]{0.644pt}{0.350pt}}
\put(540,281){\rule[-0.175pt]{0.644pt}{0.350pt}}
\put(542,282){\rule[-0.175pt]{0.644pt}{0.350pt}}
\put(545,283){\rule[-0.175pt]{0.644pt}{0.350pt}}
\put(548,284){\rule[-0.175pt]{0.644pt}{0.350pt}}
\put(550,285){\rule[-0.175pt]{0.644pt}{0.350pt}}
\put(553,286){\rule[-0.175pt]{0.644pt}{0.350pt}}
\put(556,287){\rule[-0.175pt]{0.644pt}{0.350pt}}
\put(558,288){\rule[-0.175pt]{0.644pt}{0.350pt}}
\put(561,289){\rule[-0.175pt]{0.644pt}{0.350pt}}
\put(564,290){\rule[-0.175pt]{0.644pt}{0.350pt}}
\put(566,291){\rule[-0.175pt]{0.644pt}{0.350pt}}
\put(569,292){\rule[-0.175pt]{0.644pt}{0.350pt}}
\put(572,293){\rule[-0.175pt]{0.644pt}{0.350pt}}
\put(574,294){\rule[-0.175pt]{0.644pt}{0.350pt}}
\put(577,295){\rule[-0.175pt]{0.644pt}{0.350pt}}
\put(580,296){\rule[-0.175pt]{0.644pt}{0.350pt}}
\put(582,297){\rule[-0.175pt]{0.644pt}{0.350pt}}
\put(585,298){\rule[-0.175pt]{0.644pt}{0.350pt}}
\put(588,299){\rule[-0.175pt]{0.644pt}{0.350pt}}
\put(590,300){\rule[-0.175pt]{0.644pt}{0.350pt}}
\put(593,301){\rule[-0.175pt]{0.644pt}{0.350pt}}
\put(596,302){\rule[-0.175pt]{0.644pt}{0.350pt}}
\put(598,303){\rule[-0.175pt]{0.644pt}{0.350pt}}
\put(601,304){\rule[-0.175pt]{0.644pt}{0.350pt}}
\put(604,305){\rule[-0.175pt]{0.644pt}{0.350pt}}
\put(606,306){\rule[-0.175pt]{0.644pt}{0.350pt}}
\put(609,307){\rule[-0.175pt]{0.644pt}{0.350pt}}
\put(612,308){\rule[-0.175pt]{0.644pt}{0.350pt}}
\put(614,309){\rule[-0.175pt]{0.644pt}{0.350pt}}
\put(617,310){\rule[-0.175pt]{0.644pt}{0.350pt}}
\put(620,311){\rule[-0.175pt]{0.644pt}{0.350pt}}
\put(622,312){\rule[-0.175pt]{0.644pt}{0.350pt}}
\put(625,313){\rule[-0.175pt]{0.644pt}{0.350pt}}
\put(628,314){\rule[-0.175pt]{0.644pt}{0.350pt}}
\put(630,315){\rule[-0.175pt]{0.644pt}{0.350pt}}
\put(633,316){\rule[-0.175pt]{0.644pt}{0.350pt}}
\put(636,317){\rule[-0.175pt]{0.644pt}{0.350pt}}
\put(638,318){\rule[-0.175pt]{0.644pt}{0.350pt}}
\put(641,319){\rule[-0.175pt]{0.644pt}{0.350pt}}
\put(644,320){\rule[-0.175pt]{0.644pt}{0.350pt}}
\put(646,321){\rule[-0.175pt]{0.644pt}{0.350pt}}
\put(649,322){\rule[-0.175pt]{0.644pt}{0.350pt}}
\put(652,323){\rule[-0.175pt]{0.644pt}{0.350pt}}
\put(654,324){\rule[-0.175pt]{0.824pt}{0.350pt}}
\put(658,325){\rule[-0.175pt]{0.824pt}{0.350pt}}
\put(661,326){\rule[-0.175pt]{0.824pt}{0.350pt}}
\put(665,327){\rule[-0.175pt]{0.824pt}{0.350pt}}
\put(668,328){\rule[-0.175pt]{0.824pt}{0.350pt}}
\put(672,329){\rule[-0.175pt]{0.824pt}{0.350pt}}
\put(675,330){\rule[-0.175pt]{0.824pt}{0.350pt}}
\put(678,331){\rule[-0.175pt]{0.824pt}{0.350pt}}
\put(682,332){\rule[-0.175pt]{0.824pt}{0.350pt}}
\put(685,333){\rule[-0.175pt]{0.824pt}{0.350pt}}
\put(689,334){\rule[-0.175pt]{0.824pt}{0.350pt}}
\put(692,335){\rule[-0.175pt]{0.824pt}{0.350pt}}
\put(696,336){\rule[-0.175pt]{0.824pt}{0.350pt}}
\put(699,337){\rule[-0.175pt]{0.824pt}{0.350pt}}
\put(702,338){\rule[-0.175pt]{0.824pt}{0.350pt}}
\put(706,339){\rule[-0.175pt]{0.824pt}{0.350pt}}
\put(709,340){\rule[-0.175pt]{0.824pt}{0.350pt}}
\put(713,341){\rule[-0.175pt]{0.824pt}{0.350pt}}
\put(716,342){\rule[-0.175pt]{0.824pt}{0.350pt}}
\put(720,343){\rule[-0.175pt]{0.824pt}{0.350pt}}
\put(723,344){\rule[-0.175pt]{0.824pt}{0.350pt}}
\put(726,345){\rule[-0.175pt]{0.824pt}{0.350pt}}
\put(730,346){\rule[-0.175pt]{0.824pt}{0.350pt}}
\put(733,347){\rule[-0.175pt]{0.824pt}{0.350pt}}
\put(737,348){\rule[-0.175pt]{0.824pt}{0.350pt}}
\put(740,349){\rule[-0.175pt]{0.824pt}{0.350pt}}
\put(743,350){\rule[-0.175pt]{0.824pt}{0.350pt}}
\put(747,351){\rule[-0.175pt]{0.824pt}{0.350pt}}
\put(750,352){\rule[-0.175pt]{0.824pt}{0.350pt}}
\put(754,353){\rule[-0.175pt]{0.824pt}{0.350pt}}
\put(757,354){\rule[-0.175pt]{0.824pt}{0.350pt}}
\put(761,355){\rule[-0.175pt]{0.824pt}{0.350pt}}
\put(764,356){\rule[-0.175pt]{0.824pt}{0.350pt}}
\put(767,357){\rule[-0.175pt]{0.824pt}{0.350pt}}
\put(771,358){\rule[-0.175pt]{0.824pt}{0.350pt}}
\put(774,359){\rule[-0.175pt]{0.824pt}{0.350pt}}
\put(778,360){\rule[-0.175pt]{0.824pt}{0.350pt}}
\put(781,361){\rule[-0.175pt]{0.824pt}{0.350pt}}
\put(785,362){\rule[-0.175pt]{1.566pt}{0.350pt}}
\put(791,363){\rule[-0.175pt]{1.566pt}{0.350pt}}
\put(798,364){\rule[-0.175pt]{1.566pt}{0.350pt}}
\put(804,365){\rule[-0.175pt]{1.566pt}{0.350pt}}
\put(811,366){\rule[-0.175pt]{1.566pt}{0.350pt}}
\put(817,367){\rule[-0.175pt]{1.566pt}{0.350pt}}
\put(824,368){\rule[-0.175pt]{1.566pt}{0.350pt}}
\put(830,369){\rule[-0.175pt]{1.566pt}{0.350pt}}
\put(837,370){\rule[-0.175pt]{1.566pt}{0.350pt}}
\put(843,371){\rule[-0.175pt]{1.566pt}{0.350pt}}
\put(850,372){\rule[-0.175pt]{1.566pt}{0.350pt}}
\put(856,373){\rule[-0.175pt]{1.566pt}{0.350pt}}
\put(863,374){\rule[-0.175pt]{1.566pt}{0.350pt}}
\put(869,375){\rule[-0.175pt]{1.566pt}{0.350pt}}
\put(876,376){\rule[-0.175pt]{1.566pt}{0.350pt}}
\put(882,377){\rule[-0.175pt]{1.566pt}{0.350pt}}
\put(889,378){\rule[-0.175pt]{1.566pt}{0.350pt}}
\put(895,379){\rule[-0.175pt]{1.566pt}{0.350pt}}
\put(902,380){\rule[-0.175pt]{1.566pt}{0.350pt}}
\put(908,381){\rule[-0.175pt]{1.566pt}{0.350pt}}
\put(915,382){\rule[-0.175pt]{1.044pt}{0.350pt}}
\put(919,383){\rule[-0.175pt]{1.044pt}{0.350pt}}
\put(923,384){\rule[-0.175pt]{1.044pt}{0.350pt}}
\put(927,385){\rule[-0.175pt]{1.044pt}{0.350pt}}
\put(932,386){\rule[-0.175pt]{1.044pt}{0.350pt}}
\put(936,387){\rule[-0.175pt]{1.044pt}{0.350pt}}
\put(940,388){\rule[-0.175pt]{1.044pt}{0.350pt}}
\put(945,389){\rule[-0.175pt]{1.044pt}{0.350pt}}
\put(949,390){\rule[-0.175pt]{1.044pt}{0.350pt}}
\put(953,391){\rule[-0.175pt]{1.044pt}{0.350pt}}
\put(958,392){\rule[-0.175pt]{1.044pt}{0.350pt}}
\put(962,393){\rule[-0.175pt]{1.044pt}{0.350pt}}
\put(966,394){\rule[-0.175pt]{1.044pt}{0.350pt}}
\put(971,395){\rule[-0.175pt]{1.044pt}{0.350pt}}
\put(975,396){\rule[-0.175pt]{1.044pt}{0.350pt}}
\put(979,397){\rule[-0.175pt]{1.044pt}{0.350pt}}
\put(984,398){\rule[-0.175pt]{1.044pt}{0.350pt}}
\put(988,399){\rule[-0.175pt]{1.044pt}{0.350pt}}
\put(992,400){\rule[-0.175pt]{1.044pt}{0.350pt}}
\put(997,401){\rule[-0.175pt]{1.044pt}{0.350pt}}
\put(1001,402){\rule[-0.175pt]{1.044pt}{0.350pt}}
\put(1005,403){\rule[-0.175pt]{1.044pt}{0.350pt}}
\put(1010,404){\rule[-0.175pt]{1.044pt}{0.350pt}}
\put(1014,405){\rule[-0.175pt]{1.044pt}{0.350pt}}
\put(1018,406){\rule[-0.175pt]{1.044pt}{0.350pt}}
\put(1023,407){\rule[-0.175pt]{1.044pt}{0.350pt}}
\put(1027,408){\rule[-0.175pt]{1.044pt}{0.350pt}}
\put(1031,409){\rule[-0.175pt]{1.044pt}{0.350pt}}
\put(1036,410){\rule[-0.175pt]{1.044pt}{0.350pt}}
\put(1040,411){\rule[-0.175pt]{1.044pt}{0.350pt}}
\put(1044,412){\rule[-0.175pt]{1.262pt}{0.350pt}}
\put(1050,413){\rule[-0.175pt]{1.262pt}{0.350pt}}
\put(1055,414){\rule[-0.175pt]{1.262pt}{0.350pt}}
\put(1060,415){\rule[-0.175pt]{1.262pt}{0.350pt}}
\put(1065,416){\rule[-0.175pt]{1.262pt}{0.350pt}}
\put(1071,417){\rule[-0.175pt]{1.262pt}{0.350pt}}
\put(1076,418){\rule[-0.175pt]{1.262pt}{0.350pt}}
\put(1081,419){\rule[-0.175pt]{1.262pt}{0.350pt}}
\put(1086,420){\rule[-0.175pt]{1.262pt}{0.350pt}}
\put(1092,421){\rule[-0.175pt]{1.262pt}{0.350pt}}
\put(1097,422){\rule[-0.175pt]{1.262pt}{0.350pt}}
\put(1102,423){\rule[-0.175pt]{1.262pt}{0.350pt}}
\put(1107,424){\rule[-0.175pt]{1.262pt}{0.350pt}}
\put(1113,425){\rule[-0.175pt]{1.262pt}{0.350pt}}
\put(1118,426){\rule[-0.175pt]{1.262pt}{0.350pt}}
\put(1123,427){\rule[-0.175pt]{1.262pt}{0.350pt}}
\put(1128,428){\rule[-0.175pt]{1.262pt}{0.350pt}}
\put(1134,429){\rule[-0.175pt]{1.262pt}{0.350pt}}
\put(1139,430){\rule[-0.175pt]{1.262pt}{0.350pt}}
\put(1144,431){\rule[-0.175pt]{1.262pt}{0.350pt}}
\put(1149,432){\rule[-0.175pt]{1.262pt}{0.350pt}}
\put(1155,433){\rule[-0.175pt]{1.262pt}{0.350pt}}
\put(1160,434){\rule[-0.175pt]{1.262pt}{0.350pt}}
\put(1165,435){\rule[-0.175pt]{1.262pt}{0.350pt}}
\put(1170,436){\rule[-0.175pt]{1.262pt}{0.350pt}}
\put(1175,437){\rule[-0.175pt]{1.160pt}{0.350pt}}
\put(1180,438){\rule[-0.175pt]{1.160pt}{0.350pt}}
\put(1185,439){\rule[-0.175pt]{1.160pt}{0.350pt}}
\put(1190,440){\rule[-0.175pt]{1.160pt}{0.350pt}}
\put(1195,441){\rule[-0.175pt]{1.160pt}{0.350pt}}
\put(1200,442){\rule[-0.175pt]{1.160pt}{0.350pt}}
\put(1204,443){\rule[-0.175pt]{1.160pt}{0.350pt}}
\put(1209,444){\rule[-0.175pt]{1.160pt}{0.350pt}}
\put(1214,445){\rule[-0.175pt]{1.160pt}{0.350pt}}
\put(1219,446){\rule[-0.175pt]{1.160pt}{0.350pt}}
\put(1224,447){\rule[-0.175pt]{1.160pt}{0.350pt}}
\put(1228,448){\rule[-0.175pt]{1.160pt}{0.350pt}}
\put(1233,449){\rule[-0.175pt]{1.160pt}{0.350pt}}
\put(1238,450){\rule[-0.175pt]{1.160pt}{0.350pt}}
\put(1243,451){\rule[-0.175pt]{1.160pt}{0.350pt}}
\put(1248,452){\rule[-0.175pt]{1.160pt}{0.350pt}}
\put(1253,453){\rule[-0.175pt]{1.160pt}{0.350pt}}
\put(1257,454){\rule[-0.175pt]{1.160pt}{0.350pt}}
\put(1262,455){\rule[-0.175pt]{1.160pt}{0.350pt}}
\put(1267,456){\rule[-0.175pt]{1.160pt}{0.350pt}}
\put(1272,457){\rule[-0.175pt]{1.160pt}{0.350pt}}
\put(1277,458){\rule[-0.175pt]{1.160pt}{0.350pt}}
\put(1281,459){\rule[-0.175pt]{1.160pt}{0.350pt}}
\put(1286,460){\rule[-0.175pt]{1.160pt}{0.350pt}}
\put(1291,461){\rule[-0.175pt]{1.160pt}{0.350pt}}
\put(1296,462){\rule[-0.175pt]{1.160pt}{0.350pt}}
\put(503,637){\circle*{12}}
\put(394,245){\circle*{12}}
\put(524,275){\circle*{12}}
\put(655,324){\circle*{12}}
\put(785,362){\circle*{12}}
\put(915,382){\circle*{12}}
\put(1045,412){\circle*{12}}
\put(1176,437){\circle*{12}}
\put(1306,464){\circle*{12}}
\put(1301,463){\rule[-0.175pt]{1.160pt}{0.350pt}}
\put(459,592){\makebox(0,0)[r]{{\tiny ${1\over 2}$}-Ne$_4$}}
\put(481,592){\rule[-0.175pt]{15.899pt}{0.350pt}}
\put(394,241){\usebox{\plotpoint}}
\put(394,241){\rule[-0.175pt]{1.118pt}{0.350pt}}
\put(398,242){\rule[-0.175pt]{1.118pt}{0.350pt}}
\put(403,243){\rule[-0.175pt]{1.118pt}{0.350pt}}
\put(407,244){\rule[-0.175pt]{1.118pt}{0.350pt}}
\put(412,245){\rule[-0.175pt]{1.118pt}{0.350pt}}
\put(417,246){\rule[-0.175pt]{1.118pt}{0.350pt}}
\put(421,247){\rule[-0.175pt]{1.118pt}{0.350pt}}
\put(426,248){\rule[-0.175pt]{1.118pt}{0.350pt}}
\put(431,249){\rule[-0.175pt]{1.118pt}{0.350pt}}
\put(435,250){\rule[-0.175pt]{1.118pt}{0.350pt}}
\put(440,251){\rule[-0.175pt]{1.118pt}{0.350pt}}
\put(445,252){\rule[-0.175pt]{1.118pt}{0.350pt}}
\put(449,253){\rule[-0.175pt]{1.118pt}{0.350pt}}
\put(454,254){\rule[-0.175pt]{1.118pt}{0.350pt}}
\put(458,255){\rule[-0.175pt]{1.118pt}{0.350pt}}
\put(463,256){\rule[-0.175pt]{1.118pt}{0.350pt}}
\put(468,257){\rule[-0.175pt]{1.118pt}{0.350pt}}
\put(472,258){\rule[-0.175pt]{1.118pt}{0.350pt}}
\put(477,259){\rule[-0.175pt]{1.118pt}{0.350pt}}
\put(482,260){\rule[-0.175pt]{1.118pt}{0.350pt}}
\put(486,261){\rule[-0.175pt]{1.118pt}{0.350pt}}
\put(491,262){\rule[-0.175pt]{1.118pt}{0.350pt}}
\put(496,263){\rule[-0.175pt]{1.118pt}{0.350pt}}
\put(500,264){\rule[-0.175pt]{1.118pt}{0.350pt}}
\put(505,265){\rule[-0.175pt]{1.118pt}{0.350pt}}
\put(510,266){\rule[-0.175pt]{1.118pt}{0.350pt}}
\put(514,267){\rule[-0.175pt]{1.118pt}{0.350pt}}
\put(519,268){\rule[-0.175pt]{1.118pt}{0.350pt}}
\put(523,269){\rule[-0.175pt]{0.657pt}{0.350pt}}
\put(526,270){\rule[-0.175pt]{0.657pt}{0.350pt}}
\put(529,271){\rule[-0.175pt]{0.657pt}{0.350pt}}
\put(532,272){\rule[-0.175pt]{0.657pt}{0.350pt}}
\put(534,273){\rule[-0.175pt]{0.657pt}{0.350pt}}
\put(537,274){\rule[-0.175pt]{0.657pt}{0.350pt}}
\put(540,275){\rule[-0.175pt]{0.657pt}{0.350pt}}
\put(543,276){\rule[-0.175pt]{0.657pt}{0.350pt}}
\put(545,277){\rule[-0.175pt]{0.657pt}{0.350pt}}
\put(548,278){\rule[-0.175pt]{0.657pt}{0.350pt}}
\put(551,279){\rule[-0.175pt]{0.657pt}{0.350pt}}
\put(554,280){\rule[-0.175pt]{0.657pt}{0.350pt}}
\put(556,281){\rule[-0.175pt]{0.657pt}{0.350pt}}
\put(559,282){\rule[-0.175pt]{0.657pt}{0.350pt}}
\put(562,283){\rule[-0.175pt]{0.657pt}{0.350pt}}
\put(564,284){\rule[-0.175pt]{0.657pt}{0.350pt}}
\put(567,285){\rule[-0.175pt]{0.657pt}{0.350pt}}
\put(570,286){\rule[-0.175pt]{0.657pt}{0.350pt}}
\put(573,287){\rule[-0.175pt]{0.657pt}{0.350pt}}
\put(575,288){\rule[-0.175pt]{0.657pt}{0.350pt}}
\put(578,289){\rule[-0.175pt]{0.657pt}{0.350pt}}
\put(581,290){\rule[-0.175pt]{0.657pt}{0.350pt}}
\put(584,291){\rule[-0.175pt]{0.657pt}{0.350pt}}
\put(586,292){\rule[-0.175pt]{0.657pt}{0.350pt}}
\put(589,293){\rule[-0.175pt]{0.657pt}{0.350pt}}
\put(592,294){\rule[-0.175pt]{0.657pt}{0.350pt}}
\put(594,295){\rule[-0.175pt]{0.657pt}{0.350pt}}
\put(597,296){\rule[-0.175pt]{0.657pt}{0.350pt}}
\put(600,297){\rule[-0.175pt]{0.657pt}{0.350pt}}
\put(603,298){\rule[-0.175pt]{0.657pt}{0.350pt}}
\put(605,299){\rule[-0.175pt]{0.657pt}{0.350pt}}
\put(608,300){\rule[-0.175pt]{0.657pt}{0.350pt}}
\put(611,301){\rule[-0.175pt]{0.657pt}{0.350pt}}
\put(614,302){\rule[-0.175pt]{0.657pt}{0.350pt}}
\put(616,303){\rule[-0.175pt]{0.657pt}{0.350pt}}
\put(619,304){\rule[-0.175pt]{0.657pt}{0.350pt}}
\put(622,305){\rule[-0.175pt]{0.657pt}{0.350pt}}
\put(624,306){\rule[-0.175pt]{0.657pt}{0.350pt}}
\put(627,307){\rule[-0.175pt]{0.657pt}{0.350pt}}
\put(630,308){\rule[-0.175pt]{0.657pt}{0.350pt}}
\put(633,309){\rule[-0.175pt]{0.657pt}{0.350pt}}
\put(635,310){\rule[-0.175pt]{0.657pt}{0.350pt}}
\put(638,311){\rule[-0.175pt]{0.657pt}{0.350pt}}
\put(641,312){\rule[-0.175pt]{0.657pt}{0.350pt}}
\put(644,313){\rule[-0.175pt]{0.657pt}{0.350pt}}
\put(646,314){\rule[-0.175pt]{0.657pt}{0.350pt}}
\put(649,315){\rule[-0.175pt]{0.657pt}{0.350pt}}
\put(652,316){\rule[-0.175pt]{0.657pt}{0.350pt}}
\put(655,317){\rule[-0.175pt]{1.080pt}{0.350pt}}
\put(659,318){\rule[-0.175pt]{1.080pt}{0.350pt}}
\put(663,319){\rule[-0.175pt]{1.080pt}{0.350pt}}
\put(668,320){\rule[-0.175pt]{1.080pt}{0.350pt}}
\put(672,321){\rule[-0.175pt]{1.080pt}{0.350pt}}
\put(677,322){\rule[-0.175pt]{1.080pt}{0.350pt}}
\put(681,323){\rule[-0.175pt]{1.080pt}{0.350pt}}
\put(686,324){\rule[-0.175pt]{1.080pt}{0.350pt}}
\put(690,325){\rule[-0.175pt]{1.080pt}{0.350pt}}
\put(695,326){\rule[-0.175pt]{1.080pt}{0.350pt}}
\put(699,327){\rule[-0.175pt]{1.080pt}{0.350pt}}
\put(704,328){\rule[-0.175pt]{1.080pt}{0.350pt}}
\put(708,329){\rule[-0.175pt]{1.080pt}{0.350pt}}
\put(713,330){\rule[-0.175pt]{1.080pt}{0.350pt}}
\put(717,331){\rule[-0.175pt]{1.080pt}{0.350pt}}
\put(722,332){\rule[-0.175pt]{1.080pt}{0.350pt}}
\put(726,333){\rule[-0.175pt]{1.080pt}{0.350pt}}
\put(731,334){\rule[-0.175pt]{1.080pt}{0.350pt}}
\put(735,335){\rule[-0.175pt]{1.080pt}{0.350pt}}
\put(740,336){\rule[-0.175pt]{1.080pt}{0.350pt}}
\put(744,337){\rule[-0.175pt]{1.080pt}{0.350pt}}
\put(749,338){\rule[-0.175pt]{1.080pt}{0.350pt}}
\put(753,339){\rule[-0.175pt]{1.080pt}{0.350pt}}
\put(758,340){\rule[-0.175pt]{1.080pt}{0.350pt}}
\put(762,341){\rule[-0.175pt]{1.080pt}{0.350pt}}
\put(767,342){\rule[-0.175pt]{1.080pt}{0.350pt}}
\put(771,343){\rule[-0.175pt]{1.080pt}{0.350pt}}
\put(776,344){\rule[-0.175pt]{1.080pt}{0.350pt}}
\put(780,345){\rule[-0.175pt]{1.080pt}{0.350pt}}
\put(785,346){\rule[-0.175pt]{1.362pt}{0.350pt}}
\put(790,347){\rule[-0.175pt]{1.362pt}{0.350pt}}
\put(796,348){\rule[-0.175pt]{1.362pt}{0.350pt}}
\put(801,349){\rule[-0.175pt]{1.362pt}{0.350pt}}
\put(807,350){\rule[-0.175pt]{1.362pt}{0.350pt}}
\put(813,351){\rule[-0.175pt]{1.362pt}{0.350pt}}
\put(818,352){\rule[-0.175pt]{1.362pt}{0.350pt}}
\put(824,353){\rule[-0.175pt]{1.362pt}{0.350pt}}
\put(830,354){\rule[-0.175pt]{1.362pt}{0.350pt}}
\put(835,355){\rule[-0.175pt]{1.362pt}{0.350pt}}
\put(841,356){\rule[-0.175pt]{1.362pt}{0.350pt}}
\put(847,357){\rule[-0.175pt]{1.362pt}{0.350pt}}
\put(852,358){\rule[-0.175pt]{1.362pt}{0.350pt}}
\put(858,359){\rule[-0.175pt]{1.362pt}{0.350pt}}
\put(864,360){\rule[-0.175pt]{1.362pt}{0.350pt}}
\put(869,361){\rule[-0.175pt]{1.362pt}{0.350pt}}
\put(875,362){\rule[-0.175pt]{1.362pt}{0.350pt}}
\put(881,363){\rule[-0.175pt]{1.362pt}{0.350pt}}
\put(886,364){\rule[-0.175pt]{1.362pt}{0.350pt}}
\put(892,365){\rule[-0.175pt]{1.362pt}{0.350pt}}
\put(898,366){\rule[-0.175pt]{1.362pt}{0.350pt}}
\put(903,367){\rule[-0.175pt]{1.362pt}{0.350pt}}
\put(909,368){\rule[-0.175pt]{1.362pt}{0.350pt}}
\put(503,592){\circle*{18}}
\put(394,241){\circle*{18}}
\put(524,269){\circle*{18}}
\put(655,317){\circle*{18}}
\put(785,346){\circle*{18}}
\put(915,369){\circle*{18}}
\put(914,369){\usebox{\plotpoint}}
\put(459,547){\makebox(0,0)[r]{{\tiny ${1\over 2}$}-Ne$_5$}}
\put(481,547){\rule[-0.175pt]{15.899pt}{0.350pt}}
\put(394,245){\usebox{\plotpoint}}
\put(394,245){\rule[-0.175pt]{5.219pt}{0.350pt}}
\put(415,246){\rule[-0.175pt]{5.219pt}{0.350pt}}
\put(437,247){\rule[-0.175pt]{5.219pt}{0.350pt}}
\put(458,248){\rule[-0.175pt]{5.219pt}{0.350pt}}
\put(480,249){\rule[-0.175pt]{5.219pt}{0.350pt}}
\put(502,250){\rule[-0.175pt]{5.219pt}{0.350pt}}
\put(523,251){\rule[-0.175pt]{1.753pt}{0.350pt}}
\put(531,252){\rule[-0.175pt]{1.753pt}{0.350pt}}
\put(538,253){\rule[-0.175pt]{1.753pt}{0.350pt}}
\put(545,254){\rule[-0.175pt]{1.753pt}{0.350pt}}
\put(553,255){\rule[-0.175pt]{1.753pt}{0.350pt}}
\put(560,256){\rule[-0.175pt]{1.753pt}{0.350pt}}
\put(567,257){\rule[-0.175pt]{1.753pt}{0.350pt}}
\put(574,258){\rule[-0.175pt]{1.753pt}{0.350pt}}
\put(582,259){\rule[-0.175pt]{1.753pt}{0.350pt}}
\put(589,260){\rule[-0.175pt]{1.753pt}{0.350pt}}
\put(596,261){\rule[-0.175pt]{1.753pt}{0.350pt}}
\put(604,262){\rule[-0.175pt]{1.753pt}{0.350pt}}
\put(611,263){\rule[-0.175pt]{1.753pt}{0.350pt}}
\put(618,264){\rule[-0.175pt]{1.753pt}{0.350pt}}
\put(625,265){\rule[-0.175pt]{1.753pt}{0.350pt}}
\put(633,266){\rule[-0.175pt]{1.753pt}{0.350pt}}
\put(640,267){\rule[-0.175pt]{1.753pt}{0.350pt}}
\put(647,268){\rule[-0.175pt]{1.753pt}{0.350pt}}
\put(654,269){\rule[-0.175pt]{1.080pt}{0.350pt}}
\put(659,270){\rule[-0.175pt]{1.080pt}{0.350pt}}
\put(663,271){\rule[-0.175pt]{1.080pt}{0.350pt}}
\put(668,272){\rule[-0.175pt]{1.080pt}{0.350pt}}
\put(672,273){\rule[-0.175pt]{1.080pt}{0.350pt}}
\put(677,274){\rule[-0.175pt]{1.080pt}{0.350pt}}
\put(681,275){\rule[-0.175pt]{1.080pt}{0.350pt}}
\put(686,276){\rule[-0.175pt]{1.080pt}{0.350pt}}
\put(690,277){\rule[-0.175pt]{1.080pt}{0.350pt}}
\put(695,278){\rule[-0.175pt]{1.080pt}{0.350pt}}
\put(699,279){\rule[-0.175pt]{1.080pt}{0.350pt}}
\put(704,280){\rule[-0.175pt]{1.080pt}{0.350pt}}
\put(708,281){\rule[-0.175pt]{1.080pt}{0.350pt}}
\put(713,282){\rule[-0.175pt]{1.080pt}{0.350pt}}
\put(717,283){\rule[-0.175pt]{1.080pt}{0.350pt}}
\put(722,284){\rule[-0.175pt]{1.080pt}{0.350pt}}
\put(726,285){\rule[-0.175pt]{1.080pt}{0.350pt}}
\put(731,286){\rule[-0.175pt]{1.080pt}{0.350pt}}
\put(735,287){\rule[-0.175pt]{1.080pt}{0.350pt}}
\put(740,288){\rule[-0.175pt]{1.080pt}{0.350pt}}
\put(744,289){\rule[-0.175pt]{1.080pt}{0.350pt}}
\put(749,290){\rule[-0.175pt]{1.080pt}{0.350pt}}
\put(753,291){\rule[-0.175pt]{1.080pt}{0.350pt}}
\put(758,292){\rule[-0.175pt]{1.080pt}{0.350pt}}
\put(762,293){\rule[-0.175pt]{1.080pt}{0.350pt}}
\put(767,294){\rule[-0.175pt]{1.080pt}{0.350pt}}
\put(771,295){\rule[-0.175pt]{1.080pt}{0.350pt}}
\put(776,296){\rule[-0.175pt]{1.080pt}{0.350pt}}
\put(780,297){\rule[-0.175pt]{1.080pt}{0.350pt}}
\put(785,298){\rule[-0.175pt]{1.424pt}{0.350pt}}
\put(790,299){\rule[-0.175pt]{1.424pt}{0.350pt}}
\put(796,300){\rule[-0.175pt]{1.424pt}{0.350pt}}
\put(802,301){\rule[-0.175pt]{1.424pt}{0.350pt}}
\put(808,302){\rule[-0.175pt]{1.424pt}{0.350pt}}
\put(814,303){\rule[-0.175pt]{1.424pt}{0.350pt}}
\put(820,304){\rule[-0.175pt]{1.424pt}{0.350pt}}
\put(826,305){\rule[-0.175pt]{1.424pt}{0.350pt}}
\put(832,306){\rule[-0.175pt]{1.424pt}{0.350pt}}
\put(838,307){\rule[-0.175pt]{1.424pt}{0.350pt}}
\put(844,308){\rule[-0.175pt]{1.424pt}{0.350pt}}
\put(850,309){\rule[-0.175pt]{1.424pt}{0.350pt}}
\put(855,310){\rule[-0.175pt]{1.424pt}{0.350pt}}
\put(861,311){\rule[-0.175pt]{1.424pt}{0.350pt}}
\put(867,312){\rule[-0.175pt]{1.424pt}{0.350pt}}
\put(873,313){\rule[-0.175pt]{1.424pt}{0.350pt}}
\put(879,314){\rule[-0.175pt]{1.424pt}{0.350pt}}
\put(885,315){\rule[-0.175pt]{1.424pt}{0.350pt}}
\put(891,316){\rule[-0.175pt]{1.424pt}{0.350pt}}
\put(897,317){\rule[-0.175pt]{1.424pt}{0.350pt}}
\put(903,318){\rule[-0.175pt]{1.424pt}{0.350pt}}
\put(503,547){\circle*{24}}
\put(394,245){\circle*{24}}
\put(524,251){\circle*{24}}
\put(655,269){\circle*{24}}
\put(785,298){\circle*{24}}
\put(915,320){\circle*{24}}
\put(909,319){\rule[-0.175pt]{1.423pt}{0.350pt}}
\end{picture}

%% file: cluster_energies.tex
\begin{tabular} {rcdddd}
\multicolumn{1}{c}{}&
\multicolumn{1}{c}{$N$} &
\multicolumn{1}{c}{$E_{\rm T}$} &
\multicolumn{1}{c}{$E_0$} \\
\tableline

Ar			&3			&-2.553335364(1)		&-2.553335375(2) \\
Ne			&			&-1.7195589(3)			&-1.7195586(5) \\
{\tiny $1\over 2$}-Ne	&			&-1.308443(2)			&-1.308444(1) \\ 
\tableline
Ar			&4			&-5.1182368(2)			&-5.1182376(4) \\
Ne			&			&-3.464174(8)			&-3.464229(13) \\
{\tiny $1\over 2$}-Ne	&			&-2.64356(3)			&-2.64383(4) \\
\tableline
Ar			&5			&-7.78598(1)			&-7.7862(5) \\
Ne			&			&-5.29948(8)			&-5.3037(3) \\
{\tiny $1\over 2$}-Ne	&			&-4.0669(1)			&-4.0755(5) \\
\end{tabular}

%% file: e.tex
% GNUPLOT: LaTeX picture
\setlength{\unitlength}{0.240900pt}
\ifx\plotpoint\undefined\newsavebox{\plotpoint}\fi
\sbox{\plotpoint}{\rule[-0.200pt]{0.400pt}{0.400pt}}%
\begin{picture}(1500,900)(0,0)
\font\gnuplot=cmr10 at 10pt
\gnuplot
\sbox{\plotpoint}{\rule[-0.200pt]{0.400pt}{0.400pt}}%
\put(220.0,113.0){\rule[-0.200pt]{0.400pt}{184.048pt}}
\put(220.0,113.0){\rule[-0.200pt]{4.818pt}{0.400pt}}
\put(198,113){\makebox(0,0)[r]{$-1.0$}}
\put(1416.0,113.0){\rule[-0.200pt]{4.818pt}{0.400pt}}
\put(220.0,266.0){\rule[-0.200pt]{4.818pt}{0.400pt}}
\put(198,266){\makebox(0,0)[r]{$-0.8$}}
\put(1416.0,266.0){\rule[-0.200pt]{4.818pt}{0.400pt}}
\put(220.0,419.0){\rule[-0.200pt]{4.818pt}{0.400pt}}
\put(198,419){\makebox(0,0)[r]{$-0.6$}}
\put(1416.0,419.0){\rule[-0.200pt]{4.818pt}{0.400pt}}
\put(220.0,571.0){\rule[-0.200pt]{4.818pt}{0.400pt}}
\put(198,571){\makebox(0,0)[r]{$-0.4$}}
\put(1416.0,571.0){\rule[-0.200pt]{4.818pt}{0.400pt}}
\put(220.0,724.0){\rule[-0.200pt]{4.818pt}{0.400pt}}
\put(198,724){\makebox(0,0)[r]{$-0.2$}}
\put(1416.0,724.0){\rule[-0.200pt]{4.818pt}{0.400pt}}
\put(220.0,877.0){\rule[-0.200pt]{4.818pt}{0.400pt}}
\put(198,877){\makebox(0,0)[r]{$0.0$}}
\put(1416.0,877.0){\rule[-0.200pt]{4.818pt}{0.400pt}}
\put(220.0,113.0){\rule[-0.200pt]{0.400pt}{4.818pt}}
\put(220,68){\makebox(0,0){0}}
\put(220.0,857.0){\rule[-0.200pt]{0.400pt}{4.818pt}}
\put(355.0,113.0){\rule[-0.200pt]{0.400pt}{4.818pt}}
\put(355,68){\makebox(0,0){0.05}}
\put(355.0,857.0){\rule[-0.200pt]{0.400pt}{4.818pt}}
\put(490.0,113.0){\rule[-0.200pt]{0.400pt}{4.818pt}}
\put(490,68){\makebox(0,0){0.1}}
\put(490.0,857.0){\rule[-0.200pt]{0.400pt}{4.818pt}}
\put(625.0,113.0){\rule[-0.200pt]{0.400pt}{4.818pt}}
\put(625,68){\makebox(0,0){0.15}}
\put(625.0,857.0){\rule[-0.200pt]{0.400pt}{4.818pt}}
\put(760.0,113.0){\rule[-0.200pt]{0.400pt}{4.818pt}}
\put(760,68){\makebox(0,0){0.2}}
\put(760.0,857.0){\rule[-0.200pt]{0.400pt}{4.818pt}}
\put(896.0,113.0){\rule[-0.200pt]{0.400pt}{4.818pt}}
\put(896,68){\makebox(0,0){0.25}}
\put(896.0,857.0){\rule[-0.200pt]{0.400pt}{4.818pt}}
\put(1031.0,113.0){\rule[-0.200pt]{0.400pt}{4.818pt}}
\put(1031,68){\makebox(0,0){0.3}}
\put(1031.0,857.0){\rule[-0.200pt]{0.400pt}{4.818pt}}
\put(1166.0,113.0){\rule[-0.200pt]{0.400pt}{4.818pt}}
\put(1166,68){\makebox(0,0){0.35}}
\put(1166.0,857.0){\rule[-0.200pt]{0.400pt}{4.818pt}}
\put(1301.0,113.0){\rule[-0.200pt]{0.400pt}{4.818pt}}
\put(1301,68){\makebox(0,0){0.4}}
\put(1301.0,857.0){\rule[-0.200pt]{0.400pt}{4.818pt}}
\put(1436.0,113.0){\rule[-0.200pt]{0.400pt}{4.818pt}}
\put(1436,68){\makebox(0,0){0.45}}
\put(1436.0,857.0){\rule[-0.200pt]{0.400pt}{4.818pt}}
\put(220.0,113.0){\rule[-0.200pt]{292.934pt}{0.400pt}}
\put(1436.0,113.0){\rule[-0.200pt]{0.400pt}{184.048pt}}
\put(220.0,877.0){\rule[-0.200pt]{292.934pt}{0.400pt}}
\put(45,495){\makebox(0,0){$- \sqrt{\frac{E}{E_{\rm cl}}}$}}
\put(828,23){\makebox(0,0){$1 \over {\sqrt m}$}}
\put(291,342){\makebox(0,0)[l]{Ar}}
\put(448,476){\makebox(0,0)[l]{Ne}}
\put(542,552){\makebox(0,0)[l]{``$\frac{1}{2}$-Ne''}}
\put(1244,610){\makebox(0,0)[l]{He}}
\put(220.0,113.0){\rule[-0.200pt]{0.400pt}{184.048pt}}
\put(291,304){\vector(0,-1){115}}
\put(448,438){\vector(0,-1){115}}
\put(542,514){\vector(0,-1){114}}
\put(1244,648){\vector(0,1){114}}
\multiput(1242.92,648.00)(-0.499,0.878){127}{\rule{0.120pt}{0.802pt}}
\multiput(1243.17,648.00)(-65.000,112.336){2}{\rule{0.400pt}{0.401pt}}
\put(1179,762){\vector(-1,2){0}}
\put(1301,419){\makebox(0,0)[r]{$N = 3$}}
\multiput(220.00,113.58)(0.600,0.491){17}{\rule{0.580pt}{0.118pt}}
\multiput(220.00,112.17)(10.796,10.000){2}{\rule{0.290pt}{0.400pt}}
\multiput(232.00,123.58)(0.652,0.491){17}{\rule{0.620pt}{0.118pt}}
\multiput(232.00,122.17)(11.713,10.000){2}{\rule{0.310pt}{0.400pt}}
\multiput(245.00,133.58)(0.543,0.492){19}{\rule{0.536pt}{0.118pt}}
\multiput(245.00,132.17)(10.887,11.000){2}{\rule{0.268pt}{0.400pt}}
\multiput(257.00,144.58)(0.600,0.491){17}{\rule{0.580pt}{0.118pt}}
\multiput(257.00,143.17)(10.796,10.000){2}{\rule{0.290pt}{0.400pt}}
\multiput(269.00,154.58)(0.600,0.491){17}{\rule{0.580pt}{0.118pt}}
\multiput(269.00,153.17)(10.796,10.000){2}{\rule{0.290pt}{0.400pt}}
\multiput(281.00,164.58)(0.652,0.491){17}{\rule{0.620pt}{0.118pt}}
\multiput(281.00,163.17)(11.713,10.000){2}{\rule{0.310pt}{0.400pt}}
\multiput(294.00,174.58)(0.600,0.491){17}{\rule{0.580pt}{0.118pt}}
\multiput(294.00,173.17)(10.796,10.000){2}{\rule{0.290pt}{0.400pt}}
\multiput(306.00,184.58)(0.600,0.491){17}{\rule{0.580pt}{0.118pt}}
\multiput(306.00,183.17)(10.796,10.000){2}{\rule{0.290pt}{0.400pt}}
\multiput(318.00,194.58)(0.652,0.491){17}{\rule{0.620pt}{0.118pt}}
\multiput(318.00,193.17)(11.713,10.000){2}{\rule{0.310pt}{0.400pt}}
\multiput(331.00,204.58)(0.600,0.491){17}{\rule{0.580pt}{0.118pt}}
\multiput(331.00,203.17)(10.796,10.000){2}{\rule{0.290pt}{0.400pt}}
\multiput(343.00,214.58)(0.600,0.491){17}{\rule{0.580pt}{0.118pt}}
\multiput(343.00,213.17)(10.796,10.000){2}{\rule{0.290pt}{0.400pt}}
\multiput(355.00,224.58)(0.600,0.491){17}{\rule{0.580pt}{0.118pt}}
\multiput(355.00,223.17)(10.796,10.000){2}{\rule{0.290pt}{0.400pt}}
\multiput(367.00,234.58)(0.652,0.491){17}{\rule{0.620pt}{0.118pt}}
\multiput(367.00,233.17)(11.713,10.000){2}{\rule{0.310pt}{0.400pt}}
\multiput(380.00,244.58)(0.600,0.491){17}{\rule{0.580pt}{0.118pt}}
\multiput(380.00,243.17)(10.796,10.000){2}{\rule{0.290pt}{0.400pt}}
\multiput(392.00,254.58)(0.600,0.491){17}{\rule{0.580pt}{0.118pt}}
\multiput(392.00,253.17)(10.796,10.000){2}{\rule{0.290pt}{0.400pt}}
\multiput(404.00,264.58)(0.652,0.491){17}{\rule{0.620pt}{0.118pt}}
\multiput(404.00,263.17)(11.713,10.000){2}{\rule{0.310pt}{0.400pt}}
\multiput(417.00,274.58)(0.600,0.491){17}{\rule{0.580pt}{0.118pt}}
\multiput(417.00,273.17)(10.796,10.000){2}{\rule{0.290pt}{0.400pt}}
\multiput(429.00,284.59)(0.669,0.489){15}{\rule{0.633pt}{0.118pt}}
\multiput(429.00,283.17)(10.685,9.000){2}{\rule{0.317pt}{0.400pt}}
\multiput(441.00,293.58)(0.600,0.491){17}{\rule{0.580pt}{0.118pt}}
\multiput(441.00,292.17)(10.796,10.000){2}{\rule{0.290pt}{0.400pt}}
\multiput(453.00,303.58)(0.652,0.491){17}{\rule{0.620pt}{0.118pt}}
\multiput(453.00,302.17)(11.713,10.000){2}{\rule{0.310pt}{0.400pt}}
\multiput(466.00,313.58)(0.600,0.491){17}{\rule{0.580pt}{0.118pt}}
\multiput(466.00,312.17)(10.796,10.000){2}{\rule{0.290pt}{0.400pt}}
\multiput(478.00,323.59)(0.669,0.489){15}{\rule{0.633pt}{0.118pt}}
\multiput(478.00,322.17)(10.685,9.000){2}{\rule{0.317pt}{0.400pt}}
\multiput(490.00,332.58)(0.652,0.491){17}{\rule{0.620pt}{0.118pt}}
\multiput(490.00,331.17)(11.713,10.000){2}{\rule{0.310pt}{0.400pt}}
\multiput(503.00,342.59)(0.669,0.489){15}{\rule{0.633pt}{0.118pt}}
\multiput(503.00,341.17)(10.685,9.000){2}{\rule{0.317pt}{0.400pt}}
\multiput(515.00,351.58)(0.600,0.491){17}{\rule{0.580pt}{0.118pt}}
\multiput(515.00,350.17)(10.796,10.000){2}{\rule{0.290pt}{0.400pt}}
\multiput(527.00,361.58)(0.600,0.491){17}{\rule{0.580pt}{0.118pt}}
\multiput(527.00,360.17)(10.796,10.000){2}{\rule{0.290pt}{0.400pt}}
\multiput(539.00,371.59)(0.728,0.489){15}{\rule{0.678pt}{0.118pt}}
\multiput(539.00,370.17)(11.593,9.000){2}{\rule{0.339pt}{0.400pt}}
\multiput(552.00,380.59)(0.669,0.489){15}{\rule{0.633pt}{0.118pt}}
\multiput(552.00,379.17)(10.685,9.000){2}{\rule{0.317pt}{0.400pt}}
\multiput(564.00,389.58)(0.600,0.491){17}{\rule{0.580pt}{0.118pt}}
\multiput(564.00,388.17)(10.796,10.000){2}{\rule{0.290pt}{0.400pt}}
\multiput(576.00,399.59)(0.669,0.489){15}{\rule{0.633pt}{0.118pt}}
\multiput(576.00,398.17)(10.685,9.000){2}{\rule{0.317pt}{0.400pt}}
\multiput(588.00,408.58)(0.652,0.491){17}{\rule{0.620pt}{0.118pt}}
\multiput(588.00,407.17)(11.713,10.000){2}{\rule{0.310pt}{0.400pt}}
\multiput(601.00,418.59)(0.669,0.489){15}{\rule{0.633pt}{0.118pt}}
\multiput(601.00,417.17)(10.685,9.000){2}{\rule{0.317pt}{0.400pt}}
\multiput(613.00,427.59)(0.669,0.489){15}{\rule{0.633pt}{0.118pt}}
\multiput(613.00,426.17)(10.685,9.000){2}{\rule{0.317pt}{0.400pt}}
\multiput(625.00,436.59)(0.728,0.489){15}{\rule{0.678pt}{0.118pt}}
\multiput(625.00,435.17)(11.593,9.000){2}{\rule{0.339pt}{0.400pt}}
\multiput(638.00,445.58)(0.600,0.491){17}{\rule{0.580pt}{0.118pt}}
\multiput(638.00,444.17)(10.796,10.000){2}{\rule{0.290pt}{0.400pt}}
\multiput(650.00,455.59)(0.669,0.489){15}{\rule{0.633pt}{0.118pt}}
\multiput(650.00,454.17)(10.685,9.000){2}{\rule{0.317pt}{0.400pt}}
\multiput(662.00,464.59)(0.669,0.489){15}{\rule{0.633pt}{0.118pt}}
\multiput(662.00,463.17)(10.685,9.000){2}{\rule{0.317pt}{0.400pt}}
\multiput(674.00,473.59)(0.728,0.489){15}{\rule{0.678pt}{0.118pt}}
\multiput(674.00,472.17)(11.593,9.000){2}{\rule{0.339pt}{0.400pt}}
\multiput(687.00,482.59)(0.669,0.489){15}{\rule{0.633pt}{0.118pt}}
\multiput(687.00,481.17)(10.685,9.000){2}{\rule{0.317pt}{0.400pt}}
\multiput(699.00,491.59)(0.669,0.489){15}{\rule{0.633pt}{0.118pt}}
\multiput(699.00,490.17)(10.685,9.000){2}{\rule{0.317pt}{0.400pt}}
\multiput(711.00,500.59)(0.728,0.489){15}{\rule{0.678pt}{0.118pt}}
\multiput(711.00,499.17)(11.593,9.000){2}{\rule{0.339pt}{0.400pt}}
\multiput(724.00,509.59)(0.758,0.488){13}{\rule{0.700pt}{0.117pt}}
\multiput(724.00,508.17)(10.547,8.000){2}{\rule{0.350pt}{0.400pt}}
\multiput(736.00,517.59)(0.669,0.489){15}{\rule{0.633pt}{0.118pt}}
\multiput(736.00,516.17)(10.685,9.000){2}{\rule{0.317pt}{0.400pt}}
\multiput(748.00,526.59)(0.669,0.489){15}{\rule{0.633pt}{0.118pt}}
\multiput(748.00,525.17)(10.685,9.000){2}{\rule{0.317pt}{0.400pt}}
\multiput(760.00,535.59)(0.728,0.489){15}{\rule{0.678pt}{0.118pt}}
\multiput(760.00,534.17)(11.593,9.000){2}{\rule{0.339pt}{0.400pt}}
\multiput(773.00,544.59)(0.758,0.488){13}{\rule{0.700pt}{0.117pt}}
\multiput(773.00,543.17)(10.547,8.000){2}{\rule{0.350pt}{0.400pt}}
\multiput(785.00,552.59)(0.669,0.489){15}{\rule{0.633pt}{0.118pt}}
\multiput(785.00,551.17)(10.685,9.000){2}{\rule{0.317pt}{0.400pt}}
\multiput(797.00,561.59)(0.824,0.488){13}{\rule{0.750pt}{0.117pt}}
\multiput(797.00,560.17)(11.443,8.000){2}{\rule{0.375pt}{0.400pt}}
\multiput(810.00,569.59)(0.669,0.489){15}{\rule{0.633pt}{0.118pt}}
\multiput(810.00,568.17)(10.685,9.000){2}{\rule{0.317pt}{0.400pt}}
\multiput(822.00,578.59)(0.758,0.488){13}{\rule{0.700pt}{0.117pt}}
\multiput(822.00,577.17)(10.547,8.000){2}{\rule{0.350pt}{0.400pt}}
\multiput(834.00,586.59)(0.669,0.489){15}{\rule{0.633pt}{0.118pt}}
\multiput(834.00,585.17)(10.685,9.000){2}{\rule{0.317pt}{0.400pt}}
\multiput(846.00,595.59)(0.824,0.488){13}{\rule{0.750pt}{0.117pt}}
\multiput(846.00,594.17)(11.443,8.000){2}{\rule{0.375pt}{0.400pt}}
\multiput(859.00,603.59)(0.758,0.488){13}{\rule{0.700pt}{0.117pt}}
\multiput(859.00,602.17)(10.547,8.000){2}{\rule{0.350pt}{0.400pt}}
\multiput(871.00,611.59)(0.758,0.488){13}{\rule{0.700pt}{0.117pt}}
\multiput(871.00,610.17)(10.547,8.000){2}{\rule{0.350pt}{0.400pt}}
\multiput(883.00,619.59)(0.824,0.488){13}{\rule{0.750pt}{0.117pt}}
\multiput(883.00,618.17)(11.443,8.000){2}{\rule{0.375pt}{0.400pt}}
\multiput(896.00,627.59)(0.758,0.488){13}{\rule{0.700pt}{0.117pt}}
\multiput(896.00,626.17)(10.547,8.000){2}{\rule{0.350pt}{0.400pt}}
\multiput(908.00,635.59)(0.758,0.488){13}{\rule{0.700pt}{0.117pt}}
\multiput(908.00,634.17)(10.547,8.000){2}{\rule{0.350pt}{0.400pt}}
\multiput(920.00,643.59)(0.758,0.488){13}{\rule{0.700pt}{0.117pt}}
\multiput(920.00,642.17)(10.547,8.000){2}{\rule{0.350pt}{0.400pt}}
\multiput(932.00,651.59)(0.824,0.488){13}{\rule{0.750pt}{0.117pt}}
\multiput(932.00,650.17)(11.443,8.000){2}{\rule{0.375pt}{0.400pt}}
\multiput(945.00,659.59)(0.758,0.488){13}{\rule{0.700pt}{0.117pt}}
\multiput(945.00,658.17)(10.547,8.000){2}{\rule{0.350pt}{0.400pt}}
\multiput(957.00,667.59)(0.758,0.488){13}{\rule{0.700pt}{0.117pt}}
\multiput(957.00,666.17)(10.547,8.000){2}{\rule{0.350pt}{0.400pt}}
\multiput(969.00,675.59)(0.824,0.488){13}{\rule{0.750pt}{0.117pt}}
\multiput(969.00,674.17)(11.443,8.000){2}{\rule{0.375pt}{0.400pt}}
\multiput(982.00,683.59)(0.874,0.485){11}{\rule{0.786pt}{0.117pt}}
\multiput(982.00,682.17)(10.369,7.000){2}{\rule{0.393pt}{0.400pt}}
\multiput(994.00,690.59)(0.758,0.488){13}{\rule{0.700pt}{0.117pt}}
\multiput(994.00,689.17)(10.547,8.000){2}{\rule{0.350pt}{0.400pt}}
\multiput(1006.00,698.59)(0.874,0.485){11}{\rule{0.786pt}{0.117pt}}
\multiput(1006.00,697.17)(10.369,7.000){2}{\rule{0.393pt}{0.400pt}}
\multiput(1018.00,705.59)(0.824,0.488){13}{\rule{0.750pt}{0.117pt}}
\multiput(1018.00,704.17)(11.443,8.000){2}{\rule{0.375pt}{0.400pt}}
\multiput(1031.00,713.59)(0.758,0.488){13}{\rule{0.700pt}{0.117pt}}
\multiput(1031.00,712.17)(10.547,8.000){2}{\rule{0.350pt}{0.400pt}}
\multiput(1043.00,721.59)(0.874,0.485){11}{\rule{0.786pt}{0.117pt}}
\multiput(1043.00,720.17)(10.369,7.000){2}{\rule{0.393pt}{0.400pt}}
\multiput(1055.00,728.59)(0.950,0.485){11}{\rule{0.843pt}{0.117pt}}
\multiput(1055.00,727.17)(11.251,7.000){2}{\rule{0.421pt}{0.400pt}}
\multiput(1068.00,735.59)(0.758,0.488){13}{\rule{0.700pt}{0.117pt}}
\multiput(1068.00,734.17)(10.547,8.000){2}{\rule{0.350pt}{0.400pt}}
\multiput(1080.00,743.59)(0.874,0.485){11}{\rule{0.786pt}{0.117pt}}
\multiput(1080.00,742.17)(10.369,7.000){2}{\rule{0.393pt}{0.400pt}}
\multiput(1092.00,750.59)(0.758,0.488){13}{\rule{0.700pt}{0.117pt}}
\multiput(1092.00,749.17)(10.547,8.000){2}{\rule{0.350pt}{0.400pt}}
\multiput(1104.00,758.59)(0.950,0.485){11}{\rule{0.843pt}{0.117pt}}
\multiput(1104.00,757.17)(11.251,7.000){2}{\rule{0.421pt}{0.400pt}}
\multiput(1117.00,765.59)(0.874,0.485){11}{\rule{0.786pt}{0.117pt}}
\multiput(1117.00,764.17)(10.369,7.000){2}{\rule{0.393pt}{0.400pt}}
\multiput(1129.00,772.59)(0.758,0.488){13}{\rule{0.700pt}{0.117pt}}
\multiput(1129.00,771.17)(10.547,8.000){2}{\rule{0.350pt}{0.400pt}}
\multiput(1141.00,780.59)(0.874,0.485){11}{\rule{0.786pt}{0.117pt}}
\multiput(1141.00,779.17)(10.369,7.000){2}{\rule{0.393pt}{0.400pt}}
\multiput(1153.00,787.59)(0.950,0.485){11}{\rule{0.843pt}{0.117pt}}
\multiput(1153.00,786.17)(11.251,7.000){2}{\rule{0.421pt}{0.400pt}}
\multiput(1166.00,794.59)(0.758,0.488){13}{\rule{0.700pt}{0.117pt}}
\multiput(1166.00,793.17)(10.547,8.000){2}{\rule{0.350pt}{0.400pt}}
\multiput(1178.00,802.59)(0.874,0.485){11}{\rule{0.786pt}{0.117pt}}
\multiput(1178.00,801.17)(10.369,7.000){2}{\rule{0.393pt}{0.400pt}}
\multiput(1190.00,809.59)(0.950,0.485){11}{\rule{0.843pt}{0.117pt}}
\multiput(1190.00,808.17)(11.251,7.000){2}{\rule{0.421pt}{0.400pt}}
\multiput(1203.00,816.59)(0.874,0.485){11}{\rule{0.786pt}{0.117pt}}
\multiput(1203.00,815.17)(10.369,7.000){2}{\rule{0.393pt}{0.400pt}}
\multiput(1215.00,823.59)(0.758,0.488){13}{\rule{0.700pt}{0.117pt}}
\multiput(1215.00,822.17)(10.547,8.000){2}{\rule{0.350pt}{0.400pt}}
\multiput(1227.00,831.59)(0.874,0.485){11}{\rule{0.786pt}{0.117pt}}
\multiput(1227.00,830.17)(10.369,7.000){2}{\rule{0.393pt}{0.400pt}}
\multiput(1239.00,838.59)(0.950,0.485){11}{\rule{0.843pt}{0.117pt}}
\multiput(1239.00,837.17)(11.251,7.000){2}{\rule{0.421pt}{0.400pt}}
\multiput(1252.00,845.59)(0.758,0.488){13}{\rule{0.700pt}{0.117pt}}
\multiput(1252.00,844.17)(10.547,8.000){2}{\rule{0.350pt}{0.400pt}}
\multiput(1264.00,853.59)(0.874,0.485){11}{\rule{0.786pt}{0.117pt}}
\multiput(1264.00,852.17)(10.369,7.000){2}{\rule{0.393pt}{0.400pt}}
\multiput(1276.00,860.59)(0.950,0.485){11}{\rule{0.843pt}{0.117pt}}
\multiput(1276.00,859.17)(11.251,7.000){2}{\rule{0.421pt}{0.400pt}}
\multiput(1289.00,867.59)(0.874,0.485){11}{\rule{0.786pt}{0.117pt}}
\multiput(1289.00,866.17)(10.369,7.000){2}{\rule{0.393pt}{0.400pt}}
\multiput(1301.00,874.61)(0.909,0.447){3}{\rule{0.767pt}{0.108pt}}
\multiput(1301.00,873.17)(3.409,3.000){2}{\rule{0.383pt}{0.400pt}}
\put(1323.0,419.0){\rule[-0.200pt]{15.899pt}{0.400pt}}
\sbox{\plotpoint}{\rule[-0.500pt]{1.000pt}{1.000pt}}%
\put(1301,374){\makebox(0,0)[r]{$N = 4$}}
\multiput(1323,374)(20.756,0.000){4}{\usebox{\plotpoint}}
\put(1389,374){\usebox{\plotpoint}}
\put(220,113){\usebox{\plotpoint}}
\put(220.00,113.00){\usebox{\plotpoint}}
\put(236.07,126.13){\usebox{\plotpoint}}
\put(252.29,139.07){\usebox{\plotpoint}}
\put(268.23,152.36){\usebox{\plotpoint}}
\multiput(269,153)(15.945,13.287){0}{\usebox{\plotpoint}}
\put(284.28,165.52){\usebox{\plotpoint}}
\put(300.52,178.44){\usebox{\plotpoint}}
\put(316.47,191.72){\usebox{\plotpoint}}
\multiput(318,193)(16.451,12.655){0}{\usebox{\plotpoint}}
\put(332.81,204.51){\usebox{\plotpoint}}
\put(348.76,217.80){\usebox{\plotpoint}}
\put(364.70,231.09){\usebox{\plotpoint}}
\multiput(367,233)(16.451,12.655){0}{\usebox{\plotpoint}}
\put(381.09,243.82){\usebox{\plotpoint}}
\put(397.47,256.56){\usebox{\plotpoint}}
\put(413.71,269.47){\usebox{\plotpoint}}
\multiput(417,272)(15.945,13.287){0}{\usebox{\plotpoint}}
\put(429.79,282.59){\usebox{\plotpoint}}
\put(446.18,295.32){\usebox{\plotpoint}}
\put(462.42,308.24){\usebox{\plotpoint}}
\multiput(466,311)(16.604,12.453){0}{\usebox{\plotpoint}}
\put(478.95,320.79){\usebox{\plotpoint}}
\put(495.24,333.63){\usebox{\plotpoint}}
\put(511.69,346.24){\usebox{\plotpoint}}
\multiput(515,349)(16.604,12.453){0}{\usebox{\plotpoint}}
\put(528.11,358.93){\usebox{\plotpoint}}
\put(544.41,371.75){\usebox{\plotpoint}}
\put(560.86,384.38){\usebox{\plotpoint}}
\multiput(564,387)(16.604,12.453){0}{\usebox{\plotpoint}}
\put(577.28,397.06){\usebox{\plotpoint}}
\put(593.59,409.87){\usebox{\plotpoint}}
\put(610.39,422.05){\usebox{\plotpoint}}
\multiput(613,424)(16.604,12.453){0}{\usebox{\plotpoint}}
\put(627.05,434.42){\usebox{\plotpoint}}
\put(643.72,446.76){\usebox{\plotpoint}}
\put(660.06,459.55){\usebox{\plotpoint}}
\multiput(662,461)(16.604,12.453){0}{\usebox{\plotpoint}}
\put(676.74,471.90){\usebox{\plotpoint}}
\put(693.62,483.97){\usebox{\plotpoint}}
\put(710.23,496.42){\usebox{\plotpoint}}
\multiput(711,497)(17.677,10.878){0}{\usebox{\plotpoint}}
\put(727.62,507.71){\usebox{\plotpoint}}
\put(744.22,520.17){\usebox{\plotpoint}}
\multiput(748,523)(17.270,11.513){0}{\usebox{\plotpoint}}
\put(761.33,531.92){\usebox{\plotpoint}}
\put(778.45,543.64){\usebox{\plotpoint}}
\put(795.31,555.73){\usebox{\plotpoint}}
\multiput(797,557)(17.677,10.878){0}{\usebox{\plotpoint}}
\put(812.81,566.88){\usebox{\plotpoint}}
\put(830.08,578.39){\usebox{\plotpoint}}
\multiput(834,581)(17.270,11.513){0}{\usebox{\plotpoint}}
\put(847.38,589.85){\usebox{\plotpoint}}
\put(864.92,600.95){\usebox{\plotpoint}}
\put(882.19,612.46){\usebox{\plotpoint}}
\multiput(883,613)(17.677,10.878){0}{\usebox{\plotpoint}}
\put(899.90,623.28){\usebox{\plotpoint}}
\put(917.47,634.31){\usebox{\plotpoint}}
\multiput(920,636)(17.928,10.458){0}{\usebox{\plotpoint}}
\put(935.26,645.00){\usebox{\plotpoint}}
\put(953.04,655.69){\usebox{\plotpoint}}
\multiput(957,658)(17.928,10.458){0}{\usebox{\plotpoint}}
\put(970.95,666.20){\usebox{\plotpoint}}
\put(988.72,676.92){\usebox{\plotpoint}}
\multiput(994,680)(17.928,10.458){0}{\usebox{\plotpoint}}
\put(1006.64,687.38){\usebox{\plotpoint}}
\put(1024.70,697.61){\usebox{\plotpoint}}
\put(1042.75,707.85){\usebox{\plotpoint}}
\multiput(1043,708)(18.564,9.282){0}{\usebox{\plotpoint}}
\put(1061.20,717.34){\usebox{\plotpoint}}
\put(1079.26,727.57){\usebox{\plotpoint}}
\multiput(1080,728)(17.928,10.458){0}{\usebox{\plotpoint}}
\put(1097.37,737.69){\usebox{\plotpoint}}
\put(1115.75,747.33){\usebox{\plotpoint}}
\multiput(1117,748)(17.928,10.458){0}{\usebox{\plotpoint}}
\put(1133.87,757.43){\usebox{\plotpoint}}
\put(1152.04,767.44){\usebox{\plotpoint}}
\multiput(1153,768)(18.845,8.698){0}{\usebox{\plotpoint}}
\put(1170.60,776.68){\usebox{\plotpoint}}
\put(1188.53,787.14){\usebox{\plotpoint}}
\multiput(1190,788)(18.845,8.698){0}{\usebox{\plotpoint}}
\put(1207.09,796.39){\usebox{\plotpoint}}
\put(1225.38,806.19){\usebox{\plotpoint}}
\multiput(1227,807)(17.928,10.458){0}{\usebox{\plotpoint}}
\put(1243.58,816.11){\usebox{\plotpoint}}
\put(1261.92,825.79){\usebox{\plotpoint}}
\multiput(1264,827)(18.564,9.282){0}{\usebox{\plotpoint}}
\put(1280.34,835.34){\usebox{\plotpoint}}
\put(1298.77,844.88){\usebox{\plotpoint}}
\multiput(1301,846)(17.928,10.458){0}{\usebox{\plotpoint}}
\put(1316.91,854.95){\usebox{\plotpoint}}
\put(1335.63,863.91){\usebox{\plotpoint}}
\multiput(1338,865)(17.928,10.458){0}{\usebox{\plotpoint}}
\put(1353.80,873.90){\usebox{\plotpoint}}
\put(1360,877){\usebox{\plotpoint}}
\sbox{\plotpoint}{\rule[-0.600pt]{1.200pt}{1.200pt}}%
\put(1301,329){\makebox(0,0)[r]{$N = 5$}}
\put(1323.0,329.0){\rule[-0.600pt]{15.899pt}{1.200pt}}
\put(220,113){\usebox{\plotpoint}}
\multiput(220.00,115.24)(0.531,0.502){10}{\rule{1.740pt}{0.121pt}}
\multiput(220.00,110.51)(8.389,10.000){2}{\rule{0.870pt}{1.200pt}}
\multiput(232.00,125.24)(0.587,0.502){10}{\rule{1.860pt}{0.121pt}}
\multiput(232.00,120.51)(9.139,10.000){2}{\rule{0.930pt}{1.200pt}}
\multiput(245.00,135.24)(0.531,0.502){10}{\rule{1.740pt}{0.121pt}}
\multiput(245.00,130.51)(8.389,10.000){2}{\rule{0.870pt}{1.200pt}}
\multiput(257.00,145.24)(0.531,0.502){10}{\rule{1.740pt}{0.121pt}}
\multiput(257.00,140.51)(8.389,10.000){2}{\rule{0.870pt}{1.200pt}}
\multiput(269.00,155.24)(0.531,0.502){10}{\rule{1.740pt}{0.121pt}}
\multiput(269.00,150.51)(8.389,10.000){2}{\rule{0.870pt}{1.200pt}}
\multiput(281.00,165.24)(0.651,0.502){8}{\rule{2.033pt}{0.121pt}}
\multiput(281.00,160.51)(8.780,9.000){2}{\rule{1.017pt}{1.200pt}}
\multiput(294.00,174.24)(0.531,0.502){10}{\rule{1.740pt}{0.121pt}}
\multiput(294.00,169.51)(8.389,10.000){2}{\rule{0.870pt}{1.200pt}}
\multiput(306.00,184.24)(0.531,0.502){10}{\rule{1.740pt}{0.121pt}}
\multiput(306.00,179.51)(8.389,10.000){2}{\rule{0.870pt}{1.200pt}}
\multiput(318.00,194.24)(0.587,0.502){10}{\rule{1.860pt}{0.121pt}}
\multiput(318.00,189.51)(9.139,10.000){2}{\rule{0.930pt}{1.200pt}}
\multiput(331.00,204.24)(0.531,0.502){10}{\rule{1.740pt}{0.121pt}}
\multiput(331.00,199.51)(8.389,10.000){2}{\rule{0.870pt}{1.200pt}}
\multiput(343.00,214.24)(0.588,0.502){8}{\rule{1.900pt}{0.121pt}}
\multiput(343.00,209.51)(8.056,9.000){2}{\rule{0.950pt}{1.200pt}}
\multiput(355.00,223.24)(0.531,0.502){10}{\rule{1.740pt}{0.121pt}}
\multiput(355.00,218.51)(8.389,10.000){2}{\rule{0.870pt}{1.200pt}}
\multiput(367.00,233.24)(0.587,0.502){10}{\rule{1.860pt}{0.121pt}}
\multiput(367.00,228.51)(9.139,10.000){2}{\rule{0.930pt}{1.200pt}}
\multiput(380.00,243.24)(0.588,0.502){8}{\rule{1.900pt}{0.121pt}}
\multiput(380.00,238.51)(8.056,9.000){2}{\rule{0.950pt}{1.200pt}}
\multiput(392.00,252.24)(0.531,0.502){10}{\rule{1.740pt}{0.121pt}}
\multiput(392.00,247.51)(8.389,10.000){2}{\rule{0.870pt}{1.200pt}}
\multiput(404.00,262.24)(0.587,0.502){10}{\rule{1.860pt}{0.121pt}}
\multiput(404.00,257.51)(9.139,10.000){2}{\rule{0.930pt}{1.200pt}}
\multiput(417.00,272.24)(0.588,0.502){8}{\rule{1.900pt}{0.121pt}}
\multiput(417.00,267.51)(8.056,9.000){2}{\rule{0.950pt}{1.200pt}}
\multiput(429.00,281.24)(0.531,0.502){10}{\rule{1.740pt}{0.121pt}}
\multiput(429.00,276.51)(8.389,10.000){2}{\rule{0.870pt}{1.200pt}}
\multiput(441.00,291.24)(0.588,0.502){8}{\rule{1.900pt}{0.121pt}}
\multiput(441.00,286.51)(8.056,9.000){2}{\rule{0.950pt}{1.200pt}}
\multiput(453.00,300.24)(0.587,0.502){10}{\rule{1.860pt}{0.121pt}}
\multiput(453.00,295.51)(9.139,10.000){2}{\rule{0.930pt}{1.200pt}}
\multiput(466.00,310.24)(0.588,0.502){8}{\rule{1.900pt}{0.121pt}}
\multiput(466.00,305.51)(8.056,9.000){2}{\rule{0.950pt}{1.200pt}}
\multiput(478.00,319.24)(0.531,0.502){10}{\rule{1.740pt}{0.121pt}}
\multiput(478.00,314.51)(8.389,10.000){2}{\rule{0.870pt}{1.200pt}}
\multiput(490.00,329.24)(0.651,0.502){8}{\rule{2.033pt}{0.121pt}}
\multiput(490.00,324.51)(8.780,9.000){2}{\rule{1.017pt}{1.200pt}}
\multiput(503.00,338.24)(0.588,0.502){8}{\rule{1.900pt}{0.121pt}}
\multiput(503.00,333.51)(8.056,9.000){2}{\rule{0.950pt}{1.200pt}}
\multiput(515.00,347.24)(0.531,0.502){10}{\rule{1.740pt}{0.121pt}}
\multiput(515.00,342.51)(8.389,10.000){2}{\rule{0.870pt}{1.200pt}}
\multiput(527.00,357.24)(0.588,0.502){8}{\rule{1.900pt}{0.121pt}}
\multiput(527.00,352.51)(8.056,9.000){2}{\rule{0.950pt}{1.200pt}}
\multiput(539.00,366.24)(0.651,0.502){8}{\rule{2.033pt}{0.121pt}}
\multiput(539.00,361.51)(8.780,9.000){2}{\rule{1.017pt}{1.200pt}}
\multiput(552.00,375.24)(0.588,0.502){8}{\rule{1.900pt}{0.121pt}}
\multiput(552.00,370.51)(8.056,9.000){2}{\rule{0.950pt}{1.200pt}}
\multiput(564.00,384.24)(0.588,0.502){8}{\rule{1.900pt}{0.121pt}}
\multiput(564.00,379.51)(8.056,9.000){2}{\rule{0.950pt}{1.200pt}}
\multiput(576.00,393.24)(0.531,0.502){10}{\rule{1.740pt}{0.121pt}}
\multiput(576.00,388.51)(8.389,10.000){2}{\rule{0.870pt}{1.200pt}}
\multiput(588.00,403.24)(0.651,0.502){8}{\rule{2.033pt}{0.121pt}}
\multiput(588.00,398.51)(8.780,9.000){2}{\rule{1.017pt}{1.200pt}}
\multiput(601.00,412.24)(0.657,0.503){6}{\rule{2.100pt}{0.121pt}}
\multiput(601.00,407.51)(7.641,8.000){2}{\rule{1.050pt}{1.200pt}}
\multiput(613.00,420.24)(0.588,0.502){8}{\rule{1.900pt}{0.121pt}}
\multiput(613.00,415.51)(8.056,9.000){2}{\rule{0.950pt}{1.200pt}}
\multiput(625.00,429.24)(0.651,0.502){8}{\rule{2.033pt}{0.121pt}}
\multiput(625.00,424.51)(8.780,9.000){2}{\rule{1.017pt}{1.200pt}}
\multiput(638.00,438.24)(0.588,0.502){8}{\rule{1.900pt}{0.121pt}}
\multiput(638.00,433.51)(8.056,9.000){2}{\rule{0.950pt}{1.200pt}}
\multiput(650.00,447.24)(0.588,0.502){8}{\rule{1.900pt}{0.121pt}}
\multiput(650.00,442.51)(8.056,9.000){2}{\rule{0.950pt}{1.200pt}}
\multiput(662.00,456.24)(0.657,0.503){6}{\rule{2.100pt}{0.121pt}}
\multiput(662.00,451.51)(7.641,8.000){2}{\rule{1.050pt}{1.200pt}}
\multiput(674.00,464.24)(0.651,0.502){8}{\rule{2.033pt}{0.121pt}}
\multiput(674.00,459.51)(8.780,9.000){2}{\rule{1.017pt}{1.200pt}}
\multiput(687.00,473.24)(0.588,0.502){8}{\rule{1.900pt}{0.121pt}}
\multiput(687.00,468.51)(8.056,9.000){2}{\rule{0.950pt}{1.200pt}}
\multiput(699.00,482.24)(0.657,0.503){6}{\rule{2.100pt}{0.121pt}}
\multiput(699.00,477.51)(7.641,8.000){2}{\rule{1.050pt}{1.200pt}}
\multiput(711.00,490.24)(0.651,0.502){8}{\rule{2.033pt}{0.121pt}}
\multiput(711.00,485.51)(8.780,9.000){2}{\rule{1.017pt}{1.200pt}}
\multiput(724.00,499.24)(0.657,0.503){6}{\rule{2.100pt}{0.121pt}}
\multiput(724.00,494.51)(7.641,8.000){2}{\rule{1.050pt}{1.200pt}}
\multiput(736.00,507.24)(0.657,0.503){6}{\rule{2.100pt}{0.121pt}}
\multiput(736.00,502.51)(7.641,8.000){2}{\rule{1.050pt}{1.200pt}}
\multiput(748.00,515.24)(0.588,0.502){8}{\rule{1.900pt}{0.121pt}}
\multiput(748.00,510.51)(8.056,9.000){2}{\rule{0.950pt}{1.200pt}}
\multiput(760.00,524.24)(0.732,0.503){6}{\rule{2.250pt}{0.121pt}}
\multiput(760.00,519.51)(8.330,8.000){2}{\rule{1.125pt}{1.200pt}}
\multiput(773.00,532.24)(0.657,0.503){6}{\rule{2.100pt}{0.121pt}}
\multiput(773.00,527.51)(7.641,8.000){2}{\rule{1.050pt}{1.200pt}}
\multiput(785.00,540.24)(0.657,0.503){6}{\rule{2.100pt}{0.121pt}}
\multiput(785.00,535.51)(7.641,8.000){2}{\rule{1.050pt}{1.200pt}}
\multiput(797.00,548.24)(0.732,0.503){6}{\rule{2.250pt}{0.121pt}}
\multiput(797.00,543.51)(8.330,8.000){2}{\rule{1.125pt}{1.200pt}}
\multiput(810.00,556.24)(0.657,0.503){6}{\rule{2.100pt}{0.121pt}}
\multiput(810.00,551.51)(7.641,8.000){2}{\rule{1.050pt}{1.200pt}}
\multiput(822.00,564.24)(0.657,0.503){6}{\rule{2.100pt}{0.121pt}}
\multiput(822.00,559.51)(7.641,8.000){2}{\rule{1.050pt}{1.200pt}}
\multiput(834.00,572.24)(0.657,0.503){6}{\rule{2.100pt}{0.121pt}}
\multiput(834.00,567.51)(7.641,8.000){2}{\rule{1.050pt}{1.200pt}}
\multiput(846.00,580.24)(0.835,0.505){4}{\rule{2.529pt}{0.122pt}}
\multiput(846.00,575.51)(7.752,7.000){2}{\rule{1.264pt}{1.200pt}}
\multiput(859.00,587.24)(0.657,0.503){6}{\rule{2.100pt}{0.121pt}}
\multiput(859.00,582.51)(7.641,8.000){2}{\rule{1.050pt}{1.200pt}}
\multiput(871.00,595.24)(0.657,0.503){6}{\rule{2.100pt}{0.121pt}}
\multiput(871.00,590.51)(7.641,8.000){2}{\rule{1.050pt}{1.200pt}}
\multiput(883.00,603.24)(0.835,0.505){4}{\rule{2.529pt}{0.122pt}}
\multiput(883.00,598.51)(7.752,7.000){2}{\rule{1.264pt}{1.200pt}}
\multiput(896.00,610.24)(0.657,0.503){6}{\rule{2.100pt}{0.121pt}}
\multiput(896.00,605.51)(7.641,8.000){2}{\rule{1.050pt}{1.200pt}}
\multiput(908.00,618.24)(0.738,0.505){4}{\rule{2.357pt}{0.122pt}}
\multiput(908.00,613.51)(7.108,7.000){2}{\rule{1.179pt}{1.200pt}}
\multiput(920.00,625.24)(0.657,0.503){6}{\rule{2.100pt}{0.121pt}}
\multiput(920.00,620.51)(7.641,8.000){2}{\rule{1.050pt}{1.200pt}}
\multiput(932.00,633.24)(0.835,0.505){4}{\rule{2.529pt}{0.122pt}}
\multiput(932.00,628.51)(7.752,7.000){2}{\rule{1.264pt}{1.200pt}}
\multiput(945.00,640.24)(0.738,0.505){4}{\rule{2.357pt}{0.122pt}}
\multiput(945.00,635.51)(7.108,7.000){2}{\rule{1.179pt}{1.200pt}}
\multiput(957.00,647.24)(0.738,0.505){4}{\rule{2.357pt}{0.122pt}}
\multiput(957.00,642.51)(7.108,7.000){2}{\rule{1.179pt}{1.200pt}}
\multiput(969.00,654.24)(0.835,0.505){4}{\rule{2.529pt}{0.122pt}}
\multiput(969.00,649.51)(7.752,7.000){2}{\rule{1.264pt}{1.200pt}}
\multiput(982.00,661.24)(0.738,0.505){4}{\rule{2.357pt}{0.122pt}}
\multiput(982.00,656.51)(7.108,7.000){2}{\rule{1.179pt}{1.200pt}}
\multiput(994.00,668.24)(0.738,0.505){4}{\rule{2.357pt}{0.122pt}}
\multiput(994.00,663.51)(7.108,7.000){2}{\rule{1.179pt}{1.200pt}}
\multiput(1006.00,675.24)(0.738,0.505){4}{\rule{2.357pt}{0.122pt}}
\multiput(1006.00,670.51)(7.108,7.000){2}{\rule{1.179pt}{1.200pt}}
\multiput(1018.00,682.24)(0.835,0.505){4}{\rule{2.529pt}{0.122pt}}
\multiput(1018.00,677.51)(7.752,7.000){2}{\rule{1.264pt}{1.200pt}}
\multiput(1031.00,689.24)(0.738,0.505){4}{\rule{2.357pt}{0.122pt}}
\multiput(1031.00,684.51)(7.108,7.000){2}{\rule{1.179pt}{1.200pt}}
\multiput(1043.00,696.24)(0.792,0.509){2}{\rule{2.700pt}{0.123pt}}
\multiput(1043.00,691.51)(6.396,6.000){2}{\rule{1.350pt}{1.200pt}}
\multiput(1055.00,702.24)(0.835,0.505){4}{\rule{2.529pt}{0.122pt}}
\multiput(1055.00,697.51)(7.752,7.000){2}{\rule{1.264pt}{1.200pt}}
\multiput(1068.00,709.24)(0.738,0.505){4}{\rule{2.357pt}{0.122pt}}
\multiput(1068.00,704.51)(7.108,7.000){2}{\rule{1.179pt}{1.200pt}}
\multiput(1080.00,716.24)(0.792,0.509){2}{\rule{2.700pt}{0.123pt}}
\multiput(1080.00,711.51)(6.396,6.000){2}{\rule{1.350pt}{1.200pt}}
\multiput(1092.00,722.24)(0.738,0.505){4}{\rule{2.357pt}{0.122pt}}
\multiput(1092.00,717.51)(7.108,7.000){2}{\rule{1.179pt}{1.200pt}}
\multiput(1104.00,729.24)(0.962,0.509){2}{\rule{2.900pt}{0.123pt}}
\multiput(1104.00,724.51)(6.981,6.000){2}{\rule{1.450pt}{1.200pt}}
\multiput(1117.00,735.24)(0.738,0.505){4}{\rule{2.357pt}{0.122pt}}
\multiput(1117.00,730.51)(7.108,7.000){2}{\rule{1.179pt}{1.200pt}}
\multiput(1129.00,742.24)(0.792,0.509){2}{\rule{2.700pt}{0.123pt}}
\multiput(1129.00,737.51)(6.396,6.000){2}{\rule{1.350pt}{1.200pt}}
\multiput(1141.00,748.24)(0.792,0.509){2}{\rule{2.700pt}{0.123pt}}
\multiput(1141.00,743.51)(6.396,6.000){2}{\rule{1.350pt}{1.200pt}}
\multiput(1153.00,754.24)(0.835,0.505){4}{\rule{2.529pt}{0.122pt}}
\multiput(1153.00,749.51)(7.752,7.000){2}{\rule{1.264pt}{1.200pt}}
\multiput(1166.00,761.24)(0.792,0.509){2}{\rule{2.700pt}{0.123pt}}
\multiput(1166.00,756.51)(6.396,6.000){2}{\rule{1.350pt}{1.200pt}}
\multiput(1178.00,767.24)(0.792,0.509){2}{\rule{2.700pt}{0.123pt}}
\multiput(1178.00,762.51)(6.396,6.000){2}{\rule{1.350pt}{1.200pt}}
\multiput(1190.00,773.24)(0.962,0.509){2}{\rule{2.900pt}{0.123pt}}
\multiput(1190.00,768.51)(6.981,6.000){2}{\rule{1.450pt}{1.200pt}}
\multiput(1203.00,779.24)(0.738,0.505){4}{\rule{2.357pt}{0.122pt}}
\multiput(1203.00,774.51)(7.108,7.000){2}{\rule{1.179pt}{1.200pt}}
\multiput(1215.00,786.24)(0.792,0.509){2}{\rule{2.700pt}{0.123pt}}
\multiput(1215.00,781.51)(6.396,6.000){2}{\rule{1.350pt}{1.200pt}}
\multiput(1227.00,792.24)(0.792,0.509){2}{\rule{2.700pt}{0.123pt}}
\multiput(1227.00,787.51)(6.396,6.000){2}{\rule{1.350pt}{1.200pt}}
\multiput(1239.00,798.24)(0.962,0.509){2}{\rule{2.900pt}{0.123pt}}
\multiput(1239.00,793.51)(6.981,6.000){2}{\rule{1.450pt}{1.200pt}}
\multiput(1252.00,804.24)(0.738,0.505){4}{\rule{2.357pt}{0.122pt}}
\multiput(1252.00,799.51)(7.108,7.000){2}{\rule{1.179pt}{1.200pt}}
\multiput(1264.00,811.24)(0.792,0.509){2}{\rule{2.700pt}{0.123pt}}
\multiput(1264.00,806.51)(6.396,6.000){2}{\rule{1.350pt}{1.200pt}}
\multiput(1276.00,817.24)(0.962,0.509){2}{\rule{2.900pt}{0.123pt}}
\multiput(1276.00,812.51)(6.981,6.000){2}{\rule{1.450pt}{1.200pt}}
\multiput(1289.00,823.24)(0.738,0.505){4}{\rule{2.357pt}{0.122pt}}
\multiput(1289.00,818.51)(7.108,7.000){2}{\rule{1.179pt}{1.200pt}}
\multiput(1301.00,830.24)(0.792,0.509){2}{\rule{2.700pt}{0.123pt}}
\multiput(1301.00,825.51)(6.396,6.000){2}{\rule{1.350pt}{1.200pt}}
\multiput(1313.00,836.24)(0.792,0.509){2}{\rule{2.700pt}{0.123pt}}
\multiput(1313.00,831.51)(6.396,6.000){2}{\rule{1.350pt}{1.200pt}}
\multiput(1325.00,842.24)(0.835,0.505){4}{\rule{2.529pt}{0.122pt}}
\multiput(1325.00,837.51)(7.752,7.000){2}{\rule{1.264pt}{1.200pt}}
\multiput(1338.00,849.24)(0.792,0.509){2}{\rule{2.700pt}{0.123pt}}
\multiput(1338.00,844.51)(6.396,6.000){2}{\rule{1.350pt}{1.200pt}}
\multiput(1350.00,855.24)(0.792,0.509){2}{\rule{2.700pt}{0.123pt}}
\multiput(1350.00,850.51)(6.396,6.000){2}{\rule{1.350pt}{1.200pt}}
\multiput(1362.00,861.24)(0.835,0.505){4}{\rule{2.529pt}{0.122pt}}
\multiput(1362.00,856.51)(7.752,7.000){2}{\rule{1.264pt}{1.200pt}}
\multiput(1375.00,868.24)(0.792,0.509){2}{\rule{2.700pt}{0.123pt}}
\multiput(1375.00,863.51)(6.396,6.000){2}{\rule{1.350pt}{1.200pt}}
\put(1387,872.01){\rule{2.168pt}{1.200pt}}
\multiput(1387.00,869.51)(4.500,5.000){2}{\rule{1.084pt}{1.200pt}}
\end{picture}

%% file: mc.tex
% GNUPLOT: LaTeX picture
\setlength{\unitlength}{0.240900pt}
\ifx\plotpoint\undefined\newsavebox{\plotpoint}\fi
\sbox{\plotpoint}{\rule[-0.200pt]{0.400pt}{0.400pt}}%
\begin{picture}(1500,900)(0,0)
\font\gnuplot=cmr10 at 10pt
\gnuplot
\sbox{\plotpoint}{\rule[-0.200pt]{0.400pt}{0.400pt}}%
\put(220.0,113.0){\rule[-0.200pt]{0.400pt}{184.048pt}}
\put(220.0,113.0){\rule[-0.200pt]{4.818pt}{0.400pt}}
\put(198,113){\makebox(0,0)[r]{$3$}}
\put(1416.0,113.0){\rule[-0.200pt]{4.818pt}{0.400pt}}
\put(220.0,266.0){\rule[-0.200pt]{4.818pt}{0.400pt}}
\put(198,266){\makebox(0,0)[r]{$4$}}
\put(1416.0,266.0){\rule[-0.200pt]{4.818pt}{0.400pt}}
\put(220.0,419.0){\rule[-0.200pt]{4.818pt}{0.400pt}}
\put(198,419){\makebox(0,0)[r]{$5$}}
\put(1416.0,419.0){\rule[-0.200pt]{4.818pt}{0.400pt}}
\put(220.0,571.0){\rule[-0.200pt]{4.818pt}{0.400pt}}
\put(198,571){\makebox(0,0)[r]{$6$}}
\put(1416.0,571.0){\rule[-0.200pt]{4.818pt}{0.400pt}}
\put(220.0,724.0){\rule[-0.200pt]{4.818pt}{0.400pt}}
\put(198,724){\makebox(0,0)[r]{$7$}}
\put(1416.0,724.0){\rule[-0.200pt]{4.818pt}{0.400pt}}
\put(220.0,877.0){\rule[-0.200pt]{4.818pt}{0.400pt}}
\put(198,877){\makebox(0,0)[r]{$8$}}
\put(1416.0,877.0){\rule[-0.200pt]{4.818pt}{0.400pt}}
\put(220.0,113.0){\rule[-0.200pt]{0.400pt}{4.818pt}}
\put(220,68){\makebox(0,0){$\infty$}}
\put(220.0,857.0){\rule[-0.200pt]{0.400pt}{4.818pt}}
\put(625.0,113.0){\rule[-0.200pt]{0.400pt}{4.818pt}}
\put(625,68){\makebox(0,0){$5$}}
\put(625.0,857.0){\rule[-0.200pt]{0.400pt}{4.818pt}}
\put(727.0,113.0){\rule[-0.200pt]{0.400pt}{4.818pt}}
\put(727,68){\makebox(0,0){$4$}}
\put(727.0,857.0){\rule[-0.200pt]{0.400pt}{4.818pt}}
\put(896.0,113.0){\rule[-0.200pt]{0.400pt}{4.818pt}}
\put(896,68){\makebox(0,0){$3$}}
\put(896.0,857.0){\rule[-0.200pt]{0.400pt}{4.818pt}}
\put(1233.0,113.0){\rule[-0.200pt]{0.400pt}{4.818pt}}
\put(1233,68){\makebox(0,0){$2$}}
\put(1233.0,857.0){\rule[-0.200pt]{0.400pt}{4.818pt}}
\put(220.0,113.0){\rule[-0.200pt]{292.934pt}{0.400pt}}
\put(1436.0,113.0){\rule[-0.200pt]{0.400pt}{184.048pt}}
\put(220.0,877.0){\rule[-0.200pt]{292.934pt}{0.400pt}}
\put(45,495){\makebox(0,0){$m_{\rm c}$}}
\put(828,23){\makebox(0,0){$N$}}
\put(220.0,113.0){\rule[-0.200pt]{0.400pt}{184.048pt}}
\put(1233,732){\circle*{12}}
\put(896,601){\circle*{12}}
\put(727,513){\circle*{12}}
\put(625,462){\circle*{12}}
\end{picture}

%% file: O1ScalingPlot.tex
% GNUPLOT: LaTeX picture
\setlength{\unitlength}{0.240900pt}
\ifx\plotpoint\undefined\newsavebox{\plotpoint}\fi
\sbox{\plotpoint}{\rule[-0.175pt]{0.350pt}{0.350pt}}%
\begin{picture}(1500,1499)(0,0)
\tenrm
\sbox{\plotpoint}{\rule[-0.175pt]{0.350pt}{0.350pt}}%
\put(264,158){\rule[-0.175pt]{4.818pt}{0.350pt}}
\put(242,158){\makebox(0,0)[r]{0}}
\put(1416,158){\rule[-0.175pt]{4.818pt}{0.350pt}}
\put(264,363){\rule[-0.175pt]{4.818pt}{0.350pt}}
\put(242,363){\makebox(0,0)[r]{2}}
\put(1416,363){\rule[-0.175pt]{4.818pt}{0.350pt}}
\put(264,567){\rule[-0.175pt]{4.818pt}{0.350pt}}
\put(242,567){\makebox(0,0)[r]{4}}
\put(1416,567){\rule[-0.175pt]{4.818pt}{0.350pt}}
\put(264,772){\rule[-0.175pt]{4.818pt}{0.350pt}}
\put(242,772){\makebox(0,0)[r]{6}}
\put(1416,772){\rule[-0.175pt]{4.818pt}{0.350pt}}
\put(264,977){\rule[-0.175pt]{4.818pt}{0.350pt}}
\put(242,977){\makebox(0,0)[r]{8}}
\put(1416,977){\rule[-0.175pt]{4.818pt}{0.350pt}}
\put(264,1181){\rule[-0.175pt]{4.818pt}{0.350pt}}
\put(242,1181){\makebox(0,0)[r]{10}}
\put(1416,1181){\rule[-0.175pt]{4.818pt}{0.350pt}}
\put(264,1386){\rule[-0.175pt]{4.818pt}{0.350pt}}
\put(242,1386){\makebox(0,0)[r]{12}}
\put(1416,1386){\rule[-0.175pt]{4.818pt}{0.350pt}}
\put(264,158){\rule[-0.175pt]{0.350pt}{4.818pt}}
\put(264,113){\makebox(0,0){-1}}
\put(264,1366){\rule[-0.175pt]{0.350pt}{4.818pt}}
\put(381,158){\rule[-0.175pt]{0.350pt}{4.818pt}}
\put(381,113){\makebox(0,0){-0.8}}
\put(381,1366){\rule[-0.175pt]{0.350pt}{4.818pt}}
\put(498,158){\rule[-0.175pt]{0.350pt}{4.818pt}}
\put(498,113){\makebox(0,0){-0.6}}
\put(498,1366){\rule[-0.175pt]{0.350pt}{4.818pt}}
\put(616,158){\rule[-0.175pt]{0.350pt}{4.818pt}}
\put(616,113){\makebox(0,0){-0.4}}
\put(616,1366){\rule[-0.175pt]{0.350pt}{4.818pt}}
\put(733,158){\rule[-0.175pt]{0.350pt}{4.818pt}}
\put(733,113){\makebox(0,0){-0.2}}
\put(733,1366){\rule[-0.175pt]{0.350pt}{4.818pt}}
\put(850,158){\rule[-0.175pt]{0.350pt}{4.818pt}}
\put(850,113){\makebox(0,0){0}}
\put(850,1366){\rule[-0.175pt]{0.350pt}{4.818pt}}
\put(967,158){\rule[-0.175pt]{0.350pt}{4.818pt}}
\put(967,113){\makebox(0,0){0.2}}
\put(967,1366){\rule[-0.175pt]{0.350pt}{4.818pt}}
\put(1084,158){\rule[-0.175pt]{0.350pt}{4.818pt}}
\put(1084,113){\makebox(0,0){0.4}}
\put(1084,1366){\rule[-0.175pt]{0.350pt}{4.818pt}}
\put(1202,158){\rule[-0.175pt]{0.350pt}{4.818pt}}
\put(1202,113){\makebox(0,0){0.6}}
\put(1202,1366){\rule[-0.175pt]{0.350pt}{4.818pt}}
\put(1319,158){\rule[-0.175pt]{0.350pt}{4.818pt}}
\put(1319,113){\makebox(0,0){0.8}}
\put(1319,1366){\rule[-0.175pt]{0.350pt}{4.818pt}}
\put(1436,158){\rule[-0.175pt]{0.350pt}{4.818pt}}
\put(1436,113){\makebox(0,0){1}}
\put(1436,1366){\rule[-0.175pt]{0.350pt}{4.818pt}}
\put(264,158){\rule[-0.175pt]{282.335pt}{0.350pt}}
\put(1436,158){\rule[-0.175pt]{0.350pt}{295.825pt}}
\put(264,1386){\rule[-0.175pt]{282.335pt}{0.350pt}}
\put(45,772){\makebox(0,0)[l]{\shortstack{$\Sigma$}}}
\put(850,68){\makebox(0,0){$L^{y_T}u$}}
\put(264,158){\rule[-0.175pt]{0.350pt}{295.825pt}}
\put(674,1028){\makebox(0,0)[r]{fit}}
\put(696,1028){\rule[-0.175pt]{15.899pt}{0.350pt}}
\put(264,162){\usebox{\plotpoint}}
\put(264,162){\rule[-0.175pt]{2.891pt}{0.350pt}}
\put(276,161){\rule[-0.175pt]{5.782pt}{0.350pt}}
\put(300,160){\rule[-0.175pt]{5.541pt}{0.350pt}}
\put(323,161){\rule[-0.175pt]{11.563pt}{0.350pt}}
\put(371,162){\rule[-0.175pt]{11.322pt}{0.350pt}}
\put(418,163){\rule[-0.175pt]{8.431pt}{0.350pt}}
\put(453,164){\rule[-0.175pt]{8.672pt}{0.350pt}}
\put(489,165){\rule[-0.175pt]{8.431pt}{0.350pt}}
\put(524,166){\rule[-0.175pt]{2.891pt}{0.350pt}}
\put(536,167){\rule[-0.175pt]{5.782pt}{0.350pt}}
\put(560,168){\rule[-0.175pt]{2.891pt}{0.350pt}}
\put(572,169){\rule[-0.175pt]{5.541pt}{0.350pt}}
\put(595,170){\rule[-0.175pt]{2.891pt}{0.350pt}}
\put(607,171){\rule[-0.175pt]{2.891pt}{0.350pt}}
\put(619,172){\rule[-0.175pt]{2.891pt}{0.350pt}}
\put(631,173){\rule[-0.175pt]{2.891pt}{0.350pt}}
\put(643,174){\rule[-0.175pt]{2.891pt}{0.350pt}}
\put(655,175){\rule[-0.175pt]{2.891pt}{0.350pt}}
\put(667,176){\rule[-0.175pt]{1.325pt}{0.350pt}}
\put(672,177){\rule[-0.175pt]{1.325pt}{0.350pt}}
\put(678,178){\rule[-0.175pt]{2.891pt}{0.350pt}}
\put(690,179){\rule[-0.175pt]{1.445pt}{0.350pt}}
\put(696,180){\rule[-0.175pt]{1.445pt}{0.350pt}}
\put(702,181){\rule[-0.175pt]{1.445pt}{0.350pt}}
\put(708,182){\rule[-0.175pt]{1.445pt}{0.350pt}}
\put(714,183){\rule[-0.175pt]{1.445pt}{0.350pt}}
\put(720,184){\rule[-0.175pt]{1.445pt}{0.350pt}}
\put(726,185){\rule[-0.175pt]{1.445pt}{0.350pt}}
\put(732,186){\rule[-0.175pt]{1.445pt}{0.350pt}}
\put(738,187){\rule[-0.175pt]{1.325pt}{0.350pt}}
\put(743,188){\rule[-0.175pt]{1.325pt}{0.350pt}}
\put(749,189){\rule[-0.175pt]{0.964pt}{0.350pt}}
\put(753,190){\rule[-0.175pt]{0.964pt}{0.350pt}}
\put(757,191){\rule[-0.175pt]{0.964pt}{0.350pt}}
\put(761,192){\rule[-0.175pt]{0.964pt}{0.350pt}}
\put(765,193){\rule[-0.175pt]{0.964pt}{0.350pt}}
\put(769,194){\rule[-0.175pt]{0.964pt}{0.350pt}}
\put(773,195){\rule[-0.175pt]{0.964pt}{0.350pt}}
\put(777,196){\rule[-0.175pt]{0.964pt}{0.350pt}}
\put(781,197){\rule[-0.175pt]{0.964pt}{0.350pt}}
\put(785,198){\rule[-0.175pt]{0.964pt}{0.350pt}}
\put(789,199){\rule[-0.175pt]{0.964pt}{0.350pt}}
\put(793,200){\rule[-0.175pt]{0.964pt}{0.350pt}}
\put(797,201){\rule[-0.175pt]{0.723pt}{0.350pt}}
\put(800,202){\rule[-0.175pt]{0.723pt}{0.350pt}}
\put(803,203){\rule[-0.175pt]{0.723pt}{0.350pt}}
\put(806,204){\rule[-0.175pt]{0.723pt}{0.350pt}}
\put(809,205){\rule[-0.175pt]{0.662pt}{0.350pt}}
\put(811,206){\rule[-0.175pt]{0.662pt}{0.350pt}}
\put(814,207){\rule[-0.175pt]{0.662pt}{0.350pt}}
\put(817,208){\rule[-0.175pt]{0.662pt}{0.350pt}}
\put(820,209){\rule[-0.175pt]{0.578pt}{0.350pt}}
\put(822,210){\rule[-0.175pt]{0.578pt}{0.350pt}}
\put(824,211){\rule[-0.175pt]{0.578pt}{0.350pt}}
\put(827,212){\rule[-0.175pt]{0.578pt}{0.350pt}}
\put(829,213){\rule[-0.175pt]{0.578pt}{0.350pt}}
\put(832,214){\rule[-0.175pt]{0.578pt}{0.350pt}}
\put(834,215){\rule[-0.175pt]{0.578pt}{0.350pt}}
\put(836,216){\rule[-0.175pt]{0.578pt}{0.350pt}}
\put(839,217){\rule[-0.175pt]{0.578pt}{0.350pt}}
\put(841,218){\rule[-0.175pt]{0.578pt}{0.350pt}}
\put(844,219){\rule[-0.175pt]{0.578pt}{0.350pt}}
\put(846,220){\rule[-0.175pt]{0.578pt}{0.350pt}}
\put(848,221){\rule[-0.175pt]{0.578pt}{0.350pt}}
\put(851,222){\rule[-0.175pt]{0.578pt}{0.350pt}}
\put(853,223){\rule[-0.175pt]{0.578pt}{0.350pt}}
\put(856,224){\rule[-0.175pt]{0.482pt}{0.350pt}}
\put(858,225){\rule[-0.175pt]{0.482pt}{0.350pt}}
\put(860,226){\rule[-0.175pt]{0.482pt}{0.350pt}}
\put(862,227){\rule[-0.175pt]{0.482pt}{0.350pt}}
\put(864,228){\rule[-0.175pt]{0.482pt}{0.350pt}}
\put(866,229){\rule[-0.175pt]{0.482pt}{0.350pt}}
\put(868,230){\rule[-0.175pt]{0.482pt}{0.350pt}}
\put(870,231){\rule[-0.175pt]{0.482pt}{0.350pt}}
\put(872,232){\rule[-0.175pt]{0.482pt}{0.350pt}}
\put(874,233){\rule[-0.175pt]{0.482pt}{0.350pt}}
\put(876,234){\rule[-0.175pt]{0.482pt}{0.350pt}}
\put(878,235){\rule[-0.175pt]{0.482pt}{0.350pt}}
\put(880,236){\rule[-0.175pt]{0.379pt}{0.350pt}}
\put(881,237){\rule[-0.175pt]{0.379pt}{0.350pt}}
\put(883,238){\rule[-0.175pt]{0.379pt}{0.350pt}}
\put(884,239){\rule[-0.175pt]{0.379pt}{0.350pt}}
\put(886,240){\rule[-0.175pt]{0.379pt}{0.350pt}}
\put(887,241){\rule[-0.175pt]{0.379pt}{0.350pt}}
\put(889,242){\rule[-0.175pt]{0.379pt}{0.350pt}}
\put(890,243){\rule[-0.175pt]{0.361pt}{0.350pt}}
\put(892,244){\rule[-0.175pt]{0.361pt}{0.350pt}}
\put(894,245){\rule[-0.175pt]{0.361pt}{0.350pt}}
\put(895,246){\rule[-0.175pt]{0.361pt}{0.350pt}}
\put(897,247){\rule[-0.175pt]{0.361pt}{0.350pt}}
\put(898,248){\rule[-0.175pt]{0.361pt}{0.350pt}}
\put(900,249){\rule[-0.175pt]{0.361pt}{0.350pt}}
\put(901,250){\rule[-0.175pt]{0.361pt}{0.350pt}}
\put(903,251){\usebox{\plotpoint}}
\put(904,252){\usebox{\plotpoint}}
\put(905,253){\usebox{\plotpoint}}
\put(906,254){\usebox{\plotpoint}}
\put(908,255){\usebox{\plotpoint}}
\put(909,256){\usebox{\plotpoint}}
\put(910,257){\usebox{\plotpoint}}
\put(912,258){\usebox{\plotpoint}}
\put(913,259){\usebox{\plotpoint}}
\put(914,260){\usebox{\plotpoint}}
\put(916,261){\usebox{\plotpoint}}
\put(917,262){\usebox{\plotpoint}}
\put(918,263){\usebox{\plotpoint}}
\put(920,264){\usebox{\plotpoint}}
\put(921,265){\usebox{\plotpoint}}
\put(922,266){\usebox{\plotpoint}}
\put(924,267){\usebox{\plotpoint}}
\put(925,268){\usebox{\plotpoint}}
\put(926,269){\usebox{\plotpoint}}
\put(928,270){\usebox{\plotpoint}}
\put(929,271){\usebox{\plotpoint}}
\put(930,272){\usebox{\plotpoint}}
\put(931,273){\usebox{\plotpoint}}
\put(933,274){\usebox{\plotpoint}}
\put(934,275){\usebox{\plotpoint}}
\put(935,276){\usebox{\plotpoint}}
\put(936,277){\usebox{\plotpoint}}
\put(937,278){\usebox{\plotpoint}}
\put(939,279){\usebox{\plotpoint}}
\put(940,280){\usebox{\plotpoint}}
\put(941,281){\usebox{\plotpoint}}
\put(942,282){\usebox{\plotpoint}}
\put(943,283){\usebox{\plotpoint}}
\put(944,284){\usebox{\plotpoint}}
\put(945,285){\usebox{\plotpoint}}
\put(946,286){\usebox{\plotpoint}}
\put(947,287){\usebox{\plotpoint}}
\put(948,288){\usebox{\plotpoint}}
\put(949,289){\usebox{\plotpoint}}
\put(950,290){\usebox{\plotpoint}}
\put(951,290){\usebox{\plotpoint}}
\put(952,291){\usebox{\plotpoint}}
\put(953,292){\usebox{\plotpoint}}
\put(954,293){\usebox{\plotpoint}}
\put(955,294){\usebox{\plotpoint}}
\put(956,295){\usebox{\plotpoint}}
\put(957,296){\usebox{\plotpoint}}
\put(958,297){\usebox{\plotpoint}}
\put(959,298){\usebox{\plotpoint}}
\put(960,299){\usebox{\plotpoint}}
\put(961,300){\usebox{\plotpoint}}
\put(962,302){\usebox{\plotpoint}}
\put(963,303){\usebox{\plotpoint}}
\put(964,304){\usebox{\plotpoint}}
\put(965,305){\usebox{\plotpoint}}
\put(966,306){\usebox{\plotpoint}}
\put(967,307){\usebox{\plotpoint}}
\put(968,308){\usebox{\plotpoint}}
\put(969,309){\usebox{\plotpoint}}
\put(970,310){\usebox{\plotpoint}}
\put(971,311){\usebox{\plotpoint}}
\put(972,312){\usebox{\plotpoint}}
\put(973,313){\usebox{\plotpoint}}
\put(974,315){\usebox{\plotpoint}}
\put(975,316){\usebox{\plotpoint}}
\put(976,317){\usebox{\plotpoint}}
\put(977,318){\usebox{\plotpoint}}
\put(978,319){\usebox{\plotpoint}}
\put(979,320){\usebox{\plotpoint}}
\put(980,321){\usebox{\plotpoint}}
\put(981,323){\usebox{\plotpoint}}
\put(982,324){\usebox{\plotpoint}}
\put(983,325){\usebox{\plotpoint}}
\put(984,326){\usebox{\plotpoint}}
\put(985,327){\usebox{\plotpoint}}
\put(986,328){\usebox{\plotpoint}}
\put(987,330){\usebox{\plotpoint}}
\put(988,331){\usebox{\plotpoint}}
\put(989,332){\usebox{\plotpoint}}
\put(990,334){\usebox{\plotpoint}}
\put(991,335){\usebox{\plotpoint}}
\put(992,336){\usebox{\plotpoint}}
\put(993,337){\usebox{\plotpoint}}
\put(994,339){\usebox{\plotpoint}}
\put(995,340){\usebox{\plotpoint}}
\put(996,341){\usebox{\plotpoint}}
\put(997,342){\usebox{\plotpoint}}
\put(998,344){\usebox{\plotpoint}}
\put(999,345){\usebox{\plotpoint}}
\put(1000,346){\usebox{\plotpoint}}
\put(1001,347){\usebox{\plotpoint}}
\put(1002,349){\usebox{\plotpoint}}
\put(1003,350){\usebox{\plotpoint}}
\put(1004,351){\usebox{\plotpoint}}
\put(1005,352){\usebox{\plotpoint}}
\put(1006,354){\usebox{\plotpoint}}
\put(1007,355){\usebox{\plotpoint}}
\put(1008,356){\usebox{\plotpoint}}
\put(1009,357){\usebox{\plotpoint}}
\put(1010,359){\usebox{\plotpoint}}
\put(1011,360){\usebox{\plotpoint}}
\put(1012,361){\usebox{\plotpoint}}
\put(1013,363){\usebox{\plotpoint}}
\put(1014,364){\usebox{\plotpoint}}
\put(1015,366){\usebox{\plotpoint}}
\put(1016,367){\usebox{\plotpoint}}
\put(1017,368){\usebox{\plotpoint}}
\put(1018,370){\usebox{\plotpoint}}
\put(1019,371){\usebox{\plotpoint}}
\put(1020,373){\usebox{\plotpoint}}
\put(1021,374){\usebox{\plotpoint}}
\put(1022,375){\rule[-0.175pt]{0.350pt}{0.394pt}}
\put(1023,377){\rule[-0.175pt]{0.350pt}{0.394pt}}
\put(1024,379){\rule[-0.175pt]{0.350pt}{0.394pt}}
\put(1025,380){\rule[-0.175pt]{0.350pt}{0.394pt}}
\put(1026,382){\rule[-0.175pt]{0.350pt}{0.394pt}}
\put(1027,384){\rule[-0.175pt]{0.350pt}{0.394pt}}
\put(1028,385){\rule[-0.175pt]{0.350pt}{0.394pt}}
\put(1029,387){\rule[-0.175pt]{0.350pt}{0.394pt}}
\put(1030,389){\rule[-0.175pt]{0.350pt}{0.394pt}}
\put(1031,390){\rule[-0.175pt]{0.350pt}{0.394pt}}
\put(1032,392){\rule[-0.175pt]{0.350pt}{0.394pt}}
\put(1033,393){\rule[-0.175pt]{0.350pt}{0.361pt}}
\put(1034,395){\rule[-0.175pt]{0.350pt}{0.361pt}}
\put(1035,397){\rule[-0.175pt]{0.350pt}{0.361pt}}
\put(1036,398){\rule[-0.175pt]{0.350pt}{0.361pt}}
\put(1037,400){\rule[-0.175pt]{0.350pt}{0.361pt}}
\put(1038,401){\rule[-0.175pt]{0.350pt}{0.361pt}}
\put(1039,403){\rule[-0.175pt]{0.350pt}{0.361pt}}
\put(1040,404){\rule[-0.175pt]{0.350pt}{0.361pt}}
\put(1041,406){\rule[-0.175pt]{0.350pt}{0.361pt}}
\put(1042,407){\rule[-0.175pt]{0.350pt}{0.361pt}}
\put(1043,409){\rule[-0.175pt]{0.350pt}{0.361pt}}
\put(1044,410){\rule[-0.175pt]{0.350pt}{0.361pt}}
\put(1045,412){\rule[-0.175pt]{0.350pt}{0.401pt}}
\put(1046,413){\rule[-0.175pt]{0.350pt}{0.401pt}}
\put(1047,415){\rule[-0.175pt]{0.350pt}{0.401pt}}
\put(1048,416){\rule[-0.175pt]{0.350pt}{0.401pt}}
\put(1049,418){\rule[-0.175pt]{0.350pt}{0.401pt}}
\put(1050,420){\rule[-0.175pt]{0.350pt}{0.401pt}}
\put(1051,421){\rule[-0.175pt]{0.350pt}{0.401pt}}
\put(1052,423){\rule[-0.175pt]{0.350pt}{0.401pt}}
\put(1053,425){\rule[-0.175pt]{0.350pt}{0.401pt}}
\put(1054,426){\rule[-0.175pt]{0.350pt}{0.401pt}}
\put(1055,428){\rule[-0.175pt]{0.350pt}{0.401pt}}
\put(1056,430){\rule[-0.175pt]{0.350pt}{0.401pt}}
\put(1057,431){\rule[-0.175pt]{0.350pt}{0.422pt}}
\put(1058,433){\rule[-0.175pt]{0.350pt}{0.422pt}}
\put(1059,435){\rule[-0.175pt]{0.350pt}{0.422pt}}
\put(1060,437){\rule[-0.175pt]{0.350pt}{0.422pt}}
\put(1061,439){\rule[-0.175pt]{0.350pt}{0.422pt}}
\put(1062,440){\rule[-0.175pt]{0.350pt}{0.422pt}}
\put(1063,442){\rule[-0.175pt]{0.350pt}{0.422pt}}
\put(1064,444){\rule[-0.175pt]{0.350pt}{0.422pt}}
\put(1065,446){\rule[-0.175pt]{0.350pt}{0.422pt}}
\put(1066,447){\rule[-0.175pt]{0.350pt}{0.422pt}}
\put(1067,449){\rule[-0.175pt]{0.350pt}{0.422pt}}
\put(1068,451){\rule[-0.175pt]{0.350pt}{0.422pt}}
\put(1069,453){\rule[-0.175pt]{0.350pt}{0.422pt}}
\put(1070,454){\rule[-0.175pt]{0.350pt}{0.422pt}}
\put(1071,456){\rule[-0.175pt]{0.350pt}{0.422pt}}
\put(1072,458){\rule[-0.175pt]{0.350pt}{0.422pt}}
\put(1073,460){\rule[-0.175pt]{0.350pt}{0.422pt}}
\put(1074,461){\rule[-0.175pt]{0.350pt}{0.422pt}}
\put(1075,463){\rule[-0.175pt]{0.350pt}{0.422pt}}
\put(1076,465){\rule[-0.175pt]{0.350pt}{0.422pt}}
\put(1077,467){\rule[-0.175pt]{0.350pt}{0.422pt}}
\put(1078,468){\rule[-0.175pt]{0.350pt}{0.422pt}}
\put(1079,470){\rule[-0.175pt]{0.350pt}{0.422pt}}
\put(1080,472){\rule[-0.175pt]{0.350pt}{0.422pt}}
\put(1081,474){\rule[-0.175pt]{0.350pt}{0.442pt}}
\put(1082,475){\rule[-0.175pt]{0.350pt}{0.442pt}}
\put(1083,477){\rule[-0.175pt]{0.350pt}{0.442pt}}
\put(1084,479){\rule[-0.175pt]{0.350pt}{0.442pt}}
\put(1085,481){\rule[-0.175pt]{0.350pt}{0.442pt}}
\put(1086,483){\rule[-0.175pt]{0.350pt}{0.442pt}}
\put(1087,485){\rule[-0.175pt]{0.350pt}{0.442pt}}
\put(1088,486){\rule[-0.175pt]{0.350pt}{0.442pt}}
\put(1089,488){\rule[-0.175pt]{0.350pt}{0.442pt}}
\put(1090,490){\rule[-0.175pt]{0.350pt}{0.442pt}}
\put(1091,492){\rule[-0.175pt]{0.350pt}{0.442pt}}
\put(1092,494){\rule[-0.175pt]{0.350pt}{0.442pt}}
\put(1093,496){\rule[-0.175pt]{0.350pt}{0.462pt}}
\put(1094,497){\rule[-0.175pt]{0.350pt}{0.462pt}}
\put(1095,499){\rule[-0.175pt]{0.350pt}{0.462pt}}
\put(1096,501){\rule[-0.175pt]{0.350pt}{0.462pt}}
\put(1097,503){\rule[-0.175pt]{0.350pt}{0.462pt}}
\put(1098,505){\rule[-0.175pt]{0.350pt}{0.462pt}}
\put(1099,507){\rule[-0.175pt]{0.350pt}{0.462pt}}
\put(1100,509){\rule[-0.175pt]{0.350pt}{0.462pt}}
\put(1101,511){\rule[-0.175pt]{0.350pt}{0.462pt}}
\put(1102,513){\rule[-0.175pt]{0.350pt}{0.462pt}}
\put(1103,515){\rule[-0.175pt]{0.350pt}{0.462pt}}
\put(1104,517){\rule[-0.175pt]{0.350pt}{0.462pt}}
\put(1105,519){\rule[-0.175pt]{0.350pt}{0.526pt}}
\put(1106,521){\rule[-0.175pt]{0.350pt}{0.526pt}}
\put(1107,523){\rule[-0.175pt]{0.350pt}{0.526pt}}
\put(1108,525){\rule[-0.175pt]{0.350pt}{0.526pt}}
\put(1109,527){\rule[-0.175pt]{0.350pt}{0.526pt}}
\put(1110,529){\rule[-0.175pt]{0.350pt}{0.526pt}}
\put(1111,532){\rule[-0.175pt]{0.350pt}{0.526pt}}
\put(1112,534){\rule[-0.175pt]{0.350pt}{0.526pt}}
\put(1113,536){\rule[-0.175pt]{0.350pt}{0.526pt}}
\put(1114,538){\rule[-0.175pt]{0.350pt}{0.526pt}}
\put(1115,540){\rule[-0.175pt]{0.350pt}{0.526pt}}
\put(1116,543){\rule[-0.175pt]{0.350pt}{0.482pt}}
\put(1117,545){\rule[-0.175pt]{0.350pt}{0.482pt}}
\put(1118,547){\rule[-0.175pt]{0.350pt}{0.482pt}}
\put(1119,549){\rule[-0.175pt]{0.350pt}{0.482pt}}
\put(1120,551){\rule[-0.175pt]{0.350pt}{0.482pt}}
\put(1121,553){\rule[-0.175pt]{0.350pt}{0.482pt}}
\put(1122,555){\rule[-0.175pt]{0.350pt}{0.482pt}}
\put(1123,557){\rule[-0.175pt]{0.350pt}{0.482pt}}
\put(1124,559){\rule[-0.175pt]{0.350pt}{0.482pt}}
\put(1125,561){\rule[-0.175pt]{0.350pt}{0.482pt}}
\put(1126,563){\rule[-0.175pt]{0.350pt}{0.482pt}}
\put(1127,565){\rule[-0.175pt]{0.350pt}{0.482pt}}
\put(1128,567){\rule[-0.175pt]{0.350pt}{0.502pt}}
\put(1129,569){\rule[-0.175pt]{0.350pt}{0.502pt}}
\put(1130,571){\rule[-0.175pt]{0.350pt}{0.502pt}}
\put(1131,573){\rule[-0.175pt]{0.350pt}{0.502pt}}
\put(1132,575){\rule[-0.175pt]{0.350pt}{0.502pt}}
\put(1133,577){\rule[-0.175pt]{0.350pt}{0.502pt}}
\put(1134,579){\rule[-0.175pt]{0.350pt}{0.502pt}}
\put(1135,581){\rule[-0.175pt]{0.350pt}{0.502pt}}
\put(1136,583){\rule[-0.175pt]{0.350pt}{0.502pt}}
\put(1137,585){\rule[-0.175pt]{0.350pt}{0.502pt}}
\put(1138,587){\rule[-0.175pt]{0.350pt}{0.502pt}}
\put(1139,589){\rule[-0.175pt]{0.350pt}{0.502pt}}
\put(1140,591){\rule[-0.175pt]{0.350pt}{0.522pt}}
\put(1141,594){\rule[-0.175pt]{0.350pt}{0.522pt}}
\put(1142,596){\rule[-0.175pt]{0.350pt}{0.522pt}}
\put(1143,598){\rule[-0.175pt]{0.350pt}{0.522pt}}
\put(1144,600){\rule[-0.175pt]{0.350pt}{0.522pt}}
\put(1145,602){\rule[-0.175pt]{0.350pt}{0.522pt}}
\put(1146,605){\rule[-0.175pt]{0.350pt}{0.522pt}}
\put(1147,607){\rule[-0.175pt]{0.350pt}{0.522pt}}
\put(1148,609){\rule[-0.175pt]{0.350pt}{0.522pt}}
\put(1149,611){\rule[-0.175pt]{0.350pt}{0.522pt}}
\put(1150,613){\rule[-0.175pt]{0.350pt}{0.522pt}}
\put(1151,615){\rule[-0.175pt]{0.350pt}{0.522pt}}
\put(1152,618){\rule[-0.175pt]{0.350pt}{0.522pt}}
\put(1153,620){\rule[-0.175pt]{0.350pt}{0.522pt}}
\put(1154,622){\rule[-0.175pt]{0.350pt}{0.522pt}}
\put(1155,624){\rule[-0.175pt]{0.350pt}{0.522pt}}
\put(1156,626){\rule[-0.175pt]{0.350pt}{0.522pt}}
\put(1157,628){\rule[-0.175pt]{0.350pt}{0.522pt}}
\put(1158,631){\rule[-0.175pt]{0.350pt}{0.522pt}}
\put(1159,633){\rule[-0.175pt]{0.350pt}{0.522pt}}
\put(1160,635){\rule[-0.175pt]{0.350pt}{0.522pt}}
\put(1161,637){\rule[-0.175pt]{0.350pt}{0.522pt}}
\put(1162,639){\rule[-0.175pt]{0.350pt}{0.522pt}}
\put(1163,641){\rule[-0.175pt]{0.350pt}{0.522pt}}
\put(1164,644){\rule[-0.175pt]{0.350pt}{0.542pt}}
\put(1165,646){\rule[-0.175pt]{0.350pt}{0.542pt}}
\put(1166,648){\rule[-0.175pt]{0.350pt}{0.542pt}}
\put(1167,650){\rule[-0.175pt]{0.350pt}{0.542pt}}
\put(1168,653){\rule[-0.175pt]{0.350pt}{0.542pt}}
\put(1169,655){\rule[-0.175pt]{0.350pt}{0.542pt}}
\put(1170,657){\rule[-0.175pt]{0.350pt}{0.542pt}}
\put(1171,659){\rule[-0.175pt]{0.350pt}{0.542pt}}
\put(1172,662){\rule[-0.175pt]{0.350pt}{0.542pt}}
\put(1173,664){\rule[-0.175pt]{0.350pt}{0.542pt}}
\put(1174,666){\rule[-0.175pt]{0.350pt}{0.542pt}}
\put(1175,668){\rule[-0.175pt]{0.350pt}{0.542pt}}
\put(1176,671){\rule[-0.175pt]{0.350pt}{0.613pt}}
\put(1177,673){\rule[-0.175pt]{0.350pt}{0.613pt}}
\put(1178,676){\rule[-0.175pt]{0.350pt}{0.613pt}}
\put(1179,678){\rule[-0.175pt]{0.350pt}{0.613pt}}
\put(1180,681){\rule[-0.175pt]{0.350pt}{0.613pt}}
\put(1181,683){\rule[-0.175pt]{0.350pt}{0.613pt}}
\put(1182,686){\rule[-0.175pt]{0.350pt}{0.613pt}}
\put(1183,688){\rule[-0.175pt]{0.350pt}{0.613pt}}
\put(1184,691){\rule[-0.175pt]{0.350pt}{0.613pt}}
\put(1185,693){\rule[-0.175pt]{0.350pt}{0.613pt}}
\put(1186,696){\rule[-0.175pt]{0.350pt}{0.613pt}}
\put(1187,699){\rule[-0.175pt]{0.350pt}{0.582pt}}
\put(1188,701){\rule[-0.175pt]{0.350pt}{0.582pt}}
\put(1189,703){\rule[-0.175pt]{0.350pt}{0.582pt}}
\put(1190,706){\rule[-0.175pt]{0.350pt}{0.582pt}}
\put(1191,708){\rule[-0.175pt]{0.350pt}{0.582pt}}
\put(1192,711){\rule[-0.175pt]{0.350pt}{0.582pt}}
\put(1193,713){\rule[-0.175pt]{0.350pt}{0.582pt}}
\put(1194,715){\rule[-0.175pt]{0.350pt}{0.582pt}}
\put(1195,718){\rule[-0.175pt]{0.350pt}{0.582pt}}
\put(1196,720){\rule[-0.175pt]{0.350pt}{0.582pt}}
\put(1197,723){\rule[-0.175pt]{0.350pt}{0.582pt}}
\put(1198,725){\rule[-0.175pt]{0.350pt}{0.582pt}}
\put(1199,728){\rule[-0.175pt]{0.350pt}{0.582pt}}
\put(1200,730){\rule[-0.175pt]{0.350pt}{0.582pt}}
\put(1201,732){\rule[-0.175pt]{0.350pt}{0.582pt}}
\put(1202,735){\rule[-0.175pt]{0.350pt}{0.582pt}}
\put(1203,737){\rule[-0.175pt]{0.350pt}{0.582pt}}
\put(1204,740){\rule[-0.175pt]{0.350pt}{0.582pt}}
\put(1205,742){\rule[-0.175pt]{0.350pt}{0.582pt}}
\put(1206,744){\rule[-0.175pt]{0.350pt}{0.582pt}}
\put(1207,747){\rule[-0.175pt]{0.350pt}{0.582pt}}
\put(1208,749){\rule[-0.175pt]{0.350pt}{0.582pt}}
\put(1209,752){\rule[-0.175pt]{0.350pt}{0.582pt}}
\put(1210,754){\rule[-0.175pt]{0.350pt}{0.582pt}}
\put(1211,757){\rule[-0.175pt]{0.350pt}{0.602pt}}
\put(1212,759){\rule[-0.175pt]{0.350pt}{0.602pt}}
\put(1213,762){\rule[-0.175pt]{0.350pt}{0.602pt}}
\put(1214,764){\rule[-0.175pt]{0.350pt}{0.602pt}}
\put(1215,767){\rule[-0.175pt]{0.350pt}{0.602pt}}
\put(1216,769){\rule[-0.175pt]{0.350pt}{0.602pt}}
\put(1217,772){\rule[-0.175pt]{0.350pt}{0.602pt}}
\put(1218,774){\rule[-0.175pt]{0.350pt}{0.602pt}}
\put(1219,777){\rule[-0.175pt]{0.350pt}{0.602pt}}
\put(1220,779){\rule[-0.175pt]{0.350pt}{0.602pt}}
\put(1221,782){\rule[-0.175pt]{0.350pt}{0.602pt}}
\put(1222,784){\rule[-0.175pt]{0.350pt}{0.602pt}}
\put(1223,787){\rule[-0.175pt]{0.350pt}{0.602pt}}
\put(1224,789){\rule[-0.175pt]{0.350pt}{0.602pt}}
\put(1225,792){\rule[-0.175pt]{0.350pt}{0.602pt}}
\put(1226,794){\rule[-0.175pt]{0.350pt}{0.602pt}}
\put(1227,797){\rule[-0.175pt]{0.350pt}{0.602pt}}
\put(1228,799){\rule[-0.175pt]{0.350pt}{0.602pt}}
\put(1229,802){\rule[-0.175pt]{0.350pt}{0.602pt}}
\put(1230,804){\rule[-0.175pt]{0.350pt}{0.602pt}}
\put(1231,807){\rule[-0.175pt]{0.350pt}{0.602pt}}
\put(1232,809){\rule[-0.175pt]{0.350pt}{0.602pt}}
\put(1233,812){\rule[-0.175pt]{0.350pt}{0.602pt}}
\put(1234,814){\rule[-0.175pt]{0.350pt}{0.602pt}}
\put(1235,817){\rule[-0.175pt]{0.350pt}{0.642pt}}
\put(1236,819){\rule[-0.175pt]{0.350pt}{0.642pt}}
\put(1237,822){\rule[-0.175pt]{0.350pt}{0.642pt}}
\put(1238,825){\rule[-0.175pt]{0.350pt}{0.642pt}}
\put(1239,827){\rule[-0.175pt]{0.350pt}{0.642pt}}
\put(1240,830){\rule[-0.175pt]{0.350pt}{0.642pt}}
\put(1241,833){\rule[-0.175pt]{0.350pt}{0.642pt}}
\put(1242,835){\rule[-0.175pt]{0.350pt}{0.642pt}}
\put(1243,838){\rule[-0.175pt]{0.350pt}{0.642pt}}
\put(1244,841){\rule[-0.175pt]{0.350pt}{0.642pt}}
\put(1245,843){\rule[-0.175pt]{0.350pt}{0.642pt}}
\put(1246,846){\rule[-0.175pt]{0.350pt}{0.642pt}}
\put(1247,849){\rule[-0.175pt]{0.350pt}{0.701pt}}
\put(1248,851){\rule[-0.175pt]{0.350pt}{0.701pt}}
\put(1249,854){\rule[-0.175pt]{0.350pt}{0.701pt}}
\put(1250,857){\rule[-0.175pt]{0.350pt}{0.701pt}}
\put(1251,860){\rule[-0.175pt]{0.350pt}{0.701pt}}
\put(1252,863){\rule[-0.175pt]{0.350pt}{0.701pt}}
\put(1253,866){\rule[-0.175pt]{0.350pt}{0.701pt}}
\put(1254,869){\rule[-0.175pt]{0.350pt}{0.701pt}}
\put(1255,872){\rule[-0.175pt]{0.350pt}{0.701pt}}
\put(1256,875){\rule[-0.175pt]{0.350pt}{0.701pt}}
\put(1257,878){\rule[-0.175pt]{0.350pt}{0.701pt}}
\put(1258,881){\rule[-0.175pt]{0.350pt}{0.662pt}}
\put(1259,883){\rule[-0.175pt]{0.350pt}{0.662pt}}
\put(1260,886){\rule[-0.175pt]{0.350pt}{0.662pt}}
\put(1261,889){\rule[-0.175pt]{0.350pt}{0.662pt}}
\put(1262,892){\rule[-0.175pt]{0.350pt}{0.662pt}}
\put(1263,894){\rule[-0.175pt]{0.350pt}{0.662pt}}
\put(1264,897){\rule[-0.175pt]{0.350pt}{0.662pt}}
\put(1265,900){\rule[-0.175pt]{0.350pt}{0.662pt}}
\put(1266,903){\rule[-0.175pt]{0.350pt}{0.662pt}}
\put(1267,905){\rule[-0.175pt]{0.350pt}{0.662pt}}
\put(1268,908){\rule[-0.175pt]{0.350pt}{0.662pt}}
\put(1269,911){\rule[-0.175pt]{0.350pt}{0.662pt}}
\put(1270,914){\rule[-0.175pt]{0.350pt}{0.662pt}}
\put(1271,916){\rule[-0.175pt]{0.350pt}{0.662pt}}
\put(1272,919){\rule[-0.175pt]{0.350pt}{0.662pt}}
\put(1273,922){\rule[-0.175pt]{0.350pt}{0.662pt}}
\put(1274,925){\rule[-0.175pt]{0.350pt}{0.662pt}}
\put(1275,927){\rule[-0.175pt]{0.350pt}{0.662pt}}
\put(1276,930){\rule[-0.175pt]{0.350pt}{0.662pt}}
\put(1277,933){\rule[-0.175pt]{0.350pt}{0.662pt}}
\put(1278,936){\rule[-0.175pt]{0.350pt}{0.662pt}}
\put(1279,938){\rule[-0.175pt]{0.350pt}{0.662pt}}
\put(1280,941){\rule[-0.175pt]{0.350pt}{0.662pt}}
\put(1281,944){\rule[-0.175pt]{0.350pt}{0.662pt}}
\put(1282,947){\rule[-0.175pt]{0.350pt}{0.683pt}}
\put(1283,949){\rule[-0.175pt]{0.350pt}{0.683pt}}
\put(1284,952){\rule[-0.175pt]{0.350pt}{0.683pt}}
\put(1285,955){\rule[-0.175pt]{0.350pt}{0.683pt}}
\put(1286,958){\rule[-0.175pt]{0.350pt}{0.683pt}}
\put(1287,961){\rule[-0.175pt]{0.350pt}{0.683pt}}
\put(1288,963){\rule[-0.175pt]{0.350pt}{0.683pt}}
\put(1289,966){\rule[-0.175pt]{0.350pt}{0.683pt}}
\put(1290,969){\rule[-0.175pt]{0.350pt}{0.683pt}}
\put(1291,972){\rule[-0.175pt]{0.350pt}{0.683pt}}
\put(1292,975){\rule[-0.175pt]{0.350pt}{0.683pt}}
\put(1293,978){\rule[-0.175pt]{0.350pt}{0.683pt}}
\put(1294,980){\rule[-0.175pt]{0.350pt}{0.703pt}}
\put(1295,983){\rule[-0.175pt]{0.350pt}{0.703pt}}
\put(1296,986){\rule[-0.175pt]{0.350pt}{0.703pt}}
\put(1297,989){\rule[-0.175pt]{0.350pt}{0.703pt}}
\put(1298,992){\rule[-0.175pt]{0.350pt}{0.703pt}}
\put(1299,995){\rule[-0.175pt]{0.350pt}{0.703pt}}
\put(1300,998){\rule[-0.175pt]{0.350pt}{0.703pt}}
\put(1301,1001){\rule[-0.175pt]{0.350pt}{0.703pt}}
\put(1302,1004){\rule[-0.175pt]{0.350pt}{0.703pt}}
\put(1303,1007){\rule[-0.175pt]{0.350pt}{0.703pt}}
\put(1304,1010){\rule[-0.175pt]{0.350pt}{0.703pt}}
\put(1305,1013){\rule[-0.175pt]{0.350pt}{0.703pt}}
\put(1306,1016){\rule[-0.175pt]{0.350pt}{0.703pt}}
\put(1307,1018){\rule[-0.175pt]{0.350pt}{0.703pt}}
\put(1308,1021){\rule[-0.175pt]{0.350pt}{0.703pt}}
\put(1309,1024){\rule[-0.175pt]{0.350pt}{0.703pt}}
\put(1310,1027){\rule[-0.175pt]{0.350pt}{0.703pt}}
\put(1311,1030){\rule[-0.175pt]{0.350pt}{0.703pt}}
\put(1312,1033){\rule[-0.175pt]{0.350pt}{0.703pt}}
\put(1313,1036){\rule[-0.175pt]{0.350pt}{0.703pt}}
\put(1314,1039){\rule[-0.175pt]{0.350pt}{0.703pt}}
\put(1315,1042){\rule[-0.175pt]{0.350pt}{0.703pt}}
\put(1316,1045){\rule[-0.175pt]{0.350pt}{0.703pt}}
\put(1317,1048){\rule[-0.175pt]{0.350pt}{0.703pt}}
\put(1318,1050){\rule[-0.175pt]{0.350pt}{0.767pt}}
\put(1319,1054){\rule[-0.175pt]{0.350pt}{0.766pt}}
\put(1320,1057){\rule[-0.175pt]{0.350pt}{0.766pt}}
\put(1321,1060){\rule[-0.175pt]{0.350pt}{0.766pt}}
\put(1322,1063){\rule[-0.175pt]{0.350pt}{0.766pt}}
\put(1323,1066){\rule[-0.175pt]{0.350pt}{0.766pt}}
\put(1324,1070){\rule[-0.175pt]{0.350pt}{0.766pt}}
\put(1325,1073){\rule[-0.175pt]{0.350pt}{0.766pt}}
\put(1326,1076){\rule[-0.175pt]{0.350pt}{0.766pt}}
\put(1327,1079){\rule[-0.175pt]{0.350pt}{0.766pt}}
\put(1328,1082){\rule[-0.175pt]{0.350pt}{0.766pt}}
\put(1329,1085){\rule[-0.175pt]{0.350pt}{0.683pt}}
\put(1330,1088){\rule[-0.175pt]{0.350pt}{0.683pt}}
\put(1331,1091){\rule[-0.175pt]{0.350pt}{0.683pt}}
\put(1332,1094){\rule[-0.175pt]{0.350pt}{0.683pt}}
\put(1333,1097){\rule[-0.175pt]{0.350pt}{0.683pt}}
\put(1334,1100){\rule[-0.175pt]{0.350pt}{0.683pt}}
\put(1335,1103){\rule[-0.175pt]{0.350pt}{0.683pt}}
\put(1336,1105){\rule[-0.175pt]{0.350pt}{0.683pt}}
\put(1337,1108){\rule[-0.175pt]{0.350pt}{0.683pt}}
\put(1338,1111){\rule[-0.175pt]{0.350pt}{0.683pt}}
\put(1339,1114){\rule[-0.175pt]{0.350pt}{0.683pt}}
\put(1340,1117){\rule[-0.175pt]{0.350pt}{0.682pt}}
\put(1341,1120){\rule[-0.175pt]{0.350pt}{0.703pt}}
\put(1342,1122){\rule[-0.175pt]{0.350pt}{0.703pt}}
\put(1343,1125){\rule[-0.175pt]{0.350pt}{0.703pt}}
\put(1344,1128){\rule[-0.175pt]{0.350pt}{0.703pt}}
\put(1345,1131){\rule[-0.175pt]{0.350pt}{0.703pt}}
\put(1346,1134){\rule[-0.175pt]{0.350pt}{0.703pt}}
\put(1347,1137){\rule[-0.175pt]{0.350pt}{0.703pt}}
\put(1348,1140){\rule[-0.175pt]{0.350pt}{0.703pt}}
\put(1349,1143){\rule[-0.175pt]{0.350pt}{0.703pt}}
\put(1350,1146){\rule[-0.175pt]{0.350pt}{0.703pt}}
\put(1351,1149){\rule[-0.175pt]{0.350pt}{0.703pt}}
\put(1352,1152){\rule[-0.175pt]{0.350pt}{0.703pt}}
\put(1353,1154){\rule[-0.175pt]{0.350pt}{0.683pt}}
\put(1354,1157){\rule[-0.175pt]{0.350pt}{0.683pt}}
\put(1355,1160){\rule[-0.175pt]{0.350pt}{0.683pt}}
\put(1356,1163){\rule[-0.175pt]{0.350pt}{0.683pt}}
\put(1357,1166){\rule[-0.175pt]{0.350pt}{0.683pt}}
\put(1358,1169){\rule[-0.175pt]{0.350pt}{0.683pt}}
\put(1359,1172){\rule[-0.175pt]{0.350pt}{0.683pt}}
\put(1360,1174){\rule[-0.175pt]{0.350pt}{0.683pt}}
\put(1361,1177){\rule[-0.175pt]{0.350pt}{0.683pt}}
\put(1362,1180){\rule[-0.175pt]{0.350pt}{0.683pt}}
\put(1363,1183){\rule[-0.175pt]{0.350pt}{0.683pt}}
\put(1364,1186){\rule[-0.175pt]{0.350pt}{0.682pt}}
\put(1365,1189){\rule[-0.175pt]{0.350pt}{0.683pt}}
\put(1366,1191){\rule[-0.175pt]{0.350pt}{0.683pt}}
\put(1367,1194){\rule[-0.175pt]{0.350pt}{0.683pt}}
\put(1368,1197){\rule[-0.175pt]{0.350pt}{0.683pt}}
\put(1369,1200){\rule[-0.175pt]{0.350pt}{0.683pt}}
\put(1370,1203){\rule[-0.175pt]{0.350pt}{0.683pt}}
\put(1371,1206){\rule[-0.175pt]{0.350pt}{0.683pt}}
\put(1372,1208){\rule[-0.175pt]{0.350pt}{0.683pt}}
\put(1373,1211){\rule[-0.175pt]{0.350pt}{0.683pt}}
\put(1374,1214){\rule[-0.175pt]{0.350pt}{0.683pt}}
\put(1375,1217){\rule[-0.175pt]{0.350pt}{0.683pt}}
\put(1376,1220){\rule[-0.175pt]{0.350pt}{0.682pt}}
\put(1377,1223){\rule[-0.175pt]{0.350pt}{0.662pt}}
\put(1378,1225){\rule[-0.175pt]{0.350pt}{0.662pt}}
\put(1379,1228){\rule[-0.175pt]{0.350pt}{0.662pt}}
\put(1380,1231){\rule[-0.175pt]{0.350pt}{0.662pt}}
\put(1381,1234){\rule[-0.175pt]{0.350pt}{0.662pt}}
\put(1382,1236){\rule[-0.175pt]{0.350pt}{0.662pt}}
\put(1383,1239){\rule[-0.175pt]{0.350pt}{0.662pt}}
\put(1384,1242){\rule[-0.175pt]{0.350pt}{0.662pt}}
\put(1385,1245){\rule[-0.175pt]{0.350pt}{0.662pt}}
\put(1386,1247){\rule[-0.175pt]{0.350pt}{0.662pt}}
\put(1387,1250){\rule[-0.175pt]{0.350pt}{0.662pt}}
\put(1388,1253){\rule[-0.175pt]{0.350pt}{0.662pt}}
\put(1389,1256){\rule[-0.175pt]{0.350pt}{0.766pt}}
\put(1390,1259){\rule[-0.175pt]{0.350pt}{0.766pt}}
\put(1391,1262){\rule[-0.175pt]{0.350pt}{0.766pt}}
\put(1392,1265){\rule[-0.175pt]{0.350pt}{0.766pt}}
\put(1393,1268){\rule[-0.175pt]{0.350pt}{0.766pt}}
\put(1394,1271){\rule[-0.175pt]{0.350pt}{0.766pt}}
\put(1395,1275){\rule[-0.175pt]{0.350pt}{0.766pt}}
\put(1396,1278){\rule[-0.175pt]{0.350pt}{0.766pt}}
\put(1397,1281){\rule[-0.175pt]{0.350pt}{0.766pt}}
\put(1398,1284){\rule[-0.175pt]{0.350pt}{0.766pt}}
\put(1399,1287){\rule[-0.175pt]{0.350pt}{0.766pt}}
\put(1400,1290){\rule[-0.175pt]{0.350pt}{0.723pt}}
\put(1401,1294){\rule[-0.175pt]{0.350pt}{0.723pt}}
\put(1402,1297){\rule[-0.175pt]{0.350pt}{0.723pt}}
\put(1403,1300){\rule[-0.175pt]{0.350pt}{0.723pt}}
\put(1404,1303){\rule[-0.175pt]{0.350pt}{0.723pt}}
\put(1405,1306){\rule[-0.175pt]{0.350pt}{0.723pt}}
\put(1406,1309){\rule[-0.175pt]{0.350pt}{0.723pt}}
\put(1407,1312){\rule[-0.175pt]{0.350pt}{0.723pt}}
\put(1408,1315){\rule[-0.175pt]{0.350pt}{0.723pt}}
\put(1409,1318){\rule[-0.175pt]{0.350pt}{0.723pt}}
\put(1410,1321){\rule[-0.175pt]{0.350pt}{0.723pt}}
\put(1411,1324){\rule[-0.175pt]{0.350pt}{0.723pt}}
\put(1412,1327){\rule[-0.175pt]{0.350pt}{0.823pt}}
\put(1413,1330){\rule[-0.175pt]{0.350pt}{0.823pt}}
\put(1414,1333){\rule[-0.175pt]{0.350pt}{0.823pt}}
\put(1415,1337){\rule[-0.175pt]{0.350pt}{0.823pt}}
\put(1416,1340){\rule[-0.175pt]{0.350pt}{0.823pt}}
\put(1417,1344){\rule[-0.175pt]{0.350pt}{0.823pt}}
\put(1418,1347){\rule[-0.175pt]{0.350pt}{0.823pt}}
\put(1419,1350){\rule[-0.175pt]{0.350pt}{0.823pt}}
\put(1420,1354){\rule[-0.175pt]{0.350pt}{0.823pt}}
\put(1421,1357){\rule[-0.175pt]{0.350pt}{0.823pt}}
\put(1422,1361){\rule[-0.175pt]{0.350pt}{0.823pt}}
\put(1423,1364){\rule[-0.175pt]{0.350pt}{0.823pt}}
\put(1424,1367){\rule[-0.175pt]{0.350pt}{0.867pt}}
\put(1425,1371){\rule[-0.175pt]{0.350pt}{0.867pt}}
\put(1426,1375){\rule[-0.175pt]{0.350pt}{0.867pt}}
\put(1427,1378){\rule[-0.175pt]{0.350pt}{0.867pt}}
\put(1428,1382){\rule[-0.175pt]{0.350pt}{0.867pt}}
\put(1429,1385){\usebox{\plotpoint}}
\sbox{\plotpoint}{\rule[-0.350pt]{0.700pt}{0.700pt}}%
\put(674,983){\makebox(0,0)[r]{$L=\ \ 4$}}
\put(718,983){\raisebox{-1.2pt}{\makebox(0,0){$\Diamond$}}}
\put(836,214){\raisebox{-1.2pt}{\makebox(0,0){$\Diamond$}}}
\put(850,220){\raisebox{-1.2pt}{\makebox(0,0){$\Diamond$}}}
\put(850,220){\raisebox{-1.2pt}{\makebox(0,0){$\Diamond$}}}
\put(850,220){\raisebox{-1.2pt}{\makebox(0,0){$\Diamond$}}}
\put(850,220){\raisebox{-1.2pt}{\makebox(0,0){$\Diamond$}}}
\put(850,220){\raisebox{-1.2pt}{\makebox(0,0){$\Diamond$}}}
\put(850,220){\raisebox{-1.2pt}{\makebox(0,0){$\Diamond$}}}
\put(855,223){\raisebox{-1.2pt}{\makebox(0,0){$\Diamond$}}}
\put(862,226){\raisebox{-1.2pt}{\makebox(0,0){$\Diamond$}}}
\put(873,231){\raisebox{-1.2pt}{\makebox(0,0){$\Diamond$}}}
\put(876,233){\raisebox{-1.2pt}{\makebox(0,0){$\Diamond$}}}
\put(888,240){\raisebox{-1.2pt}{\makebox(0,0){$\Diamond$}}}
\put(901,249){\raisebox{-1.2pt}{\makebox(0,0){$\Diamond$}}}
\put(913,257){\raisebox{-1.2pt}{\makebox(0,0){$\Diamond$}}}
\put(926,267){\raisebox{-1.2pt}{\makebox(0,0){$\Diamond$}}}
\put(937,277){\raisebox{-1.2pt}{\makebox(0,0){$\Diamond$}}}
\put(950,288){\raisebox{-1.2pt}{\makebox(0,0){$\Diamond$}}}
\put(831,212){\raisebox{-1.2pt}{\makebox(0,0){$\Diamond$}}}
\put(862,226){\raisebox{-1.2pt}{\makebox(0,0){$\Diamond$}}}
\sbox{\plotpoint}{\rule[-0.500pt]{1.000pt}{1.000pt}}%
\put(674,938){\makebox(0,0)[r]{$L=\ \ 6$}}
\put(718,938){\makebox(0,0){$+$}}
\put(639,174){\makebox(0,0){$+$}}
\put(666,176){\makebox(0,0){$+$}}
\put(683,178){\makebox(0,0){$+$}}
\put(747,189){\makebox(0,0){$+$}}
\put(772,195){\makebox(0,0){$+$}}
\put(799,202){\makebox(0,0){$+$}}
\put(824,210){\makebox(0,0){$+$}}
\put(850,218){\makebox(0,0){$+$}}
\put(850,221){\makebox(0,0){$+$}}
\put(850,221){\makebox(0,0){$+$}}
\put(850,221){\makebox(0,0){$+$}}
\put(850,221){\makebox(0,0){$+$}}
\put(850,221){\makebox(0,0){$+$}}
\put(860,226){\makebox(0,0){$+$}}
\put(874,233){\makebox(0,0){$+$}}
\put(893,245){\makebox(0,0){$+$}}
\put(899,248){\makebox(0,0){$+$}}
\put(922,265){\makebox(0,0){$+$}}
\put(947,287){\makebox(0,0){$+$}}
\put(970,310){\makebox(0,0){$+$}}
\put(994,338){\makebox(0,0){$+$}}
\put(1016,367){\makebox(0,0){$+$}}
\put(1039,403){\makebox(0,0){$+$}}
\put(813,207){\makebox(0,0){$+$}}
\put(874,233){\makebox(0,0){$+$}}
\sbox{\plotpoint}{\rule[-0.250pt]{0.500pt}{0.500pt}}%
\put(674,893){\makebox(0,0)[r]{$L=\ \ 8$}}
\put(718,893){\raisebox{-1.2pt}{\makebox(0,0){$\Box$}}}
\put(518,166){\raisebox{-1.2pt}{\makebox(0,0){$\Box$}}}
\put(560,168){\raisebox{-1.2pt}{\makebox(0,0){$\Box$}}}
\put(587,170){\raisebox{-1.2pt}{\makebox(0,0){$\Box$}}}
\put(688,179){\raisebox{-1.2pt}{\makebox(0,0){$\Box$}}}
\put(728,185){\raisebox{-1.2pt}{\makebox(0,0){$\Box$}}}
\put(770,194){\raisebox{-1.2pt}{\makebox(0,0){$\Box$}}}
\put(808,205){\raisebox{-1.2pt}{\makebox(0,0){$\Box$}}}
\put(850,223){\raisebox{-1.2pt}{\makebox(0,0){$\Box$}}}
\put(850,221){\raisebox{-1.2pt}{\makebox(0,0){$\Box$}}}
\put(850,221){\raisebox{-1.2pt}{\makebox(0,0){$\Box$}}}
\put(850,221){\raisebox{-1.2pt}{\makebox(0,0){$\Box$}}}
\put(851,221){\raisebox{-1.2pt}{\makebox(0,0){$\Box$}}}
\put(851,221){\raisebox{-1.2pt}{\makebox(0,0){$\Box$}}}
\put(866,229){\raisebox{-1.2pt}{\makebox(0,0){$\Box$}}}
\put(887,241){\raisebox{-1.2pt}{\makebox(0,0){$\Box$}}}
\put(918,263){\raisebox{-1.2pt}{\makebox(0,0){$\Box$}}}
\put(927,269){\raisebox{-1.2pt}{\makebox(0,0){$\Box$}}}
\put(964,304){\raisebox{-1.2pt}{\makebox(0,0){$\Box$}}}
\put(1003,350){\raisebox{-1.2pt}{\makebox(0,0){$\Box$}}}
\put(1039,402){\raisebox{-1.2pt}{\makebox(0,0){$\Box$}}}
\put(1077,465){\raisebox{-1.2pt}{\makebox(0,0){$\Box$}}}
\put(1111,532){\raisebox{-1.2pt}{\makebox(0,0){$\Box$}}}
\put(1148,609){\raisebox{-1.2pt}{\makebox(0,0){$\Box$}}}
\put(792,200){\raisebox{-1.2pt}{\makebox(0,0){$\Box$}}}
\put(887,241){\raisebox{-1.2pt}{\makebox(0,0){$\Box$}}}
\put(674,848){\makebox(0,0)[r]{$L=10$}}
\put(718,848){\makebox(0,0){$\times$}}
\put(377,162){\makebox(0,0){$\times$}}
\put(436,163){\makebox(0,0){$\times$}}
\put(476,164){\makebox(0,0){$\times$}}
\put(619,172){\makebox(0,0){$\times$}}
\put(676,178){\makebox(0,0){$\times$}}
\put(736,187){\makebox(0,0){$\times$}}
\put(791,200){\makebox(0,0){$\times$}}
\put(850,208){\makebox(0,0){$\times$}}
\put(850,223){\makebox(0,0){$\times$}}
\put(850,221){\makebox(0,0){$\times$}}
\put(850,221){\makebox(0,0){$\times$}}
\put(851,221){\makebox(0,0){$\times$}}
\put(851,222){\makebox(0,0){$\times$}}
\put(873,232){\makebox(0,0){$\times$}}
\put(903,251){\makebox(0,0){$\times$}}
\put(947,287){\makebox(0,0){$\times$}}
\put(960,300){\makebox(0,0){$\times$}}
\put(1012,361){\makebox(0,0){$\times$}}
\put(1068,453){\makebox(0,0){$\times$}}
\put(1119,546){\makebox(0,0){$\times$}}
\put(1173,666){\makebox(0,0){$\times$}}
\put(1222,785){\makebox(0,0){$\times$}}
\put(1274,925){\makebox(0,0){$\times$}}
\put(768,193){\makebox(0,0){$\times$}}
\put(903,251){\makebox(0,0){$\times$}}
\put(674,803){\makebox(0,0)[r]{$L=12$}}
\put(718,803){\makebox(0,0){$\triangle$}}
\put(298,161){\makebox(0,0){$\triangle$}}
\put(350,161){\makebox(0,0){$\triangle$}}
\put(542,167){\makebox(0,0){$\triangle$}}
\put(617,172){\makebox(0,0){$\triangle$}}
\put(698,180){\makebox(0,0){$\triangle$}}
\put(771,194){\makebox(0,0){$\triangle$}}
\put(849,222){\makebox(0,0){$\triangle$}}
\put(851,223){\makebox(0,0){$\triangle$}}
\put(851,222){\makebox(0,0){$\triangle$}}
\put(881,236){\makebox(0,0){$\triangle$}}
\put(921,266){\makebox(0,0){$\triangle$}}
\put(997,335){\makebox(0,0){$\triangle$}}
\put(1067,454){\makebox(0,0){$\triangle$}}
\put(1141,595){\makebox(0,0){$\triangle$}}
\put(1208,754){\makebox(0,0){$\triangle$}}
\put(1280,937){\makebox(0,0){$\triangle$}}
\put(1346,1136){\makebox(0,0){$\triangle$}}
\put(1416,1340){\makebox(0,0){$\triangle$}}
\put(741,187){\makebox(0,0){$\triangle$}}
\put(980,325){\makebox(0,0){$\triangle$}}
\put(921,262){\makebox(0,0){$\triangle$}}
\end{picture}

%% file: review.bbl
\begin{thebibliography}{99}

\bibitem{reviews}
M.H. Kalos and P.A. Whitlock, \it {Monte Carlo Methods,} Vol. 1,
(Wiley, 1986) ; B.L. Hammond, W.A. Lester Jr. and P.J. Reynolds, {\it
Monte Carlo Methods in Ab Initio Quantum Chemistry} (World Scientific
Pub., 1994); D.  Ceperley and L. Mit\'{a}\v{s} in {\it Advances in
Chemical Physics, Vol. XCIII}, edited by I. Prigogine and S.A. Rice
(Wiley, NY 1996); J .B. Anderson, Int. Rev. Phys. Chem. {\bf 14}, 85
(1995).  J. B.  Anderson, in {\it Quantum Mechanical Electronic
Structure Calculations with Chemical Accuracy,} edited by S. R.
Langhoff (Kluwer Academic Publishers, Dordrecht, 1995).

\bibitem{M}
J.D. Morgan in {\it Numerical Determination of the Electronic
Structure of Atoms, Diatomic and Polyatomic Molecules}, edited by
M. Defranceschi and J. Delhalle (Kluwer Academic Publishers, 1989).

\bibitem{CyrusPRL88}
C.J. Umrigar, K.G. Wilson and J.W. Wilkins, \prl {\bf 60}, 1719
(1988).

\bibitem{CyrusAthens88}
C.J. Umrigar, K.G. Wilson and J.W. Wilkins, in \it Computer Simulation
Studies in Condensed Matter Physics, Recent Developments, \rm edited
by D.P. Landau K.K. Mon and H.B. Sch\"uttler, Springer Proc. Phys.
(Springer, Berlin, 1988).

\bibitem{georgia.88}
M. P. Nightingale and R. G. Caflisch, in {\it Computer Simulation
Studies in Condensed Matter Physics,} edited by D.  P. Landau and
H. B.  Sch"ttler (eds.),p. 208-213 (Springer, Berlin 1988).

\bibitem{boston.88}
M. P. Nightingale, in {\it Proceedings of the Third International
Conference on Supercomputing,} edited by L. P. Kartashev and S. I.
Kartashev Vol. I, 427-436 (1988).

\bibitem{NB.PRL.60} M.P. Nightingale and H.W.J. Bl\"ote,
\prl {\bf 60}, 1662 (1988).

\bibitem{GraNigh93} E. Granato and M.P. Nightingale,
\prb {\bf 48}, 7438 (1993).

\bibitem{MushNigh.94}
A. Mushinski and M.P. Nightingale, J. Chem. Phys. {\bf 101}, 8831
(1994).

\bibitem{NighGraKos95}
M.P. Nightingale, E. Granato and J.M. Kosterlitz,
\prb {\bf 52}, 7402 (1995).

\bibitem{MeiMushNigh96}
M. Meierovich, A. Mushinski and M.P. Nightingale
\jcp, in print, and URL http://xxx.lanl.gov/abs/chem-ph/9512001.

\bibitem{NighBloeprl.96}
M.P. Nightingale and H.W.J. Bl\"ote, \prl {\bf 76} 4548, 1996; Also
see URL http://xxx.lanl.gov/abs/cond-mat/9601059.

\bibitem{NighBloeprb.96}
M.P. Nightingale and H.W.J. Bl\"ote, \prb {\bf 54} 1001, 1996; also
see URL http://xxx.lanl.gov/abs/cond-mat/9602089.

\bibitem{FilippiUmrigar96}
Claudia Filippi and C.J. Umrigar, \jcp {\bf 105}, 213 (1996).

\bibitem{NighUmrLM}
M.P. Nightingale and C.J. Umrigar, unpublished.  For a test version of
the program contact us at cyrus@tc.cornell.edu or
nigh@repeter.phys.uri.edu.

\bibitem{CyrusPeterKarl93}
C.J. Umrigar, M. P. Nightingale and K.J. Runge, \jcp {\bf 99}, 2865
(1993).

\bibitem{BH}
S.F. Boys and N.C. Handy, Proc. R. Soc. London Ser. A {\bf 309}, 209
(1969).

\bibitem{SM}
K.E. Schmidt and J.W. Moskowitz, J. Chem. Phys. {\bf 93}, 4172 (1990).

\bibitem{SXM}
K.E. Schmidt, J. Xiang and J.W. Moskowitz in {\it Recent Progress in
Many-Body Theories}, edited by T.L. Ainsworth, C.E.  Campbell, B.E.
Clements and E. Krotsheck (Plenum Press, New York, 1992).

\bibitem{HU}
C.J. Huang, C.J. Umrigar and M.P. Nightingale, unpublished.

\bibitem{MUSM} C.R. Myers, C.J. Umrigar, J.P. Sethna and J.D. Morgan,
Phys. Rev. A {\bf 44}, 5537 (1991).

\bibitem{Fisher}
The Hartree Fock energies were calculated using the program described
in {\it The Hartree Fock Method for Atoms}, Charlotte Froese Fischer,
(Wiley, New York 1977).

\bibitem{Freund}
The exact energies were calculated using a modified version of the
program described in D.E. Freund, B.D. Huxtable and J.D. Morgan, Phys.
Rev. A., {\bf 29}, 980 (1984).

\bibitem{Drake}
G.F.W. Drake and Zong-Chao Yan, Chem. Phys. Lett.  {\bf 229}, 486
(1994).

\bibitem{Fisher93} Charlotte Froese Fischer, J. Phys. B., {\bf 26}, 855 (1993).

\bibitem{Davidson} E.R. Davidson, S.A. Hagstrom, S.J. Chakravorty, V.M. Umar
and C.F. Fischer, Phys. Rev. A {\bf 44}, 7071 (1991).

\bibitem{Chakravor}
S.J. Chakravorty, S.R. Gwaltney, E.R. Davidson, F.A. Parpia and
C.F. Fischer, Phys. Rev. A {\bf 47}, 3649 (1993).

\bibitem{wetting} For reviews see, S. Dietrich in
{\it Phase Transitions and Critical Phenomena,} edited by C. Domb and
J.L. Lebowitz, Academic, London 1988) Vol. {\bf 12}; J.O. Indekeu,
Int. J. Mod. Phys. B {\bf 8}, 309 (1994).

\bibitem{fermion.unbinding}
V.R. Pandharipande, S.C. Pieper and R.B. Wieringa, \prb, {\bf 34},
4571 (1986); S. Stringari, {\it Proceedings of the International
School of Physics ``Enrico Fermi'', Course CVII,} edited by G. Scoles
(North-Holland, 1990).

\bibitem{Marks.internet.visualization}
See URL
http://www.phys.uri.edu/people/mark\_meierovich/visual/Main.html for
an informal presentation.

\bibitem{BLH} H.W.J. Bl\"ote, E. Luijten and J.R. Heringa,
J. Phys. A {\bf 28}, 6289 (1995).

\bibitem{BHHSM}
H.W.J. Bl\"ote, J.R. Heringa, A. Hoogland, E.W. Meyer and T.S. Smit,
preprint (1996).

\bibitem{GT}
R. Gupta and P. Tamayo, preprint; to appear in Int. J. Mod. Phys
(1996).

\bibitem{FerLan} A.M. Ferrenberg and D.P. Landau,
\prb {\bf 44}, 5081 (1991).

\bibitem{LGZJ}
 J.C. Le Guillou and J. Zinn-Justin, \prb {\bf 21} 3976 (1980).

\bibitem{Yang}
 C.P. Yang,
 Proc. Symp. Appl. Math. {\bf 15}, 351 (1963).

\end{thebibliography}
